\documentclass{aa}    

\newcount\arxivmode
\ifnum\arxivmode=0 
  \def\deffigs{deffigs}
\fi
\ifnum\arxivmode=1
  \def\deffigs{.}
\fi

\usepackage{graphicx,url,twoopt,natbib}
\usepackage[varg]{txfonts}           
\usepackage{hyperref}                
\usepackage[usenames]{color}
\usepackage{acronym}      
\hypersetup{
  colorlinks=true,  
  urlcolor=rrblue,    
  citecolor=rrblue,   
  linkcolor=rrblue    
}

\usepackage[titles]{tocloft}
\usepackage{stdclsdv} 
\definecolor{bllsht}{rgb}{0.40, 0.05, 0.01} 
\long\def\@ #1@{\par\vspace{0.5ex}\noindent
  {\Large \textcolor{red}{@~}}\color{bllsht}{#1}\color{black}\\[0.5ex]}

\definecolor{rrblue}{rgb}{0.15,0.0,0.8} 

\def\rrref#1#2{\hyperref[#2]{#1~}\ref{#2}}

\makeatletter
\newcommand{\bibnote}[2]{\global\@namedef{#1note}{#2~}}
\newcommand{\biblink}[2]{\global\@namedef{#1link}{#2}}
\makeatother

\definecolor{amber}{rgb}{1.0, 0.49, 0.0}

\bibpunct{(}{)}{;}{a}{}{,}    
\makeatletter
  \protected\def\sppresslink{\def\hyper@linkstart##1##2{}\let\hyper@linkend\@empty}
  \newcommandtwoopt{\citeads}[3][][]{%
   \href{http://ui.adsabs.harvard.edu/abs/#3/abstract}%
        {\sppresslink \citealp[#1][#2]{#3}}
   \biblink{#3}{\href{http://ui.adsabs.harvard.edu/abs/#3/abstract}{ADS}}}
 \newcommandtwoopt{\citepads}[3][][]{%
   \href{http://ui.adsabs.harvard.edu/abs/#3/abstract}%
        {\sppresslink \citep[#1][#2]{#3}}
   \biblink{#3}{\href{http://ui.adsabs.harvard.edu/abs/#3/abstract}{ADS}}}
 \newcommandtwoopt{\citetads}[3][][]{%
   \href{http://ui.adsabs.harvard.edu/abs/#3/abstract}%
        {\sppresslink \citet[#1][#2]{#3}}
  \biblink{#3}{\href{http://ui.adsabs.harvard.edu/abs/#3/abstract}{ADS}}}
 \newcommandtwoopt{\citeyearads}[3][][]{%
   \href{http://ui.adsabs.harvard.edu/abs/#3/abstract}%
        {\sppresslink \citeyear[#1][#2]{#3}}
   \biblink{#3}{\href{http://ui.adsabs.harvard.edu/abs/#3/abstract}{ADS}}}
\makeatother

\def\linkadspage#1#2#3{\href{http://adsabs.harvard.edu/cgi-bin/nph-data_query?bibcode=#1\&link_type=ARTICLE\&db_key=AST\#page=#2}{#3 (pdf\,#2)}}

\def\linkpdfpage#1#2#3{\href{#1\#page=#2}{#3 (pdf\,#2)}}

\newcount\longrefs
\def\adv{\ifnum\longrefs=1 {Adv.\ Space Res.} \else 
                           {Adv.\ Sp'\ Res.}\fi}
\def\aap{\ifnum\longrefs=1 {Astron.\ Astrophys.}\else 
                           {A\hbox{\rm \&}A}\fi}
\def\aapr{\ifnum\longrefs=1 {Astron.\ Astrophys.\ Rev.}\else 
                            {A\hbox{\rm \&}AR}\fi}
\def\aaps{\ifnum\longrefs=1 {Astron.\ Astrophys.\ Suppl.}\else 
                            {A\hbox{\rm \&}A Suppl.}\fi}
\def\actaa{\ifnum\longrefs=1 {Acta Astronomica}\else
                            {Acta Astron.}\fi}
\def\aipcs{\ifnum\longrefs=1 {Am.\ Inst.\ Phys.\ Conf.\ Series}\else
                             {AIP Conf.\ Ser.}\fi}
\def\aj{\ifnum\longrefs=1 {Astron.\ J.}\else 
                          {AJ}\fi} 
\def\ao{\ifnum\longrefs=1 {Applied Optics}\else 
                           {Appl.\ Opt.}\fi} 
\def\aspcs{\ifnum\longrefs=1 {Astron.\ Soc.\ Pacific Conf.\ Series}\else 
                           {ASP Conf.\ Ser.}\fi} 
\def\apj{\ifnum\longrefs=1 {Astrophys.\ J.}\else 
                           {ApJ}\fi} 
\def\apjl{\ifnum\longrefs=1 {Astrophys.\ J. Lett.}\else 
                            {ApJL}\fi} 
\def\aplett{\ifnum\longrefs=1 {Astrophys.\ J. Lett.}\else 
                            {ApJ}\fi} 
\def\apjs{\ifnum\longrefs=1 {Astrophys.\ J. Suppl.}\else 
                            {ApJS}\fi}
\def\apss{\ifnum\longrefs=1 {Astrophys.\ Space Sci.}\else 
                            {Astrophys.\ Space Sci.}\fi}
\def\araa{\ifnum\longrefs=1 {Ann.\ Rev.\ Astron.\ Astrophys.}\else 
                            {ARA\hbox{\rm \&}A}\fi}
\def\azh{\ifnum\longrefs=1 {Astronomicheskii Zhurnal}\else 
                            {Astron.\ Zhur.}\fi}
\def\baas{\ifnum\longrefs=1 {Bull.\ Am.\ Astron.\ Soc.}\else 
                            {BAAS}\fi}
\def\bain{\ifnum\longrefs=1 {Bull.\ Astronom.\ Institutes Netherlands}\else
                            {Bull.\ Astr.\ Inst.\ Neth.}\fi}
\def\cjaa{\ifnum\longrefs=1 {Chin.\ J.\ Astron.\ Astrophys.}\else 
                            {Chin.\ J.\ A\&A}\fi}
\def\gca{\ifnum\longrefs=1 {Geochim.\ Cosmochim.\ Acta}\else 
                           {Geochim.\ Cosmochim.\ Acta}\fi}
\def\grl{\ifnum\longrefs=1 {Geophys.\ Res.\ Lett.}\else 
                           {Geoph.\ Res.\ Lett.}\fi}
\def\iaucirc{\ifnum\longrefs=1 {IAU Circulars}\else 
                          {IAU Circ.}\fi}
\def\icarus{\ifnum\longrefs=1 {Icarus}\else 
                          {Icarus}\fi}
\def\ip{\ifnum\longrefs=1 {in press}\else 
                          {in press}\fi}
\def\jcap{\ifnum\longrefs=1 {Jour.\ Cosmology Astropart.\ Phys.}\else 
                          {JCAP}\fi}
\def\jgr{\ifnum\longrefs=1 {J.\ Geophys.\ Res.}\else 
                           {J.\ Geophys.\ Res.}\fi}  
\def\jrasc{\ifnum\longrefs=1 {J.\ Royal Astron.\ Soc.\ Canada}\else 
                             {JRAS Can.}\fi}  
\def\memsai{\ifnum\longrefs=1 {Mem.~Soc.~Astron.~Italiana}\else
                              {MmSAI}\fi}
\def\mnras{\ifnum\longrefs=1 {Mon.\ Not.\ Roy.\ Astron.\ Soc.}\else 
                             {MNRAS}\fi} 
\def\na{\ifnum\longrefs=1 {New Astronomy}\else 
                          {New Astron.}\fi}
\def\nar{\ifnum\longrefs=1 {New Astronomy rev.}\else 
                           {New Astron.\ Rev.}\fi}
\def\nat{\ifnum\longrefs=1 {Nature}\else 
                           {Nat}\fi}
\def\pasa{\ifnum\longrefs=1 {Pub.\ Astron.\ Soc.\ Australia}\else 
                            {PASA}\fi} 
\def\pasj{\ifnum\longrefs=1 {Pub.\ Astron.\ Soc.\ Japan}\else 
                            {PASJ}\fi} 
\def\pasp{\ifnum\longrefs=1 {Pub.\ Astron.\ Soc.\ Pacific}\else 
                            {PASP}\fi} 
\def\physscr{\ifnum\longrefs=1 {Physica Scripta}\else 
                               {Phys.\ Scrip.}\fi} 
\def\planss{\ifnum\longrefs=1 {Planetary \& Space Science}\else 
                              {Plan. \& Space Sci.}\fi} 
\def\pre{\ifnum\longrefs=1 {Phys.\ Rev.\ E}\else
                           {Phys.\ Rev.\ E}\fi}
\def\procspie{\ifnum\longrefs=1 {Proc.\ SPIE}\else 
                                {Proc.\ SPIE}\fi} 
\def\qjras{\ifnum\longrefs=1 {Quarterly J.\ Royal Astron.\ Soc.}\else 
                             {QJRAS}\fi} 
\def\rmxaa{\ifnum\longrefs=1 {Revista Mexicana de Astron.\ y Astrofys.}\else 
                             {RMxAA}\fi} 
\def\sa{\ifnum\longrefs=1 {Soviet Astron..}\else 
                          {Sov.\ Astron.}\fi}
\def\skytel{\ifnum\longrefs=1 {Sky \& Telescope}\else 
                              {Sky \& Tel.}\fi} 
\def\solphys{\ifnum\longrefs=1 {Solar Phys.}\else 
                               {SoPh}\fi}
\def\sovast{\ifnum\longrefs=1 {Soviet Astron.}\else 
                              {Sov.\ Ast.}\fi}
\def\ssr{\ifnum\longrefs=1 {Space Sci.\ Rev.}\else 
                           {Space Sci.\ Rev.}\fi}
\def\zap{\ifnum\longrefs=1 {Zeitschr.\ f.\ Astrophysik}\else
                               {Z.\ Astrophys.}\fi}

\newacro{AA}{Astronomy \& Astrophysics}  
\newacro{ADS}{Astrophysics Data System}
\newacro{AIA}{Atmospheric Imaging Assembly}
\newacro{ALMA}{Atacama Large Millimeter/submillimeter Array}
\newacro{AO}{adaptive optics}
\newacro{ApJ}{Astrophysical Journal}
\newacro{AR}{active region}
\newacro{bb}{bound-bound}
\newacro{bf}{bound-free}
\newacro{BFI}{Broad-band Filter Imager}
\newacro{CE}{coronal equilibrium}
\newacro{CfA}{Center for Astrophysics}
\newacro{CME}{coronal mass ejection}
\newacro{CRD}{complete redistribution}
\newacro{CRISP}{CRisp Imaging SpectroPolarimeter}
\newacro{CRISPEX}{CRisp SPectral EXplorer}
\newacro{CS}{coherent scattering}
\newacro{DEM}{Differential Emission Measure}
\newacro{DF}{dynamic fibril}
\newacro{DKIST}{Daniel K. Inouye Solar Telescope}
\newacro{DLR}{Deutsches Zentrum f\"ur Luft- und Raumfahrt}
\newacro{DOT}{Dutch Open Telescope}
\newacro{DST}{Richard B. Dunn Solar Telescope}   
\newacro{EB}{Ellerman bomb}
\newacro{EDP}{\'{E}dition Diffusion Presse}  
\newacro{EIT}{Extreme ultraviolet Imaging Telescope}
\newacro{EPIC}{European participation in Solar-C}
\newacro{ERC}{European Research Council}
\newacro{ESA}{European Space Agency}
\newacro{EST}{European Solar Telescope}
\newacro{EUV}{extreme ultraviolet}
\newacro{FAF}{flaring active-region fibril}
\newacro{ff}{free-free}
\newacro{FITS}{Flexible Image Transport System}
\newacro{FOV}{field of view}
\newacro{fov}{field of view}
\newacro{FWHM}{full width at half maximum}
\newacro{HAO}{High Altitude Observatory}
\newacro{HD}{hydrodynamics}
\newacro{Hi-C}{High Resolution Coronal Imager Sounding Rocket}
\newacro{HMI}{Helioseismic and Magnetic Imager}
\newacro{IAA}{Instituto de Astrof\'{i}sica de Andaluc\'{i}a}
\newacro{IAC}{Instituto de Astrof\'{i}sica de Canarias}
\newacro{IAS}{Institut d'Astrophysique Spatiale}
\newacro{IAU}{International Astronomical Union}
\newacro{IBIS}{Interferometric Bi-dimensional Spectrometer}
\newacro{IDL}{Interactive Data Language}
\newacro{IMaX}{Imaging Magnetograph eXperiment}
\newacro{INAF}{Istituto Nazionale di Astrofisica}
\newacro{IB}{IRIS bomb}
\newacro{IR}{infrared}
\newacro{IRIS}{Interface Region Imaging Spectrograph}
\newacro{ISAS}{Institute of Space and Astronautical Science}
\newacro{ISP}{Institute for Solar Physics}
\newacro{ISS}{International Space Station}
\newacro{ISSI}{International Space Science Institute}
\newacro{ITA}{Institute for Theoretical Astrophysics}
\newacro{JAXA}{Japan Aerospace Exploration Agency}
\newacro{JSOC}{Joint Science Operations Center}
\newacro{KIS}{Kiepenheuer--Institut f\"{u}r Sonnenphysik}
\newacro{KPNO}{Kitt Peak National Observatory}
\newacro{LASP}{Laboratory for Atmospheric and Space Physics}
\newacro{LC}{liquid cristal}
\newacro{LMSAL}{Lockheed Martin Solar and Astrophysics Labratory}
\newacro{LOS}{line of sight}
\newacro{LTE}{local thermodynamic equilibrium}
\newacro{MC}{magnetic concentration}
\newacro{MCAO}{multi-conjugate adaptive optics} 
\newacro{MDI}{Michelson Doppler Imager}
\newacro{ME}{Milne-Eddington} 
\newacro{MHD}{magnetohydrodynamics}
\newacro{MOMFBD}{Multi-Object Multi-Frame Blind Deconvolution}
\newacro{MPE}{Max--Planck--Institut f\"ur extraterrestrische Physik}
\newacro{MPG}{Max--Planck--Gesellschaft}
\newacro{MPS}{Max Planck Institute for Solar System Research}
\newacro{MSSL}{Mullard Space Science Laboratory}
\newacro{MTF}{modulation transfer function}
\newacro{NAOJ}{National Astronomical Observatory of Japan}
\newacro{NASA}{National Aeronautics and Space Administration}
\newacro{NIST}{National Institute of Standards and Technology}
\newacro{NLTE}{non-local thermodynamic equilibrium}
\newacro{NLFFF}{non-linear force-free field}
\newacro{NOAA}{National Oceanic and Atmospheric Administration}
\newacro{non-E}{non-equilibrium}
\newacro{NSO}{National Solar Observatory}
\newacro{NWO}{Netherlands Organisation for Scientific Research}
\newacro{PHE}{propagating heating event}
\newacro{PRD}{partial redistribution}
\newacro{PROBA2}{PRoject for OnBoard Autonomy}
\newacro{PSBE}{post Saha-Boltzmann extinction}
\newacro{PSF}{point spread function}
\newacro{QS}{quiet Sun}
\newacro{QSEB}{quiet-Sun Ellerman-like brightening} 
\newacro{RAL}{Rutherford Appleton Laboratory}
\newacro{RBE}{rapid blue-shifted excursion}
\newacro{R-MHD}{radiation hydrodynamics}
\newacro{rms}{root mean square}
\newacro{RMS}{root mean square}
\newacro{ROB}{Royal Observatory of Belgium}
\newacro{ROI}{region of interest}
\newacro{RRE}{rapid red-shifted excursion}
\newacro{RTE}{radiative transfer equation}
\newacro{RTSA}{Radiative Transfer in Stellar Atmospheres}
\newacro{SCF}{slender \CaIIH\ fibril}
\newacro{SE}{statistical equilibrium}
\newacro{SB}{Saha Boltzmann}
\newacro{SDO}{Solar Dynamics Observatory}
\newacro{SJI}{slit-jaw image}
\newacro{SLI}{slit image}
\newacro{SNR}{signal-to-noise ratio}
\newacro{SO}{Solar Orbiter}
\newacro{SoHO}{Solar and Heliospheric Observatory}
\newacro{SP}{Spectropolarimeter}
\newacro{SST}{Swedish 1-m Solar Telescope}
\newacro{SUMER}{Solar Ultraviolet Measurements of Emitted Radiation}
\newacro{SUFI}{Sunrise Filter Imager}
\newacro{SVD}{singular value decomposition}
\newacro{SVST}{Swedish Vacuum Solar Telescope}
\newacro{STX}{Solar Telescope X}
\newacro{THEMIS}{T\'{e}lescope H\'{e}liographique pour l'Etude du 
   Magn\'{e}tisme et des Instabilit\'{e} Solaires}     
\newacro{TR}{transition region}
\newacro{TRACE}{Transition Region and Coronal Explorer}
\newacro{TSI}{total solar irradiance}
\newacro{UT}{Universal Time}
\newacro{UV}{ultraviolet}
\newacro{VAULT}{Very high angular resolution ultraviolet telescope}
\newacro{VIRGO}{Variability of solar IRradiance and Gravity Oscillations}
\newacro{VTT}{Vacuum Tower Telescope}    
\newacro{XRT}{X-Ray Telescope}

\long\def\startignore #1\stopignore{}   

\def\rmit#1{{\it #1}}              
           
\def\ie{\rmit{i.e.,}}              
\def\eg{\rmit{e.g.,}}              

\def\specchar#1{\uppercase{#1}}    
\def\specand{ and }                
\def\specand{\,\&\,}               

\def\CIV{\mbox{C\,\specchar{iv}}}

\def\CaII{\mbox{Ca\,\specchar{ii}}}

\def\FeI{\mbox{Fe\,\specchar{i}}}

\def\FeVIII{\mbox{Fe\,\specchar{viii}}}

\def\FeXII{\mbox{Fe\,\specchar{xii}}}

\def\HI{\mbox{H\,\specchar{i}}} 
 
\def\HeI{\mbox{He\,\specchar{i}}} 
\def\HeII{\mbox{He\,\specchar{ii}}}

\def\MgII{\mbox{Mg\,\specchar{ii}}}

\def\MnI{\mbox{Mn\,\specchar{i}}}

\def\SiI{\mbox{Si\,\specchar{i}}}

\def\SiIV{\mbox{Si\,\specchar{iv}}}

\def\Halpha{\mbox{H\hspace{0.1ex}$\alpha$}} 

\def\Hbeta{\mbox{H\hspace{0.2ex}$\beta$}}

\def\Lyalpha{\mbox{Ly$\hspace{0.2ex}\alpha$}}

\def\HeIDthree{\mbox{He\,\specchar{i}\,\,D$_3$}}

\def\NaID{\mbox{Na\,\specchar{i}\,\,D}}

\def\MgIb{\mbox{Mg\,\specchar{i}\,b}}

\def\CaIIK{\mbox{Ca\,\specchar{ii}\,\,K}}       
\def\CaIIH{\mbox{Ca\,\specchar{ii}\,\,H}}
\def\CaIIHK{\mbox{Ca\,\specchar{ii}\,\,H{\specand}K}}
\def\HK{\mbox{H{\specand}K}}

\def\KtwoV{\mbox{K$_{2V}$}}

\def\HtwoV{\mbox{H$_{2V}$}}

\def\CaIR{\mbox{Ca\,\specchar{ii}\,8542\,\AA}} 

\def\hk{\mbox{h{\specand}k}}

\def\level #1 #2#3#4{$#1 \; ^{#2} \mbox{#3} ^{#4}$}   

\def\tis{\!=\!\!}                          

\def\rmit#1{#1}                 

\ifnum\arxivmode=0
\fi
\def\subsectionrr#1{\vspace{-1.5ex} \subsection*{#1} 
  \vspace{-1ex} 
  \addcontentsline{toc}{subsection}{#1}}
\def\addtocline#1{\addcontentsline{toc}{subsection}{#1}}

\begin{document}  


\twocolumn[{%
  \includegraphics[width=3cm]{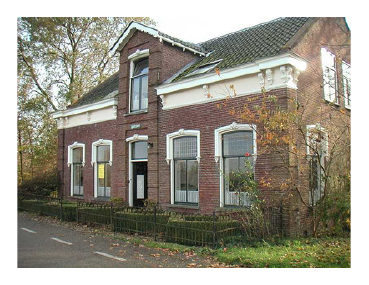}\\[-20mm]
  \hspace*{35mm} 
  {\large {\sf Lingezicht Astrophysics Reports} 1 
             ~~--~~ \hyperref[sec:history]{\today}
             ~~--~~
  \href{https://webspace.science.uu.nl/~rutte101/rrweb/rjr-pubs/2020LingAstRep...1R.pdf}{newest version}}

\vspace*{5ex} 
\begin{center}
  {\LARGE \bf SolO campfires in SDO images}\\[3ex] 
  {\large \bf Robert J. Rutten$^{1, 2, 3}$}\\[3ex]
  \begin{minipage}[t]{16cm} \small        
       \mbox{~~~~~~~~~} $^1$ 
\href{https://webspace.science.uu.nl/~rutte101/Lingezicht_Astrophysics.html}
{Lingezicht Astrophysics}, 
             Deil, The Netherlands\\
       \mbox{~~~~~~~~~} $^2$ 
\href{https://www.mn.uio.no/astro/english}
{Institute of Theoretical Astrophysics},
             University in Oslo, Oslo, Norway\\
       \mbox{~~~~~~~~~} $^3$ 
\href{https://www.mn.uio.no/rocs/english}
{Rosseland Centre for Solar Physics},
             University in Oslo, Oslo, Norway\\[2ex]

\label{sec:abstract}
{\bf Abstract.~} I present the appearance of ``Solar Orbiter
campfires'' in simultaneous images from the Solar Dynamics
Observatory where most are visible although less sharp. 
I also show such features elsewhere in the SDO database.
I show some in detail and discuss their nature.  
\vspace*{2ex}
  \end{minipage}
\end{center}
}] 

\begin{minipage}[t]{0.9\columnwidth}  
  \tableofcontents
  \label{sec:contents}
\end{minipage}
\newpage

\parskip=1ex
\section{Introduction}     \label{sec:introduction}
On July 16, 2020 a friend with interest in science alerted me to an
on-line ESA--NASA press conference that afternoon to announce first
images from Solar Orbiter (SoLO): ``closer than ever before''.

I had not expected this so soon after launch, but found a
\href{https://commons.wikimedia.org/wiki/File:Animation_of_Solar_Orbiter%27s_trajectory_-_polar_view.webm}{video movie}
of the ecliptic-projected orbit and saw that SolO dropped behind Earth
so much at launch that it fell considerably from Earth orbit
towards this relatively close pass (0.5~AU) then already.

A newspaper journalist then informed me embargoedly that the press
release would boast ``many little flames''.  

Naturally I speculated about familiar solar little-flame phenomena:
Ellerman bombs (EB,
\citeads{2013JPhCS.440a2007R}), 
quiet-Sun Ellerman-like brightenings (QSEB,
\citeads{2016A&A...592A.100R}), 
flaring active-region fibrils (FAF,
\citeads{2016A&A...590A.124R}), 
IRIS bombs (IB, \citeads{2014Sci...346C.315P}), 
UV bursts (\citeads{2018SSRv..214..120Y}), 
tips of spicules-II in on-disk appearance as rapid blue excursion
(RBE,
\citeads{2008ApJ...679L.167L}) 
or rapid red excursion (RRE,
\citeads{2013ApJ...769...44S}), 
and (ALMA) mm bursts
(\citeads{2020A&A...643A..41D}). 

Attending the press conference made clear that the little flames,
coyly called ``campfires'', sat in images from the EUI 174\,\AA\ HRI
telescope implying million-K temperature and hence excluding EBs and
QSEBs since these reconnection flames remain photospheric, not
reaching EUV temperatures and even not breaking through the \Halpha\
fibril canopy (which also hides them for ALMA,
\citeads{2017A&A...598A..89R}). 

The tips of RBEs and RREs do reach out of the canopy and often reach
EUV temperatures
(\citeads{2016ApJ...820..124H}), 
but these occur ubiquitously around network including quiet unipolar
network, much less scarce than the dozen or so campfires in the
approximately $0.5 \times 0.5$~R$_{\rm Sun}$ SolO image that was
shown.

The SolO campfires looked like FAFs to me. 

I have inspected many FAFs in specific {\tt EBFAF} detection movies
multiplying SDO/AIA 1700\,\AA\ and 1600\,\AA\ after
normalization, but the only detailed FAF descriptions are in
\citetads{2015ApJ...812...11V}. 
More in \rrref{appendix}{sec:16001700}.

Here I report on my eventual finding these SolO campfires in SDO
images. 
To cut the long story short: they are not FAFs. 
I speculate what instead. 

The contents table serves as clickable outline; I therefore refrain
from adding a descriptive contents outline here.

I keep the figures at the end to facilitate parallel text and figure
inspection using a second pdf-reader instance (or printing the text if
you prefer paper). 
For many images zoom-in to detail may be useful.
Most are full-page to enable blinking by page flipping
(\hyperref[sec:blinking]{how-to per viewer}).

\section{Finding SolO campfires}    \label{sec:finding}
Immediately after the press conference I shared my FAF suspicion and
asked whether the images would be public and whether I might have
access to them for checking per SDO with my EBFAF detection
technique. 
The quick answer from EUI PI David Berghmans was ``yes'' -- but that
it would take some months.

For a first look, being curious and impatient, I therefore saved
the press-release high-resolution 174\,\AA\ image
(\rrref{figure}{fig:soloim}), noting that the others shown were cutouts
of it, asked for information where on the Sun and when it was taken,
and downloaded the SDO/AIA 171\,\AA\ daily movie for the stipulated
May 30, 2020 date to locate the SolO image scene.   

The SolO EUI/HRI telescopes have 2048$\times$2048 0.5~arcsec pixels,
the same angular pixel size as SDO/HMI (4096$\times$4096) and TRACE
(1025$\times$1024) and close to the 0.6~arcsec of the four SDO/AIA
telescopes to which SSW's {\tt aia\_prep.pro} rescales HMI products.
Hence, from about 0.5~AU the campfires should be twice sharper with
SolO, but since they extend multiple pixels in the press-release image
they should be visible in AIA 171\,\AA\ also.

The orbit movie suggested that SolO was about 40 degrees in advance of
the Earth along the ecliptic, so I concentrated on pattern recognition
of the SolO scene in \rrref{figure}{fig:soloim} towards the West limb in
the AIA 171\,\AA\ movie, hoping that the EUI/HRI telescope pointed
near sub-SolO disk center and not towards the SolO West limb where the
scene would already be backside for SDO.\footnote{And for IRIS,
Hinode, SST etc.\ if co-pointing had been requested -- which I hoped
since multi-telescope multi-diagnostic observation is the proper way
to do solar physics in my view, motivating my efforts
in SDO--other telescope co-alignment and before that in running
\href{https://webspace.science.uu.nl/~rutte101/esmn/ESMN_home.html}
{EU networks}.}

I did not find the scene, also not for SolO's full-disk 304\,\AA\ images
in the press release.

I got no answer on my when and where question\footnote{Nor a reaction
on this report although I shared a link while writing it. 
The admonition now under \rrref{figure}{fig:soloim} came only
post-posting.
The current live link is
\href{https://webspace.science.uu.nl/~rutte101/rrweb/rjr-pubs/2020LingAstRep...1R.pdf}{this
website post} updating the
\href{https://arxiv.org/pdf/2009.00376.pdf}{arXiv post} and also
retaining an
\linkpdfpage{https://webspace.science.uu.nl/~rutte101/rrweb/rjr-pubs/2020LingAstRep...1R.pdf}{14}{epilogue} 
on starting this report series.}
and I also realized that the SolO image might have been taken anywhere
on the Sun with any orientation, perhaps even flipped or mirrored, and
possibly beyond the SDO limb.
Then Greg Slater (LMSAL) pointed out that the zoom movie shown in the
press conference morphed between unrelated, disjoint scenes and so
could not be used in location finding; the 304\,\AA\ full-disk view I
had taken as guide might be from another date and have other
orientation. 
He also suggested that the high-resolution image might have been taken
during June at yet smaller SolO distance to the Sun but increasing
SolO--Earth sight-line difference. 

I therefore collected daily SD0\,171\,\AA\ movies for many days and
played them endlessly against the SolO image in many diverse
orientations, trying visual pattern recognition -- very confusing, the
proverbial needle in a haystack. 
After many days of increasing frustration I gave up.

Eventually, on July 28 ESA mission scientist Daniel M\"uller
informed me that the image was actually taken on May 30 as stipulated,
around 14:54~UT. 
This enabled full-resolution full-cadence full-disk AIA sequence
downloads and inspection at more detail than the SDO daily movies.

It took me a few days more but then I finally recognized the SolO
scene in SDO\,171\,\AA. 
After all my far-too-wide casting I found that the image was actually
taken somewhat east of sub-SolO disk center, not flipped or mirrored,
and had only small-angle rotation from solar North up as seen from the
Earth and SDO.

\section{SDO data collection}   \label{sec:gathering}
Over the past decade I developed an extensive IDL pipeline to
cross-align 
\href{http://jsoc.stanford.edu}{JSOC}
 ``{\tt im\_patch}'' cutouts precisely between all SDO
diagnostics and then co-align the results with small fields from other
telescopes, in particular the Swedish 1-m Solar Telescope (SST).
The first part is used here.

I aim to present this pipeline in another report in this series, but
some detail is given already in my brief manual in the IDL directory
where I make this software
available.\footnote{\href{https://webspace.science.uu.nl/~rutte101/Recipes_IDL.html}{\url{https://webspace.science.uu.nl/~rutte101/Recipes_IDL.html}}}

In a nutshell, a single call of
\href{https://webspace.science.uu.nl/~rutte101/rridl/sdolib/sdo_getdata_rr.pro}{\tt
sdo\_getdata.pro} requests, collects, and cross-aligns SDO cutouts
from \href{http://jsoc.stanford.edu}{JSOC}, small ones at full cadence
for the target area and large ones at lower cadence around disk
center. The latter are used to find spatial offsets between the SDO
diagnostics (``channels'') by cross-correlation of many small
subfields, usually 30$\times$30~arcsec each, applying apparent
height-of-formation differences and iterative removing outliers, as
function of time during the requested sequence duration. 
Examples before and after are shown in \rrref{figure}{fig:drifts}.
These ``driftscenter'' results are stored and used for the target
cutouts, usually yielding cutout cross-alignment precision about
0.1~arcsec (an order better than the start-off co-registration
by {\tt aia\_prep.pro}). 

With {\tt sdo\_getdata\_rr,'2020.05.30\_14:50',15,375,148,
xsize=700,ysize=700} I targeted a wider area around the SolO field
during 15 minutes. 
The program took 42~min\footnote{Deep Thought computation.}
to deliver the material used here. 

In visual comparison I found that the best match with the SolO image
in \rrref{figure}{fig:soloim} occurred around time step 46 or
14:58:46~UT. 

In addition I made full-disk images and charts at this best-match time
with my
\href{https://webspace.science.uu.nl/~rutte101/rridl/sdolib/sdo_diskfigs.pro}
{\tt sdo\_diskfigs.pro} which may be used for any non-eclipsed SDO
moment and includes optional GONG \Halpha\ download. 
It yielded the full-disk SDO overviews in
\rrref{figures}{fig:fullmag}--\ref{fig:full193} with the JSOC target
cutout outlined.   

There were two minor active regions near the East limb. 
The outlined SolO target shows nothing active beyond quiet network on
the surface, but diffuse and also concentrated patches of EUV
brightness -- formerly ``bright points'' but ``fires'' now.

Normally, I obtain precise co-alignment with the ``other'' telescope
(called STX for Solar Telescope X in my 
\href{https://webspace.science.uu.nl/~rutte101/Recipes_IDL.html}
{software} and
\href{https://webspace.science.uu.nl/~rutte101/rridl/00-README/sdo-manual.html}{manual})
which includes removal of small SDO wobbles left over from the JSOC
whole-pixel cutout selection.  
Precise pixel-by-pixel SDO--STX comparisons are then possible and
easily done with my versatile
\href{https://webspace.science.uu.nl/~rutte101/rridl/imagelib/showex.pro}
{\tt showex.pro}\footnote{The underlying engine is
\href{https://webspace.science.uu.nl/~rutte101/rridl/imagelib/movex.pro}
{\tt movex.pro}, built on SSW's {\tt ximovie.pro} written by {\O}yvind
Wikst{\o}l and Viggo Hansteen for pre-Hinode {\em Solar-B\/}.}
sequence browser and blinker which can load very many different files
and also cube parameters in memory together with jpg images or mpg
movies and then zoom-in to pixel level.
 
In this case, however, I could not use my
\href{https://webspace.science.uu.nl/~rutte101/rridl/sdolib/sdo_stx_align.pro}{\tt sdo\_stx\_align.pro} because the SDO scene is foreshortened by its
limbward viewing, differential across its large field, with respect to
the SolO scene nearer disk center (probably with opposite
foreshortening).
 
No pixel-by-pixel co-registration therefore, but it was easy to
manually co-locate SolO campfires and larger
``brushfires''\,\footnote{Ron Moore corrected my English by writing
``brushfire'' in a reaction. \label{foot:moore} 
I had used ``bushfire'' for a fire larger than a ``campfire'' but smaller
than ``burning prairie'' (obviously sardonic), but I did not
appreciate that ``bushfire'' is Australian dialect for any wildfire
including giant forest fires.
Now I call the larger-than-campfire fires ``brushfire''.} precisely in
SDO/AIA images with {\tt showex} and then zoom-in to pixel detail to
obtain pixel-clicked joint location output for matching ROI = ``region
of interest'' cutouts of the SolO image and the SDO images.

These double coordinate pairs became the cutout centers for the 17 ROI
cutout assemblies shown in
\rrref{figures}{fig:cfroilocs}--\ref{fig:sdoroiloc304x131}.  Their fields
are sufficiently small to permit de-foreshortening and
height-difference corrections detailed below.

\section{SolO--SDO scene} \label{sec:SDO}
\rrref{Figure}{fig:soloim} shows the SolO press-release high-resolution
image from the 174\,\AA\ EUI HRI telescope.\footnote{The SolO/HRI
pixels are 0.5~arcsec just as for SDO/HMI but twice smaller in km on
the Sun at 0.5~AU distance. 
I could not add $(X,Y)$ axes to \rrref{figure}{fig:soloim} because I do
not know the sub-SolO XCEN and YCEN pointing values nor the precise
pixel size and image orientation.
The derotation applied here is a visual estimate.}  
I see about a dozen campfires in subjective selection of tiny bright
features. 
The 2048$\times$2048~px image measures about 402$\times$402~Mm$^2$ on
the Sun, 
suggesting campfire density about $10^{-4}$~Mm$^{-2}$ 
or less than a thousand on the Sun if the
remaining surface -- in this quiet cycle phase -- shows them
likewise. 
Too scarce for a significant role in coronal heating.

\rrref{Figures}{fig:fieldmag}--\ref{fig:field193} show corresponding but
somewhat larger SDO cutouts (plus a GONG \Halpha\ cutout) at the
best-match time in a selection that is diagnostically ordered
bottom-to-top or cool-to-hot in na\"{\i}ve interpretation. 
For each SDO image the greyscaling is defined by the entire 15-min
downloaded sequence to have common greyscales per diagnostic for the
ROI cutouts in \rrref{figures}{fig:roi1}--\ref{fig:roiE}. 
The axes specify standard solar $(X,Y)$ coordinates with the origin at
sub-Earth disk center, $Y$ pointing to the solar North pole, in
arcsec\footnote{Here called ``SDO arcsec'' for viewing from Earth.}.

IDL {\tt showex} inspection with zoom-in and blinking is the best
method for detailed comparison of these image sequences; the reader
may do so by installing my pipeline, duplicate the above {\tt
sdo\_getdata} command and inspect with {\tt showex}. 
Flipping the below figure pages is a poor blinking emulation; zoom-in
to detail is emulated as poorly in the ROI cutout assemblies in
\rrref{figures}{fig:roi1}--\ref{fig:roiE}.

Unfortunately, the SolO image cannot be blinked at the pixel level to
the SDO ones due to the considerable morphing by different and varying
foreshortening. 
This also inhibits scatter correlations with Strous diagrams defined
in \linkadspage{2019A&A...632A..96R}{2}{Section~2}\footnote{Direct pdf
page links as this one open the cited page on your screen with all pdf
viewers I know -- but not reliably under macOS.
Mac users may instead be shunted to the first page and must then
manually find the cited page. 
For the Mac-challenged I add the pdf page number in these links. 
Acrobat may require undoing security settings disabling the opening of
web pages.} of \citetads{2019A&A...632A..96R} 
and used here in \rrref{figure}{fig:scats}.

Visual comparison with \rrref{figure}{fig:soloim} shows that many
campfires are also visible in the hotter AIA diagnostics, not only in
AIA 171\,\AA\ (\rrref{figure}{fig:field171}). 
In AIA 193\,\AA\ (\rrref{figure}{fig:field193}) especially in dark areas
where they stand out clearer.

Upshot: SolO was not required for noticing these small flames,
although it does show them better.
The gratifying good news is that the EUI HRI 174\,\AA\ telescope
functions well.\footnote{Just as the granulation and magnetic bright
points in the press-released first DKIST images show nothing not
already known from \eg\ SST imaging, but do demonstrate promising
telescope functioning. 
But also the same story -- for lack of asked-for pointing information
I did not find their scene in HMI images.}

The SolO image appears clipped at the brightest levels. 
I therefore made the histograms in \rrref{figure}{fig:hists}. 
The SolO histogram shows no sign of clipping but an extended
highest-brightness tail not present in the AIA 171\,\AA\ histogram. 
I therefore added ``cooler'' and ``hotter'' AIA 131 and 193\,\AA\
histograms. 
The latter shows a similar tail, suggesting either nonlinear response
in the SolO image or that its 174\,\AA\ bandpass includes hotter
contributions than AIA's 171\,\AA\ bandpass. 
The visibility of many SolO campfires in \rrref{figure}{fig:field193}
suggests the latter.

\section{Campfires in other SDO scenes} \label{sec:otherSDO}

Obviously the next question is whether the area imaged by SolO was
somehow fortuitously special and lucky in uniquely harboring campfires
not visible anywhere else or anytime before. 
The press release claimed ``omnipresent miniature solar flares'' but
showed only the field of \rrref{figure}{fig:soloim}.

I therefore repeated {\tt sdo\_getdata} sequence collection and
processing again for the best-match time but at sign-reversed $X$ and/or $Y$
to sample all four disk quadrants likewise, and also for the SolO
pointing at the same date and time in the past three years. 
\rrref{Figures}{fig:sdo-minx-193}--\ref{fig:sdo2017-mag} show the
results. 
They are combined in ``triples' per location, arranged for easy
blinking by page flipping (\hyperref[sec:blinking]{how-to viewer hints}).

The first figure in each triple is the AIA~193\,\AA\ image, selecting
this wavelength because it shows SolO campfires clearest of
\rrref{figures}{fig:fieldmag}--\ref{fig:field193}.

The second per triple shows my ``SDO fire detector'' construct to
detect campfires and brushfires distinct from more ubiquitous
chromospheric heating. 
It was inspired by {\tt EBFAF} detection
(\rrref{appendix}{sec:16001700}) and \rrref{figure}{fig:scats} and
proved useful in \rrref{figure}{fig:sdoroiloc304x131} of which the
caption describes the construction. 
The greyscale clip and color threshold are defined for quiet network.

The detector construct appears to supply an amazingly good spatial
separator between chromospheric heating and coronal heating in
quiet-Sun regions. 
Chromospheric heating shows up as ubiquitous grey network patches,
similarly present nearly everywhere.
Coronal heating shows up as sparser small campfires and larger
loop-bundle brushfires of which the cyan detector pixels emphasize the
feet. 
More diffuse coronal heating with large-scale connectivity seen in
193\,\AA\ spreads around brushfires and may well be caused by these or
preceding instances.

The third per triple shows the corresponding HMI magnetogram, clipped
as described in the caption of \rrref{figure}{fig:fieldmag}.

Triple blinking is best done pairwise: 193\,\AA\,--\,detector and
detector\,--\,magnetogram. 

Blinking 193\,\AA\,--\,detector pairs in
\rrref{figures}{fig:sdo-minx-193}--\ref{fig:sdo2017-mag} shows that
the detector does a good job in locating small campfires.
Some of these fields contained active regions around which the thick
haze of extended coronal loops in 193\,\AA\ hides small fires
underneath, but the detector unveils some.
Elsewhere the 193\,\AA\ and detector images in these six other-scene
triples display similar tiny campfires at roughly similar (scarce)
density as in \rrref{figures}{fig:soloim} and
\ref{fig:sdoroiloc304x131}. 

Upshot: campfires seem indeed a sparse but omnipresent solar
phenomenon.
If SolO had been targeted elsewhere or launched earlier we would have
had the same press conference.

Blinking detector\,--\,magnetogram pairs in
\rrref{figures}{fig:sdo-minx-193}--\ref{fig:sdo2017-mag} shows
directly and unequivocally that all larger brushfires are located
above or between mixed-polarity patches of multiple MCs (magnetic
concentrations), suggesting EUV-visible heating due to bipolar MC
happenings on the surface. 
All lie in extended grey patches in the detector images that
represent dynamically heated chromosphere above and around network
including ``quiet'' network (\rrref{appendix}{sec:ha304}).

Only the smallest campfires seem to not always share bipolar magnetic
feet (at HMI resolution and sensitivity), but these still lie
preferentially in heated-chromosphere network patches.

Blinking 193\,\AA\,--\,detector pairs shows nearly 1:1 correspondence
between campfire presence in 193\,\AA\ and cyan detector pixels,
suggesting common heating to lower and higher temperatures.

The larger brushfires generally consist of close-packed short bright
arches in 193\,\AA\ that connect opposite-polarity MC clusters in the
magnetograms. 
In many the corresponding cyan detector pixels are grouped at the arch
feet.

The campfires generally differ from the brushfires by not showing
multiple-arch topography but just small single flames. 

Another difference lies in their time behavior seen in {\tt showex}
viewing each downloaded time sequence: the campfire flames appear
momentarily, mostly without repeat at the same location, whereas the
brushfires live longer, usually throughout these 20-min sequence
durations.  

I also collected long-duration SDO sequences following the SolO scene
six hours before and six hours after the SolO moment and found that
most brushfire sites kept brushfiring continuously, irregularly but
generally present during this time span.

The next question is whether campfires are a global quiet-Sun
phenomenon still present during minima, or instead an activity
phenomenon restricted to the activity belts and occurring with cycle
modulation.

I therefore repeated {\tt sdo\_getdata} sequence collection and
processing for similar-size North-pole and South-pole target areas
during the maximum of cycle 24 and during the subsequent present
minimum.
\rrref{Figures}{fig:sdo-np-max-193}--\ref{fig:sdo-sp-min-mag} show the
results, again grouped together in 193\,\AA\ -- detector --
magnetogram triples for pairwise blinking.

These four polar areas also show campfires in quiet areas, including
polar holes. 
Blinking 193\,\AA\,--\,detector pairs suggests that the smaller ones
lie roughly as deep in the atmosphere in these diagnostics: they jump
less limbward than the hazier 193\,\AA\ protrusions above them. 
I tried ${\tt heightdiff} = 1000$~km for their construction
(\rrref{appendix}{sec:16001700}) but it made no difference in fire
detection. 
Campfires are a chromospheric phenomenon\footnote{Qua height, but not qua
\citetads{1868RSPS...17..131L} 
since not seen in \Halpha.}.

The North-pole at maximum in \rrref{figure}{fig:sdo-np-max-fire} shows
fewer fires towards the limb than the others -- but there may be
blocking along slanted lines-of sight to deep-lying fires, more at
larger activity.
Overall campfires seem a global phenomenon.

Blinking detector\,--\,magnetogram polar pairs shows increasing
limbward offsets, to the extent that near the limb fires and MC
patterns are hard to match. 
Expected since the average height of the 304\,\AA\ chromosphere is
multiple thousands of km, translating to projected shifts of multiple
arcseconds there.\footnote{My pipeline presently uses limb value
${\tt heightdiff} = 3600$~km for AIA 304\,\AA\ to HMI magnetogram
cross-alignment, twice as much as in plane-parallel standard models
but likely on the small side. 
An unpublished limb spectrum taken early in the IRIS mission by Bart
De Pontieu and reduced by Han Uitenbroek shows \MgII\ \hk\ sticking
out beyond 5000~km, whereas \MgII\ ionizes away at temperature 20~kK
(coronal equilibrium) or lower (Saha-Boltzmann) about where \HeII\
comes up, so should appear ``below'' or near that unless non-E applies
(\rrref{appendix}{sec:ha304}).
See also \citetads{2019SoPh..294..161A}.} 
 
Thus, fires lie relatively deep for EUV phenomena, but not below
the chromosphere as photospheric EBs and QSEBs do.

The ten-year full-disk high-cadence SDO database potentially permits
an immensely rich harvest of campfires and brushfires, including
appearance in other diagnostics as in these triples and also scene
evolution before and after as in the ROI figures below. 
The fire detector offers an easy way to find SDO fires. 
The first projects suggested in \rrref{section}{sec:conclusion} are
obvious strategies to elaborate on these few first-look trial samples
shown here.

However, the 20 blink pairs of this section already furnish a vivid
tell-tale overview of fire occurrence and properties.
The tentative conclusion is that all or nearly all fires represent
small-scale heating caused by bipolar MC interactions on the surface,
likely occurring as globally as somewhat active mixed-polarity network
does.  
 
The larger, \ie\ brushfires, clearly heat the corona locally above
them and possibly elsewhere via long-loop connectivity.
The smallest, \ie\ campfires, are located at chromosphere heights and
may possibly contribute coronal heating in minor fashion.
 
Chromosphere heating (of the ubiquitous grey network patches in the
detector images, see \rrref{appendix}{sec:ha304}) is instead wide-spread and 
not due to fires.

The next sections enlarge fires in the SolO field of
\rrref{figure}{fig:soloim} for more detailed inspection.

\section{SolO--SDO campfire cutouts} \label{sec:campfires}

\rrref{Figures}{fig:roi1}--\ref{fig:roi12} compare ROI = ``region of
interest'' cutouts of SolO 174\,\AA\ campfires and corresponding
co-located SDO diagnostics. 
These cutouts measure 24$\times$24~arcsec. 
\rrref{Figure}{fig:cfroilocs} shows their locations in the SolO 174\,\AA\
image.

The SDO cutout panels are de-foreshortened to mitigate slanted
limbward viewing. 
This means that the original cutout pixels, which sample compressed
solar surface in the limb direction with respect to transverse and at
sub-SDO disk center, are increased in number in the limbward direction
to stretch the surface they sample to the extent they would have in
sub-SDO viewing from above.\footnote{But not obtaining the
actual view from above because the projection and blocking in slanted
viewing remain. 
Remapping cannot repair that the Sun is 3D non-spherical 
although we zoom her 2D flat and describe her plane-parallel.}

Since I do not know the sub-SolO $(X,Y)$ pointing I cannot apply
deforeshortening to the SolO cutout at the top of these ROI figures.
It is likely less but opposite.

The de-foreshortened SDO cutouts are ordered in time-delay columns,
respectively 5, 2 and 1~min before the best-match time and then 1, 2,
and 5~min later. 
These samplings emulate {\tt showex} time-sliding while blinking.

From bottom to top the diagnostic order is as for
\rrref{figures}{fig:fieldmag}--\ref{fig:field193}, but adding the HMI
continuum showing granulation at the bottom and replacing AIA
1600\,\AA\ with a construct called 16001700 and based on the
1600/1700\,\AA\ ratio to display excess AIA 1600\,\AA\ brightenings.

I switched to this construct instead of my usual {\tt EBFAF} detection
described in \rrref{appendix}{sec:16001700} when I saw the scene in the
AIA images. 
I would not have made my FAF suggestion if I had immediately been able
to inspect SDO UV images for the correct date, time and disk location. 
When I finally could do so with
\rrref{figures}{fig:field1700}--\ref{fig:field1600} I saw no indication
of any FAF whatsoever. 
Also no activity whatsoever in \rrref{figure}{fig:fieldmag} -- whereas
the A in FAF stands for active region.

The new 16001700 construct therefore serves to note excess 1600\,\AA\
brightenings less obvious than FAFs. 
The brightness range in these ratio panels is set to $1.5-2.5$ based
on inspection of EB- and FAF-rich data. 
The lower threshold excludes excess 1600/1700 brightening in ordinary
magnetic concentrations (MC) masquerading as pseudo-EB
(\rrref{appendix}{sec:16001700}). 
The upper threshold does accommodate EBs and FAFs but such large
enhancements are not reached here.

A complication in the construction of these ratio panels is that they
need correction for the apparent limbward shifts of MC brightenings.
This is also detailed in \rrref{appendix}{sec:16001700}.

The yellow plus signs mark the locations that I clicked manually in
zoom-in {\tt showex} inspection of \rrref{figure}{fig:soloim} and in
parallel of \rrref{figures}{fig:field171}--\ref{fig:field193}, blinking
the latter pair to find the best match, to select each ROI location.

The small yellow arrow to the upper right in the leftmost continuum
panel at the bottom shows the local limb direction. 
Its length corresponds to the projection of a 3600~km tall upright
structure (my {\tt heightdiff} value for the 304\,\AA\ chromosphere).  
Roughly this arrow indicates the position shift one may expect between
the surface and EUV samplings.

The axis scales are SDO-size arcsec for both, corresponding to
441.4~km on the Sun. 
For the SolO cutout they refer to the center of the full image. 
For the SDO cutouts the arcseconds are ``stretched'' by
de-foreshortening but the cutout centers are non-stretched sub-SDO
$(X,Y)$ values.

Per diagnostic the greyscale of each cutout is set by bytescaling the
whole-field sequence, making brightenings comparable between different
ROI figures.

My inspections of these assemblies were unusual for me because in all
my SDO--STX (usually SST but also DST and IRIS) co-alignments I have
always searched for bottom-up cause-effect order. 
My searches concerned happenings in the photosphere or chromosphere
that would or did not cause hotter and higher EUV response.

The SolO campfires are dense and hot features of which the cause is
the issue here -- but the disturbance causing them may also come from
above, as in the blobby coronal rain of
\citetads{2012SoPh..280..457A} 
and flocculent flows of
\citetads{2012ApJ...750...22V}. 
The assembly rows are ordered top down from AIA~193\,\AA\ to the HMI
continuum, but so comparisons should be made in both directions. 
At least the arrow of time from left to right should be unique.

Let me comment on these ROI figures one-by-one:

{\em ROI-1\/}.~ This is the arrow-marked campfire in the annotated
version of the high-resolution image in the SolO press
release.\footnote{I think that (as usual) the selection was not a
``typical'' (= average) example but rather the nicest.}

The SolO cutout on top shows an upward arc that is not present at the
same time in AIA 171\,\AA\ while weakly present in AIA 193\,\AA\ and
seen best in AIA 304\,\AA\ where it stays a few minutes.

The four AIA EUVs are all also bright at $\Delta t \tis -5$~min and
then re-brighten during two minutes from $\Delta t \tis -1$~min,
suggesting recurrence. 
Inspection with {\tt showex} indeed showed larger previous
brightening, maximal 7~min before the SolO moment.

There is a bipolar MC pair at the surface with significant excess
brightening in AIA 1700\,\AA\ and also momentary 1600\,\AA\ excess
brightening in the 16001700 construct that peaks at
$\Delta t \tis -2$~min and then decays.

The granules in the bottom row are as uninteresting as always. 
I hoped to see vorticity around the campfire site but don't see this.
There may be granular convergence to the yellow cross but this is hard
to tell at HMI image quality.

Top-down or bottom-up?

The MC pair at the exact location (account for the projection arrow in
the first bottom panel), its brightening in 1700\,\AA\ and its
larger-than-usual excess brightening in 1600\,\AA\ suggest bottom-up
with the MCs as agent.

However, the 16001700 brightening is weaker than EBs which do not make
it to hot AIA wavelengths, and if they did one would not expect the
precise co-spatiality of the bright grain in 171\,\AA\ and 193\,\AA\
seen here. 
FAFs do show effects in these high-temperature diagnostics but as
distant expanding arcs
(\citeads{2015ApJ...812...11V}). 
Apparent co-spatiality is also seen in the 193\,\AA\,--\,detector
blink pairs in the limb triples
(\rrref{figures}{fig:sdo-np-max-193}--\ref{fig:sdo-sp-min-mag}).

Top-down supposedly means for such a small disturbance propagating
down that it follows field lines and so naturally ends up in a surface
MC since all field lines are rooted in such. 
The earlier 1600 and 1700 brightenings may then follow from a similar
earlier disturbance dropping from above. 
Indeed, in {\tt showex} they show the 7-min earlier brightening too,
reaching maximum at small (about 12\,s) delay.\footnote{{\tt Showex}
can also plot timelines per pixel. 
The AIA UV exposures sampled at 24~s cadence are interpolated with all
others to the 12-s 171\,\AA\ timings in my pipeline. 
All samplings along columns are synchronous, as are the triples 
in \rrref{figures}{fig:sdo-minx-193}--\ref{fig:sdo-sp-min-mag}.}

The AIA~304\,\AA\ row inspired me to add \rrref{appendix}{sec:ha304}.
The campfire feature is most extended and complex here, and there are
larger fuzzy patches around it. 
In both bottom-up and top-down scenarios one would expect it to show
up between the UVs and the hotter EUVs in size and appearance. 
The simultaneous visibility of the arc in the SolO image and in
304\,\AA\ might then suggest that SolO includes a cooler line in its
174\,\AA\ passband.

However, to me the fuzzy long-lived AIA 304\,\AA\ appearance suggests
response to irradiation from above followed by non-E visibility
extension as described in \rrref{appendix}{sec:ha304}. 


{\em ROI-2\/}.~ Not a single flame but a more complex multi-feature
campfire in the SolO cutout at the top. 
AIA 193\,\AA\ mimics it best.
Its continuous presence and the presence of mixed-polarity MCs on the
surface suggest that this is a bottom-up feature of the type I call
brushfire and show in the next section. 
The agent causing it happened before these sequences.

{\em ROI-3\/}.~ Weak brightening at the center of a fuzzy
diabolo-shaped feature. 
Both are clearest in the SolO cutout.
Nothing in the 16001700 panel, weak monopolar MCs on the surface. 
The diffuse surrounding 304\,\AA\ brightness again suggests hot
irradiation from above.

{\em ROI-4\/}.~ Rather like a weak version of ROI-1 but without
precursor. 
A nearby MC on the surface surrounded by weak opposite-polarity MCs. 
Again the feature is most extended and fuzzy in 304\,\AA.

{\em ROI-5\/}.~ Similar to ROI-4. 
Brightest in AIA 131\,\AA\ at just the SolO moment.

{\em ROI-6\/}.~ Similar again but embedded in a longer rather
persistent feature .

{\em ROI-7\/}.~ Local brightening in the SolO cutout but not standing
out in any SDO panel. 
Perhaps foreshortening blocking by the fuzzy surroundings best seen in
the 304\,\AA\ panel.

{\em ROI-8\/}.~ Present in the SDO EUV panels, also earlier, then
fading. 
Again 304\,\AA\ shows the most extended surroundings. 
The 16001700 panel has an excess feature but dislocated.

{\em ROI-9\/}.~ Already present at the start, with some mixed-polarity field
on the surface. 
Perhaps a small brushfire.

{\em ROI-10\/}.~ Weak but also persistent from the start, with some
mixed-polarity MCs on the surface.

{\em ROI-11\/}.~ Truly a SolO campfire because there is nothing in any
AIA diagnostic including the fire detector in
\rrref{figure}{fig:sdoroiloc304x131}. 
AIA 171\,\AA\ shows weak streaking in the same direction, that's all. 
Nothing at its location in the HMI magnetogram.
The bright SolO streak may be just that, higher up and may be blocked
by foreground opacity in the slanted AIA viewing through the gas above
the large brushfire of ROI-D.

{\em ROI-12\/}.~ Local brightening in a small brushfire.

\section{SolO--SDO brushfire cutouts} \label{sec:brushfires}
In this section I add similar cutout figures for some of the larger
bright patches in \rrref{figure}{fig:soloim}. 
I call these ``brushfire'' after language correction by Ron Moore
(\rrref{footnote}{foot:moore}). 
They likely all are regular ``coronal bright points'' but I keep to
my phenomenological name in this purely observational inventory.

I selected five called ROI-A -- ROI-E. 
Their locations and the corresponding ROI cutouts in the SolO
174\,\AA\ image are shown in \rrref{Figure}{fig:bfroilocs}. 
These cutouts measure 36$\times$36~arcsec$^2$.

\rrref{Figures}{fig:roiA}--\ref{fig:roiE} again show a selection of SDO
diagnostics. 
The column timings are now respectively 9, 6, and 3~min before the
best-match time in the last column.

All five EUV-bright patches remain more or less the same during these
nine minutes, so that discussing their origin is literally beyond this
study: for each it took place or started before my downloaded SDO sequences.

However, all five sit above locations with somewhat enhanced mixed-polarity
network. 
\rrref{Figures}{fig:sdoroilocmag}--\ref{fig:sdoroiloc304x131} show the
HMI magnetogram of \rrref{figure}{fig:fieldmag} and various EUV
diagnostics with all ROI boxes superimposed. 
Blinking shows that also all other bright EUV patches of brushfire
size sit above similar mixed-polarity network concentrations, whereas
campfires, as SolO prototype number 1, can be at quieter monopolar
locations.

It therefore seems likely that the brushfires are all bottom-up cases
resulting from magnetodynamic opposite-polarity interactions on the
surface. 
The same is suggested by the high degree of correspondence between
larger fires and somewhat denser mixed-polarity MC assemblies in the
``other'' SDO scenes in
\rrref{figures}{fig:sdo-minx-193}--\ref{fig:sdo-sp-min-mag}.

\section{Discussion}

\subsectionrr{Campfire nature} \label{sec:nature}
By definition the SolO campfires are ``EUV bursts'' or ``SolO
bursts''.

Initially I did not check the extensive literature on coronal (X-ray)
bright points\footnote{SolO campfires are surely ``bright points'' at
lower resolution. 
Solar physicists have a bad habit of calling unresolved features
``points'' -- in
\citetads{1991SoPh..134...15R} 
we went from \CaII\ \KtwoV\ and \HtwoV\ internetwork ``bright point''
to ``grain''; for ``magnetic bright point'' I use ``magnetic
concentration''. 
At least umbral dots are not points.} 
as to whether these little SolO critters have earlier been noted,
described, analyzed in this bright-point context. 
This might well be the case since I found them also in the SDO images
and archive -- but perhaps they were just too small, scarce and
inconsequential to excite interest.

Classical coronal bright points, recently reviewed by
\citetads{2019LRSP...16....2M}, 
extend 10--60\,arcsec, are bundled short loops bright at EUV and
soft-X-ray wavelengths, and typically live multiple to many hours. 
They are the larger fires here that I phenomenologically called
brushfires.

However, after my initial arXiv post
(\href{https://arxiv.org/pdf/2009.00376.pdf}{2009.00376}) Ron Moore
pointed me to
\citetads{1998ApJ...501..386F}. 
This beautiful study compared quiet-Sun coronal activity with
underlying network much as displayed here in triple
\rrref{figures}{fig:sdo-minx-193}--\ref{fig:sdo-sp-min-mag}.
Their example \linkadspage{1998ApJ...501..386F}{2}{Fig.~1} from
SOHO/EIT is similar to \rrref{figure}{fig:soloim} and the ten AIA
193\,\AA\ triple members here. 
Their KPNO magnetograms (\linkadspage{1998ApJ...501..386F}{4}{their
Fig.~2}) appear less sharp but of higher sensitivity than the HMI
magnetograms here. 
They detected hundreds coronal bright points in their six combined
data sets. 
Most were supergranular-cell size and therefore classical ones,
but there were also smaller ones, even down to the
2.6\,arcsec EIT pixel size\footnote{SOHO arcsec, not as close as SolO
but a teeny bit closer than SDO.}. 
These one-pixel EIT brightenings were likely campfires -- long avant
la lettre. 

\citetads{1998ApJ...501..386F} 
concluded that most if not all quiet-Sun coronal heating is driven by
bipolar network interactions. 
The canonical view of coronal bright points (or Skylab X-ray bright
points) is indeed that these bright small-scale loop systems connect
network MCs with opposite polarities
(\citeads{2019LRSP...16....2M}). 
Blinking the triples in
\rrref{figures}{fig:sdo-minx-193}--\ref{fig:sdo-sp-min-mag} vividly
confirms this.

The exception to bipolar MC footpoint driving may be campfires as in
ROI-3 in \rrref{figure}{fig:roi3} and ROI-6 in
\rrref{figure}{fig:roi6}.
They show only minor underlying MCs in their HMI magnetogram cutouts
-- but still some and the low HMI sensitivity sets a severe detection
limit. 
If strong bipolar MC canceling occurred at campfire sites one should
observe EBs; their absence already requires weaker interactions only
whereas even EB cancelations are already difficult to find at HMI
quality (see comparisons in
\linkadspage{2013JPhCS.440a2007R}{6}{figure 4} of
\citeads{2013JPhCS.440a2007R}). 

So what are campfires?

Their apparent co-spatiality in the 193\,\AA\,--\,detector limb
triples (\rrref{figures}{fig:sdo-np-max-193}--\ref{fig:sdo-sp-min-mag})
suggests chromospheric location.

They are not EBs or FAFs: not in active regions and no excess
1600\,\AA\ brightenings.

They are not quiet-Sun QSEBs: no excess 1600\,\AA\ brightenings, and
QSEBs do not become EUV-bright.

They are not spicule-II tips at the end of RREs or RBEs: far too
scarce and not emanating from and around network.
 
What else as bottom-up disturbance candidate? 

Vortices come to mind (``swirls'', ``tornadoes'', ``cyclones'').
The original granulation-flow vortex of
\citetads{1988Natur.335..238B} 
extended 7~arcsec; later detections went smaller (\eg\
\citeads{2009A&A...493L..13A}, 
\citeads{2010ApJ...723L.139B}, 
\citeads{2011MNRAS.416..148V}, 
\citeads{2012Natur.486..505W}, 
\citeads{2019NatCo..10.3504L}, 
\citeads{2019A&A...632A..97L}). 
Mesogranular-scale vortices occur at about ten times higher
density then campfires while
\linkadspage{2018A&A...618A..51T}{5}{Figure 3} of
\citetads{2018A&A...618A..51T} 
does not show bright EUV response to a long-lived one.
Maybe campfires mark rare unusually large multi-granule vortices.

Yet larger prominence tornadoes detected in AIA~171\,\AA\ (\eg\
\citeads{2013ApJ...774..123W}, 
but see
\citeads{2014SoPh..289..603P}) 
aren't campfire candidates since there are too few filaments in the
SDO fields here.

Granular-size swirls occur at least a hundred times more than
campfires. 
Obviously no campfire agent, but since spicules-II show multiple wave
signatures including torsion
(\citeads{2012ApJ...752L..12D}, 
\citeads{2013ASPC..470...49R}) 
while emanating from network (mono- or bipolar), it may well be that
granular swirls affecting embedded MCs contribute to or cause the
quiet chromosphere.

I {\tt showex}-inspected the HMI granulation around the campfires
searching for apparent vorticity in time-sliding, but the
HMI-continuum granulation seems too low in quality for such visual
detection or for reliable small-scale flow mapping and vorticity
charting. 

What rests are top-down candidates.

Particle beam impacts are a familiar flare phenomenon but are also
suggested to occur in small-scale nanoflare context from higher-up
filamentary reconnection near the apex of thin loops. 
Such reconnection may generate beams of energetic particles that
travel along field lines down to their footpoints and heat the
underlying chromosphere there (\eg\
\citeads{2018A&A...620L...5B}, 
\citeads{2020A&A...643A..27F}). 
This seems a good way to explain small-scale and deeply-located
cospatial 304\,\AA\ and 131\,\AA\ brightening up to cyan intensities
in the fire detector images as coming from above, with particle beams
penetrating deeper than conduction in partially neutral chromospheric
gas (\linkadspage{2020A&A...643A..27F}{14}{Fig.~13} of
\citeads{2020A&A...643A..27F}). 

In this scheme cyan detector footpoints of short-loop brushfires in
193\,\AA\ mark higher-up intermittent reconnection imposed by bipolar
network dynamics affecting their feet, with recurrent brushfire
longevity imposed by continuous network replenishment as described by
\citetads{1997ApJ...487..424S}. 
The field-topology reconnective cause for brushfires is then bottom-up
into the corona while concurrent local chromospheric foot heating
results top-down from there. 

The small momentary non-arch campfires may then mark yet smaller and
more incidental high-up reconnection events with a resulting particle
beam lighting up the chromosphere directly underneath into brightness
in all AIA EUVs.

A very different heating-from-above option: axion quark nuggets are
claimed to impact the Sun and so not only explain SolO campfires but
also all solar EUV emission, solar impulsive radio events, coronal
heating and hence the solar wind, and then also all dark matter
(search ``axion campfires'' on arXiv,
not cited here because I won't).

Instead I wondered 
about more mundane Sun-impacting meteors and consulted
AIA-comet-impact expert Karel Schrijver.
He suggested instead CME left-over return into the Sun and sent me a
dramatic AIA 193\,\AA\ movie taken 2012-12-14 as an example.
It shows very blobby (``flocculent'') downpour back onto the lower
atmosphere. 
Karel suggested that SolO campfires may represent small versions of
such blobby return splashes, and that perhaps AIA difference movies
might indicate such even at their coarse resolution.  

In the ROI cutouts the brushfires are most probably bottom-up
cases originating in bipolar field interactions on the surface. 
Also the campfires that appear as small brushfires (ROI-2, 9, 10, 12
-- unattended campfires?).
But for the prototype ROI-1 campfire in \rrref{figure}{fig:roi1} and
similar others (ROI 3 -- 8) a small post-CME splash-down blob may be a
viable mechanism.   

If so, the ROI-1 blob splashed down into a surface MC.
Blinking \rrref{figure}{fig:sdoroiloc193} against
\rrref{figure}{fig:sdoroilocmag} shows that there is not a single ROI,
campfire or brushfire, without MCs underneath -- while there is as
much non-magnetic grey internetwork area to splash down in for
non-discriminating splashers.
However, impacting post-CME blobs likely remain charged and follow
field lines to roots on the surface, just as post-reconnection
particle beams do. 
The smallest cyan-colored fires in the fire detector scenes in triple
\rrref{figures}{fig:sdo-minx-193}--\ref{fig:sdo-sp-min-mag} are indeed
all located in grey network-chromosphere patches.

\subsectionrr{Quiet-Sun heating}

The campfires obviously display heated atmosphere, but the AIA images
in \rrref{figures}{fig:field304}--\ref{fig:field193} show no obvious
effect on their surroundings. 
In cutout \rrref{figures}{fig:roi1}--\ref{fig:roi12} diffuse AIA
304\,\AA\ brightness around them suggests wider spreading, but I
attribute that to EUV irradiation from above
(\rrref{appendix}{sec:ha304}). 

The larger brushfires in the AIA 193\,\AA\ images seem to initiate
larger-scale loops and diffuse coronal connectivity around them, but
the tiny campfires do not.
As smallest instance of the coronal bright point phenomenon they also
seem the least important in coronal heating. 

A non-expected result of this study is the remarkable efficacy of the
fire detector 304$\times$131 product in separating quiet-Sun
chromospheric and coronal heating. 
The detector images show very distinct appearances between these,
suggesting very different mechanisms.  

Heated quiet-Sun chromosphere appears in the detector images as grey
patches everywhere. 
These have the same surface pattern as reversed GONG-\Halpha\
brightness patches because both spread around quiet network. 
They occupy the same space with the same gas but in different phases
of small-scale dynamic heating and cooling which occurs ubiquitously
around quiet network, whether monopolar or bipolar.
It is intermittent, recurrent and frequent, consisting of propagating
heating events with spicules-II as principal candidate. 
In these hydrogen and helium (\ie\ almost all gas) ionize with
\HeII\ 304\,\AA\ emissivity emphasizing their tips. 
Most gas then retracts along the track of the heating event, cooling
and recombining, with \Halpha-core darkness retardedly emphasizing the
launch sites nearer the magnetic
network. \rrref{Appendix}{sec:ha304} gives background and detail.

Heated quiet-Sun corona appears in the AIA 193\,\AA\ images as coronal
bright points (brushfires and campfires) and in diffuse
wider-connectivity patterns around brushfires. 
The fire detector images pinpoint their chromospheric feet as cyan
pixels, co-spatially heated in AIA 304\,\AA\ and 131\,\AA\ but not
remarkable in \Halpha\ (\rrref{Appendix}{sec:ha304}).
In quiet-Sun areas brushfires result from minor bipolar MC
interactions and do not occur at monopolar network (``minor''
because there are no EBs or FAFs in the UV sequences; such fierce
low-altitude reconnection occurs only in active regions). 

Thus, the fire detector images suggest quiet-Sun heating that
intrinsically differs between chromospheric and coronal heating. 
My initial question whether cause-effect relations are bottom-up or
top-down has both as answer for both in the scenarios proposed in
\rrref{Appendix}{sec:ha304}.
In brief: in quiet chromosphere spicule-II heating and ionization
brighten 304 and 131\,\AA\ upward while subsequent downward cooling
and recombination of the same gas cause the fibrilar \Halpha\
chromosphere. 
In quiet corona bipolar MC footpoint dynamics impose high-up
small-scale filamentary reconnection while resulting downward particle
beams cause 304 and 131\,\AA\ feet brightenings in the chromosphere.

Whether the vaguer diffuse coronal connection patterns in the AIA 171
and 193\,\AA\ images were made exclusively by previous brushfires or
by additional other agents cannot be addressed with the short
sequences I downloaded for
\rrref{figures}{fig:sdo-minx-193}--\ref{fig:sdo-sp-min-mag}. 
However, the continued brushfire presence and activity in my 12-hour SDO
sequences of the SolO field around the SolO moment suggest continuous
bipolar MC renewal through newly arriving network replenishment
(\citeads{1997ApJ...487..424S}) 
and indeed long-term production of the observed larger-scale
connectivity.


\subsectionrr{Campfire prospects} \label{sec:prospects}
IRIS PI Bart De Pontieu and the SST's Peter S\"utterlin informed me
that on May 30 there was no coordinated co-targeting with SolO.

I hope there will be in future, but the
\href{https://commons.wikimedia.org/wiki/File:Animation_of_Solar_Orbiter%27s_trajectory_-_polar_view.webm}{SolO orbit movie}
suggests that in the upcoming SolO perihelia the mission will be
between 70 and 110 degrees in advance or behind the Earth in ecliptic
projection.\footnote{Because SolO will store its 10-day campaign data
taken during its perihelia in spring and autumn far from Earth
awaiting delivery during its subsequent mid-winter aphelia when we
also get thereabouts for closer download, hence with Earth a quarter
orbit off during the taking.}
This is promising for STEREO-type dual-sightline studies,
but complicates multi-diagnostic same-scene co-targeting because
co-pointing Earth-based and near-Earth telescopes will suffer severe
foreshortening offset from the SolO target scenes, worse than in this
study (if not farside from here).

The best time for down-the-throat conjunction co-targeting seems
around 2022-03-15 when SolO first crosses Mercury's orbit with Earth
only about 10~degrees behind in solar-center direction.\footnote{I
would love co-pointed SolO EUI/HRI \Lyalpha\ and \HeII\ 304\,\AA\ with
SST/CHROMIS \Hbeta, SST/CRISP \Halpha\ and IRIS in view of
\rrref{appendix}{sec:ha304}, but this is early in the year for best
seeing on La Palma. 
Maybe DKIST by then.}
In June 2027 SolO in its perihelion and Earth will again be close in
ecliptic direction, but then SolO will be intentionally kicked
poleward by Venus for a better view of polar areas sampled here in
\rrref{figures}{fig:sdo-np-max-193}--\ref{fig:sdo-sp-min-mag}.

\section{Conclusion} \label{sec:conclusion}

Most coronal fires in this report seem controlled bottom-up by MC
interactions of quiet mixed-polarity network on the surface, as
suggested for X-ray/coronal bright points since Skylab (\eg\
\citeads{1971IAUS...43..397K}). 
They seem to furnish most quiet-Sun coronal heating. 

Coronal fire heating is very distinct from ubiquitous chromospheric
heating around all network in the fire detector images that separate
these classes remarkably well by fire feet marking.

Whether coronal fires mark MC reconnection that may be diagnosed from
cancelation, fly-by reconnection that may be diagnosed from MC
trajectories, reconnective twisting that may be diagnosed from
swirling, reconnective or ion-neutral slingshotting that may be
diagnosed from simulation, or Alv\'enic wave heating that may be
diagnosed from propagation is beyond this study.\footnote{Or all
contribute adhering to the
\linkadspage{1990IAUS..138..501R}{9}{``Principle of Solar
Communicativity''} formulated at the only solar IAU symposium ever in
the USSR (\citeads{1990IAUS..138..501R}).} 

Campfires represent the smallest, scarcest, briefest, least important
specimens of the coronal bright point phenomenon.
The one-pixel EIT \FeXII\,195\,\AA\ coronal bright points of
\citetads{1998ApJ...501..386F} 
were a decades-earlier campfire detection. 
SDO shows them daily in multitudes for a decade already.
Blinking the triples in
\rrref{figures}{fig:sdo-minx-193}--\ref{fig:sdo-sp-min-mag} suggests
that they reside at chromospheric heights and do not contribute
noticeable coronal heating.

Some smallest campfires, including some original SolO campfires, may
represent exceptions to bipolar footpoint interaction and perhaps mark
larger-scale surface vortices or impacting field-guided CME return
blobs. 

I suggest that coronal brushfire heating corresponds to sets 1 and 2
in \linkadspage{2020A&A...643A..27F}{13}{Fig.~12} of
\citetads{2020A&A...643A..27F} with particle-beam foot heating giving
the fire pinpoint functionality of the chromospheric 304$\times$131
detector product.

For the campfires I suggest that most represent small localized
chromosphere heating as in set 3 of this figure.
It comes electrically from above, is quickly radiated away, and does
not affect the \Halpha\ chromosphere -- making ``St.~Elmo's fires'' a
better nickname for these harmless little electric flames.
They may be diagnosed by charting IRIS \SiIV\ Dopplershifts at AIA
campfires
(\citeads{2014Sci...346B.315T}, 
\citeads{2018ApJ...856..178P}). 

The obvious recommendation from this study is to not await properly
validated SolO images or future SolO campfire observations but to
study them in the SDO database using fire detection as in
\rrref{figure}{fig:sdoroiloc304x131} to find them.

If my institute would offer access to studentships and
students\footnote{If a student of this report desires to use material
or techniques used here I likely will cooperate.}
I would propose as projects:

\vspace*{-2ex}
\begin{enumerate} \parskip=1ex 

\item obtain longer SDO sequences than here and use ROI cutout
timeline analysis to study fire recurrence and doubling where multiple
feet heating occurs;

\item apply the SDO fire detector of \rrref{figure}{fig:sdoroiloc304x131}
to the entire SDO database to assemble campfire statistics as function
of the cycle, latitude, open/closed field geometry, etcetera;

\item do the same for more interesting brushfires in relation to
larger-scale coronal connectivity patterns. 
Also produce latitude--time fire activity diagrams in the style of
David Hathaway's beautiful
\href{http://solarcyclescience.com/bin/magbfly.png}{magnetic butterfly
diagram};

\item try to find campfires that follow on CMEs with 
subsequent flocculent rain, first for trial demonstration and when
successful with machine learning for the entire SDO database;

\item search for granular and/or MC vortices at fire sites.
For SolO use simultaneous SolO/PHI granulation imaging. 
For SDO use simultaneous Hinode/SOT or groundbased granulation
imaging. 
Expand to chromospheric vortices where chromospheric diagnostics are
available. 
Expand if possible to the SDO database when successful;

\item use Bifrost or MURaM simulations to replicate and demonstrate
the apparent brightening differences and limb shifts of MCs between
different ultraviolet wavelengths, including the AIA, TRACE and
IRIS/SLI ones;

\item use SST/CHROMIS observations in the extended \CaII\ \HK\ wings
to study MC stratifications following
\citetads{2005A&A...437.1069S}, 
extend these to larger height with IRIS \MgII\ \hk\ scans, do so for
quiet network and brushfire-feet mixed-polarity network, also cover
center-to-limb sampling, and compare with Bifrost or MURaM
mixed-network simulations with \HK\ and \hk\ wing synthesis using
RH\,1.5D (\citeads{2015A&A...574A...3P}); 

\item develop a {\tt FAFDETECT} algorithm and search the SDO
database for slow wind sources as in 
\citetads{2015NatCo...6.5947B}; 

\item use machine learning to search the SDO database for retarded
correlation of AIA 304\,\AA\ brightenings after AIA 1600\,\AA\
brightenings, and also of AIA 304\,\AA\ brightenings after flocculent
CME return flows in the hotter AIA's.  Best done with early SDO data when
AIA~304\,\AA\ had higher signal-to-noise;

\item use automation on the SDO and GONG \Halpha\ databases to select
the best \Halpha\ images, de-stretch and co-align these precisely with
AIA 304\,\AA\, and quantify pattern equality including relative time
delay. 
Better would be to find sharper joint data sets for these two lines
and use those. 
Yet better would be to add ALMA imaging.


\end{enumerate}


\begin{appendix}
\parindent=0ex
\parskip=1ex
\renewcommand\thefigure{\arabic{figure}} 

\section{AIA 1600 and 1700\, \AA\ comparison} \label{sec:16001700}

My initial impression from watching the press conference was that SolO
campfires might be FAFs. 
I therefore elaborate on these here. 
The way to find them is by comparing AIA 1600\,\AA\ images to
1700\,\AA\ images which is also a good way to find EBs beyond the blue
wing of \Halpha. 
Comparing these UV images brings intricacies also detailed here.

I start this explanation with EBs because ultraviolet EB and FAF
detections are coupled, although it was evident already during the
press conference that campfires are not EBs or QSEBs because neither
type of low-atmosphere reconnection reaches EUV visibility.

\subsectionrr{EBs and FAFs}
EBs\footnote{EB naming: \citetads{1917ApJ....46..298E} 
called them ``hydrogen bombs''. 
``Ellerman bombs'' came from \citetads{1960PNAS...46..165M} 
but I now prefer ``bursts'' since the border-police-challenging b-word
cost me a laptop.}
were discovered and defined by
\citetads{1917ApJ....46..298E} 
as sudden small brightenings in active regions in the outer wings of
the Balmer lines. 
By now it is well-established that EBs mark strong-field reconnection
in the low photosphere that does not break through the overlying
chromospheric canopy of \Halpha\ fibrils. 
``Strong field'' means kilogauss ``fluxtube'' magnetic concentrations
(MC) in the \citetads{1977PhDT.......237S} 
sense, but they don't have to cancel completely, only partially, which
means that in coarse SDO/HMI magnetograms one does not observe bipolar
feature pairs vanishing against each other at EB sites. 
At higher resolution, as from the SST in the third row of
\linkadspage{2013JPhCS.440a2007R}{6}{figure 4} of
\citetads{2013JPhCS.440a2007R} 
as compared to the fourth HMI row there, one does see partial
vanishing. 
I believe that EB cancelations have not yet been studied at the so-far
highest resolution and magnetic sensitivity (both needed) of Hinode's
SOT/SP, perhaps because Hinode's \Halpha\ imaging first suffered bubbles
and then died.

EBs are observed per Ellerman definition in the outer wings of the
Balmer lines and also in \CaIIHK, but not in the \NaID\ and \MgIb\
lines (\citeads{1917ApJ....46..298E}, 
\citeads{2015ApJ...808..133R}, 
\citeads{2016A&A...590A.124R}). 

A decisive characteristic is bright-flame appearance when observed
towards the limb (\citeads{2011ApJ...736...71W}). 

They also stand out in AIA 1700 and yet more in 1600\,\AA\ images
through metal ionization leaving only the Balmer continuum and
Rayleigh scattering as opacity agents
(\citeads{2016A&A...590A.124R}). 

The best way to spot potential magnetic cancelations at EB sites with
SDO is not searching for MC cancelation in HMI magnetograms but detecting
fast convergence of magnetic bright points in AIA 1700\,\AA\ that mark
opposite-polarity MCs. 

\citetads{2016A&A...592A.100R} 
reported QSEBs = ``quiet-Sun Ellerman-like brightenings''.
These are similar partial MC reconnections but in quiet network away
from active regions. 
\citetads{2017A&A...601A.122D} 
simulated them with MURaM;
\citetads{2020A&A...641L...5J} 
found more and more detail with \Hbeta\ images from SST/CHROMIS, twice
sharper than SST/CRISP \Halpha\ (like SolO over SDO).

Neither EBs nor QSEBs are of interest here because their reconnective
heating doesn't reach EUV-line temperatures.  Their role here is their
AIA UV detectability. 
 
FAFs\footnote{FAF naming: problematic. 
In \citetads{2013JPhCS.440a2007R} 
we noted them as ``small flaring arch filaments and microflares'', I
think following a report by Brigitte Schmieder -- but I don't remember
which and she has too many non-open-access for easy search.  
\citetads{2009ApJ...701.1911P} 
noted them in 1600\,\AA\ images from TRACE, called them ``transient
loops''. and reported them as a new phenomenon -- but probably \CaIIK\
``microflares'' (\eg\
\citeads{2002ApJ...574.1074S}) 
described similar outbursts.  
We used FAF = “flaring arch filament” in
\citetads{2015ApJ...812...11V} 
but in \citetads{2016A&A...590A.124R} 
I proposed ``flaring active-region fibril'' to avoid confusion with
the larger and stabler structures making up ``arch filament systems''
in emerging and flaring active regions.}
are also easily identified through enhanced brightness in the
ultraviolet continua sampled by SDO. 
In contrast to EBs they are less round, move fast along filamentary
tracks, and appear more enhanced in 1600\,\AA, presumably from \CIV\
contributions.
They start as similar photospheric partial MC reconnection events, but
their reconnection proceeds upwards to above the \Halpha\ canopy as
emulated in numerical MHD simulations by
\citetads{2017ApJ...839...22H}. 
They may leave signatures in the hotter AIA EUV diagnostics but these
appear as rapidly expanding arcs (perhaps shells) that are hard to
detect (\citeads{2015ApJ...812...11V}). 

\subsectionrr{EBFAF movies} \label{sec:ebfaf}
I started on {\tt EBFAF} detection while reviewing the EB literature
for \citetads{2013JPhCS.440a2007R}. 
I found that many older but also recent publications erroneously
addressed ordinary MCs as EBs although
\citetads{1917ApJ....46..298E} 
already warned against this -- we then called these ``pseudo-EBs''.
 
I then wrote
\href{https://webspace.science.uu.nl/~rutte101/rridl/sdolib/sdo_makeeblocmovie.pro}
{\tt sdo\_makeeblocmovie.pro} which produces {\tt EBFAF} movies
comparing AIA 1600 to 1700\,\AA. 
I experimented with subtraction, division and multiplication of the
two after normalizing each to its mean, and settled on multiplication.
In the 1600$\times$1700 panels of these movies EBs stand out
dramatically by being very bright while roundish and stationary during
a few minutes; FAFs are as bright or yet brighter but have elongated
shape and move very fast along filamentary paths. 

Triggered by EB-manuscript referee requests I then got some years into
the habit of checking any new EB publication by producing and
inspecting corresponding SDO {\tt EBFAF} movies, including
\citetads{2013SoPh..283..307N}, 
\citetads{2013ApJ...774...32V}, 
\citetads{2013A&A...557A.102B}, 
\citetads{2013SoPh..288...39Y}, 
\citetads{2013ApJ...779..125N}, 
\citetads{2014ApJ...792...13H}, 
\citetads{2014Sci...346C.315P}, 
\citetads{2015ApJ...798...19N}, 
\citetads{2015ApJ...812...11V}, 
\citetads{2015ApJ...810...38K}, 
\citetads{2015A&A...582A.104R} 
and more that I don't remember.\footnote{The worst identified over
3000 EBs ``radiating enough excess energy to heat the corona'' -- but
they were all ordinary MCs, pseudo-EBs, that do not obtain excess
brightness from heating and anyhow radiate that away.
The best was where my SDO inspections reversed the conclusion of the
manuscript.}
I showed these at various meetings.

\subsectionrr{EBDETECT}
\citetads{2019A&A...626A...4V} 
perfected EB finding in AIA UV images combining ten SST--SDO data sets
for evaluating different options. 
The resulting \href{https://github.com/grviss/ebdetect}{\tt EBDETECT}
program uses AIA 1700\,\AA\ to avoid FAFs and finds most of the stronger
\Halpha\ EBs by setting a severe brightness threshold (over 5$\sigma$ above
mean) as well as lifetime and size requirements. 

With this recipe the entire SDO database is accessible to study the
occurrence of strong EBs, whereas all EB studies before were limited
to small fields sampled briefly in \Halpha\ with groundbased
telescopes.

\subsectionrr{FAFDETECT}
\citetads{2019A&A...626A...4V} 
did not develop an analogous {\tt FAFDETECT} program, but it would be
a similar effort and similarly contain a severe brightness threshold
but then require elongated shape and fast motion along
filament-shaped tracks.  

I suspect it would be worthwhile to let also such a detector loose on
the SDO database. 
For example, when a \linkadspage{2015NatCo...6E5947B}{26}{slow-wind
source map} was published by
\citetads{2015NatCo...6.5947B} 
I made corresponding preceding {\tt EBFAF} movies and noted
that the wind-producing active regions were also rich in FAFs,
\ie\ in having canopy-piercing reconnection events. 

\subsectionrr{AIA 1600/1700 with limbshift correction} \label{sec:shift}
Inspection of \rrref{figures}{fig:field1600} and \ref{fig:field1700} (or
rather the corresponding cube files using {\tt showex}) showed
immediately that running {\tt sdo\_makeeblocmovie} or {\tt
EBDETECT} makes no sense for this quiet area. 

Nevertheless, I wanted some indication of excess AIA 1600\,\AA\
brightening over 1700\,\AA\ brightening while lower than for EB or FAF
localizing.
Renewed experimentation brought me to division instead of
multiplication.

A complication arose for such weaker excess detection: the apparent
limbward shifts of MCs in AIA 1600 versus 1700\,\AA. 
It is immediately obvious per {\tt showex} when zooming in to
near-limb areas, showing displacements up to half an AIA 0.6~arcsec
pixel.  

\rrref{Figure}{fig:tileshifts-1617-0} displays them as shift vector
chart.
Such figures are an optional byproduct of my SDO cross-alignment
pipeline for many years already, but I have not published any.

\rrref{Figure}{fig:zonalshifts-1617-0}, also a regular product, shows that
these apparent shifts increase linearly limbward and that radial
components dominate them.

In \rrref{figure}{fig:tileshifts-1617-220} each subfield ``tile''
(similar to those in which the pipeline splits the disk-center JSOC
cutouts for SDO cross-alignments) has been shifted back with radial
counter-shift increasing to 220~km at the limb before
cross-correlation.
The vectors in this residual shift chart are much smaller but show
large-scale patterns. 
I then made such residue charts for the first minute of every month of
the 10 full SDO years so far and found to my surprise that this
pattern is roughly the same on all.
I don't know whether it comes from errors in my programs or from fixed
small-scale imaging difference between the two bands, but the
amplitudes are only of order 0.1~px ($\sim$ 40~km) and negligible in
my cross-alignment averaging over many tiles.

Since last year apparent limbshift corrections as these are applied in
my SDO cross-alignment pipeline; they improved it. 
In particular, they enabled using AIA 304\,\AA\ versus HMI
magnetograms as ``anchor'' pair to couple the EUV channels
directly to HMI, bypassing the UV channels I used before in an
intermediate step. 
This gives better {\tt driftscenter} results because magnetograms
sample a thinner atmospheric layer and so suffer less from such
apparent shifts and from blocking by foreshortening. 
The result for the present SDO download is shown at left in
\rrref{figure}{fig:drifts}.

\subsectionrr{MCs in AIA 1600 and 1700\,\AA}
I conclude this UV appendix by discussing the reason for the apparent
1600 versus 1700\,\AA\ brightening and limbward shifts of
MCs. 

Plane-parallel colleagues attribute both to larger opacity at
1600\,\AA\ due to increased \SiI\ photoionization, and possibly larger
\CIV\ contributions. 
I doubt the latter because the tile chart in
\rrref{figure}{fig:zonalshifts-1617-0} is regular and does not reflect
EUV brightness patterns. 
Only near and at the limb do I note sight-line integrated \CIV\
contributions. 

Higher 1600\,\AA\ formation is correct for idealized gas in the
VALIIIC plane-parallel solar analogon star of
\citetads{1981ApJS...45..635V} 
and indeed evidenced in \linkadspage{1981ApJS...45..635V}{39}{these
panels of their informative Fig.~36}. 
A plane-parallel plage model with outward increasing facular
temperature excess then also explains facular 1600 over 1700\,\AA\
brightening. 

Such plane-parallel construct was the basis for often-followed facular
SATIRE irradiance parametrization in
\citetads{1999A&A...345..635U} 
which uses the FALP model of
\citetads{1993ApJ...406..319F} 
although discarding its chromospheric temperature rise to avoid
strong-line core reversals from assuming LTE line formation (the
classical trick of \citetads{1967ZA.....65..365H} 
underlying the formerly popular but equally unrealistic model of
\citeads{1974SoPh...39...19H}). 

However, in non-plane-parallel solar reality this explanation fails
entirely.
MC brightening and enhanced MC brightening are {\em not} due to extra
heating with upward increasing excess but to non-plane structure
viewing.
MCs appear bright because they are holes in the surface sampling
higher temperatures underneath, brighter when seen deeper.
The observed 1600 versus 1700\,\AA\ brightness enhancements and
limbward offsets are due to {\em smaller\/} opacity at 1600\,\AA,
resulting in more apparent MC transparency.
It gives deeper hole viewing from above and deeper penetration beyond
holes in slanted viewing.
\rrref{Figure}{fig:uvcartoons} and its caption review how and why.
Old stuff but ignored in irradiance modeling.

For continuum bright points, G-band bright points and faculae
non-heated multi-D hole viewing has been properly reproduced with numerical
simulations by \citetads{2004ApJ...607L..59K}, 
\citetads{2004ApJ...610L.137C}, 
\citetads{2005A&A...430..691S} 
and \citetads{2009A&A...499..301V}. 
The dark limbward MC foot was emphasized by
\citetads{2005A&A...430..691S} 
and is clearly observed near the limb in AIA 1600--1700\,\AA\ zoom-in
{\tt showex} blinking. 
Together with the growth of the bright stalk sampling the granule
behind, these combined morphology changes explain the apparent
limbward shifts in \rrref{figure}{fig:tileshifts-1617-0}.

Such MURaM and Bifrost simulations might easily be extended to
1600\,\AA\ versus 1700\,\AA\ formation comparison and furnish a
sounder network/plage recipe for irradiance modeling. 
Relatively easy because non-E and ion-neutral separation
play no role so deep in the atmosphere; the hardest part is accounting
for the non-LTE ultraviolet line haze
(\citeads{2019arXiv190804624R}). 

\section{AIA 304\, \AA\ and GONG \Halpha\ comparison} 
\label{sec:ha304}

\subsectionrr{Chromosphere in \Halpha\ and \HeII\,304\,\AA}
\rrref{Figures}{fig:fieldharev} and \ref{fig:field304}\footnote{I
co-aligned these two images with
\href{https://webspace.science.uu.nl/~rutte101/rridl/imagelib/findalignimages.pro}
{\tt findalignimages.pro} which is my engine for SDO--STX
co-alignments. 
It uses iterative best-fit determination of relative scales, shifts
and rotation with Tom Metcalf's {\tt auto\_align\_images.pro} in SSW. 
I applied the results per
\href{https://webspace.science.uu.nl/~rutte101/rridl/imagelib/reformimage.pro}{\tt
reformimage.pro}.
In this case the trick was to blur both images considerably.
When blinking this pair the scenes appear to jump, but detailed {\tt
showex} inspection shows close alignment.}
show coarsely similar patterns in greyscale-reversed \Halpha\ and
\HeII\,304\,\AA.  
\rrref{Figure}{fig:scats} demonstrates this similarity statistically.

One would expect that observing some atmospheric domain or structure
in the one line would exclude observing the same in the other line,
since \HI\ should be fully ionized where \HeI\ is ionized at its twice
larger ionization threshold. 

Indeed in plane-parallel standard models \Halpha\ samples the middle
chromosphere\footnote{The so-called ``chromosphere'' with temperatures
around 7000~K in standard models as VALIIIC of
\citetads{1981ApJS...45..635V}, 
FALC of \citetads{1993ApJ...406..319F} 
and ALC7 of \citetads{2008ApJS..175..229A} 
is wrongly named. 
There it is primarily defined by fitting apparent ultraviolet
brightness temperature maxima reached in acoustic shocks (Carlsson \&
Stein
\citeyearads{Carlsson+Stein1994}, 
\citeyearads{1995ApJ...440L..29C}), 
emphasized because linearly attributing mean brightness intensity to
mean temperature doesn't hold in the Wien part of the spectrum. 
These clapotispheric shocks sit under the \Halpha\ fibril canopies
that are the on-disk Balmer-line counterpart to Lockyer's
(\citeyearads{1868RSPS...17..131L}) 
off-limb line-colored ring which is made magnetodynamically, not
hydrodynamically.}, whereas \HeII\ 304\,\AA\ samples the much hotter
``transition region'' in these models.
Observed pattern similarities would then imply that these discordant
(mutually exclusive) regimes vary jointly in mapping local
opacity/emissivity variations that similarly affect the chromosphere
and overlying transition region.

I think that this is seriously misleading oversimplification.
To me the reversed \Halpha\ and 304\,\AA\ scenes look too
similar.\footnote{This struck me first when watching the marvelous AIA
video wall at LMSAL with Marc DeRosa in 2011. 
All other EUVs showed the corona in various disguises but this
diagnostic showed the chromosphere more or less as \Halpha\ does with
its extended fibril canopies, itself unique in chromosphere rendering
in the visible (bar \Hbeta). 
That splendid view inspired my work since. 
I concentrated on \Halpha\ and some \Lyalpha, but it is time for
\HeI~10830, \HeIDthree, \HeII\,304\,\AA. 
While missing \HeI~584\,\AA\ I welcome SolO's EUI/HRI \Lyalpha\
capability.}
I pose that both lines show the ``chromosphere'' defined by
\citetads{1868RSPS...17..131L} 
as what one sees in Balmer lines and \HeIDthree. 

I suggest that assuming statistical equilibrium (SE) is wrong for
\Halpha\ and possibly wrong for \HeII\,304\,\AA\ in situations where
previously heated gas cools and recombines -- and I suggest that this
occurs always and everywhere in chromospheric canopies covering the
clapotisphere.
The latter is already shockingly dynamic as proven by
\citetads{1997ApJ...481..500C} 
and its low-altitude post-shock-cooling internetwork gas gets very far
out of SE for hydrogen
(\citeads{2007A&A...473..625L}). 

Thus, I see the chromosphere around quiet network, \ie\ the \Halpha\
fibril canopies that figure so prominently in any \Halpha\ filtergram,
as a domain that is continually pervaded and renewed by ``propagating
heating events'' (PHE), far from static or statistical equilibria; a
domain in which most dark \Halpha\ features result from opacity
boosting in dynamic cooling events with hot onsets (Rutten
\citeyearads{2016A&A...590A.124R}, 
\citeyearads{2017A&A...598A..89R}, 
\citeyearads{2017IAUS..327....1R}; 
\citeads{2017A&A...597A.138R}, 
\citeads{2019A&A...632A..96R}). 

In this view the similarity and co-correlation of the quiet \Halpha\
chromosphere and the quiet \HeII\,304\,\AA\ chromosphere result
because both represent dynamism that is patterned by underlying
magnetic network: dynamic heating for 304\,\AA, subsequent dynamic
cooling for \Halpha. 
They do not show the same chromosphere instantaneously but they do
obey the same underlying network patterning in showing gas in
different stages of intermittent ionization and recombination.

A corollary of dynamic non-equilibrium \Halpha\ opacity boosting is
that \Halpha\ fibril canopies should also be opaque at the ALMA
wavelengths and hide the clapotisphere from its solar view, in
conflict with SE-based numerical predictions.
ALMA is often advertised as LTE thermometer, hence obey SE to the
extreme, but its intensities represent temperature directly only for
optically thick features whereas the ALMA opacities defining these
thicknesses, hence feature visibility, get very far from SE in cooling
gas.
Non-equilibrium cool-after-hot retardance affects the whole top of the
hydrogen term diagram, similarly boosting the \Halpha\ and \HI\
free-free opacities by orders of magnitude. 
Since clapotisphere shocks (\CaII\ \HtwoV\ and \KtwoV\ internetwork
grains) are nowhere visible in canopy-dominated \Halpha-core images
they cannot be seen with ALMA either
(\citeads{2017A&A...598A..89R}), 

This prediction is getting confirmed
(\citeads{2019ApJ...881...99M}, 
\citeads{2020A&A...643A..41D}, 
\citeads{2020arXiv200512717C}) 
and simulated
(\citeads{2020ApJ...891L...8M}), 
implying indirect confirmation of the dynamic non-equilibrium nature
of the chromosphere.

\subsectionrr{Dark in \Halpha\ versus bright in \HeII\,304\,\AA}
The \Halpha\ greyscale reversal that helps to obtain pattern match
with \HeII\,304\,\AA\ between \rrref{figures}{fig:fieldharev} and
\ref{fig:field304} (quantified in \rrref{figure}{fig:scats}) also
supports my view. 
With static and statistic equilibria it is hard to explain that a dark
feature in the one should match a bright feature in the other as these
figures show.

\Halpha\ is a heavily scattering line in which fibrils that are very
optically thick become very dark from the 
\citetads{1965SAOSR.174..101A} 
$\sqrt{\varepsilon}$ scattering law, lowering their line source
function towards the feature surface (see
\linkadspage{2003rtsa.book.....R}{112}{Section~4.3} of
\citeads{2003rtsa.book.....R} 
or the summary in \linkadspage{2019arXiv190804624R}{7}{Section~3} of
\citeads{2019arXiv190804624R}). 
Cool-after-hot non-E opacity enhancement darkens these \Halpha\
features considerably.
Such enhancements reached even factors 10$^{12}$ for \Halpha\ in
cool-down after internetwork shocks in the simulation of
\citetads{2007A&A...473..625L}, 
shown in the last panel of
\href{https://webspace.science.uu.nl/~rutte101/rrweb/rjr-movies/hion2_fig1_movie.mov}{this
movie of their Fig.~1}.

On the contrary, most hot coronal features are optically thin or at
least effectively thin in the AIA EUV diagnostics; the greyish
\HeII\,304\,\AA\ patches in \rrref{figure}{fig:field304} likely also.
Then one doesn't talk source function but just emissivity.
In lines the first is set by the ratio of upper and lower level
populations, the second by the upper-level population alone
(\linkadspage{2003rtsa.book.....R}{44}{equations~2.71 and 2.69} of
\citeads{2003rtsa.book.....R}). 
Avrett $\sqrt{\varepsilon}$ darkening requires feature thickness
beyond thermalization lengths.  
\Halpha\ fibrils are thick enough that the population ratio is set by
$\sqrt{\varepsilon}$ darkening, more for larger opacity = larger
lower-level population.
For \HeII\ 304\,\AA\ larger upper-level population gives larger
brightness to thin or thinnish features.\footnote{Filaments are also
dark in \Halpha\ from scattering photospheric radiation (often with
underneath brightening by backscattering as observed by
\citeads{1975SoPh...45..119K}) 
and may well be non-E-enhanced opaque and dark since they appear
ridden through by frequent disturbances. 
When filaments are also observed darkly in \HeII\,304\,\AA\ this is
more likely due to frequency-redistributing bound-free scattering of
chromospheric radiation in the \HI\ Lyman and \HeI\ continua
(\linkadspage{1999ASPC..184..181R}{12}{Figure 10} of
\citeads{1999ASPC..184..181R}) 
in which every interaction darkens the
intensity, whereas resonance scattering needs many interactions for
$\sqrt{\varepsilon}$ darkening.} 

Thus, observing the same gas bright in \HeII\ 304\,\AA\ and dark in
\Halpha\ requires substantial emissivity in the first, large opacity
in the second. 
In SE these are mutually exclusive for given temperature, as in
standard models, but in cooling-after-hot gas they may coexist
sequentially with non-E retardation.

Another ALMA corollary is the prediction that ALMA images of the quiet
chromosphere show patterns coarsely similar to reversed \Halpha\ and
AIA 304\,\AA\ images. 
Direct ALMA -- 304\,\AA\ cross-correlation will be hampered by the
small ALMA field size; it is likely better to use GONG \Halpha\ as
intermediary since that can first be co-aligned with AIA 304\,\AA\
using large center cutouts as in my
\href{https://webspace.science.uu.nl/~rutte101/rridl/gonglib/gong_sdo.pro}
{\tt gong\_sdo.pro}.


\subsectionrr{Non-E fibril canopies in \Halpha\ and \HeII\,304\,\AA}
For \Halpha\
\citetads{2019A&A...632A..96R} 
showed that many dark fibrils constituting dense canopies around
network are made by spicules-II appearing as \Halpha-wing RBEs and
RREs on the disk. 
The latter are not detected in GONG wide-band images, but their
subsequent fibrilar products show up as unresolved grey patches in reversed
\rrref{figure}{fig:fieldharev}.

The dark fibrils follow a few minutes after the spicule-II heating
jets, as cooling backflow gas maintaining the large \Halpha\ opacity
gotten in the hot onset through non-E retardation during subsequent
minutes.
Their opacities then decline gradually but their opacity excess over
actual-temperature SE estimation (non-LTE population departure
coefficient $b_2$ of \Halpha's lower level) increases rapidly. 
Even if these fibrils cool enough to show strong CO lines they can
still be prominently opaque and dark in \Halpha. 

The contrail of
\citetads{2017A&A...597A.138R} 
was a prototype of dark-fibril production. 
Its subsequent retraction was evident from the later \Halpha\ Doppler
profiles in \linkadspage{2017A&A...597A.138R}{5}{their Figure 4}. 
In the statistical follow-up for many fibrils by
\citetads{2019A&A...632A..96R} 
subsequent down-drafting dark-fibril presence is evident in \eg\
\linkadspage{2019A&A...632A..96R}{6}{Fig.~4} and
\linkadspage{2019A&A...632A..96R}{10}{Fig.~8}. 

The physical reason for non-E \Halpha\ retardation is the 10-eV size of
the \Lyalpha\ jump inhibiting collisional settling
(\citeads{2002ApJ...572..626C}). 
For \HeII\,304\,\AA\ the underlying \HeI\ jump from $n \tis 1$ to
$n \tis 2$ is twice larger and may cause similar non-E retardation for
the \HeI\ top and ion populations in cooling after heating.

The fact that \HeII\,304\,\AA\ similarly shows greyish blobs around
surface network with coarse pattern correspondence to reversed
\Halpha\ may mean only that the \Halpha\ scene effectively portrays
previous \HeII\ temperatures, but it is also possible that also
\HeII\,304\,\AA\ itself stays overpopulated in gas that would already
be too cool to show it per SE estimation. 
In the AIA 304\,\AA\ column of
\linkadspage{2017A&A...597A.138R}{3}{Fig.~2} of
\citetads{2017A&A...597A.138R} 
the heating-jet brightness lingers longer than in the 193\,\AA\ images
when the \Halpha-wing RBE is over and the very dark \Halpha-core
contrail takes its place. 

So, possibly both \Halpha\ and \HeII\,304\,\AA\ live above their
station.
Of course they cannot match precisely. 
In propagating heating events such as spicules-II seen on the disk
bright \HeII\ 304 will extend further than the dark RBEs and RREs in
the outer \Halpha\ wings (not seen with GONG), beyond where hydrogen
ionizes. 
In the subsequent return phases bright \HeII\,304\,\AA\ features
likely sample hotter parts of cool-down recombination tracks also
earlier and further from network than dark \Halpha\ fibrils. 
The latter darken most close to the network launch sites.

Blinking \rrref{figures}{fig:fieldharev} and \ref{fig:field304} indeed
suggests, even at their low resolution and large noise, that the grey
304\,\AA\ patches generally extend well beyond the grey
reversed-\Halpha\ patches.
Also, per {\tt showex}\footnote{Blinker {\tt showex} also accepts (e)ps
or pdf figure files; I run it also from the command line piping to IDL
with a script shown in my
\href{https://webspace.science.uu.nl/~rutte101/rridl/00-README/sdo-manual.html}{pipeline
manual}.}
zoom-in many reversed-\Halpha\ patches show bright grains at their
centers, \ie\ darkest fibril parts at the end of their retraction.
The \Halpha\ blurring applied in \rrref{figure}{fig:scats} increased
the overall anticorrelation shown there.

{\tt Showex} timesliding of the AIA 304\,\AA\ sequence shows that the
304\,\AA\ grey-patch patterns are stable over 15~min, with rapid
small-scale fluctuations (making me use the temporal mean in
\rrref{figure}{fig:scats} and fire detector construction).
The rapid recurrence of on-disk spicules-II found by
\citetads{2013ApJ...764..164S} 
and
\citetads{2019A&A...632A..96R} 
indeed suggests continual maintenance of heating-around-network
patterns. 

\subsectionrr{Chromosphere in AIA 131\,\AA}
The next EUV diagnostic is AIA~131\,\AA\ in \rrref{figure}{fig:field131}.
Blinking against 304\,\AA\ in \rrref{figure}{fig:field304} shows
remarkable similarity for the grey patches, quantified in the
rightmost Strous diagram in \rrref{figure}{fig:scats}.
These patches should be dominated by \FeVIII\ emissivity in this
passband.
 
\linkadspage{2017A&A...597A.138R}{7}{Figure~6} of
\citetads{2017A&A...597A.138R} 
shows that the characteristic coronal-equilibrium (CE) formation
temperatures for \Halpha, \HeII\,304\,\AA\ and \FeVIII\,131\AA\ are
about 12, 50 and 400~kK, respectively, 
while for LTE Saha-Boltzmann (SB) equilibrium (at chromospheric
density) they are only about 8, 16 and 56~kK, respectively.

The actual values must lie between these simplistic limits, with CE
gaining validity over SB for increasing temperature and decreasing
density, but in any case these SB--CE estimation ranges are enormous
-- whereas all three lines visibly conform in rendering the network
chromosphere, the last two lines the closest but \Halpha\ coarsely
also in its network-imposed patterning.

This suggests overlap. 
My non-E suggestion is that PHEs (as spicules-II) heat and ionize the
chromosphere around network to \FeVIII\,131\,\AA\ emissivity, followed
by along-the-track retraction cooling and recombination showing emission
in \HeII\,304\,\AA\ that may have non-E retardance, and then retarded
$\sqrt{\varepsilon}$ darkening in \Halpha.

The quantification of 304--131\,\AA\ similarity in
\rrref{figure}{fig:scats} inspired the 304$\times$131\,\AA\ image
multiplication of my SDO fire detector, detailed in the caption of
\rrref{figure}{fig:sdoroiloc304x131}. 
 
\subsectionrr{Chromosphere around quiet network}
In summary: I suggest that the rough similarity of the greyish scenes
in \Halpha\ (\rrref{figure}{fig:fieldharev}) and in \HeII\,304\,\AA\
(\rrref{figure}{fig:field304}) is because both lines sample essentially
the same highly dynamic chromosphere around network, also seen in
\FeVIII\,131\,\AA\ (\rrref{figure}{fig:field131}).

I conclude that the dynamic chromosphere around quiet network cannot
be characterized by a single temperature, certainly not the SE
equilibrium temperature for \Halpha\ in a standard model. 
Everywhere around network PHE's continually heat gas which then mostly
flows back while cooling.
The most likely agents are spicules-II observed as RBEs and RREs in
\Halpha. 
These dynamic heating agents and their dark \Halpha-core products
constitute the grey patches making up most of the scenes in
\rrref{figures}{fig:fieldharev}--\ref{fig:field131} and seen best in
\rrref{figure}{fig:sdoroiloc304x131} and in the fire detector triple
members of
\rrref{figures}{fig:sdo-minx-193}--\ref{fig:sdo-sp-min-mag}.

The earlier Bifrost simulations of \Halpha\ fibrils
(\citeads{2012ApJ...749..136L}, 
\citeads{2015ApJ...802..136L}) 
and accompanying ALMA scenes
(\citeads{2015A&A...575A..15L}0 
lacked spicules-II and their fibrilar aftermaths around network, but
recent addition of non-E retardation as well as ion-neutral ambipolar
separation yields better numerical analogs
(\citeads{2020ApJ...889...95M}, 
\citeads{2020arXiv200512717C}).

I expect that these will confirm that generally \Halpha\ lives
furthest above its station, in the form of dark cooling but non-E
recombining fibrils producing opaque canopies around network that
chart remembrance of much hotter things past.
\citetads{1868RSPS...17..131L} 
named the chromosphere after the dark horse (helium his white one?).

\subsectionrr{Dark EUV features}
The darkest patches between grey chromospheric network in the fire
detector images match dark features in most of the AIA EUV images
(blink \rrref{figures}{fig:field304}--\ref{fig:field193}). 
The AIA 335, 304, 221, 131, and 94\,\AA\ images show them similarly.
They are sharper and best defined in AIA~171\,\AA\ but they
correspond least in 193\,\AA\ (a brushfire example is specified in the
caption of \rrref{figure}{fig:sdoroiloc304x131}). 
They are not present (dark or bright) in the GONG \Halpha\ image.

EUV darkness is either due to lack of EUV emissivity (as in coronal
holes) or due to extinction (not ``absorption'') from bound-free
out-of-the-passband scattering in overlying neutral gas
(\linkadspage{1999ASPC..184..181R}{12}{Figure 10} of
\citeads{1999ASPC..184..181R}). 
In such scattering the re-emitted photon is not only redistributed in
direction but also in frequency, most likely to near the bound-free
edge threshold far from the AIA passband, so that virtually every
scattered photon diminishes the detected intensity for radiation
originating deeper along the line of sight.
Hence, dark features of this type require overlying gas containing
sufficient neutral hydrogen and/or neutral helium (their visibility in
335\,\AA\ excludes ionized helium).
The discordant non-visibility or flip-into-bright of these features in
the 193\,\AA\ images then implies that the hot gas contributing noticeable
193\,\AA\ emissivity overlies cooler gas seen in the others.

\subsectionrr{Chromosphere under brushfires}
\Halpha\ in \rrref{figure}{fig:fieldharev} differs markedly from the
subsequent EUV images in that AIA 304\,\AA\ already shows bright
brushfires as the hotter EUV diagnostics do: when blinking
\rrref{figure}{fig:field304} with
\rrref{figures}{fig:field131}--\ref{fig:field193} these are remarkably
co-spatial at least in bright footpoints -- but GONG \Halpha\ does not
show them, also not in non-reversed \rrref{figure}{fig:fieldha}.

If this footpoint heating was bottom-up one might expect conspicuous
features in \Halpha, but the scatter diagrams in
\rrref{figure}{fig:scats} display only slight-above-average GONG
\Halpha\ brightening where AIA 304\,\AA\ is brightest.
Top-down particle beams may have less penetration into or heating of
high-density \Halpha\ gas, yet denser where it gets neutral with low
electron pressure.

In addition, most of the campfire ROI cutouts in
\rrref{figures}{fig:roi1}--\ref{fig:roi12} show extended diffuse
304\,\AA\ brightness around the small campfire that roughly duplicates
diffuse brightness patterns in the hotter AIA cutouts, in particular
the 193\,\AA\ ones.  
I attribute this difference with \Halpha\ to how irradiation
contributes ionization.

Hydrogen is mainly ionized by photoionization in the Balmer continuum
from the deep photosphere, a very bland irradiator from below because
the granulation pattern gets erased in 3D scattering through the upper
photosphere
(\citeads{2012A&A...540A..86R}, 
\citeads{2012ApJ...749..136L}). 

In very hot instances collisional ionization takes over, but elsewhere
the degree of hydrogen ionization is set by the $n \tis 2$ population
controlled by \Lyalpha, with Balmer continuum NLTE as instantaneous
SE-obeying modifier. 
It is the retarded settling of \Lyalpha\ that governs retarded non-E
hydrogen recombination usually called ``non-E ionization''. 
Where \Halpha\ lives far above its station from slow \Lyalpha\
settling, so does \HI\ ionization and with it the \HI\ free-free
continuum dominating the ALMA mm passbands
(\linkadspage{2017A&A...598A..89R}{4}{Figure 1} of
\citeads{2017A&A...598A..89R}). 

In contrast, \HeI\ ionization also senses irradiation from above,
making coronal holes visible in \HeI~10830\,\AA\ and \HeII\,304\,\AA.
This may occur via downward EUV irradiation ionizing \HeI\ (edge at
504\,\AA) and exciting \HeII\,304\,\AA, or from ionizing \HeII\ (edge
at 228\,\AA) with cascade-recombination through \HeII\,304\,\AA.

When such EUV irradiance patches ionize \HeI\ in cool or cooling gas
with slow \HeI\ $n\tis1 - 2$ settling to the Boltzmann ratio, the
resulting \HeII\,304\,\AA\ emissivities may also maintain
non-E-retarded memorial boosting.

  

\subsectionrr{\HeII\,304\,\AA\ in a well-sampled brushfire}
\linkadspage{2017A&A...597A.138R}{2}{Figure 1} of
\citetads{2017A&A...597A.138R} 
targeted active bipolar network with a patch of overlying small
EUV-bright loops constituting a brushfire best seen in the
last 193\,\AA\ image.
This is one of the combined SST--IRIS datasets made available by
\citetads{2020A&A...641A.146R}. 

This brushfire appeared as a double set of 15-arcsec EUV-bright arches
connecting multiple patches of active bipolar network underneath. 
Most brushfires in the triple-member 193\,\AA\ images here show single
sets of such arches.

The 304\,\AA\ and 193\,\AA\ images in that figure and the
corresponding cutout sequences in the subsequent
\linkadspage{2017A&A...597A.138R}{3}{Figure 2} show that the 304\,\AA\
bright patch conforms roughly to the 193\,\AA\ bright patch but with
sharper bright and dark features, many with filamentary shape.

The IRIS 1400\,\AA\ slitjaw image in the first figure and cutouts in
the third column of the second figure show roughly similar bright and
dark patterns as 304\,\AA, less bright but sharper.

With 1400\,\AA\ ``bright'' I mean the grey patches, excluding the
brighter roundish grains which are surface MCs that are
hole-brightened in the 1400\,\AA\ continuum as in
\rrref{figure}{fig:uvcartoons} and also seen as bright points in the
\Halpha\ blue-wing cutouts in the first column, unless covered by the
dark RBE-like jet that is the topic of that study. 
Where these shine through the grey 1400\,\AA\ patches called bright
here the latter must be optically thin.

For the times of rows 4 and 6 in the
\linkadspage{2017A&A...597A.138R}{3}{second figure} the
\linkadspage{2017A&A...597A.138R}{4}{next figure} adds panels sampling
\CaIR\ and 171\,\AA.  
The \CaIR\ blue-wing panels shows the MCs also, hole-brightened by
less collisional broadening and superposed on reversed granulation
without shielding.

The 1400\,\AA\ similarity to the 304\,\AA\ scene comes from the
contribution of the \SiIV\ lines in the IRIS passband. 
The \HeII\ and \SiIV\ ion occurrence temperatures overlap pairwise in
the CE and SB limits in \linkadspage{2017A&A...597A.138R}{7}{Figure~6}
of \citetads{2017A&A...597A.138R}. 
The adjacent \linkadspage{2017A&A...597A.138R}{7}{Figure~7} holds for
the SB limit which is better for hot dense gas. 
The \HeII\ and \SiIV\ humps there occupy the same temperatures but
with $10^3$ extinction ratio from the abundance ratio. 
Indeed, the 1400\,\AA\ and 304\,\AA\ features in the quoted
\linkadspage{2017A&A...597A.138R}{2}{Figure 1} and
\linkadspage{2017A&A...597A.138R}{3}{Figure 2} agree when regarding
the former as much optically thinner than the latter.

The bright fine structuring, seen best in the better-mapping
1400\,\AA\ images, does not duplicate the more diffuse 193\,\AA\ scene
so that EUV irradiation is not the only agent.
Part of the fine-scale brightening is heating from below as evident in
the extending RBE-like jet in the upper rows which is dark in the first
column, bright in 1400\,\AA, and produced the subsequent
dark \Halpha-core contrail in the second column. 
There are a few more such upward-heating features that are dark in the
first column and bright in 1400\,\AA, but most smaller grey 1400\,\AA\
patches in the lower rows may well be due to downward particle
heating.
They appear smaller than in larger-opacity 304\,\AA\ and indeed lie
around the yet brighter grains that mark footpoint MCs into which the
accelerated particles should funnel down.


\ifnum\arxivmode=0  
\section{Epilogue on publishing} \label{sec:epilogue}

During this effort I wondered how to publish it.
Formally, in a regular refereed ``journal''?
The most fitting seemed {\em Solar Physics\/} but last year Springer
put me into a depression by mishandling
\citetads{2019arXiv190804624R}. 
I wrote that away by posting my grudges in a dozen bitter
\href{https://webspace.science.uu.nl/~rutte101/Recipes_publications.html}
{Springer nasturtia}.
 
Also, ``publish or perish'' is no longer an issue for me -- {\em alea iacta 
est\/}.\footnote{Six years Latin in high school did not forward my career, 
nor did five years classical Greek in which Caesar probably said this.}

I then wondered about starting my own institute series, in line with
starting my own institute when Utrecht University 
\href{https://webspace.science.uu.nl/~rutte101/Closure_Utrecht.html}
{killed 370-year-old Utrecht astronomy} -- including Dutch solar physics.
I arrived too late in this 370-year time\-span to publish in French in
the {\em Recherches Astronomiques de l'Observatoire d'Utrecht} and
also for the subsequent switch to German as principal science
language\footnote{For a brilliant example see Kees de Jager's
\linkadspage{1962ZA.....55...66T}{1}{scathing review} of ``Physics of
the solar chromosphere'' of ADS-misattributed
\citetads{1961psc..book.....A} 
in ADS-mislabeled \citetads{1962ZA.....55...66T},
\linkadspage{1962ZA.....55...70W}{1}{ending here} in ADS-mislabeled
\citetads{1962ZA.....55...70W}.},
but in time to
\href{https://ui.adsabs.harvard.edu/abs/1967BAN....19..254F/abstract}
{publish} in the {\em Bulletin of the Astronomical Institutes of the
Netherlands} when institute publications where still mainstream
platforms\footnote{Until new BAN
editor Stuart Pottasch joined national journals into A\&A.}. 

I am also much impressed by the 250 citations of Mats Carlsson's
\citeyearads{1986UppOR..33.....C} exemplary Uppsala report describing
{\tt MULTI}, more than for our most boring (since nonsolar) {\tt
MULTI}-based \citetads{1994A&A...288..860C} which remains my
most-cited publication.

Another inspiring example are Edsger Dijkstra's
\href{http://www.cs.utexas.edu/users/EWD/indexBibTeX.html}{personal
writings} found when I wondered why IDL
\href{https://webspace.science.uu.nl/~rutte101/stuff/dijkstra_EWD1019.pdf}
{counts fingers from 0 to 9}.\footnote{Needing {\tt nx/(nx-1)} and
{\tt nx+1} corrections in SDO image axis labeling as in {\tt
xaxis=(indgen(nx)*float(nx)/(nx-1)-(nx+1)/2.)*CDELT1+XCEN} for solar
X. 
Do you?}

I then decided on this institute-report format for the present
analysis, and hopefully more.

In hindsight my recent single-author publications might also have been
Lingezicht reports without further ado.
However, I must concede that the outstanding referee and the attentive
scientific editor of Springer-mishandled
\citetads{2019arXiv190804624R} 
significantly improved that conference\footnote{IAU GA in 2018 at
Vienna -- my last pre-covid non-line meeting. 
Since then I became the only solar physicist always working
in the institute office.} 
write-up.

Likewise, I must add that I much appreciate A\&A--EDP's efficient
publication handling, including helpful language editing and not
molesting my hyperlink macros as other astronomy publishers including
ApJ do.
I also like the {\tt aa.cls} latex style and used it here by starting
with my
\href{https://webspace.science.uu.nl/~rutte101/Report_recipe.html}{report
example} for students as template.
Naturally I also appreciate A\&A's fee waiver for Lingezicht
Astrophysics since Utrecht's demise.
However, I judged the material presented here too provincial, too
detailed and too inconclusive for submission to A\&A -- too much honor
for a single non-validated image even though press-released.

I intend to post these reports on my website\footnote{I did
with this one
\href{https://webspace.science.uu.nl/~rutte101/rrweb/rjr-pubs/2020LingAstRep...1R.pdf}{here} and update it since  as specified in
the \hyperref[sec:history]{version history}.}.
They are in good company there since my most-read publication
(\href{https://webspace.science.uu.nl/~rutte101/rrweb/rjr-pubs/2006-Manicouagan-report.pdf}{Rutten
2006})
\biblink{2006-Manicouagan-report}
{\href{https://webspace.science.uu.nl/~rutte101/rrweb/rjr-pubs/2006-Manicouagan-report.pdf}{URL}}
and my most-used publication
(\citeads{2003rtsa.book.....R}) 
were both only self-posted, successfully relying on Google's indexing
(plus ADS for the second) for spreading. 
Such posts may be kept ``living'' with updates.
I did for
\cite{2006-Manicouagan-report} but servicing
\citetads{2003rtsa.book.....R} remains overdue. 
 
I may also post these reports at arXiv for ADS access, citing, and
persistence beyond my website and me.\footnote{In our bitter dispute
last year Springer's manager of {\em Solar Physics\/} wrote that
arXiv-only ``publication'' is viable but undesirable. 
Boycotting hybrid double-dipping predator publishing now seemed so
desirable to me that I forsook my principle of archiving refereed
only. 
Colleagues who saw the present report as manuscript advised arxiving;
I did (\href{https://arxiv.org/abs/2009.00376}{arXiv 2009.00376})
but without this epilogue.}  
Such posts may also be kept alive.\footnote{I soon revised the arXiv
upload.
Differences between arXiv versions can be inspected by downloading
their source materials and running {\tt latexdiff}.}

Obvious advantages of self-publication are that I may embellish,
pontificate, expostulate, footnote-comment, caption-didact, hyperlink,
joke and be personal (first names) just as I like without fearing
editors, that I may skip reviewing literature without fearing
citation-desiring referees, that all my fancy hyperlinks keep working
unmolested, that I don't have to minimize figure sizes and numbers in
fear of page limits, and that I can use formats I
like\footnote{Including Eugene Avrett's one-point-per-paragraph style
in dog-eared VAL (\citeyearads{1973ApJ...184..605V}, 
\citeyearads{1976ApJS...30....1V}, 
\citeyearads{1981ApJS...45..635V}) 
preprints, the only paper ones still on my desk.}
as well as keep figures separate at the end (for parallel viewing) and
full-page (for blinking).

An obvious disadvantage is that my Hirsch index growth may suffer, but
at the median of senior colleagues in
\linkadspage{2018arXiv180408709R}{8}{Fig.~1} of
\citetads{2018arXiv180408709R} 
I cherish mediocrity.

After I decided on its format this report grew considerably in
content, hyperlinks, references, links to published figures,
footnotes, appendices (new to me), and very many figure
pages\footnote{Hoping that easy blinking makes them actually get
inspected. 
In a formal publication the 75 figures here, most full-page and all
adorning the single press-release image in the first, would probably
editorially be made ``online material'' (oldfashionedly since printing
cost and volume thickness are no longer issues for publications that
are themselves online material). 
I rarely inspect such if it requires more than one-click access --
which I implemented for figure panel blinkers in
\citetads{2019A&A...632A..96R}, 
similar to my direct
\href{https://webspace.science.uu.nl/~rutte101/Turning_citations.html}{cited-page
openers} also used here (for the non-Mac-challenged). 
I also suspect that in a formal publication these \thefootnote\
footnotes would be editorially scrapped.}, notwithstanding considerable
shortening by shunting detailed descriptions of my
\href{https://webspace.science.uu.nl/~rutte101/rridl/00-README/sdo-manual.html}{SDO--STX
alignment pipeline} and my
\href{https://webspace.science.uu.nl/~rutte101/rridl/imagelib/showex.pro}
{\tt showex.pro} browser to future Lingezicht reports.

\fi 


\end{appendix}

\begin{footnotesize}
\subsubsection*{Acknowledgments}
\addcontentsline{toc}{section}{Acknowledgments}
\parindent=0ex
\parskip=0.5ex
\noindent
I am indebted to Alan Ferguson for alerting me to the SolO press
conference and to Ron Moore for correcting bushfire to brushfire.

SDO is a NASA mission of which all data are Alan-Title-style exemplary
public, without any reservation whatsoever, and made easily accessible
and ready-for-science through extensive and commendable efforts by the
Stanford (HMI, JSOC) and LMSAL (AIA) teams. 
I regularly and reliably obtain and appreciate additional help from
Phil Scherrer, Arthur Amezcua, Mark Cheung, Greg Slater, John Serafin,
Sam Freeland.

Solar Orbiter is a space mission of international collaboration
between ESA and NASA. 

As always I made much use of the splendid SolarSoft and ADS libraries.

\subsubsection*{Version history}
\label{version} 
\addcontentsline{toc}{section}{Version history}
\parindent=0ex
\parskip=0.5ex
\label{sec:history} 
{\em July 16, 2020\/} -- Solar Orbiter press conference.
Started scene search in SDO.

{\em August 3, 2020\/} -- Eureka: found SolO campfires in SDO/AIA
EUVs. 

{\em August 31, 2020\/} -- initial version posted on my website and on
arXiv; \href{https://arxiv.org/pdf/2009.00376v1.pdf}{arXiv version}
without epilogue.
Since then the
\href{https://webspace.science.uu.nl/~rutte101/rrweb/rjr-pubs/2020LingAstRep...1R.pdf}{website version} is ``live'' latest.

{\em September 14, 2020\/} --
\href{https://arxiv.org/pdf/2009.00376.pdf}{arXiv version 2}: SolO caveat
under \rrref{figure}{fig:soloim}, twenty new ``other scene'' blink pairs
in \rrref{figures}{fig:sdo-minx-193}--\ref{fig:sdo-sp-min-mag},
``bushfire'' corrected to ``brushfire'' (\rrref{footnote}{foot:moore}),
better fire detector by basing thresholds on quiet network
(\rrref{figure}{fig:sdoroiloc304x131}), AIA
304\,--\,131\,\AA\ driftplot added
(\rrref{figure}{fig:drifts}), reference to
\citetads{1998ApJ...501..386F}, 
much text re-written and added.

{\em September 15, 2020\/} -- reference to
  \citetads{2019SoPh..294..161A}. 

{\em September 20, 2020\/} -- improvements, this history,
$\backslash$ref{} links blue for dichromats.

{\em September 25, 2020\/} -- more fluxtube review below
\rrref{figure}{fig:uvcartoons}.

{\em October 19, 2020\/} -- brushfire detail at the end of
\rrref{appendix}{sec:ha304}.

{\em November 25, 2020\/} -- \Halpha\ sequence diagram in
\rrref{figure}{fig:scats}.
 
{\em December 4, 2020\/} -- more references, particle beams after
\citetads{2020A&A...643A..27F}. 

{\em December 22, 2020\/} -- St.~Elmo's fires.

{\em \today\/} -- Third arXiv upload.

\end{footnotesize}

\addcontentsline{toc}{section}{References}
\bibliographystyle{aa-note} 
\bibliography{rjrfiles,adsfiles} 

\begin{thebibliography}{117}
\expandafter\ifx\csname natexlab\endcsname\relax\def\natexlab#1{#1}\fi

\bibitem[{{Alissandrakis}(2019)}]{2019SoPh..294..161A}
{Alissandrakis}, C.~E. 2019, \solphys, 294, 161 \csname
  2019SoPh..294..161Alink\endcsname~\csname 2019SoPh..294..161Anote\endcsname

\bibitem[{{Antolin} {et~al.}(2012){Antolin}, {Vissers}, \& {Rouppe van der
  Voort}}]{2012SoPh..280..457A}
{Antolin}, P., {Vissers}, G., \& {Rouppe van der Voort}, L. 2012, \solphys,
  280, 457 \csname 2012SoPh..280..457Alink\endcsname~\csname
  2012SoPh..280..457Anote\endcsname

\bibitem[{{Attie} {et~al.}(2009){Attie}, {Innes}, \&
  {Potts}}]{2009A&A...493L..13A}
{Attie}, R., {Innes}, D.~E., \& {Potts}, H.~E. 2009, \aap, 493, L13 \csname
  2009A&A...493L..13Alink\endcsname~\csname 2009A&A...493L..13Anote\endcsname

\bibitem[{{Avrett}(1965)}]{1965SAOSR.174..101A}
{Avrett}, E.~H. 1965, SAO Special Report, 174, 101 \csname
  1965SAOSR.174..101Alink\endcsname~\csname 1965SAOSR.174..101Anote\endcsname

\bibitem[{{Avrett} \& {Loeser}(2008)}]{2008ApJS..175..229A}
{Avrett}, E.~H. \& {Loeser}, R. 2008, \apjs, 175, 229 \csname
  2008ApJS..175..229Alink\endcsname~\csname 2008ApJS..175..229Anote\endcsname

\bibitem[{{Bakke} {et~al.}(2018){Bakke}, {Frogner}, \&
  {Gudiksen}}]{2018A&A...620L...5B}
{Bakke}, H., {Frogner}, L., \& {Gudiksen}, B.~V. 2018, \aap, 620, L5 \csname
  2018A&A...620L...5Blink\endcsname~\csname 2018A&A...620L...5Bnote\endcsname

\bibitem[{{Bello Gonz{\'a}lez} {et~al.}(2013){Bello Gonz{\'a}lez}, {Danilovic},
  \& {Kneer}}]{2013A&A...557A.102B}
{Bello Gonz{\'a}lez}, N., {Danilovic}, S., \& {Kneer}, F. 2013, \aap, 557, A102
  \csname 2013A&A...557A.102Blink\endcsname~\csname
  2013A&A...557A.102Bnote\endcsname

\bibitem[{{Berger} {et~al.}(2004){Berger}, {Rouppe van der Voort},
  {L{\"o}fdahl}, {Carlsson}, {Fossum}, {Hansteen}, {Marthinussen}, {Title}, \&
  {Scharmer}}]{2004A&A...428..613B}
{Berger}, T.~E., {Rouppe van der Voort}, L.~H.~M., {L{\"o}fdahl}, M.~G.,
  {et~al.} 2004, \aap, 428, 613 \csname
  2004A&A...428..613Blink\endcsname~\csname 2004A&A...428..613Bnote\endcsname

\bibitem[{{Bonet} {et~al.}(2010){Bonet}, {M{\'a}rquez}, {S{\'a}nchez Almeida},
  {Palacios}, {Mart{\'\i}nez Pillet}, {Solanki}, {del Toro Iniesta}, {Domingo},
  {Berkefeld}, {Schmidt}, {Gandorfer}, {Barthol}, \&
  {Kn{\"o}lker}}]{2010ApJ...723L.139B}
{Bonet}, J.~A., {M{\'a}rquez}, I., {S{\'a}nchez Almeida}, J., {et~al.} 2010,
  \apjl, 723, L139 \csname 2010ApJ...723L.139Blink\endcsname~\csname
  2010ApJ...723L.139Bnote\endcsname

\bibitem[{{Bose} {et~al.}(2019){Bose}, {Henriques}, {Joshi}, \& {Rouppe van der
  Voort}}]{2019A&A...631L...5B}
{Bose}, S., {Henriques}, V. M.~J., {Joshi}, J., \& {Rouppe van der Voort}, L.
  2019, \aap, 631, L5 \csname 2019A&A...631L...5Blink\endcsname~\csname
  2019A&A...631L...5Bnote\endcsname

\bibitem[{{Brandt} {et~al.}(1994){Brandt}, {Rutten}, {Shine}, \& {Trujillo
  Bueno}}]{1994ASIC..433..251B}
{Brandt}, P.~N., {Rutten}, R.~J., {Shine}, R.~A., \& {Trujillo Bueno}, J. 1994,
  in NATO Advanced Science Institutes (ASI) Series C, Vol. 433, Solar and
  stellar magnetism, ed. R.~J. {Rutten} \& C.~J. {Schrijver}, 251 \csname
  1994ASIC..433..251Blink\endcsname~\csname 1994ASIC..433..251Bnote\endcsname

\bibitem[{{Brandt} {et~al.}(1988){Brandt}, {Scharmer}, {Ferguson}, {Shine},
  {Tarbell}, \& {Title}}]{1988Natur.335..238B}
{Brandt}, P.~N., {Scharmer}, G.~B., {Ferguson}, S., {et~al.} 1988, \nat, 335,
  238 \csname 1988Natur.335..238Blink\endcsname~\csname
  1988Natur.335..238Bnote\endcsname

\bibitem[{{Brooks} {et~al.}(2015){Brooks}, {Ugarte-Urra}, \&
  {Warren}}]{2015NatCo...6.5947B}
{Brooks}, D.~H., {Ugarte-Urra}, I., \& {Warren}, H.~P. 2015, Nature
  Communications, 6, 5947 \csname 2015NatCo...6.5947Blink\endcsname~\csname
  2015NatCo...6.5947Bnote\endcsname

\bibitem[{{B{\"{u}}nte} {et~al.}(1993){B{\"{u}}nte}, {Solanki}, \&
  {Steiner}}]{1993A&A...268..736B}
{B{\"{u}}nte}, M., {Solanki}, S.~K., \& {Steiner}, O. 1993, \aap, 268, 736
  \csname 1993A&A...268..736Blink\endcsname~\csname
  1993A&A...268..736Bnote\endcsname

\bibitem[{Carlsson \& Stein(1994)}]{Carlsson+Stein1994}
Carlsson, M. \& Stein, R.~F. 1994, in Chromospheric Dynamics, ed. M.~Carlsson,
  Proc.\ Miniworkshop (Oslo: Inst.\ Theor.\ Astrophys.), 47--77 \csname
  Carlsson+Stein1994link\endcsname~\csname Carlsson+Stein1994note\endcsname

\bibitem[{{Carlsson} \& {Stein}(1995)}]{1995ApJ...440L..29C}
{Carlsson}, M. \& {Stein}, R.~F. 1995, \apjl, 440, L29 \csname
  1995ApJ...440L..29Clink\endcsname~\csname 1995ApJ...440L..29Cnote\endcsname

\bibitem[{{Carlsson} \& {Stein}(1997)}]{1997ApJ...481..500C}
{Carlsson}, M. \& {Stein}, R.~F. 1997, \apj, 481, 500 \csname
  1997ApJ...481..500Clink\endcsname~\csname 1997ApJ...481..500Cnote\endcsname

\bibitem[{{Carlsson} \& {Stein}(2002)}]{2002ApJ...572..626C}
{Carlsson}, M. \& {Stein}, R.~F. 2002, \apj, 572, 626 \csname
  2002ApJ...572..626Clink\endcsname~\csname 2002ApJ...572..626Cnote\endcsname

\bibitem[{{Carlsson} {et~al.}(2004){Carlsson}, {Stein}, {Nordlund}, \&
  {Scharmer}}]{2004ApJ...610L.137C}
{Carlsson}, M., {Stein}, R.~F., {Nordlund}, {\r{A}}., \& {Scharmer}, G.~B.
  2004, \apjl, 610, L137 \csname 2004ApJ...610L.137Clink\endcsname~\csname
  2004ApJ...610L.137Cnote\endcsname

\bibitem[{{Chintzoglou} {et~al.}(2020){Chintzoglou}, {De Pontieu},
  {Mart{\'\i}nez-Sykora}, {Hansteen}, {de la Cruz Rodr{\'\i}guez},
  {Szydlarski}, {Jafarzadeh}, {Wedemeyer}, {Bastian}, \& {Sa{\'\i}nz
  Dalda}}]{2020arXiv200512717C}
{Chintzoglou}, G., {De Pontieu}, B., {Mart{\'\i}nez-Sykora}, J., {et~al.} 2020,
  arXiv e-prints, arXiv:2005.12717 \csname
  2020arXiv200512717Clink\endcsname~\csname 2020arXiv200512717Cnote\endcsname

\bibitem[{{da Silva Santos} {et~al.}(2020){da Silva Santos}, {de la Cruz
  Rodr{\'\i}guez}, {White}, {Leenaarts}, {Vissers}, \&
  {Hansteen}}]{2020A&A...643A..41D}
{da Silva Santos}, J.~M., {de la Cruz Rodr{\'\i}guez}, J., {White}, S.~M.,
  {et~al.} 2020, \aap, 643, A41 \csname
  2020A&A...643A..41Dlink\endcsname~\csname 2020A&A...643A..41Dnote\endcsname

\bibitem[{{Danilovic}(2017)}]{2017A&A...601A.122D}
{Danilovic}, S. 2017, \aap, 601, A122 \csname
  2017A&A...601A.122Dlink\endcsname~\csname 2017A&A...601A.122Dnote\endcsname

\bibitem[{{De Pontieu} {et~al.}(2012){De Pontieu}, Carlsson, {Rouppe van der
  Voort}, Rutten, Hansteen, \& Watanabe}]{2012ApJ...752L..12D}
{De Pontieu}, B., Carlsson, M., {Rouppe van der Voort}, L. H.~M., {et~al.}
  2012, \apjl, 752, L12 1 \csname 2012ApJ...752L..12Dlink\endcsname~\csname
  2012ApJ...752L..12Dnote\endcsname

\bibitem[{{Ellerman}(1917)}]{1917ApJ....46..298E}
{Ellerman}, F. 1917, \apj, 46, 298 \csname
  1917ApJ....46..298Elink\endcsname~\csname 1917ApJ....46..298Enote\endcsname

\bibitem[{{Falconer} {et~al.}(1998){Falconer}, {Moore}, {Porter}, \&
  {Hathaway}}]{1998ApJ...501..386F}
{Falconer}, D.~A., {Moore}, R.~L., {Porter}, J.~G., \& {Hathaway}, D.~H. 1998,
  \apj, 501, 386 \csname 1998ApJ...501..386Flink\endcsname~\csname
  1998ApJ...501..386Fnote\endcsname

\bibitem[{{Fontenla} {et~al.}(1993){Fontenla}, {Avrett}, \&
  {Loeser}}]{1993ApJ...406..319F}
{Fontenla}, J.~M., {Avrett}, E.~H., \& {Loeser}, R. 1993, \apj, 406, 319
  \csname 1993ApJ...406..319Flink\endcsname~\csname
  1993ApJ...406..319Fnote\endcsname

\bibitem[{{Frogner} {et~al.}(2020){Frogner}, {Gudiksen}, \&
  {Bakke}}]{2020A&A...643A..27F}
{Frogner}, L., {Gudiksen}, B.~V., \& {Bakke}, H. 2020, \aap, 643, A27 \csname
  2020A&A...643A..27Flink\endcsname~\csname 2020A&A...643A..27Fnote\endcsname

\bibitem[{{Hansteen} {et~al.}(2017){Hansteen}, {Archontis}, {Pereira},
  {Carlsson}, {Rouppe van der Voort}, \& {Leenaarts}}]{2017ApJ...839...22H}
{Hansteen}, V.~H., {Archontis}, V., {Pereira}, T.~M.~D., {et~al.} 2017, \apj,
  839, 22 \csname 2017ApJ...839...22Hlink\endcsname~\csname
  2017ApJ...839...22Hnote\endcsname

\bibitem[{{Henriques} {et~al.}(2016){Henriques}, {Kuridze}, {Mathioudakis}, \&
  {Keenan}}]{2016ApJ...820..124H}
{Henriques}, V.~M.~J., {Kuridze}, D., {Mathioudakis}, M., \& {Keenan}, F.~P.
  2016, \apj, 820, 124 \csname 2016ApJ...820..124Hlink\endcsname~\csname
  2016ApJ...820..124Hnote\endcsname

\bibitem[{{Hill} {et~al.}(2019){Hill}, {Hammel}, {Mart{\'{\i}}nez-Pillet}, {de
  Wijn}, {Gosain}, {Burkepile}, {Henney}, {McAteer}, {Bain}, {Manchester},
  {Lin}, {Roth}, {Ichimoto}, \& {Suematsu}}]{2019BAAS...51g..74H}
{Hill}, F., {Hammel}, H., {Mart{\'{\i}}nez-Pillet}, V., {et~al.} 2019, in
  Bulletin of the American Astronomical Society, Vol.~51, 74 \csname
  2019BAAS...51g..74Hlink\endcsname~\csname 2019BAAS...51g..74Hnote\endcsname

\bibitem[{{Holweger}(1967)}]{1967ZA.....65..365H}
{Holweger}, H. 1967, \zap, 65, 365 \csname
  1967ZA.....65..365Hlink\endcsname~\csname 1967ZA.....65..365Hnote\endcsname

\bibitem[{{Holweger} \& {M{\"{u}}ller}(1974)}]{1974SoPh...39...19H}
{Holweger}, H. \& {M{\"{u}}ller}, E.~A. 1974, \solphys, 39, 19 \csname
  1974SoPh...39...19Hlink\endcsname~\csname 1974SoPh...39...19Hnote\endcsname

\bibitem[{{Hong} {et~al.}(2014){Hong}, {Ding}, {Li}, {Fang}, \&
  {Cao}}]{2014ApJ...792...13H}
{Hong}, J., {Ding}, M.~D., {Li}, Y., {Fang}, C., \& {Cao}, W. 2014, \apj, 792,
  13 \csname 2014ApJ...792...13Hlink\endcsname~\csname
  2014ApJ...792...13Hnote\endcsname

\bibitem[{{Joshi} {et~al.}(2020){Joshi}, {Rouppe van der Voort}, \& {de la Cruz
  Rodr{\'\i}guez}}]{2020A&A...641L...5J}
{Joshi}, J., {Rouppe van der Voort}, L. H.~M., \& {de la Cruz Rodr{\'\i}guez},
  J. 2020, \aap, 641, L5 \csname 2020A&A...641L...5Jlink\endcsname~\csname
  2020A&A...641L...5Jnote\endcsname

\bibitem[{{Keller} {et~al.}(2004){Keller}, {Sch{\"u}ssler}, {V{\"o}gler}, \&
  {Zakharov}}]{2004ApJ...607L..59K}
{Keller}, C.~U., {Sch{\"u}ssler}, M., {V{\"o}gler}, A., \& {Zakharov}, V. 2004,
  \apjl, 607, L59 \csname 2004ApJ...607L..59Klink\endcsname~\csname
  2004ApJ...607L..59Knote\endcsname

\bibitem[{{Kim} {et~al.}(2015){Kim}, {Yurchyshyn}, {Bong}, {Cho}, {Cho}, {Lee},
  {Lim}, {Park}, {Yang}, {Ahn}, {Goode}, \& {Jang}}]{2015ApJ...810...38K}
{Kim}, Y.-H., {Yurchyshyn}, V., {Bong}, S.-C., {et~al.} 2015, \apj, 810, 38
  \csname 2015ApJ...810...38Klink\endcsname~\csname
  2015ApJ...810...38Knote\endcsname

\bibitem[{{Kostik} \& {Orlova}(1975)}]{1975SoPh...45..119K}
{Kostik}, R.~I. \& {Orlova}, T.~V. 1975, \solphys, 45, 119 \csname
  1975SoPh...45..119Klink\endcsname~\csname 1975SoPh...45..119Knote\endcsname

\bibitem[{{Krieger} {et~al.}(1971){Krieger}, {Vaiana}, \& {van
  Speybroeck}}]{1971IAUS...43..397K}
{Krieger}, A.~S., {Vaiana}, G.~S., \& {van Speybroeck}, L.~P. 1971, in Solar
  Magnetic Fields, ed. R.~{Howard}, Vol.~43, 397 \csname
  1971IAUS...43..397Klink\endcsname~\csname 1971IAUS...43..397Knote\endcsname

\bibitem[{{Langangen} {et~al.}(2008){Langangen}, {De Pontieu}, {Carlsson},
  {Hansteen}, {Cauzzi}, \& {Reardon}}]{2008ApJ...679L.167L}
{Langangen}, {\O}., {De Pontieu}, B., {Carlsson}, M., {et~al.} 2008, \apjl,
  679, L167 \csname 2008ApJ...679L.167Llink\endcsname~\csname
  2008ApJ...679L.167Lnote\endcsname

\bibitem[{{Leenaarts} {et~al.}(2007){Leenaarts}, {Carlsson}, {Hansteen}, \&
  {Rutten}}]{2007A&A...473..625L}
{Leenaarts}, J., {Carlsson}, M., {Hansteen}, V., \& {Rutten}, R.~J. 2007, \aap,
  473, 625 \csname 2007A&A...473..625Llink\endcsname~\csname
  2007A&A...473..625Lnote\endcsname

\bibitem[{{Leenaarts} {et~al.}(2012){Leenaarts}, {Carlsson}, \& {Rouppe van der
  Voort}}]{2012ApJ...749..136L}
{Leenaarts}, J., {Carlsson}, M., \& {Rouppe van der Voort}, L. 2012, \apj, 749,
  136 \csname 2012ApJ...749..136Llink\endcsname~\csname
  2012ApJ...749..136Lnote\endcsname

\bibitem[{{Leenaarts} {et~al.}(2015){Leenaarts}, {Carlsson}, \& {Rouppe van der
  Voort}}]{2015ApJ...802..136L}
{Leenaarts}, J., {Carlsson}, M., \& {Rouppe van der Voort}, L. 2015, \apj, 802,
  136 \csname 2015ApJ...802..136Llink\endcsname~\csname
  2015ApJ...802..136Lnote\endcsname

\bibitem[{{Leenaarts} {et~al.}(2006{\natexlab{a}}){Leenaarts}, {Rutten},
  {Carlsson}, \& {Uitenbroek}}]{2006A&A...452L..15L}
{Leenaarts}, J., {Rutten}, R.~J., {Carlsson}, M., \& {Uitenbroek}, H.
  2006{\natexlab{a}}, \aap, 452, L15 \csname
  2006A&A...452L..15Llink\endcsname~\csname 2006A&A...452L..15Lnote\endcsname

\bibitem[{{Leenaarts} {et~al.}(2006{\natexlab{b}}){Leenaarts}, {Rutten},
  {S{\"u}tterlin}, {Carlsson}, \& {Uitenbroek}}]{2006A&A...449.1209L}
{Leenaarts}, J., {Rutten}, R.~J., {S{\"u}tterlin}, P., {Carlsson}, M., \&
  {Uitenbroek}, H. 2006{\natexlab{b}}, \aap, 449, 1209 \csname
  2006A&A...449.1209Llink\endcsname~\csname 2006A&A...449.1209Lnote\endcsname

\bibitem[{{Lites} {et~al.}(2008){Lites}, {Kubo}, {Socas-Navarro}, {Berger},
  {Frank}, {Shine}, {Tarbell}, {Title}, {Ichimoto}, {Katsukawa}, {Tsuneta},
  {Suematsu}, {Shimizu}, \& {Nagata}}]{2008ApJ...672.1237L}
{Lites}, B.~W., {Kubo}, M., {Socas-Navarro}, H., {et~al.} 2008, \apj, 672, 1237
  \csname 2008ApJ...672.1237Llink\endcsname~\csname
  2008ApJ...672.1237Lnote\endcsname

\bibitem[{{Lites} {et~al.}(1999){Lites}, {Rutten}, \&
  {Berger}}]{1999ApJ...517.1013L}
{Lites}, B.~W., {Rutten}, R.~J., \& {Berger}, T.~E. 1999, \apj, 517, 1013
  \csname 1999ApJ...517.1013Llink\endcsname~\csname
  1999ApJ...517.1013Lnote\endcsname

\bibitem[{{Liu} {et~al.}(2019{\natexlab{a}}){Liu}, {Carlsson}, {Nelson}, \&
  {Erd{\'e}lyi}}]{2019A&A...632A..97L}
{Liu}, J., {Carlsson}, M., {Nelson}, C.~J., \& {Erd{\'e}lyi}, R.
  2019{\natexlab{a}}, \aap, 632, A97 \csname
  2019A&A...632A..97Llink\endcsname~\csname 2019A&A...632A..97Lnote\endcsname

\bibitem[{{Liu} {et~al.}(2019{\natexlab{b}}){Liu}, {Nelson}, {Snow}, {Wang}, \&
  {Erd{\'e}lyi}}]{2019NatCo..10.3504L}
{Liu}, J., {Nelson}, C.~J., {Snow}, B., {Wang}, Y., \& {Erd{\'e}lyi}, R.
  2019{\natexlab{b}}, Nature Communications, 10, 3504 \csname
  2019NatCo..10.3504Llink\endcsname~\csname 2019NatCo..10.3504Lnote\endcsname

\bibitem[{{Livingston} \& {Wallace}(1987)}]{1987ApJ...314..808L}
{Livingston}, W. \& {Wallace}, L. 1987, \apj, 314, 808 \csname
  1987ApJ...314..808Llink\endcsname~\csname 1987ApJ...314..808Lnote\endcsname

\bibitem[{{Lockyer}(1868)}]{1868RSPS...17..131L}
{Lockyer}, J.~N. 1868, Proceedings of the Royal Society of London Series I, 17,
  131 \csname 1868RSPS...17..131Llink\endcsname~\csname
  1868RSPS...17..131Lnote\endcsname

\bibitem[{{Loukitcheva} {et~al.}(2015){Loukitcheva}, {Solanki}, {Carlsson}, \&
  {White}}]{2015A&A...575A..15L}
{Loukitcheva}, M., {Solanki}, S.~K., {Carlsson}, M., \& {White}, S.~M. 2015,
  \aap, 575, A15 \csname 2015A&A...575A..15Llink\endcsname~\csname
  2015A&A...575A..15Lnote\endcsname

\bibitem[{{Low}(1985)}]{1985ApJ...293...31L}
{Low}, B.~C. 1985, \apj, 293, 31 \csname
  1985ApJ...293...31Llink\endcsname~\csname 1985ApJ...293...31Lnote\endcsname

\bibitem[{{Madjarska}(2019)}]{2019LRSP...16....2M}
{Madjarska}, M.~S. 2019, Living Reviews in Solar Physics, 16, 2 \csname
  2019LRSP...16....2Mlink\endcsname~\csname 2019LRSP...16....2Mnote\endcsname

\bibitem[{{Mart{\'\i}nez-Sykora}
  {et~al.}(2020{\natexlab{a}}){Mart{\'\i}nez-Sykora}, {De Pontieu}, {de la Cruz
  Rodrigu{\'{e}}z}, \& {Chintzoglou}}]{2020ApJ...891L...8M}
{Mart{\'\i}nez-Sykora}, J., {De Pontieu}, B., {de la Cruz Rodrigu{\'{e}}z}, J.,
  \& {Chintzoglou}, G. 2020{\natexlab{a}}, \apjl, 891, L8 \csname
  2020ApJ...891L...8Mlink\endcsname~\csname 2020ApJ...891L...8Mnote\endcsname

\bibitem[{{Mart{\'\i}nez-Sykora}
  {et~al.}(2020{\natexlab{b}}){Mart{\'\i}nez-Sykora}, {Leenaarts}, {De
  Pontieu}, {N{\'o}brega-Siverio}, {Hansteen}, {Carlsson}, \&
  {Szydlarski}}]{2020ApJ...889...95M}
{Mart{\'\i}nez-Sykora}, J., {Leenaarts}, J., {De Pontieu}, B., {et~al.}
  2020{\natexlab{b}}, \apj, 889, 95 \csname
  2020ApJ...889...95Mlink\endcsname~\csname 2020ApJ...889...95Mnote\endcsname

\bibitem[{{McMath} {et~al.}(1960){McMath}, {Mohler}, \&
  {Dodson}}]{1960PNAS...46..165M}
{McMath}, R.~R., {Mohler}, O.~C., \& {Dodson}, H.~W. 1960, Proceedings of the
  National Academy of Science, 46, 165 \csname
  1960PNAS...46..165Mlink\endcsname~\csname 1960PNAS...46..165Mnote\endcsname

\bibitem[{{Molnar} {et~al.}(2019){Molnar}, {Reardon}, {Chai}, {Gary},
  {Uitenbroek}, {Cauzzi}, \& {Cranmer}}]{2019ApJ...881...99M}
{Molnar}, M.~E., {Reardon}, K.~P., {Chai}, Y., {et~al.} 2019, \apj, 881, 99
  \csname 2019ApJ...881...99Mlink\endcsname~\csname
  2019ApJ...881...99Mnote\endcsname

\bibitem[{{Muller}(1984)}]{1984apoa.conf..382M}
{Muller}, R. 1984, in Active Phenomena in the Outer Atmosphere of the Sun and
  Stars, 382 \csname 1984apoa.conf..382Mlink\endcsname~\csname
  1984apoa.conf..382Mnote\endcsname

\bibitem[{{Nelson} {et~al.}(2013{\natexlab{a}}){Nelson}, {Doyle},
  {Erd{\'e}lyi}, {Huang}, {Madjarska}, {Mathioudakis}, {Mumford}, \&
  {Reardon}}]{2013SoPh..283..307N}
{Nelson}, C.~J., {Doyle}, J.~G., {Erd{\'e}lyi}, R., {et~al.}
  2013{\natexlab{a}}, \solphys, 283, 307 \csname
  2013SoPh..283..307Nlink\endcsname~\csname 2013SoPh..283..307Nnote\endcsname

\bibitem[{{Nelson} {et~al.}(2015){Nelson}, {Scullion}, {Doyle}, {Freij}, \&
  {Erd{\'e}lyi}}]{2015ApJ...798...19N}
{Nelson}, C.~J., {Scullion}, E.~M., {Doyle}, J.~G., {Freij}, N., \&
  {Erd{\'e}lyi}, R. 2015, \apj, 798, 19 \csname
  2015ApJ...798...19Nlink\endcsname~\csname 2015ApJ...798...19Nnote\endcsname

\bibitem[{{Nelson} {et~al.}(2013{\natexlab{b}}){Nelson}, {Shelyag},
  {Mathioudakis}, {Doyle}, {Madjarska}, {Uitenbroek}, \&
  {Erd{\'e}lyi}}]{2013ApJ...779..125N}
{Nelson}, C.~J., {Shelyag}, S., {Mathioudakis}, M., {et~al.}
  2013{\natexlab{b}}, \apj, 779, 125 \csname
  2013ApJ...779..125Nlink\endcsname~\csname 2013ApJ...779..125Nnote\endcsname

\bibitem[{{November}(2004)}]{2004A&A...417..333N}
{November}, L.~J. 2004, \aap, 417, 333 \csname
  2004A&A...417..333Nlink\endcsname~\csname 2004A&A...417..333Nnote\endcsname

\bibitem[{{Panasenco} {et~al.}(2014){Panasenco}, {Martin}, \&
  {Velli}}]{2014SoPh..289..603P}
{Panasenco}, O., {Martin}, S.~F., \& {Velli}, M. 2014, \solphys, 289, 603
  \csname 2014SoPh..289..603Plink\endcsname~\csname
  2014SoPh..289..603Pnote\endcsname

\bibitem[{{Pariat} {et~al.}(2009){Pariat}, {Masson}, \&
  {Aulanier}}]{2009ApJ...701.1911P}
{Pariat}, E., {Masson}, S., \& {Aulanier}, G. 2009, \apj, 701, 1911 \csname
  2009ApJ...701.1911Plink\endcsname~\csname 2009ApJ...701.1911Pnote\endcsname

\bibitem[{{Pereira} \& {Uitenbroek}(2015)}]{2015A&A...574A...3P}
{Pereira}, T. M.~D. \& {Uitenbroek}, H. 2015, \aap, 574, A3 \csname
  2015A&A...574A...3Plink\endcsname~\csname 2015A&A...574A...3Pnote\endcsname

\bibitem[{{Peter} {et~al.}(2014){Peter}, {Tian}, {Curdt}, {Schmit}, {Innes},
  {De Pontieu}, {Lemen}, {Title}, {Boerner}, {Hurlburt}, {Tarbell}, {Wuelser},
  {Mart{\'\i}nez-Sykora}, {Kleint}, {Golub}, {McKillop}, {Reeves}, {Saar},
  {Testa}, {Kankelborg}, {Jaeggli}, {Carlsson}, \&
  {Hansteen}}]{2014Sci...346C.315P}
{Peter}, H., {Tian}, H., {Curdt}, W., {et~al.} 2014, Science, 346, 1255726
  \csname 2014Sci...346C.315Plink\endcsname~\csname
  2014Sci...346C.315Pnote\endcsname

\bibitem[{{Polito} {et~al.}(2018){Polito}, {Testa}, {Allred}, {De Pontieu},
  {Carlsson}, {Pereira}, {Go{\v{s}}i{\'c}}, \& {Reale}}]{2018ApJ...856..178P}
{Polito}, V., {Testa}, P., {Allred}, J., {et~al.} 2018, \apj, 856, 178 \csname
  2018ApJ...856..178Plink\endcsname~\csname 2018ApJ...856..178Pnote\endcsname

\bibitem[{{Rezaei} \& {Beck}(2015)}]{2015A&A...582A.104R}
{Rezaei}, R. \& {Beck}, C. 2015, \aap, 582, A104 \csname
  2015A&A...582A.104Rlink\endcsname~\csname 2015A&A...582A.104Rnote\endcsname

\bibitem[{{Rouppe van der Voort} {et~al.}(2020){Rouppe van der Voort}, {De
  Pontieu}, {Carlsson}, {de la Cruz Rodr{\'\i}guez}, {Bose}, {Chintzoglou},
  {Drews}, {Froment}, {Go{\v{s}}i{\'c}}, {Graham}, {Hansteen}, {Henriques},
  {Jafarzadeh}, {Joshi}, {Kleint}, {Kohutova}, {Leifsen},
  {Mart{\'\i}nez-Sykora}, {N{\'o}brega-Siverio}, {Ortiz}, {Pereira}, {Popovas},
  {Quintero Noda}, {Sainz Dalda}, {Scharmer}, {Schmit}, {Scullion}, {Skogsrud},
  {Szydlarski}, {Timmons}, {Vissers}, {Woods}, \&
  {Zacharias}}]{2020A&A...641A.146R}
{Rouppe van der Voort}, L.~H.~M., {De Pontieu}, B., {Carlsson}, M., {et~al.}
  2020, \aap, 641, A146 \csname 2020A&A...641A.146Rlink\endcsname~\csname
  2020A&A...641A.146Rnote\endcsname

\bibitem[{{Rouppe van der Voort} {et~al.}(2016){Rouppe van der Voort},
  {Rutten}, \& {Vissers}}]{2016A&A...592A.100R}
{Rouppe van der Voort}, L. H.~M., {Rutten}, R.~J., \& {Vissers}, G. J.~M. 2016,
  \aap, 592, A100 \csname 2016A&A...592A.100Rlink\endcsname~\csname
  2016A&A...592A.100Rnote\endcsname

\bibitem[{{Rutten}(1990)}]{1990IAUS..138..501R}
{Rutten}, R.~J. 1990, in IAU Symposium, Vol. 138, Solar Photosphere: Structure,
  Convection, and Magnetic Fields, ed. J.~O. {Stenflo}, 501--516 \csname
  1990IAUS..138..501Rlink\endcsname~\csname 1990IAUS..138..501Rnote\endcsname

\bibitem[{{Rutten}(1995)}]{1995ESASP.376a.151R}
{Rutten}, R.~J. 1995, in ESA Special Pub., Vol. 376, Helioseismology, ed. J.~T.
  Hoeksema, V.~Domingo, B.~Fleck, \& B.~Battrick, 151--163 \csname
  1995ESASP.376a.151Rlink\endcsname~\csname 1995ESASP.376a.151Rnote\endcsname

\bibitem[{{Rutten}(1999)}]{1999ASPC..184..181R}
{Rutten}, R.~J. 1999, in Astron.\ Soc.\ Pacific Conf.\ Series, Vol. 184,
  Magnetic Fields and Oscillations, ed. B.~{Schmieder}, A.~{Hofmann}, \&
  J.~{Staude}, Third Adv.\ in Solar Physics Euroconf., 181--200 \csname
  1999ASPC..184..181Rlink\endcsname~\csname 1999ASPC..184..181Rnote\endcsname

\bibitem[{{Rutten}(2003)}]{2003rtsa.book.....R}
{Rutten}, R.~J. 2003, Radiative Transfer in Stellar Atmospheres (Utrecht:
  Lecture notes Utrecht University) \csname
  2003rtsa.book.....Rlink\endcsname~\csname 2003rtsa.book.....Rnote\endcsname

\bibitem[{{Rutten}(2013)}]{2013ASPC..470...49R}
{Rutten}, R.~J. 2013, in Astron.\ Soc.\ Pacific Conf.\ Series, Vol. 470, 370
  Years of Astronomy in Utrecht, ed. G.~{Pugliese}, A.~{de Koter}, \&
  M.~{Wijburg}, 49--58 \csname 2013ASPC..470...49Rlink\endcsname~\csname
  2013ASPC..470...49Rnote\endcsname

\bibitem[{{Rutten}(2016)}]{2016A&A...590A.124R}
{Rutten}, R.~J. 2016, \aap, 590, A124 \csname
  2016A&A...590A.124Rlink\endcsname~\csname 2016A&A...590A.124Rnote\endcsname

\bibitem[{{Rutten}(2017{\natexlab{a}})}]{2017IAUS..327....1R}
{Rutten}, R.~J. 2017{\natexlab{a}}, in IAU Symposium, Vol. 327, Fine structure
  and dynamics of the solar atmosphere, ed. S.~{Vargas Dom{\'{\i}}nguez}, A.~G.
  {Kosovichev}, P.~{Antolin}, \& L.~{Harra}, 1--15 \csname
  2017IAUS..327....1Rlink\endcsname~\csname 2017IAUS..327....1Rnote\endcsname

\bibitem[{{Rutten}(2017{\natexlab{b}})}]{2017A&A...598A..89R}
{Rutten}, R.~J. 2017{\natexlab{b}}, \aap, 598, A89 \csname
  2017A&A...598A..89Rlink\endcsname~\csname 2017A&A...598A..89Rnote\endcsname

\bibitem[{{Rutten}(2019)}]{2019arXiv190804624R}
{Rutten}, R.~J. 2019, \solphys, 294, 165 1 \csname
  2019arXiv190804624Rlink\endcsname~\csname 2019arXiv190804624Rnote\endcsname

\bibitem[{{Rutten} {et~al.}(2011){Rutten}, {Leenaarts}, {Rouppe van der Voort},
  {de Wijn}, {Carlsson}, \& {Hansteen}}]{2011A&A...531A..17R}
{Rutten}, R.~J., {Leenaarts}, J., {Rouppe van der Voort}, L.~H.~M., {et~al.}
  2011, \aap, 531, A17 \csname 2011A&A...531A..17Rlink\endcsname~\csname
  2011A&A...531A..17Rnote\endcsname

\bibitem[{{Rutten} \& {Rouppe van der Voort}(2017)}]{2017A&A...597A.138R}
{Rutten}, R.~J. \& {Rouppe van der Voort}, L.~H.~M. 2017, \aap, 597, A138
  \csname 2017A&A...597A.138Rlink\endcsname~\csname
  2017A&A...597A.138Rnote\endcsname

\bibitem[{{Rutten} {et~al.}(2019){Rutten}, {Rouppe van der Voort}, \& {De
  Pontieu}}]{2019A&A...632A..96R}
{Rutten}, R.~J., {Rouppe van der Voort}, L. H.~M., \& {De Pontieu}, B. 2019,
  \aap, 632, A96 \csname 2019A&A...632A..96Rlink\endcsname~\csname
  2019A&A...632A..96Rnote\endcsname

\bibitem[{{Rutten} {et~al.}(2015){Rutten}, {Rouppe van der Voort}, \&
  {Vissers}}]{2015ApJ...808..133R}
{Rutten}, R.~J., {Rouppe van der Voort}, L.~H.~M., \& {Vissers}, G.~J.~M. 2015,
  \apj, 808, 133 1 \csname 2015ApJ...808..133Rlink\endcsname~\csname
  2015ApJ...808..133Rnote\endcsname

\bibitem[{{Rutten} \& {Uitenbroek}(1991)}]{1991SoPh..134...15R}
{Rutten}, R.~J. \& {Uitenbroek}, H. 1991, \solphys, 134, 15 \csname
  1991SoPh..134...15Rlink\endcsname~\csname 1991SoPh..134...15Rnote\endcsname

\bibitem[{{Rutten} \& {Uitenbroek}(2012)}]{2012A&A...540A..86R}
{Rutten}, R.~J. \& {Uitenbroek}, H. 2012, \aap, 540, A86 \csname
  2012A&A...540A..86Rlink\endcsname~\csname 2012A&A...540A..86Rnote\endcsname

\bibitem[{{Rutten} {et~al.}(2013){Rutten}, {Vissers}, {Rouppe van der Voort},
  {S{\"u}tterlin}, \& {Vitas}}]{2013JPhCS.440a2007R}
{Rutten}, R.~J., {Vissers}, G.~J.~M., {Rouppe van der Voort}, L.~H.~M.,
  {S{\"u}tterlin}, P., \& {Vitas}, N. 2013, in J.\ Physics Conf.\ Series, Vol.
  440, Eclipse on the Coral Sea: Cycle 24 Ascending, ed. P.~S. {Cally},
  R.~{Erd{\'{e}}lyi}, \& A.~A. {Norton}, 1--13 \csname
  2013JPhCS.440a2007Rlink\endcsname~\csname 2013JPhCS.440a2007Rnote\endcsname

\bibitem[{{Schrijver} {et~al.}(1997){Schrijver}, {Title}, {van Ballegooijen},
  {Hagenaar}, \& {Shine}}]{1997ApJ...487..424S}
{Schrijver}, C.~J., {Title}, A.~M., {van Ballegooijen}, A.~A., {Hagenaar},
  H.~J., \& {Shine}, R.~A. 1997, \apj, 487, 424 \csname
  1997ApJ...487..424Slink\endcsname~\csname 1997ApJ...487..424Snote\endcsname

\bibitem[{{Sekse} {et~al.}(2013{\natexlab{a}}){Sekse}, {Rouppe van der Voort},
  \& {De Pontieu}}]{2013ApJ...764..164S}
{Sekse}, D.~H., {Rouppe van der Voort}, L., \& {De Pontieu}, B.
  2013{\natexlab{a}}, \apj, 764, 164 \csname
  2013ApJ...764..164Slink\endcsname~\csname 2013ApJ...764..164Snote\endcsname

\bibitem[{{Sekse} {et~al.}(2013{\natexlab{b}}){Sekse}, {Rouppe van der Voort},
  {De Pontieu}, \& {Scullion}}]{2013ApJ...769...44S}
{Sekse}, D.~H., {Rouppe van der Voort}, L., {De Pontieu}, B., \& {Scullion}, E.
  2013{\natexlab{b}}, \apj, 769, 44 \csname
  2013ApJ...769...44Slink\endcsname~\csname 2013ApJ...769...44Snote\endcsname

\bibitem[{{Sheminova} {et~al.}(2005){Sheminova}, {Rutten}, \& {Rouppe van der
  Voort}}]{2005A&A...437.1069S}
{Sheminova}, V.~A., {Rutten}, R.~J., \& {Rouppe van der Voort}, L.~H.~M. 2005,
  \aap, 437, 1069 \csname 2005A&A...437.1069Slink\endcsname~\csname
  2005A&A...437.1069Snote\endcsname

\bibitem[{{Shimizu} {et~al.}(2002){Shimizu}, {Shine}, {Title}, {Tarbell}, \&
  {Frank}}]{2002ApJ...574.1074S}
{Shimizu}, T., {Shine}, R.~A., {Title}, A.~M., {Tarbell}, T.~D., \& {Frank}, Z.
  2002, \apj, 574, 1074 \csname 2002ApJ...574.1074Slink\endcsname~\csname
  2002ApJ...574.1074Snote\endcsname

\bibitem[{{Solanki}(1993)}]{1993SSRv...63....1S}
{Solanki}, S.~K. 1993, \ssr, 63, 1 \csname
  1993SSRv...63....1Slink\endcsname~\csname 1993SSRv...63....1Snote\endcsname

\bibitem[{{Spruit}(1976)}]{1976SoPh...50..269S}
{Spruit}, H.~C. 1976, \solphys, 50, 269 \csname
  1976SoPh...50..269Slink\endcsname~\csname 1976SoPh...50..269Snote\endcsname

\bibitem[{{Spruit}(1977)}]{1977PhDT.......237S}
{Spruit}, H.~C. 1977, PhD thesis, - \csname
  1977PhDT.......237Slink\endcsname~\csname 1977PhDT.......237Snote\endcsname

\bibitem[{{Steiner}(2005)}]{2005A&A...430..691S}
{Steiner}, O. 2005, \aap, 430, 691 \csname
  2005A&A...430..691Slink\endcsname~\csname 2005A&A...430..691Snote\endcsname

\bibitem[{{Stellmacher} \& {Wiehr}(1991)}]{1991A&A...251..675S}
{Stellmacher}, G. \& {Wiehr}, E. 1991, \aap, 251, 675 \csname
  1991A&A...251..675Slink\endcsname~\csname 1991A&A...251..675Snote\endcsname

\bibitem[{{Stenflo}(1984)}]{1984AdSpR...4h...5S}
{Stenflo}, J.~O. 1984, Advances in Space Research, 4, 5 \csname
  1984AdSpR...4h...5Slink\endcsname~\csname 1984AdSpR...4h...5Snote\endcsname

\bibitem[{{Stenflo} {et~al.}(1984){Stenflo}, {Solanki}, {Harvey}, \&
  {Brault}}]{1984A&A...131..333S}
{Stenflo}, J.~O., {Solanki}, S., {Harvey}, J.~W., \& {Brault}, J.~W. 1984,
  \aap, 131, 333 \csname 1984A&A...131..333Slink\endcsname~\csname
  1984A&A...131..333Snote\endcsname

\bibitem[{{Testa} {et~al.}(2014){Testa}, {De Pontieu}, {Allred}, {Carlsson},
  {Reale}, {Daw}, {Hansteen}, {Mart{\'{\i}}nez-Sykora}, {Liu}, {DeLuca},
  {Golub}, {McKillop}, {Reeves}, {Saar}, {Tian}, {Lemen}, {Title}, {Boerner},
  {Hurlburt}, {Tarbell}, {Wuelser}, {Kleint}, {Kankelborg}, \&
  {Jaeggli}}]{2014Sci...346B.315T}
{Testa}, P., {De Pontieu}, B., {Allred}, J., {et~al.} 2014, Science, 346,
  1255724 \csname 2014Sci...346B.315Tlink\endcsname~\csname
  2014Sci...346B.315Tnote\endcsname

\bibitem[{{Title} \& {Berger}(1996)}]{1996ApJ...463..797T}
{Title}, A.~M. \& {Berger}, T.~E. 1996, \apj, 463, 797 \csname
  1996ApJ...463..797Tlink\endcsname~\csname 1996ApJ...463..797Tnote\endcsname

\bibitem[{{Tziotziou} {et~al.}(2018){Tziotziou}, {Tsiropoula}, {Kontogiannis},
  {Scullion}, \& {Doyle}}]{2018A&A...618A..51T}
{Tziotziou}, K., {Tsiropoula}, G., {Kontogiannis}, I., {Scullion}, E., \&
  {Doyle}, J.~G. 2018, \aap, 618, A51 \csname
  2018A&A...618A..51Tlink\endcsname~\csname 2018A&A...618A..51Tnote\endcsname

\bibitem[{{Unruh} {et~al.}(1999){Unruh}, {Solanki}, \&
  {Fligge}}]{1999A&A...345..635U}
{Unruh}, Y.~C., {Solanki}, S.~K., \& {Fligge}, M. 1999, \aap, 345, 635 \csname
  1999A&A...345..635Ulink\endcsname~\csname 1999A&A...345..635Unote\endcsname

\bibitem[{{Vargas Dom{\'\i}nguez} {et~al.}(2011){Vargas Dom{\'\i}nguez},
  {Palacios}, {Balmaceda}, {Cabello}, \& {Domingo}}]{2011MNRAS.416..148V}
{Vargas Dom{\'\i}nguez}, S., {Palacios}, J., {Balmaceda}, L., {Cabello}, I., \&
  {Domingo}, V. 2011, \mnras, 416, 148 \csname
  2011MNRAS.416..148Vlink\endcsname~\csname 2011MNRAS.416..148Vnote\endcsname

\bibitem[{{Vernazza} {et~al.}(1981){Vernazza}, {Avrett}, \&
  {Loeser}}]{1981ApJS...45..635V}
{Vernazza}, J.~E., {Avrett}, E.~H., \& {Loeser}, R. 1981, \apjs, 45, 635
  \csname 1981ApJS...45..635Vlink\endcsname~\csname
  1981ApJS...45..635Vnote\endcsname

\bibitem[{{Vissers} \& {Rouppe van der Voort}(2012)}]{2012ApJ...750...22V}
{Vissers}, G. \& {Rouppe van der Voort}, L. 2012, \apj, 750, 22 \csname
  2012ApJ...750...22Vlink\endcsname~\csname 2012ApJ...750...22Vnote\endcsname

\bibitem[{{Vissers} {et~al.}(2013){Vissers}, {Rouppe van der Voort}, \&
  {Rutten}}]{2013ApJ...774...32V}
{Vissers}, G.~J.~M., {Rouppe van der Voort}, L.~H.~M., \& {Rutten}, R.~J. 2013,
  \apj, 774, 32 1 \csname 2013ApJ...774...32Vlink\endcsname~\csname
  2013ApJ...774...32Vnote\endcsname

\bibitem[{{Vissers} {et~al.}(2019){Vissers}, {Rouppe van der Voort}, \&
  {Rutten}}]{2019A&A...626A...4V}
{Vissers}, G. J.~M., {Rouppe van der Voort}, L. H.~M., \& {Rutten}, R.~J. 2019,
  \aap, 626, A4 \csname 2019A&A...626A...4Vlink\endcsname~\csname
  2019A&A...626A...4Vnote\endcsname

\bibitem[{{Vissers} {et~al.}(2015){Vissers}, {Rouppe van der Voort}, {Rutten},
  {Carlsson}, \& {De Pontieu}}]{2015ApJ...812...11V}
{Vissers}, G.~J.~M., {Rouppe van der Voort}, L.~H.~M., {Rutten}, R.~J.,
  {Carlsson}, M., \& {De Pontieu}, B. 2015, \apj, 812, 11 1 \csname
  2015ApJ...812...11Vlink\endcsname~\csname 2015ApJ...812...11Vnote\endcsname

\bibitem[{{Vitas} {et~al.}(2009){Vitas}, {Viticchi{\`e}}, {Rutten}, \&
  {V{\"o}gler}}]{2009A&A...499..301V}
{Vitas}, N., {Viticchi{\`e}}, B., {Rutten}, R.~J., \& {V{\"o}gler}, A. 2009,
  \aap, 499, 301 \csname 2009A&A...499..301Vlink\endcsname~\csname
  2009A&A...499..301Vnote\endcsname

\bibitem[{{Watanabe} {et~al.}(2011){Watanabe}, {Vissers}, {Kitai}, {Rouppe van
  der Voort}, \& {Rutten}}]{2011ApJ...736...71W}
{Watanabe}, H., {Vissers}, G., {Kitai}, R., {Rouppe van der Voort}, L., \&
  {Rutten}, R.~J. 2011, \apj, 736, 71 \csname
  2011ApJ...736...71Wlink\endcsname~\csname 2011ApJ...736...71Wnote\endcsname

\bibitem[{{Wedemeyer} {et~al.}(2013){Wedemeyer}, {Scullion}, {Rouppe van der
  Voort}, {Bosnjak}, \& {Antolin}}]{2013ApJ...774..123W}
{Wedemeyer}, S., {Scullion}, E., {Rouppe van der Voort}, L., {Bosnjak}, A., \&
  {Antolin}, P. 2013, \apj, 774, 123 \csname
  2013ApJ...774..123Wlink\endcsname~\csname 2013ApJ...774..123Wnote\endcsname

\bibitem[{{Wedemeyer-B{\"o}hm} {et~al.}(2012){Wedemeyer-B{\"o}hm}, {Scullion},
  {Steiner}, {Rouppe van der Voort}, {de La Cruz Rodrigu{\'{e}}z}, {Fedun}, \&
  {Erd{\'e}lyi}}]{2012Natur.486..505W}
{Wedemeyer-B{\"o}hm}, S., {Scullion}, E., {Steiner}, O., {et~al.} 2012, \nat,
  486, 505 \csname 2012Natur.486..505Wlink\endcsname~\csname
  2012Natur.486..505Wnote\endcsname

\bibitem[{{Yang} {et~al.}(2013){Yang}, {Chae}, {Lim}, {Park}, {Cho}, {Maurya},
  {Song}, {Kim}, \& {Goode}}]{2013SoPh..288...39Y}
{Yang}, H., {Chae}, J., {Lim}, E.-K., {et~al.} 2013, \solphys, 288, 39 \csname
  2013SoPh..288...39Ylink\endcsname~\csname 2013SoPh..288...39Ynote\endcsname

\bibitem[{{Young} {et~al.}(2018){Young}, {Tian}, {Peter}, {Rutten}, {Nelson},
  {Huang}, {Schmieder}, {Vissers}, {Toriumi}, {Rouppe van der Voort},
  {Madjarska}, {Danilovic}, {Berlicki}, {Chitta}, {Cheung}, {Madsen},
  {Reardon}, {Katsukawa}, \& {Heinzel}}]{2018SSRv..214..120Y}
{Young}, P.~R., {Tian}, H., {Peter}, H., {et~al.} 2018, \ssr, 214, 120 1
  \csname 2018SSRv..214..120Ylink\endcsname~\csname
  2018SSRv..214..120Ynote\endcsname

\bibitem[{{Zwaan}(1965)}]{1965smss.book.....Z}
{Zwaan}, C. 1965, Sunspot models; a study of sunspot spectra \csname
  1965smss.book.....Zlink\endcsname~\csname 1965smss.book.....Znote\endcsname

\bibitem[{{Zwaan}(1967)}]{1967SoPh....1..478Z}
{Zwaan}, C. 1967, \solphys, 1, 478 \csname
  1967SoPh....1..478Zlink\endcsname~\csname 1967SoPh....1..478Znote\endcsname

\bibitem[{{Zwaan}(1978)}]{1978SoPh...60..213Z}
{Zwaan}, C. 1978, \solphys, 60, 213 \csname
  1978SoPh...60..213Zlink\endcsname~\csname 1978SoPh...60..213Znote\endcsname

\end{thebibliography}

\clearpage  
\listoffigures
\addcontentsline{toc}{section}{Figures}
\addtocline{List}

\vfill
\begin{small}
\subsectionrr{Figure blinking by page flipping} 
Viewer-dependent; try single-page, fit-to-page, full-page, full-screen,
presentation mode, thumbnail panel, left-right, up-down, page up-down.
\vspace*{-2ex}
\begin{itemize} \itemsep=0.5ex 

\item Gnome {\tt evince}: zoom-in (CTRL +), page-blink with up-down in
the thumbnail side panel or full-page and left-right;

\item Acrobat {\tt acroread}: full-page (CTRL L), zoom-in (CTRL +),
blink left-right;

\item Ubuntu {\tt qpdfview}: full-screen, zoom-in (CTRL right), blink
left-right;

\item Chrome and Firefox pdf viewers: left-right, or full-page
and up-down or page up-down;

\item MacOS: {\tt Firefox} left-right or full-page and up-down,~
{\tt Safari} full-page up-down,~
{\tt Preview} full-page up-down or mouse scroller, zoom-in and 
left-right.
\label{sec:blinking}
\end{itemize}
\vspace{-2ex}
\end{small} \mbox{}


\begin{figure*}
  \centering
  \includegraphics[width=\textwidth]{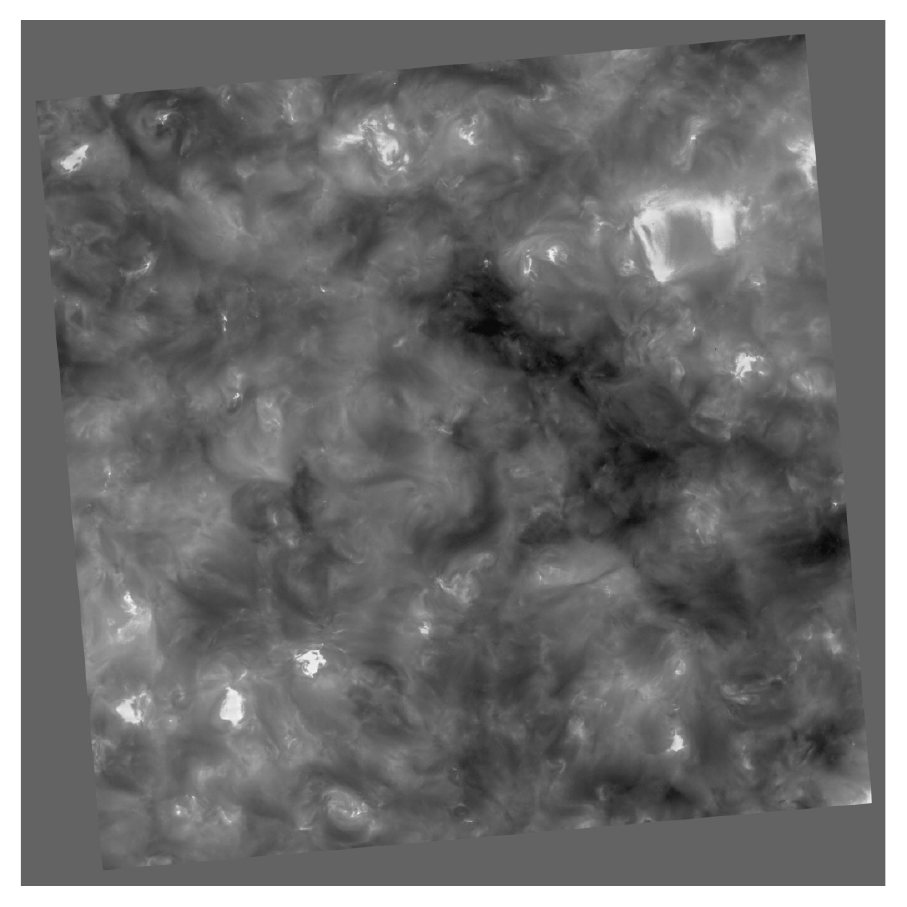}
  \caption[Press-release SolO 174\,\AA\ image]{%
  \mbox{}\label{fig:soloim}   
  \addtocline{SolO press-release image} 
  Press-release SolO 174\,\AA\
  image, roughly rotated to solar $(X,Y)$ as seen from SDO. 
  After my long search for it this scene is burned into my brain.
  ``SolO campfires'' appear as tiny bright stalks; zoom-in may help to
  spot them. 
  It is the sole SolO input for this report; all other images are from
  SDO, plus one GONG \Halpha\ image. \\
  {\bf Caveat disclaimer:} EUI PI David Berghmans reprimanded me later
  that this image ``is not properly validated'' and showing it here is
  ``unfair to the intrinsic quality of the EUI instrument'' -- but it
  serves here only to display the ``campfires'' shown and advertised
  during the SolO press conference, and as proof that I eventually did
  find this non-specified SolO scene in SDO (compare with
  \rrref{figure}{fig:field193}). 
  Proper validation may make this image yet better and enable proper
  axes, but will not affect any other image here nor SDO campfires in
  these (which by existing guarantee that the SolO ones can't validate
  away).\\
  {\em Copyright\/}: Solar Orbiter/EUI Team/ESA \& NASA; CSL, IAS, MPS,
  PMOD/WRC, ROB, UCL/MSSL.
  }\end{figure*}


\begin{figure*}
  \centering
  \includegraphics[width=\textwidth]{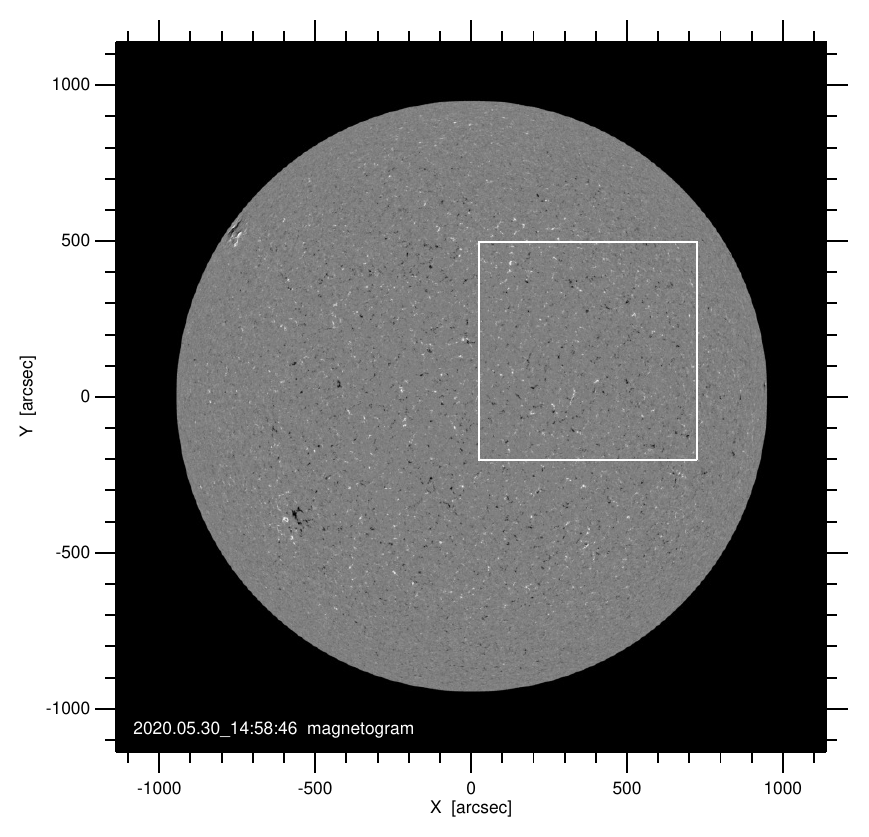}
  \caption[Full-disk HMI magnetogram]{\label{fig:fullmag} %
  \addtocline{SDO SolO field}
  HMI magnetogram at best-match time, with the selected field marked
  by the white frame. 
  It contains the SolO field. 
  It did not contain major activity, present here only towards the
  East limb (where IRIS pointed that day, still farside for SolO).
  }\end{figure*}
\begin{figure*}
  \centering
  \includegraphics[width=\textwidth]{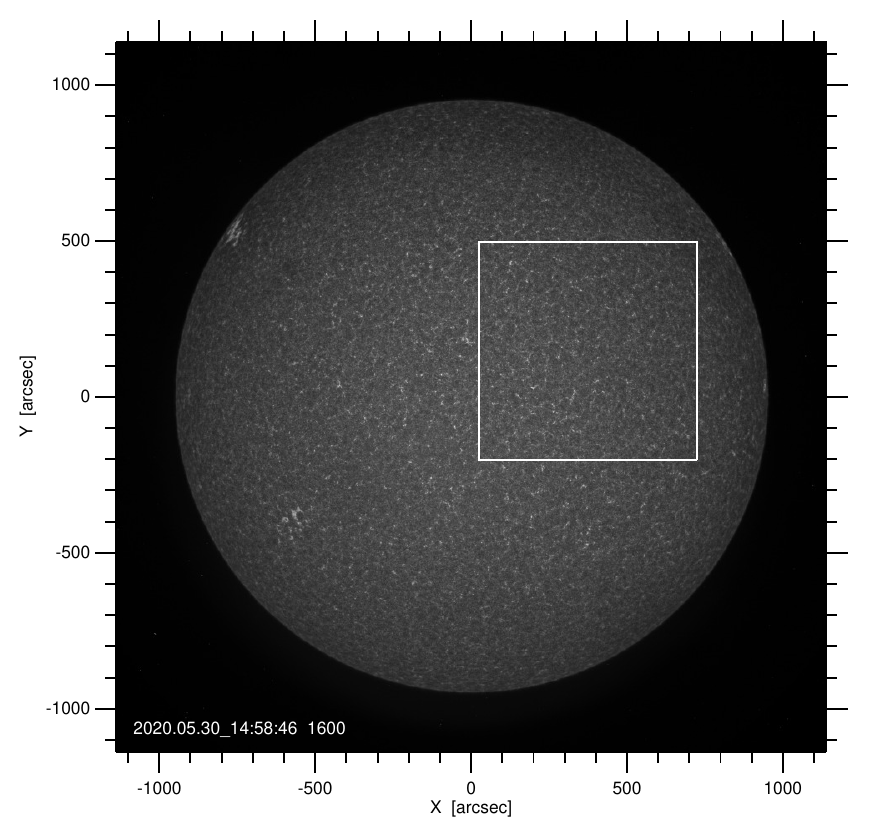}
  \caption[Full-disk AIA 1600\,\AA\ image]{\label{fig:full1600} %
  AIA 1600\,\AA\ image at best-match time. 
  It outlines the magnetic network seen in \rrref{figure}{fig:fullmag}
  very well as small bright grains obeying \rrref{figure}{fig:uvcartoons}.
  }\end{figure*}
\begin{figure*}
  \centering
  \includegraphics[width=\textwidth]{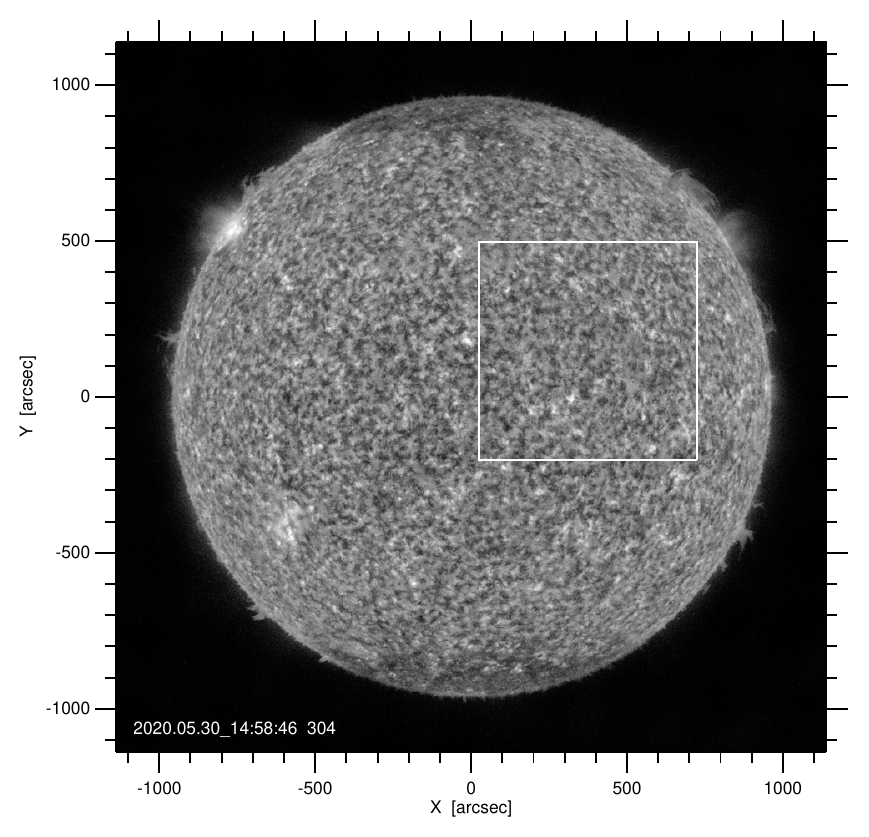}
  \caption[Full-disk AIA 304\,\AA\ image]{\label{fig:full304} AIA
  304\,\AA\ image at best-match time. 
  The dense coverage by grey patches is addressed in
  \rrref{appendix}{sec:ha304}. 
  They show the chromosphere with traditional ``transition region''
  temperatures but also visible in \Halpha. 
  Notice small brighter patches called ``brushfires'' below, as coyly
  as ``campfires'' but with language correction from Ron Moore
  (\rrref{footnote}{foot:moore}). 
  }\end{figure*}
\begin{figure*}
  \centering
  \includegraphics[width=\textwidth]{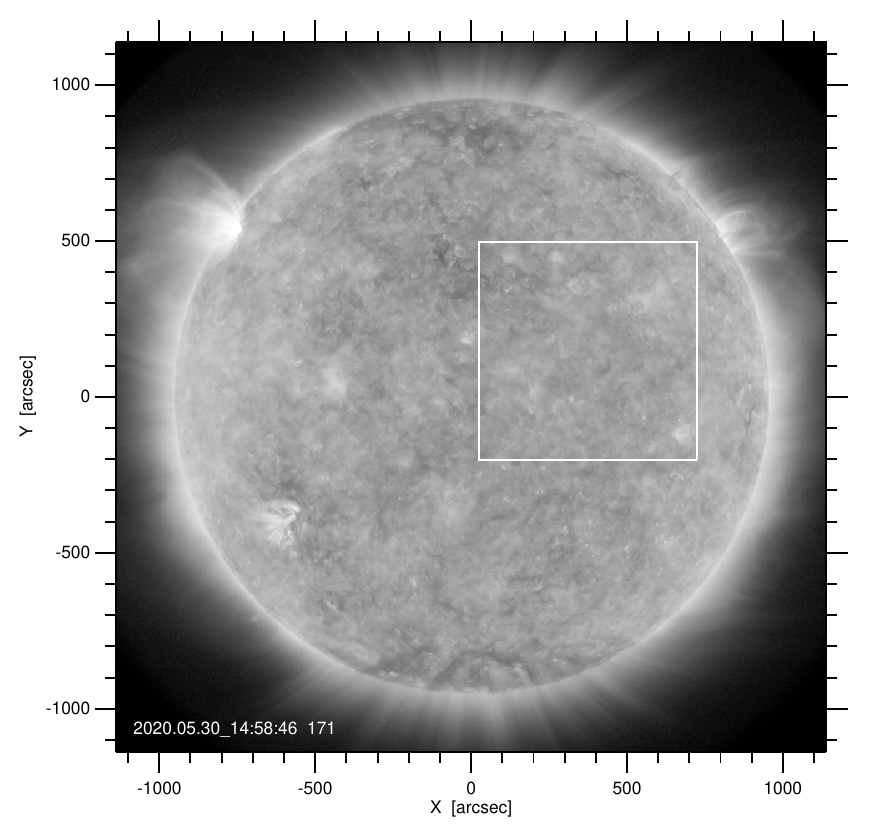}
  \caption[Full-disk AIA 171\,\AA\ image]{\label{fig:full171} %
  AIA 171\,\AA\ image at best-match time. 
  Obviously this AIA diagnostic is the best to show coronal
  connectivity and non-nearby-connectivity in polar plumes.
  }\end{figure*}
\begin{figure*}
  \centering
  \includegraphics[width=\textwidth]{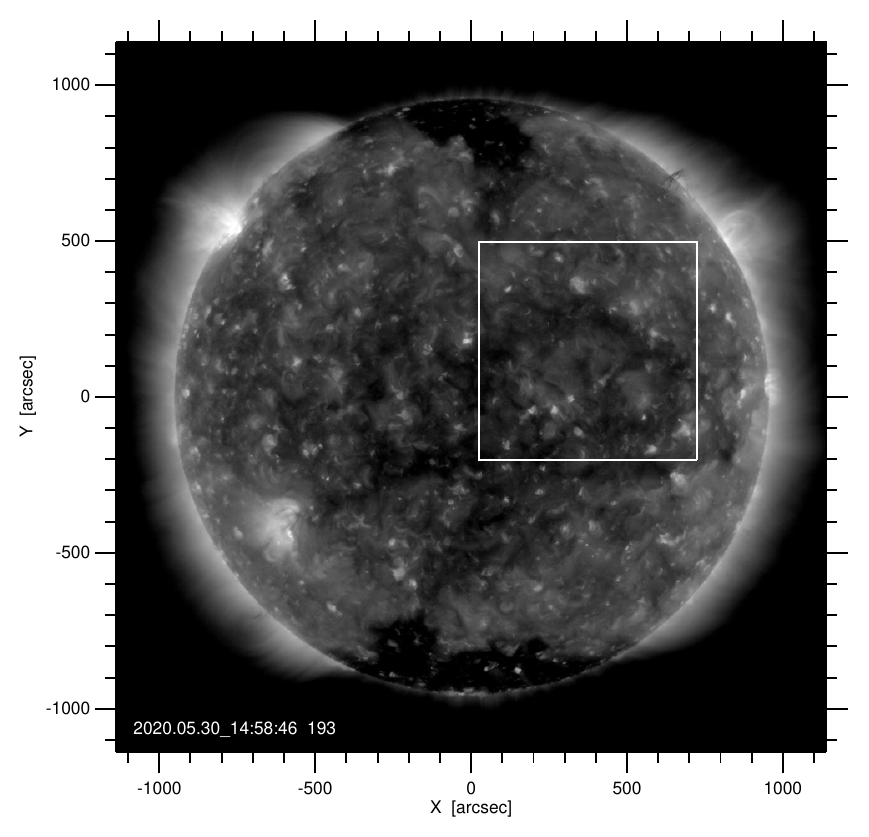}
  \caption[Full-disk AIA 193\,\AA\ image]{\label{fig:full193} %
  AIA 193\,\AA\ image at best-match time. 
  This AIA diagnostic also shows coronal connectivity but outlines
  coronal holes (here especially at the poles) and similar areas of
  less optical thickness in hot coronal emissivity clearer.
  ``Fires'' (campfires or brushfires) stand out very well.  
  If you blur your vision the pattern within the box and the scene in
  \rrref{figure}{fig:soloim} agree, demonstrating that I did find the
  latter in SDO images -- eventually. 
  In hindsight, I mistakenly searched AIA 171\,\AA\ movies that looked
  as the preceding figure; with this diagnostic I might have succeeded
  faster.
  }\end{figure*}



\begin{figure*}
  \centering
  \includegraphics[width=\textwidth]{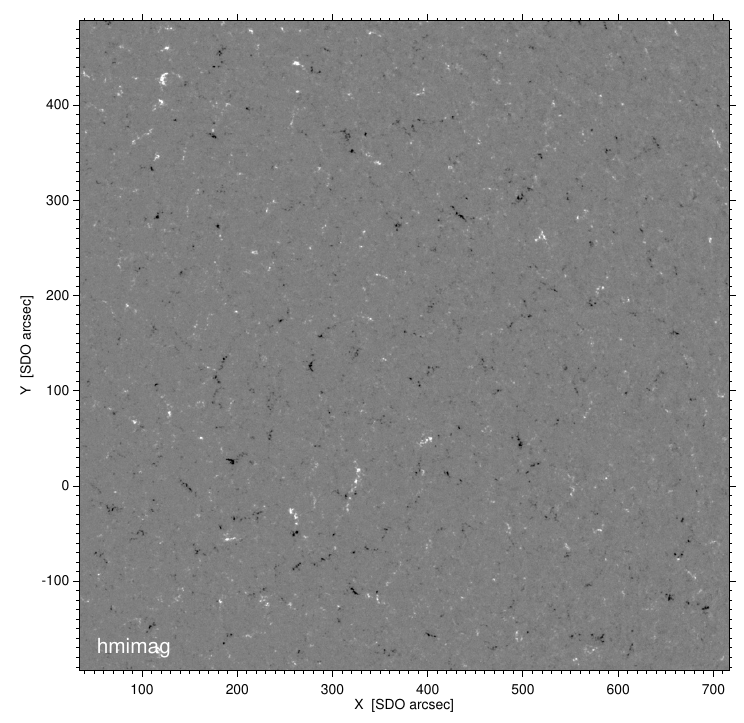}
  \caption[SolO-field HMI magnetogram]{\label{fig:fieldmag} %
  The SDO field at best-match time in the downloaded HMI magnetogram
  sequence.
  The greyscale is clipped at apparent flux density
  $B_{\rm app}^{\rm HMI} = \pm 200$~Mx\,cm$^{-2}$ to enhance the
  visibility of the network fields occupying this quiet area. 
  The subscript {\em app\/} stands for ``apparent'' following
  \citetads{1999ApJ...517.1013L} 
  to emphasize that while these units formally equal Gauss units, the
  actual intrinsic field strength in the magnetic concentrations (MC)
  charted here is much higher, of kilogauss amplitude. 
  The superscript {\em HMI\/} specifies the apparent/intrinsic ratio
  as a specific instrument property. 
  The small black and white (``bipolar'') MCs are roughly arranged in
  the apparent ``network'' incompletely outlining supergranular
  convection cells (\eg\
  \citeads{1997ApJ...487..424S}). 
  The grey internetwork in between shows just noise due to low
  sensitivity. 
  Charting internetwork fields needs higher sensitivity, as with
  Hinode/SP (\eg\
  \citeads{2008ApJ...672.1237L}). 
  The area is quiet and generally mixed-polarity. 
  On small scales there appear to be many monopolar clusters of a
  dozen or so same-color grains, but with better magnetic sensitivity
  these likely will show opposite-polarity concentrations as well.
  In higher resolution observations all such MCs are no longer
  pointlike but have fast-varying morphology following local
  intergranular-lane dynamics (\eg\ the SST ``flowers'' of
  \citeads{2004A&A...428..613B}). 
  }\end{figure*}
\begin{figure*}
  \centering
  \includegraphics[width=\textwidth]{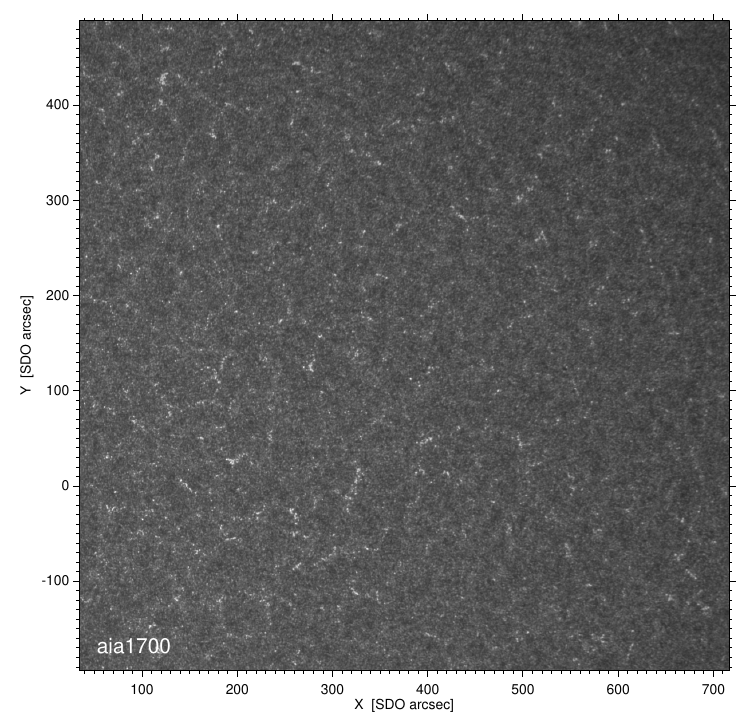}
  \caption[SolO-field AIA 1700\,\AA\ image]{\label{fig:field1700} %
  The SDO field at best-match time in the AIA 1700\,\AA\ image.
  The greyscale shows the square root of the intensity after clipping
  the brightest pixels in the 15-min sequence. 
  The scene shows ``bright points'' closely corresponding to the
  MCs in the preceding figure and roughly arranged in quasi-cellular
  supergranulation-driven network patterns with greyish internetwork
  shock interference patterns inside. 
  When playing the sequence as a movie the bright points remain fairly
  stationary while the shock patterns move around very fast in erratic
  fashion. 
  The latter are not discussed here but were brilliantly identified by
  \citetads{1997ApJ...481..500C} 
  as clapotispheric shocks
  (\linkadspage{1995ESASP.376a.151R}{8}{Figure~12} of
  \citeads{1995ESASP.376a.151R}) 
  driven and patterned by $p$-mode interference.
  }\end{figure*}
\begin{figure*}
  \centering
  \includegraphics[width=\textwidth]{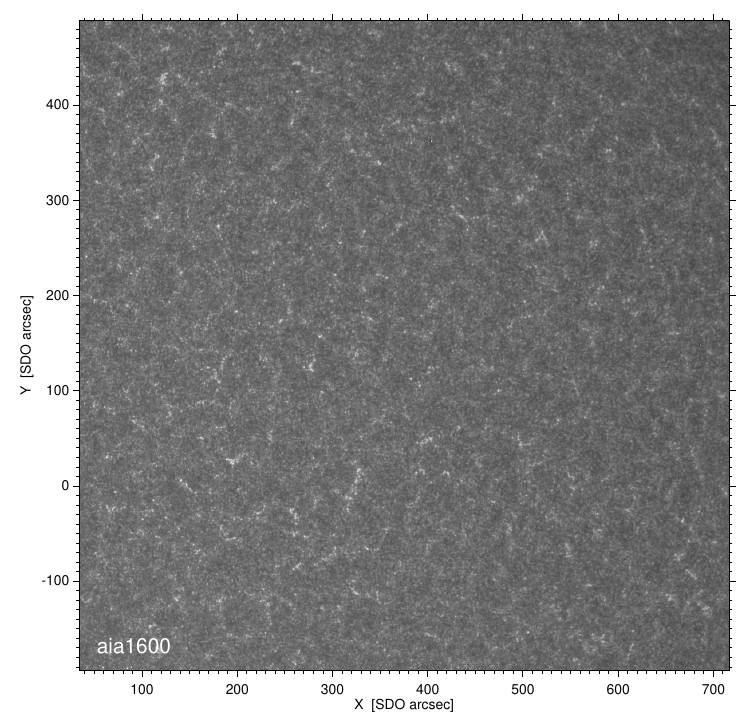}
  \caption[SolO-field AIA 1600\,\AA\ image]{\label{fig:field1600} %
  The SDO field at best-match time in the AIA 1600\,\AA\ image.
  The greyscale shows the square root of the intensity after clipping
  the brightest pixels in the 15-min sequence.
  The scene is very similar to the one in the 1700\,\AA\ image in
  preceding \rrref{figure}{fig:field1700}, but in blinking at sufficient
  zoom-in the 1600\,\AA\ bright points appear shifted limbward with
  different morphology. 
  This is detailed in \rrref{appendix}{sec:shift} and in
  \rrref{figure}{fig:uvcartoons} with a review-style caption.
  }\end{figure*}
\begin{figure*}
  \centering
  \includegraphics[width=\textwidth]{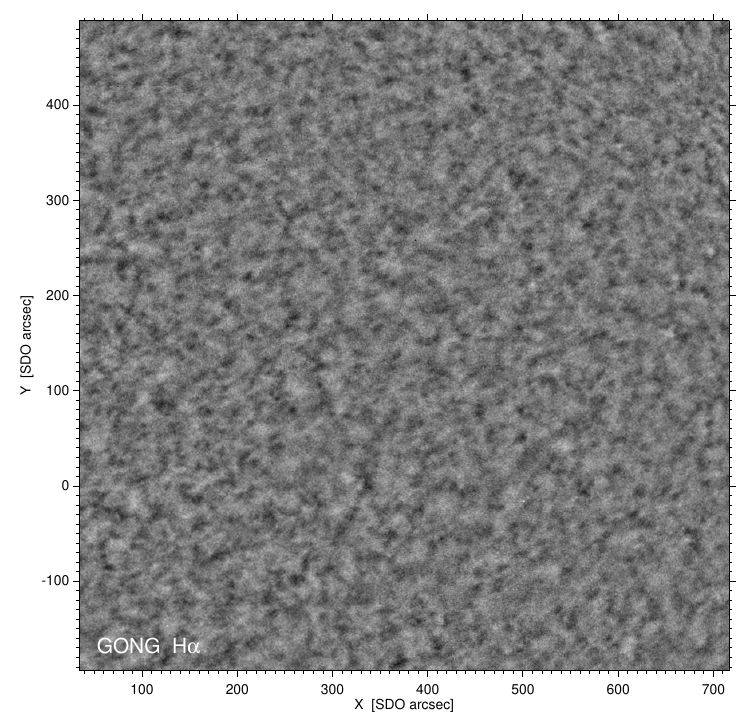}
  \caption[SolO-field in GONG \Halpha\ image] {\label{fig:fieldha} %
  The SDO field at best-match time in a simultaneous Big Bear GONG
  \Halpha\ image. 
  The cutout is bytescaled after severe clipping and limb darkening
  removal. 
  The angular resolution is too low to show that the darker, roughly
  linear, patches are made up of fibrils. 
  Higher-resolution \Halpha\ imaging would show that these emanate
  from network. 
  Yet-higher-resolution outer-wing \Halpha\ imaging would show slender
  recurrent RBEs and RREs emanating further from network
  (\linkadspage{2013ApJ...769...44S}{6}{Figure~4} of
  \citeads{2013ApJ...769...44S}). 
  \rrref{Appendix}{sec:ha304} argues that the latter produce the former.
  The plans for a DOT-clone GONG successor
  (\linkadspage{2019BAAS...51g..74H}{8}{Fig.~13} of
  \citeads{2019BAAS...51g..74H}) 
  include \Halpha\ tuning to the outer wings.
  The present GONG network takes full-disk images of this quality
  sampling the wide core of \Halpha\ at 20\,s or slower cadence with
  24/7 coverage about 90\% since 2009, before SDO.
  The resolution is nominally twice worse but worsens further by
  station-dependent and time-dependent seeing also causing
  rubber-sheet warping. 
  See the \href{https://nso.edu/data/nisp-data/h-alpha}{GONG \Halpha\
  fact sheet}.
  }\end{figure*}
\begin{figure*}
  \centering
  \includegraphics[width=\textwidth]{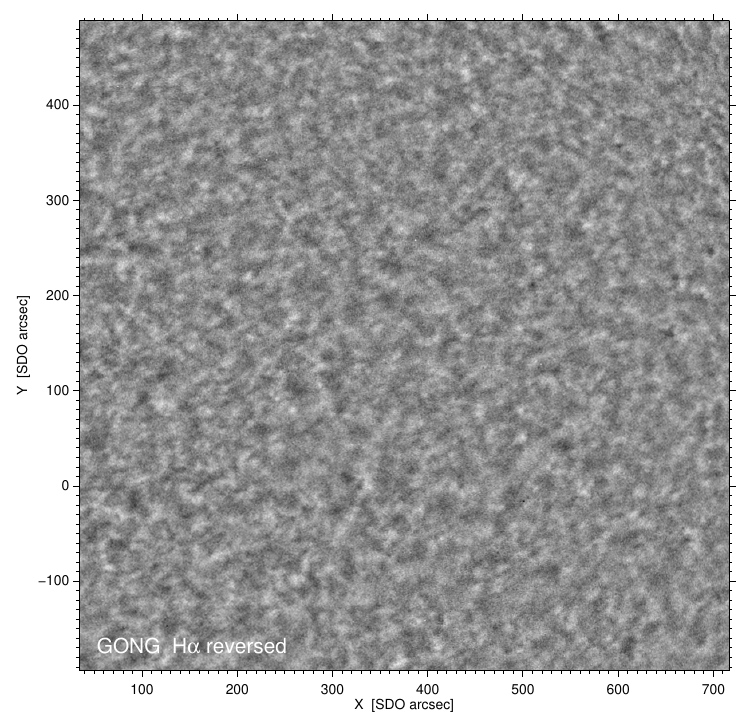}
  \caption[SolO-field in reversed GONG \Halpha\ image]
  {\label{fig:fieldharev} %
  The same GONG \Halpha\ image as in preceding
  \rrref{figure}{fig:fieldha}, but with the greyscale reversed.
  This image is inserted here to facilitate blinking to the next (AIA
  304\,\AA). 
  How to blink pdf pages in a pdf reader is treated on
  \rrref{page}{sec:blinking}. 
  When you blink them the scenes appear to jump due to your eye
  detecting patch size differences but they are actually well aligned.
  This comparison is striking and discussed in \rrref{appendix}{sec:ha304}.
  }\end{figure*}
\begin{figure*}
  \centering
  \includegraphics[width=\textwidth]{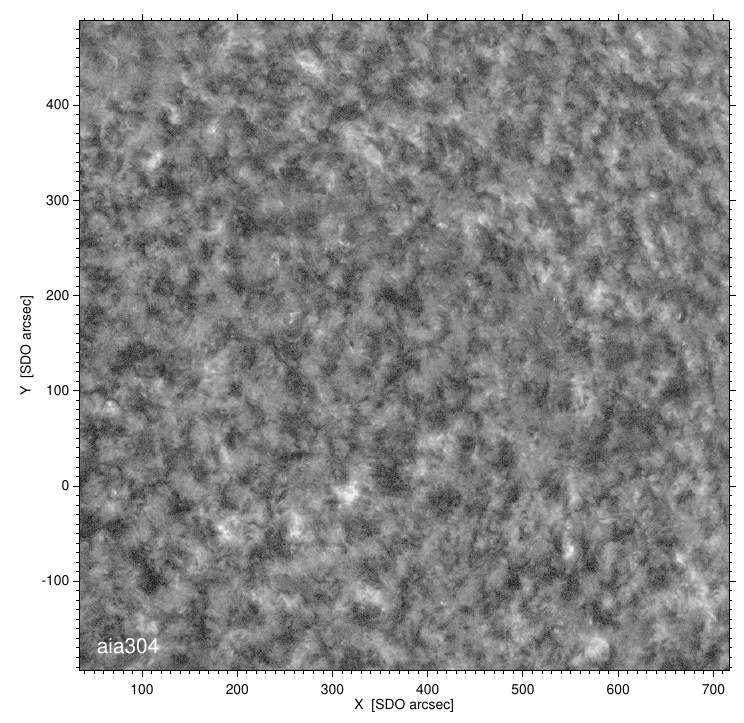}
  \caption[SolO-field AIA 304\,\AA\ image]{\label{fig:field304} %
  The SDO field at best-match time in the AIA 304\,\AA\ image.
  The greyscale shows the logarithm of the intensity in rescaling with
  \href{https://webspace.science.uu.nl/~rutte101/rridl/sdolib/sdo_intscale.pro}{\tt
  sdo\_intscale.pro} and clipping the brightest pixels in the 15-min
  sequence. 
  Unfortunately, the AIA 304\,\AA\ channel lost much sensitivity over
  the years; initially the signal-to-noise was much better. 
  The campfires detailed in \rrref{section}{sec:campfires} are not very
  obvious but blinking against the next images shows they are here
  too. 
  Their presence here and in the next AIA 131\,\AA\ image led to the
  SDO fire detector defined in \rrref{figure}{fig:sdoroiloc304x131} and
  shown as second member in the triples of
  \rrref{figures}{fig:sdo-minx-193}--\ref{fig:sdo-sp-min-mag}.
  The brighter patches mark brushfires (\rrref{section}{sec:brushfires})
  that are also seen better in the next images. 
  Blinking back to the preceding reversed \Halpha\ image in
  \rrref{figure}{fig:fieldharev} shows rough but remarkable overall
  pattern correspondences everywhere for the grainy grey chromosphere
  patches.
  This similarity is discussed in \rrref{appendix}{sec:ha304}.
  }\end{figure*}
\begin{figure*}
  \centering
  \includegraphics[width=\textwidth]{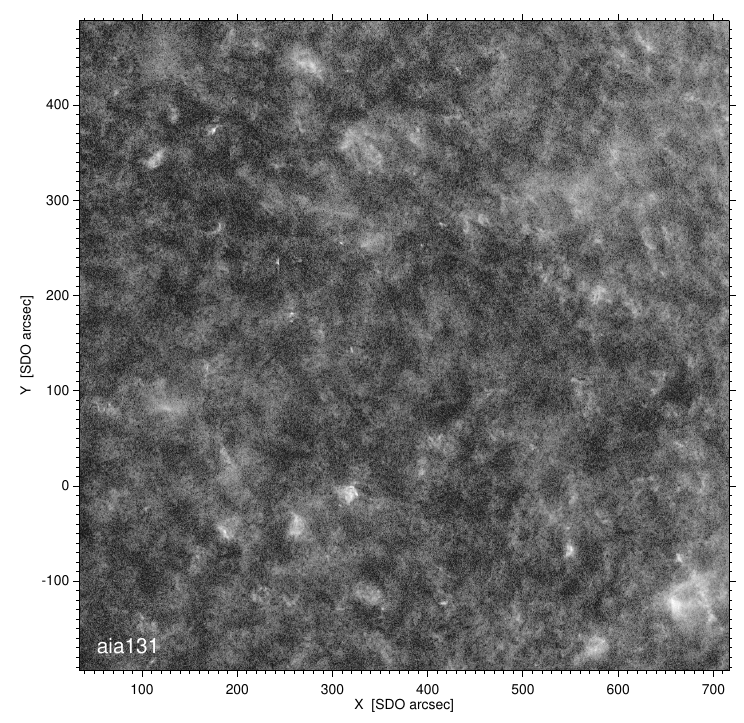}
  \caption[SolO-field AIA 131\,\AA\ image]{\label{fig:field131} %
  The SDO field at best-match time in the AIA 131\,\AA\ image.
  The greyscale shows the logarithm of the intensity in rescaling with
  \href{https://webspace.science.uu.nl/~rutte101/rridl/sdolib/sdo_intscale.pro}{\tt
  sdo\_intscale.pro} and clipping the brightest pixels in the 15-min
  sequence. 
  This image is also noisy because the scene is so quiet, but it shows
  strict grey-patch correspondence with AIA 304\,\AA\ when blinking
  against preceding \rrref{figure}{fig:field304}, plus more diffuse
  brightenings that are clearer and more extended in the next AIA
  171\,\AA\ image. 
  Some campfires already stand out as tiny bright features.
  In \rrref{figure}{fig:sdoroiloc304x131} this image is multiplied with
  the preceding 304\,\AA\ image to demonstrate feasibility of global
  SDO fire detection.
  }\end{figure*}
\begin{figure*}
  \centering
  \includegraphics[width=\textwidth]{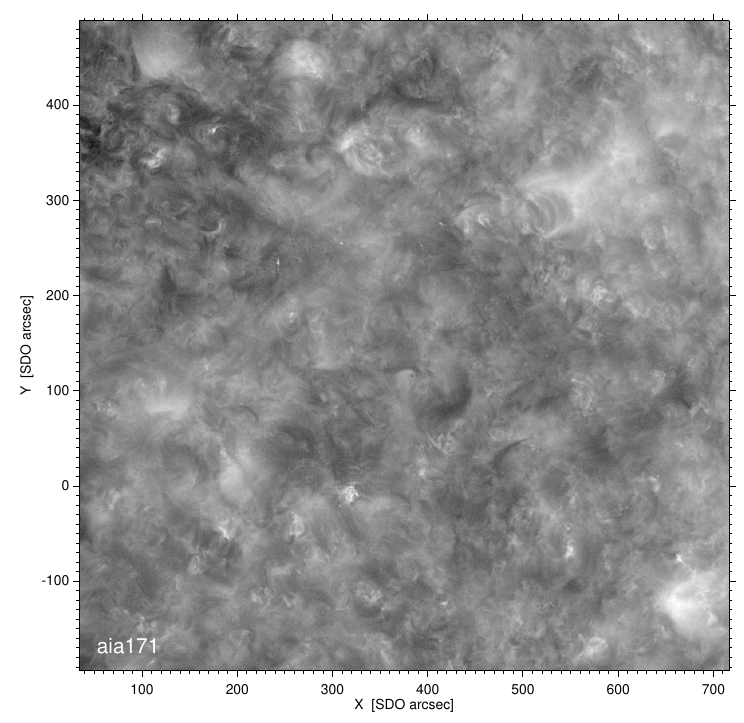}
  \caption[SolO-field AIA 171\,\AA\ image]{\label{fig:field171} %
  The SDO field at best-match time in the AIA 171\,\AA\ image.
  The greyscale shows the logarithm of the intensity in rescaling with
  \href{https://webspace.science.uu.nl/~rutte101/rridl/sdolib/sdo_intscale.pro}{\tt
  sdo\_intscale.pro} and clipping the brightest pixels in the
  15-min sequence. 
  Blinking against the 131\,\AA\ scene in \rrref{figure}{fig:field131}
  shows addition of diffuse coronal connections. 
  I expected this image to be the one to be compared to the SolO
  press-release  174\,\AA\ image in \rrref{figure}{fig:soloim}.
  They cannot be blinked directly because they differ much in
  foreshortening, but the target scene is indeed the same.
  Actually the next one suits better. 
  }\end{figure*}
\begin{figure*}
  \centering
  \includegraphics[width=\textwidth]{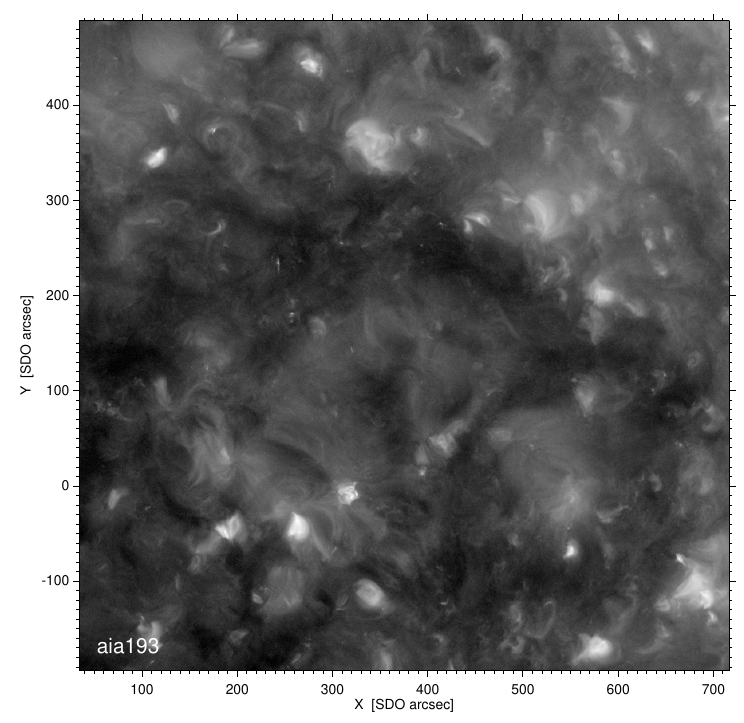}
  \caption[SolO-field AIA 193\,\AA\ image]{\label{fig:field193} %
  The SDO field at best-match time in the AIA 193\,\AA\ image.
  The greyscale shows the logarithm of the intensity in rescaling with
  \href{https://webspace.science.uu.nl/~rutte101/rridl/sdolib/sdo_intscale.pro}{\tt
  sdo\_intscale.pro} and clipping the brightest pixels in the 15-min
  sequence. 
  It shows yet more hazy coronal connectivity than the 171\,\AA\ image
  in preceding \rrref{figure}{fig:field171}.
  The tiny campfires stand out brightly where they occur in dark
  areas. 
  This image shows the best correspondence with the input
  press-release scene in \rrref{figure}{fig:soloim}, perhaps through
  the common high-intensity tails in next \rrref{figure}{fig:hists}.
  The larger bright patches are treated as brushfires in
  \rrref{section}{sec:brushfires}. 
  They represent classical coronal bright points (review by
  \citeads{2019LRSP...16....2M}). 
  They are smaller and more homogeneously bright than in 171 and
  131\,\AA\ where they partly contain darker fibrils between brighter
  feet.
  }\end{figure*}

\begin{figure*}
  \centering
  \includegraphics[width=8cm]{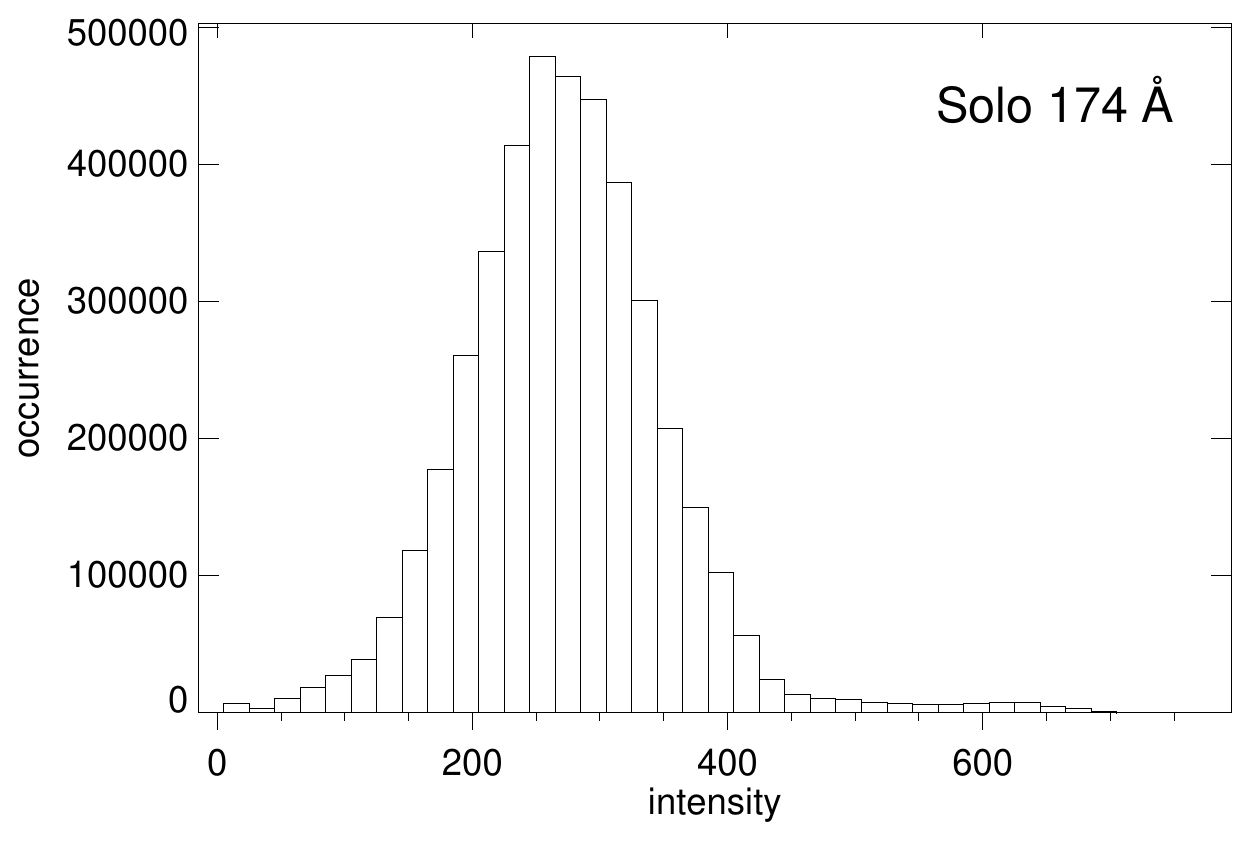}
  \includegraphics[width=8cm]{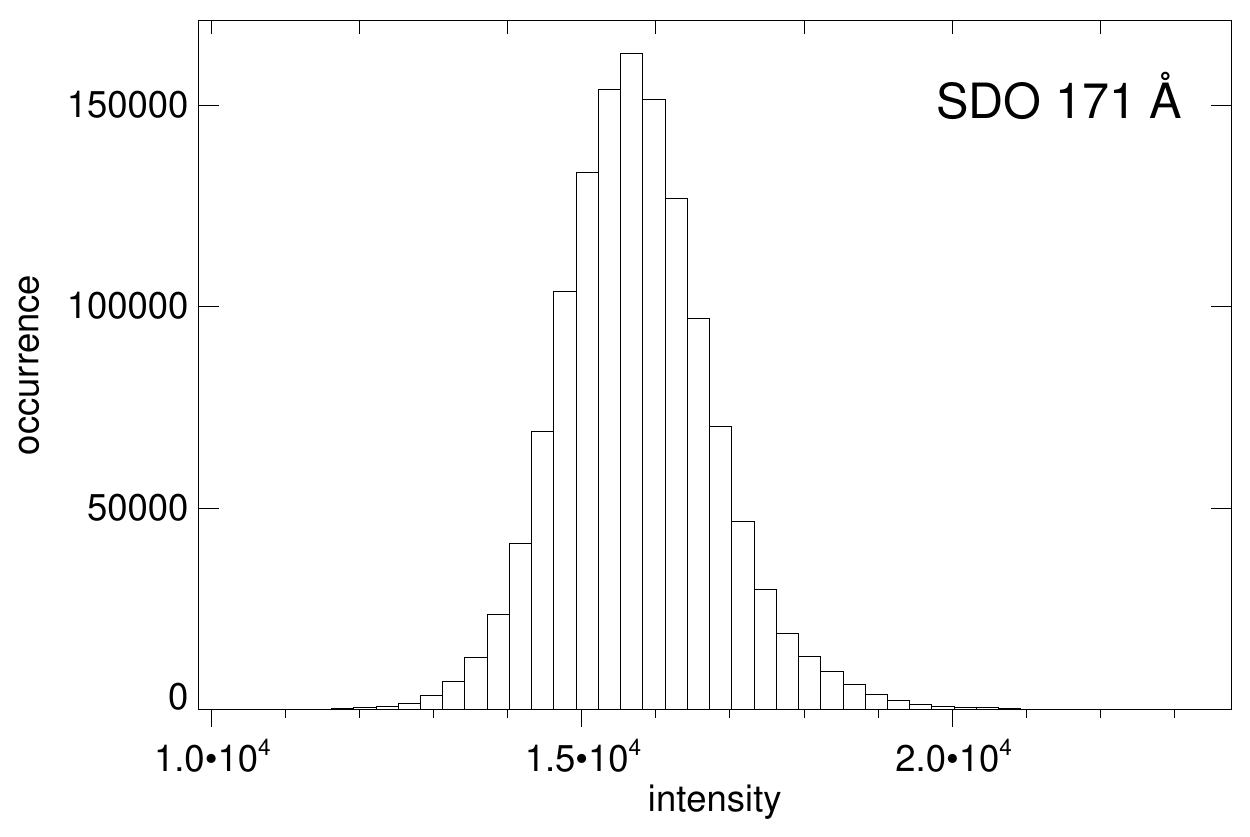}\\
  \includegraphics[width=8cm]{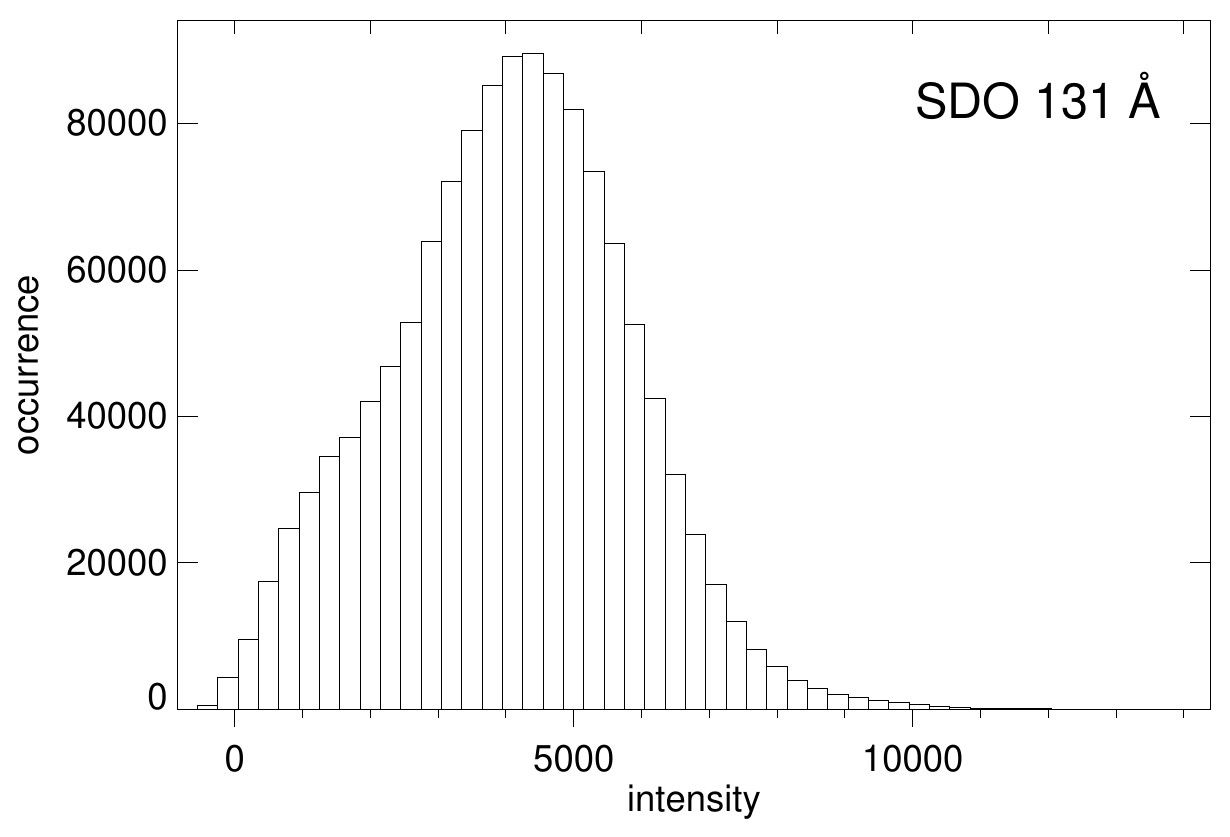}
  \includegraphics[width=8cm]{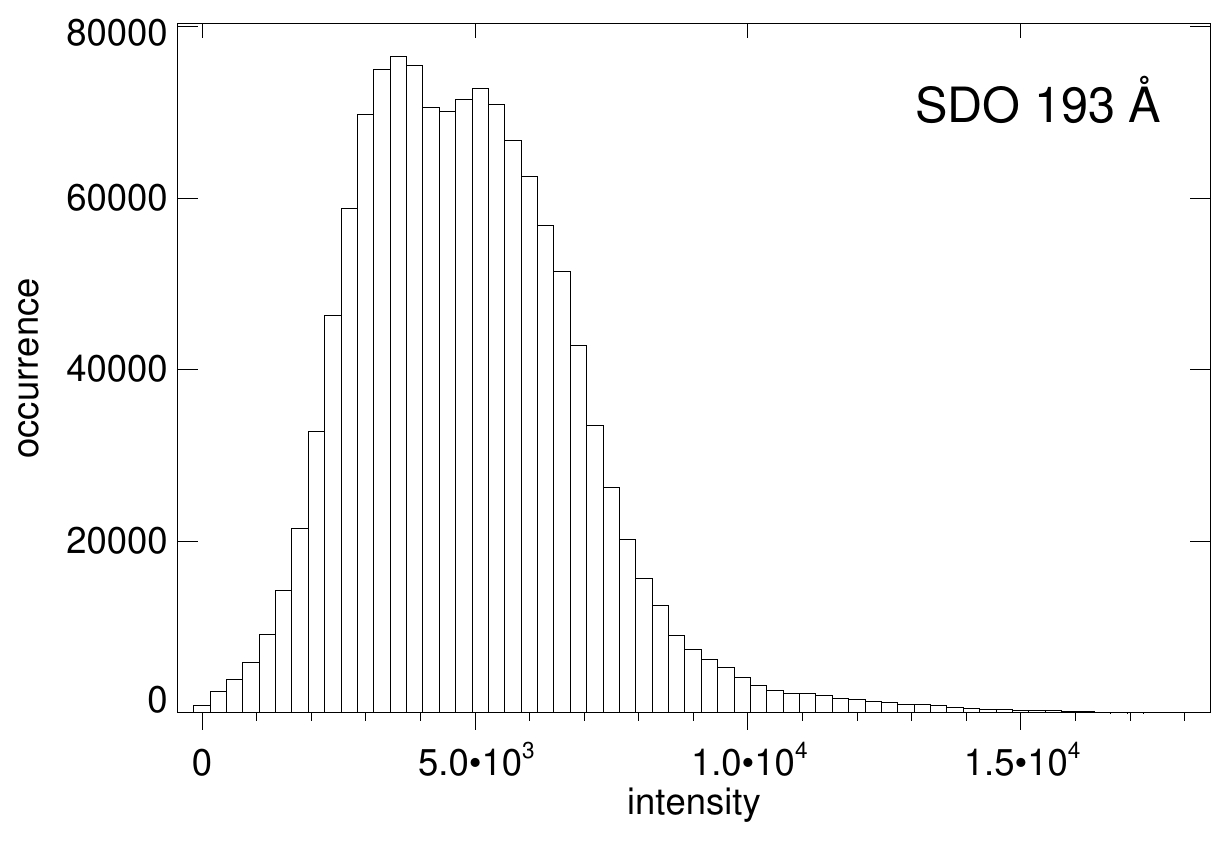}
  \caption[Pixel intensity histograms]{\label{fig:hists} %
  Histograms of the pixel intensities in the SolO 174\,\AA\ image and
  in three SDO images at the best-match time, respectively SDO
  171\,\AA, 131\,\AA\ and 193\,\AA. 
  Extended highest-brightness tails show up in all but especially in
  the SolO 174\,\AA\ and AIA 193\,\AA\ distributions. 
  The latter also shows a double-hump peak; there are indeed many dark
  areas in \rrref{figure}{fig:field193}.
  }\end{figure*}


\begin{figure*}
  \centering
  \includegraphics[width=\textwidth]{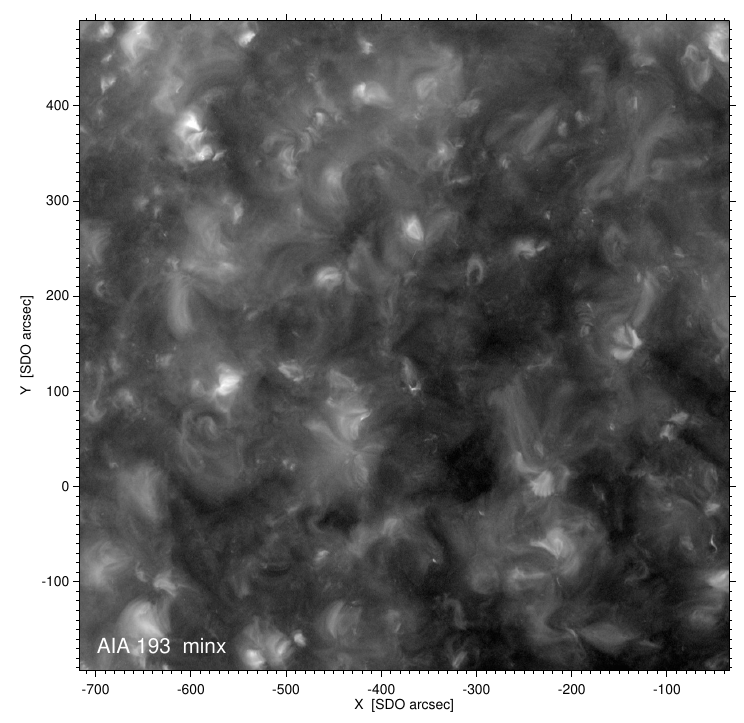}
  \caption[AIA 193\,\AA\ image in the North-East]
  {\label{fig:sdo-minx-193} %
  \addtocline{SDO other-field triples} Comparable AIA 193\,\AA\ cutout
  for May 30, 2020 14:58:46~UT as in \rrref{figure}{fig:field193} but
  East at $(X,Y) = (-375,148)$.
  Just as the SolO field this same-size other field shows about a
  dozen tiny campfires and more larger brushfires.
  Wider-connecting diffuse 193\,\AA\ brightness patterns occur mostly
  around brushfires and suggest origin in earlier or persistent
  brushfires. 
  }\end{figure*}
\begin{figure*}
  \centering
  \includegraphics[width=\textwidth]{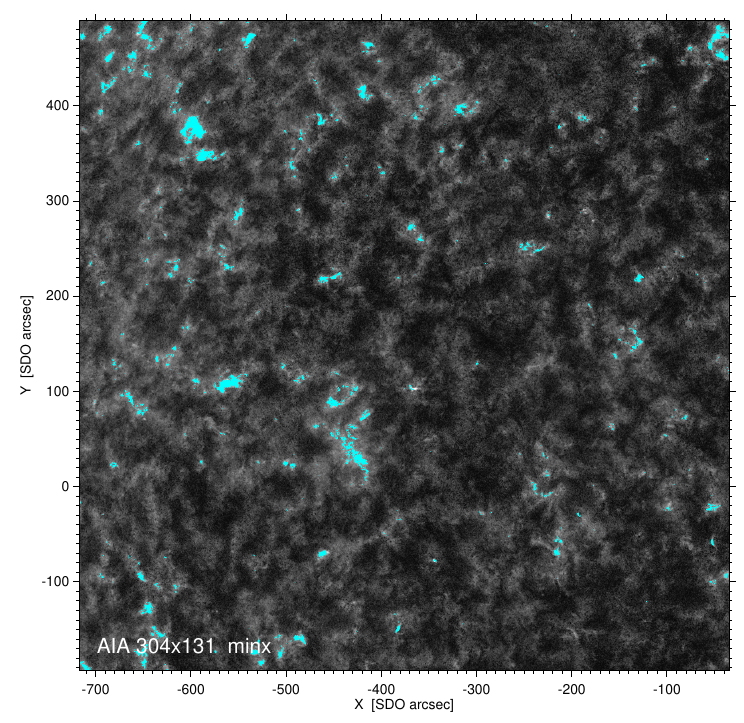}
  \caption[SDO fire detector in the North-East]
  {\label{fig:sdo-minx-fire} %
  SDO fire detector cutout for the same field as in preceding
  \rrref{figure}{fig:sdo-minx-193}.
  Its construction is described in the caption of
  \rrref{figure}{fig:sdoroiloc304x131}.
  It serves to separate chromospheric heating (ubiquitous grey
  patches) and coronal heating (small dispersed areas colored cyan).
  In these and the following ``other SDO scene'' triples these fire
  detector images are inserted between the AIA 193\,\AA\ images and
  the HMI magnetograms for blinking backward and forward with these.
  It shows that many small bright 193\,\AA\ loop bundles
  constituting larger brushfires have cyan-colored detector
  brightenings at their feet located in bipolar network.
  The smallest point-like campfires go without such closed-loop
  connectivity.
  }\end{figure*}
\begin{figure*}
  \centering
  \includegraphics[width=\textwidth]{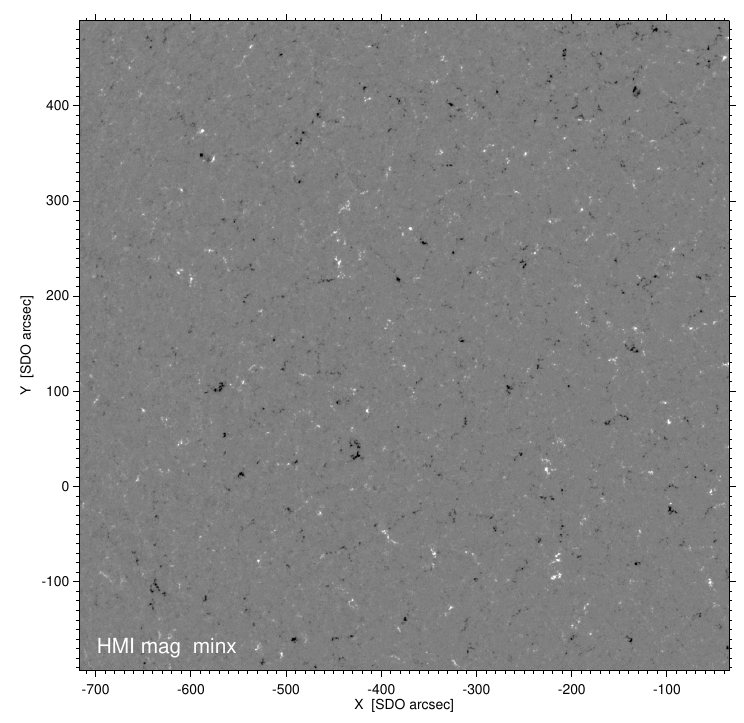}
  \caption[HMI magnetogram in the North-East]
  {\label{fig:sdo-minx-mag} %
  HMI magnetogram for the field in the two preceding figures.
  The field strengths are clipped as described for
  \rrref{figure}{fig:fieldmag}.
  Blinking with the two preceding figures shows that small
  ``brushfire'' loop bundles in 193\,\AA\ connect small groups of
  opposite-polarity MCs.
  }\end{figure*}

\begin{figure*}
  \centering
  \includegraphics[width=\textwidth]{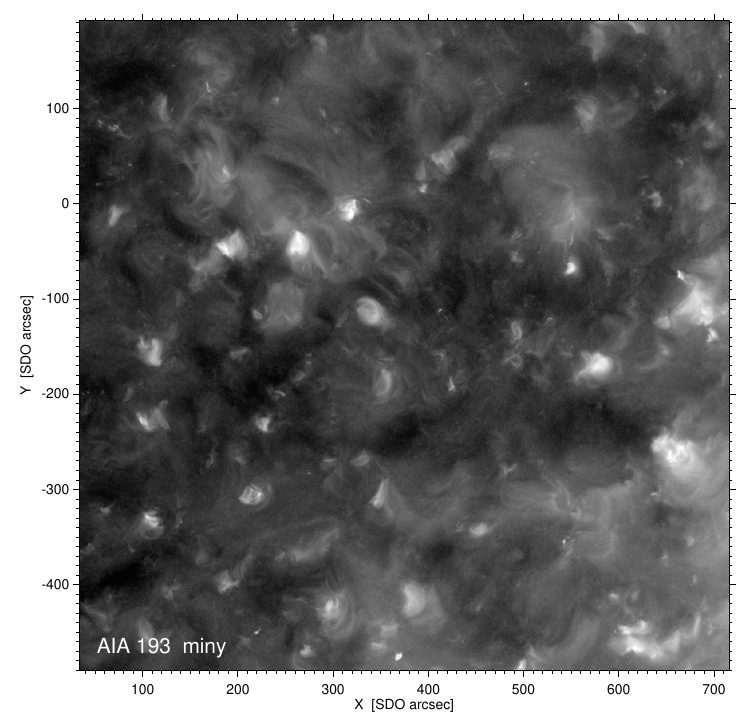}
  \caption[AIA 193\,\AA\ image in the South-West]
  {\label{fig:sdo-miny-193} %
  Comparable AIA 193\,\AA\ cutout for May 30, 2020 14:58:46~UT T as in
  \rrref{figure}{fig:field193} but South at $(X,Y) = (375,-148)$.
  Again about a dozen tiny campfires and more larger brushfires.
  }\end{figure*}
\begin{figure*}
  \centering
  \includegraphics[width=\textwidth]{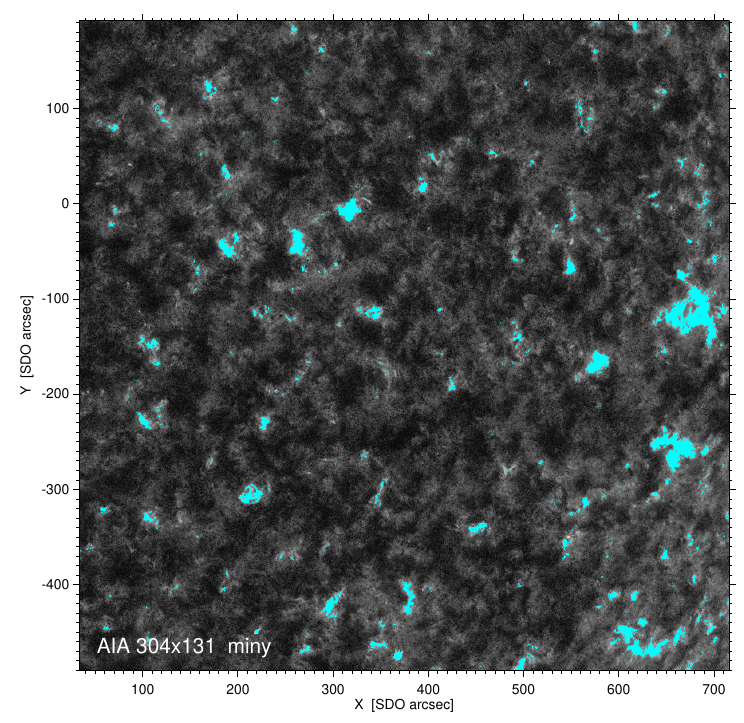}
  \caption[SDO fire detector image in the South-West]
  {\label{fig:sdo-miny-fire} %
  SDO fire detector cutout for the same field as in preceding
  \rrref{figure}{fig:sdo-miny-193}.
  }\end{figure*}
\begin{figure*}
  \centering
  \includegraphics[width=\textwidth]{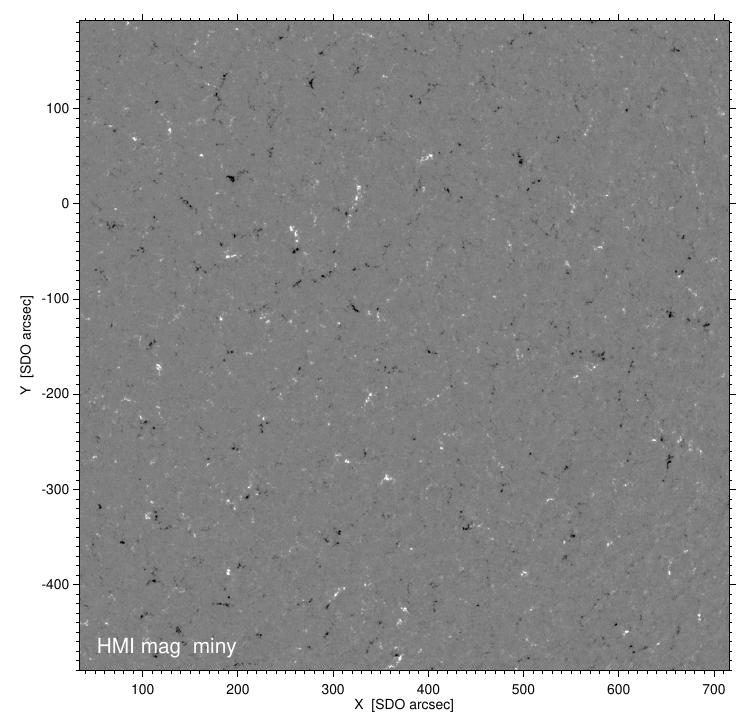}
  \caption[HMI magnetogram in the South-West]
  {\label{fig:sdo-miny-mag} %
  HMI magnetogram for the field in the two preceding figures.
  }\end{figure*}
 
\begin{figure*}
  \centering
  \includegraphics[width=\textwidth]{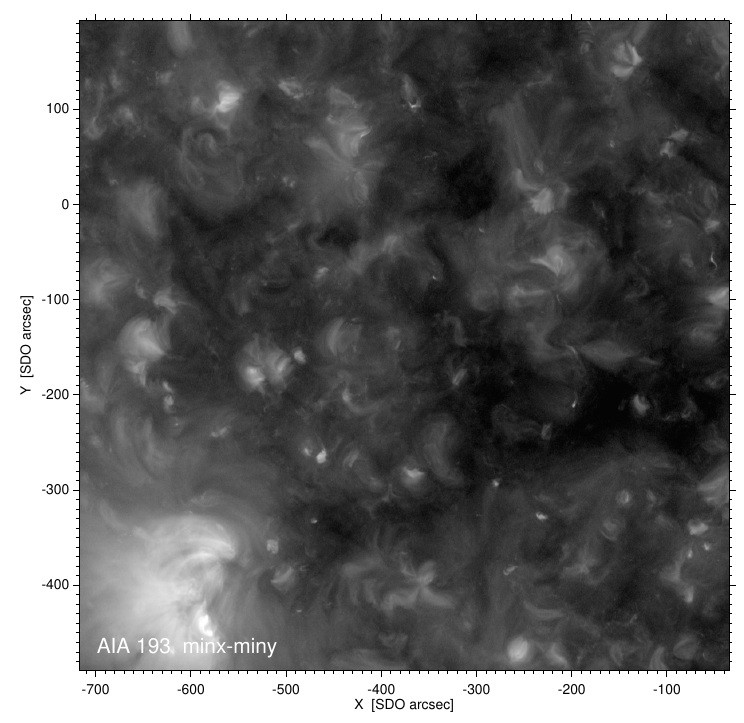}
  \caption[AIA 193\,\AA\ image in the South-East]
{\label{fig:sdo-minx-miny-193} %
  Comparable AIA 193\,\AA\ cutout for May 30, 2020 14:58:46~UT as in
  \rrref{figure}{fig:field193} but South-East at $(X,Y) = (-375,-148)$.
  }\end{figure*}
\begin{figure*}
  \centering
  \includegraphics[width=\textwidth]{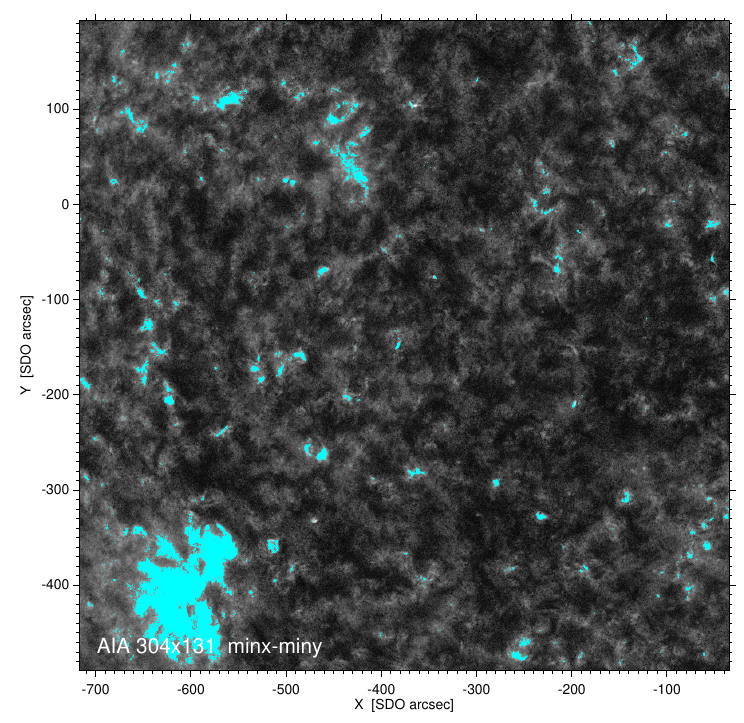}
  \caption[SDO fire detector image in the South-East]
  {\label{fig:sdo-minx-miny-fire} %
  SDO fire detector cutout for the same field as in the preceding
  figure.  
  Around the small active plage at lower-left some campfires are
  veiled by coronal loops in the preceding AIA 193\,\AA\ image
  put appear unveiled in this chart.
  }\end{figure*}
\begin{figure*}
  \centering
  \includegraphics[width=\textwidth]{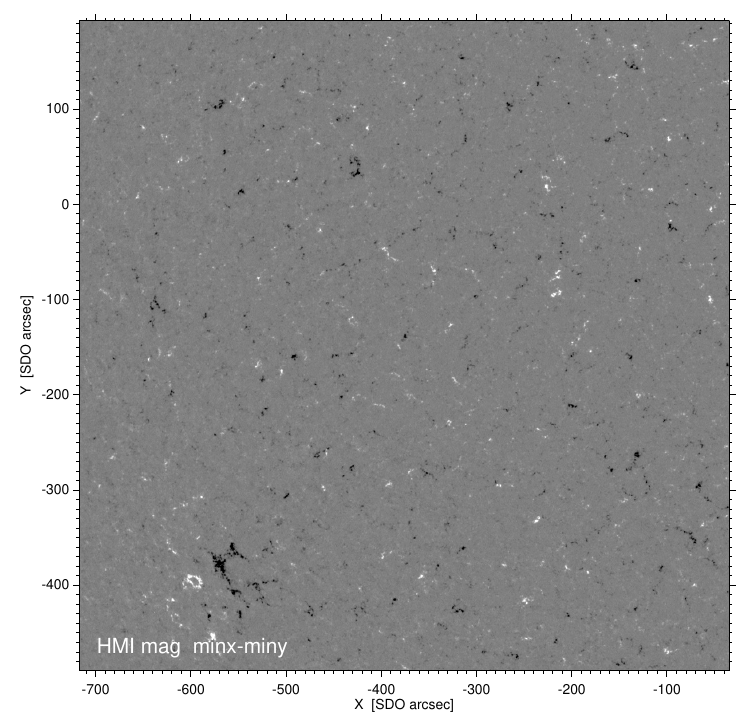}
  \caption[HMI magnetogram in the South-East]
  {\label{fig:sdo-minx-miny-mag} %
  HMI magnetogram for the field in the two preceding figures.
  }\end{figure*}


\begin{figure*}
  \centering
  \includegraphics[width=\textwidth]{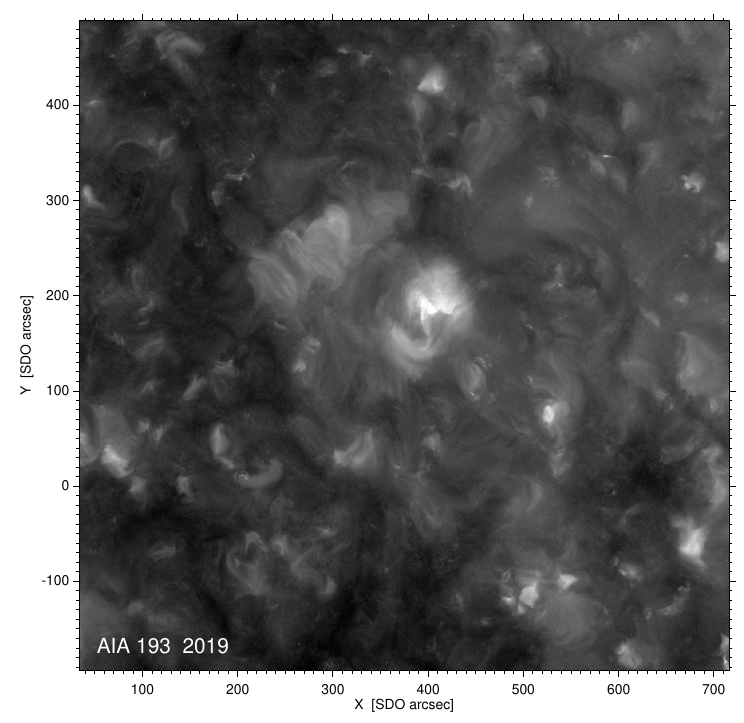}
  \caption[SolO-field AIA 193\,\AA\ image in 2019]
  {\label{fig:sdo2019-193} %
  Comparable AIA 193\,\AA\ cutout for the same $(X,Y) = (375,148)$
  arcsec location as best-match \rrref{figure}{fig:field193} and also
  on May 30 near 14:58~UT, but in 2019. 
  There is a small active plage near the center.
  }\end{figure*}
\begin{figure*}
  \centering
  \includegraphics[width=\textwidth]{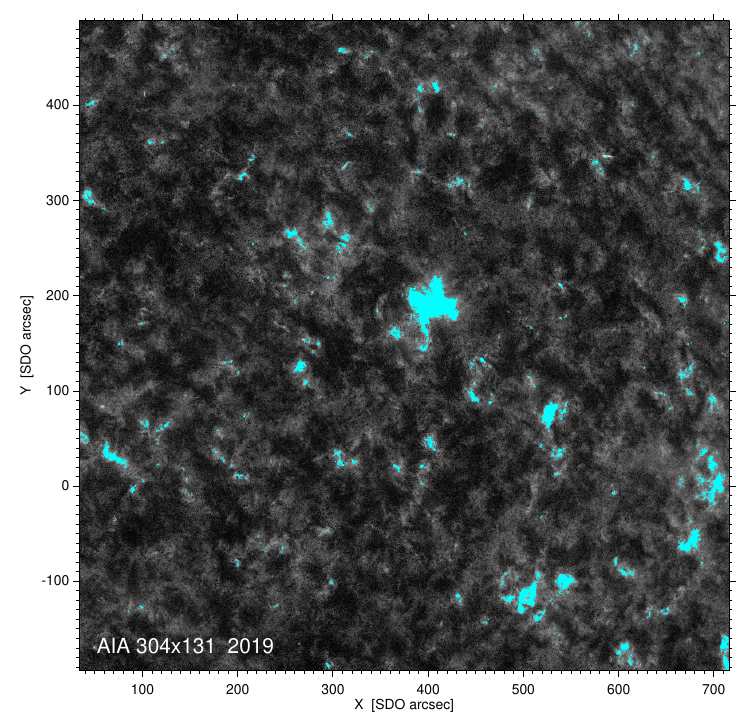}
  \caption[SolO-field SDO fire detector image in 2019]
  {\label{fig:sdo2019-fire} %
  SDO fire detector cutout for the same field as in the preceding
  figure.
  }\end{figure*}
\begin{figure*}
  \centering
  \includegraphics[width=\textwidth]{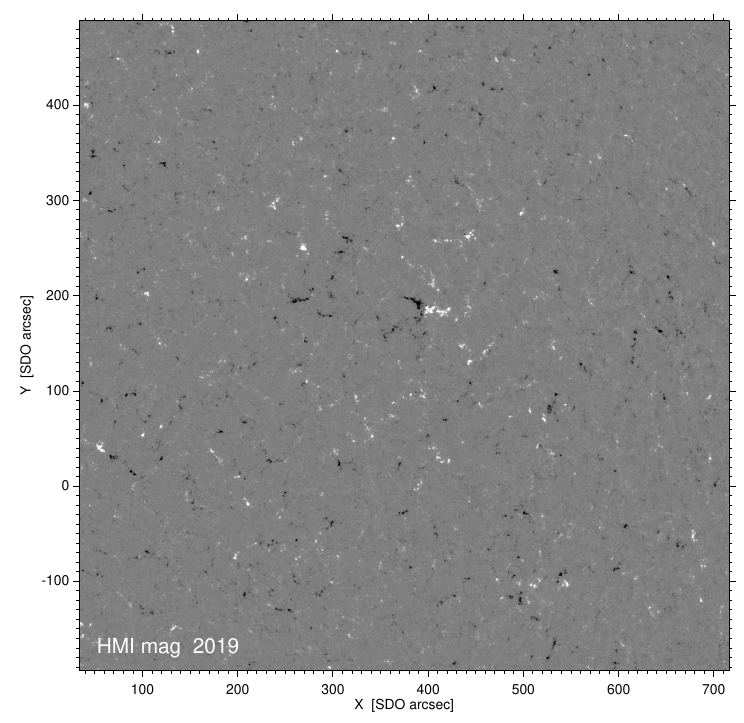}
  \caption[SolO-field HMI magnetogram in 2019]
  {\label{fig:sdo2019-mag} %
  HMI magnetogram for the field in the two preceding figures.
  }\end{figure*}

\begin{figure*}
  \centering
  \includegraphics[width=\textwidth]{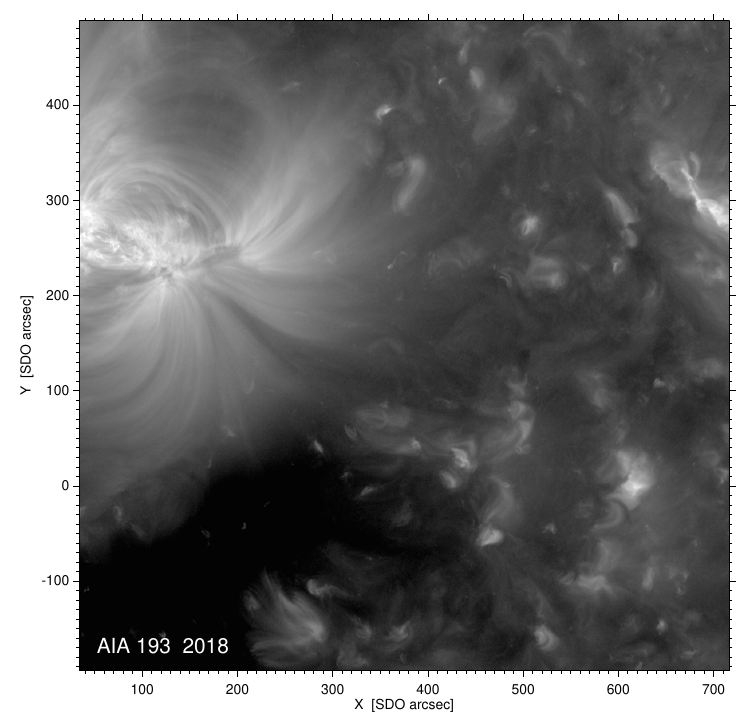}
  \caption[SolO-field AIA 193\,\AA\ image in 2018]
  {\label{fig:sdo2018-193} %
  Comparable AIA 193\,\AA\ cutout for the same $(X,Y) = (375,148)$
  arcsec location as best-match \rrref{figure}{fig:field193} and also on
  May 30 near 14:58~UT, but in 2018. 
  This scene is the most active portrayed in this report. 
  }\end{figure*}
\begin{figure*}
  \centering
  \includegraphics[width=\textwidth]{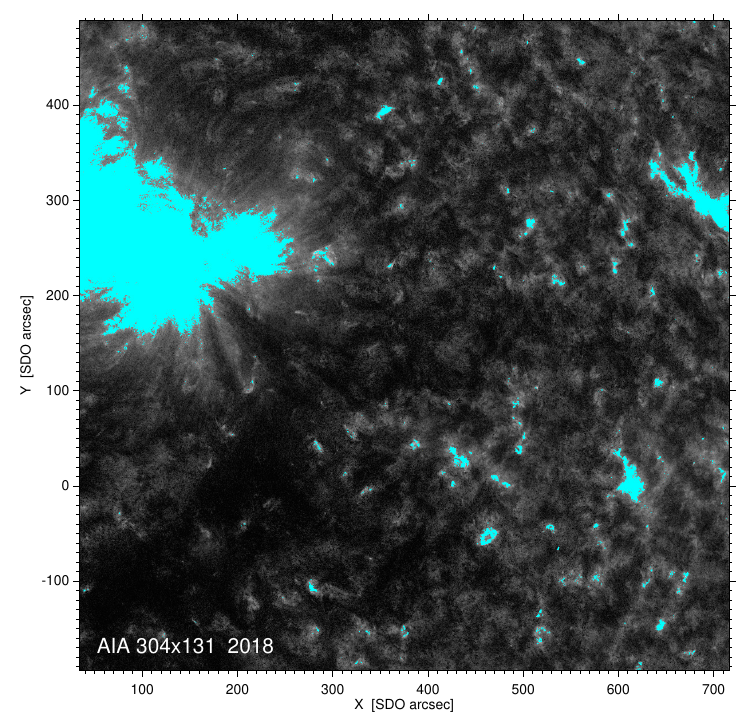}
  \caption[SolO-field SDO fire detector image in 2018]
  {\label{fig:sdo2018-fire} %
  SDO fire detector cutout for the same field as in the preceding
  figure. 
  The bytescale is clipped for the active region to maintain visibilty
  (grey) of the quiet network elsewhere, as described under
  \rrref{figure}{fig:sdoroiloc304x131}.
  }\end{figure*}
\begin{figure*}
  \centering
  \includegraphics[width=\textwidth]{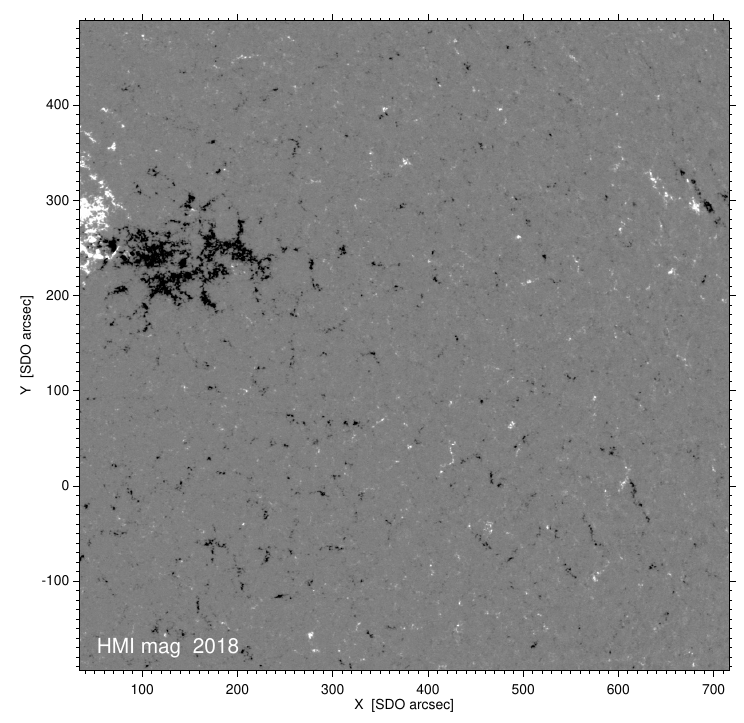}
  \caption[SolO-field HMI magnetogram in 2018]
  {\label{fig:sdo2018-mag} %
  HMI magnetogram for the field in the two preceding figures.
  }\end{figure*}

\begin{figure*}
  \centering
  \includegraphics[width=\textwidth]{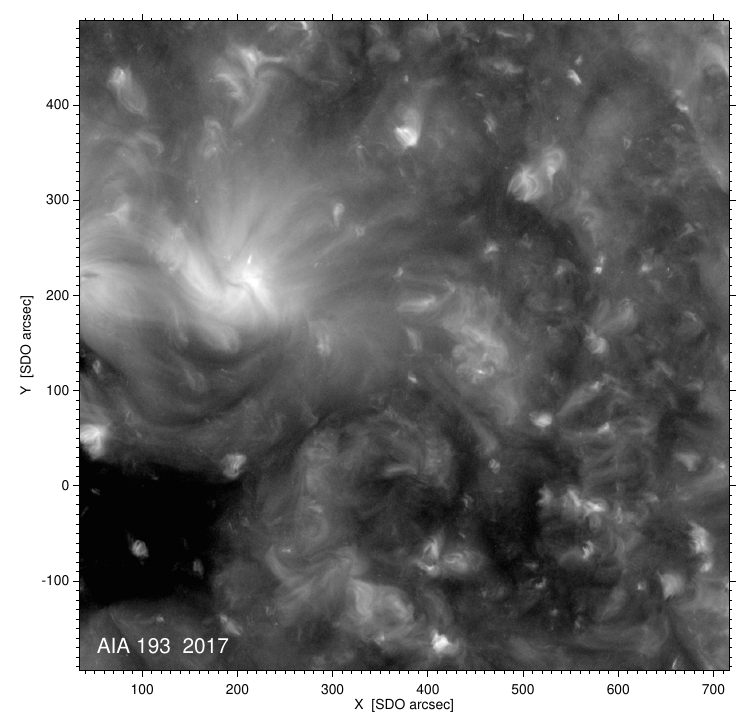}
  \caption[SolO-field AIA 193\,\AA\ image in 2017]
  {\label{fig:sdo2017-193} %
  Comparable AIA 193\,\AA\ cutout for the same $(X,Y) = (375,148)$
  arcsec location as best-match \rrref{figure}{fig:field193} and also on
  May 30 near 14:58~UT, but in 2017.
  }\end{figure*}
\begin{figure*}
  \centering
  \includegraphics[width=\textwidth]{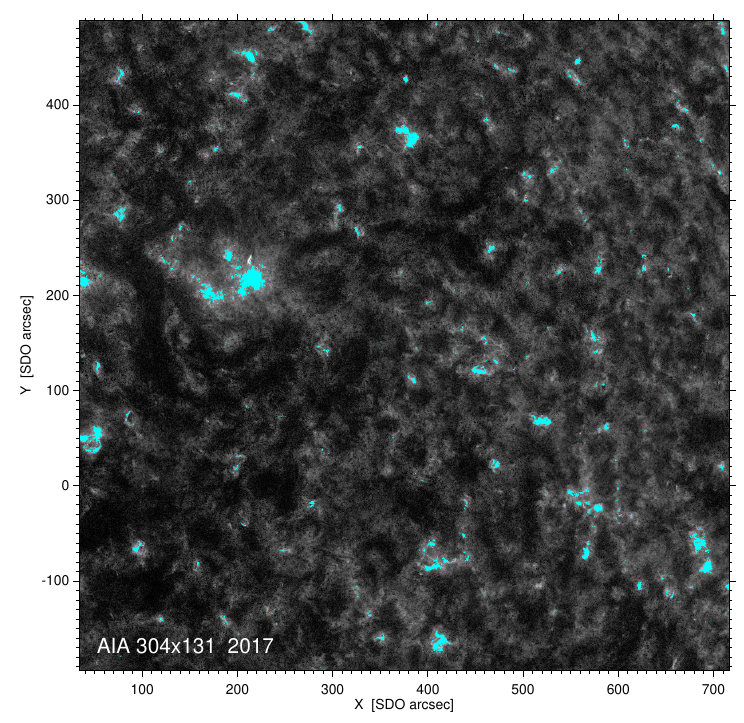}
  \caption[SolO-field SDO fire detector image in 2017]
  {\label{fig:sdo2017-fire} %
  SDO fire detector cutout for the same field as in the preceding
  figure.
  }\end{figure*}
\begin{figure*}
  \centering
  \includegraphics[width=\textwidth]{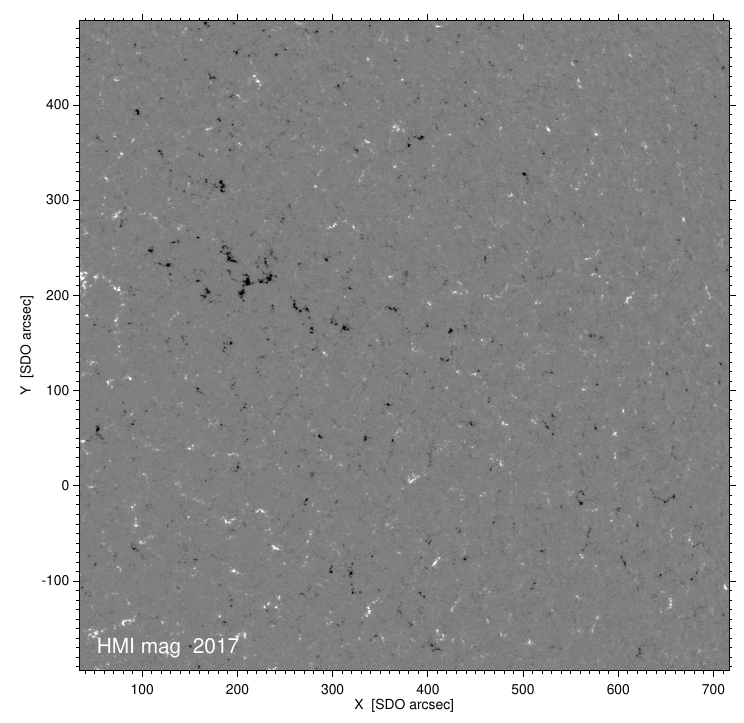}
  \caption[SolO-field HMI magnetogram in 2017]
  {\label{fig:sdo2017-mag} %
  HMI magnetogram for the field in the two preceding figures.
  The small active region is mostly monopolar but at higher
  sensitivity than offered by HMI it is rather likely that there are
  or have been canceling white-polarity MCs.
  }\end{figure*}


\begin{figure*}
  \centering
  \includegraphics[width=\textwidth]{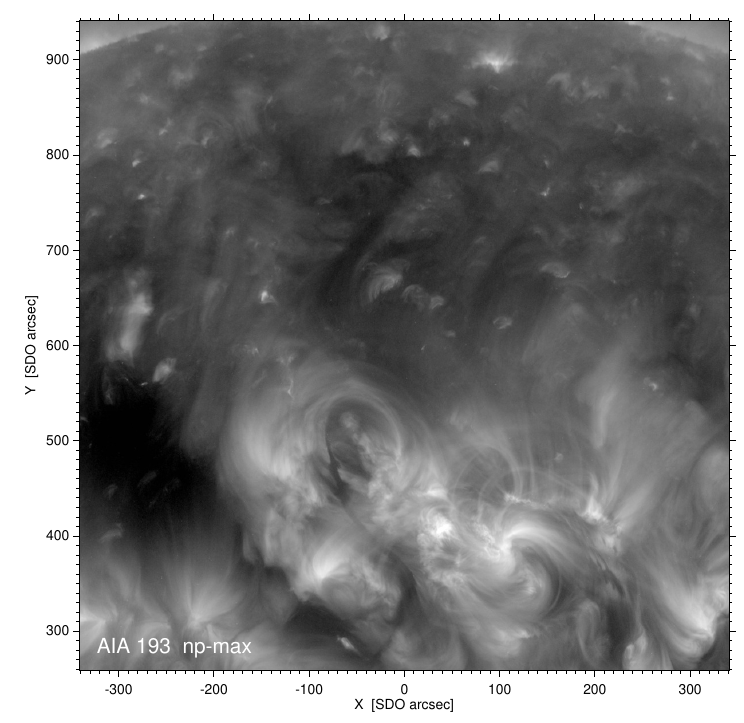}
  \caption[North-pole AIA 193\,\AA\ image at maximum]
  {\label{fig:sdo-np-max-193} %
  AIA 193\,\AA\ cutout near the solar North pole on April 1, 2014 at
  00:00~UT. 
  Much activity at lower latitudes.
  }\end{figure*}
\begin{figure*}
  \centering
  \includegraphics[width=\textwidth]{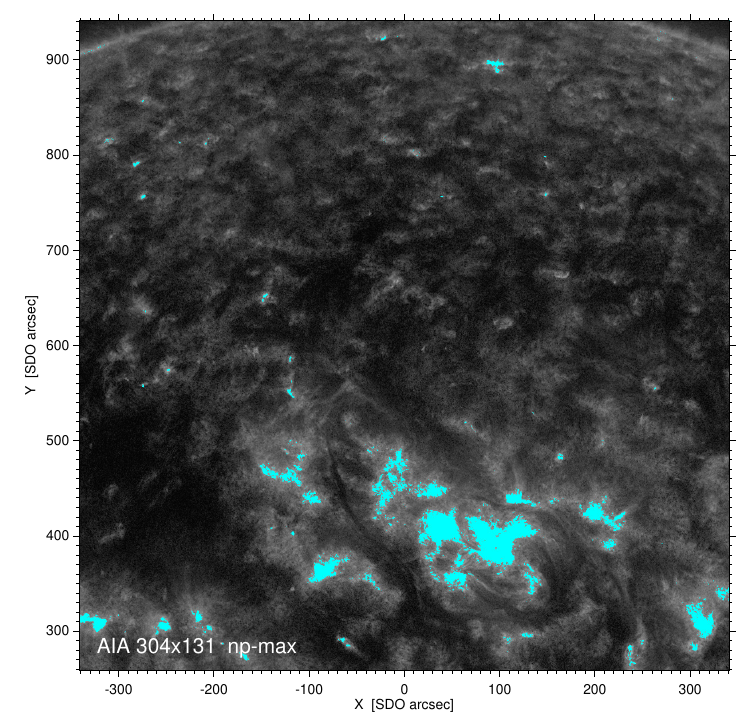}
  \caption[North-pole SDO fire detector image at maximum]
  {\label{fig:sdo-np-max-fire} %
  SDO fire detector cutout for the same field as in the preceding
  figure. 
  There are very few fires here in the upper part North of the active
  regions. 
  All the grey chromosphere patches suggest more. 
  I wonder whether you must look down their throat to see them.  
  I applied ${\tt heightdiff} = 1000$~km correction (see
  \rrref{appendix}{sec:16001700}) to the whole field but this made no
  difference. 
  There may also be too much veiling along the slanted lines of sight
  by all the widely-connected coronal haze in the preceding figure.
  }\end{figure*}
\begin{figure*}
  \centering
  \includegraphics[width=\textwidth]{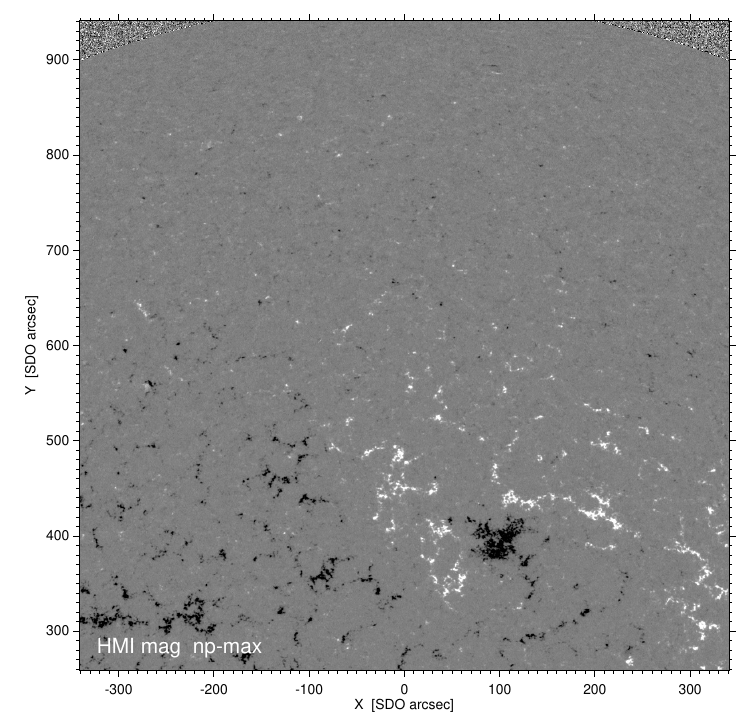}
  \caption[North-pole HMI magnetogram at maximum]
  {\label{fig:sdo-np-max-mag} %
  HMI magnetogram for the field in the two preceding figures.
  }\end{figure*}

\begin{figure*}
  \centering
  \includegraphics[width=\textwidth]{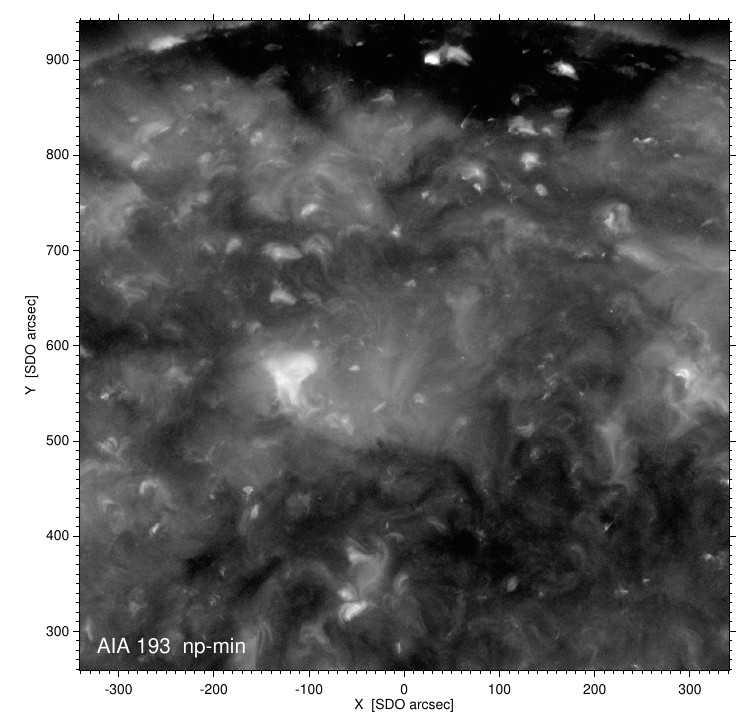}
  \caption[North-pole AIA 193\,\AA\ image at minimum]
  {\label{fig:sdo-np-min-193} %
  AIA 193\,\AA\ cutout near the solar North pole on June 1,2019 at
  00:00~UT. 
  }\end{figure*}
\begin{figure*}
  \centering
  \includegraphics[width=\textwidth]{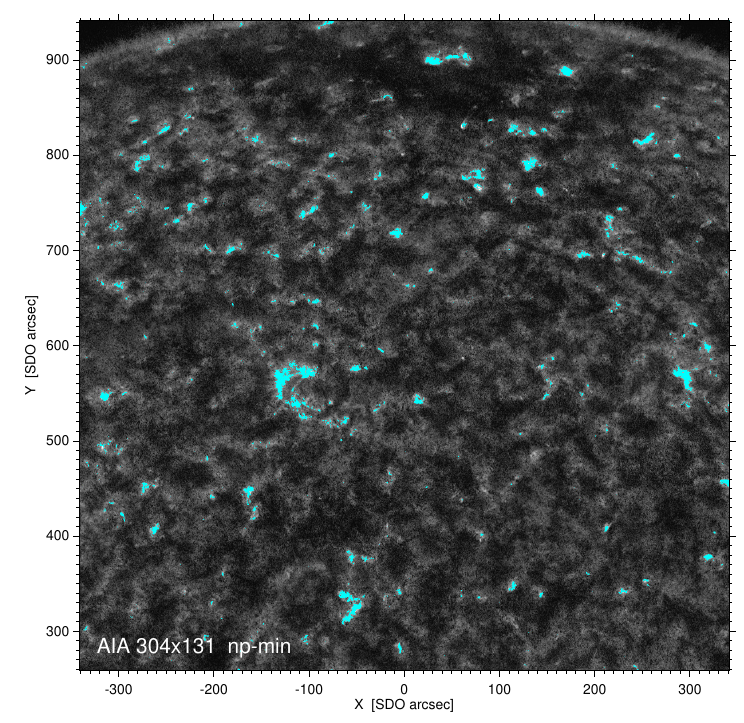}
  \caption[North-pole SDO fire detector image at minimum]
  {\label{fig:sdo-np-min-fire} %
  SDO fire detector cutout for the same field as in the preceding
  figure. 
  The small-fire density is much higher than in maximum-activity
  \rrref{figure}{fig:sdo-np-max-fire}.
  }\end{figure*}
\begin{figure*}
  \centering
  \includegraphics[width=\textwidth]{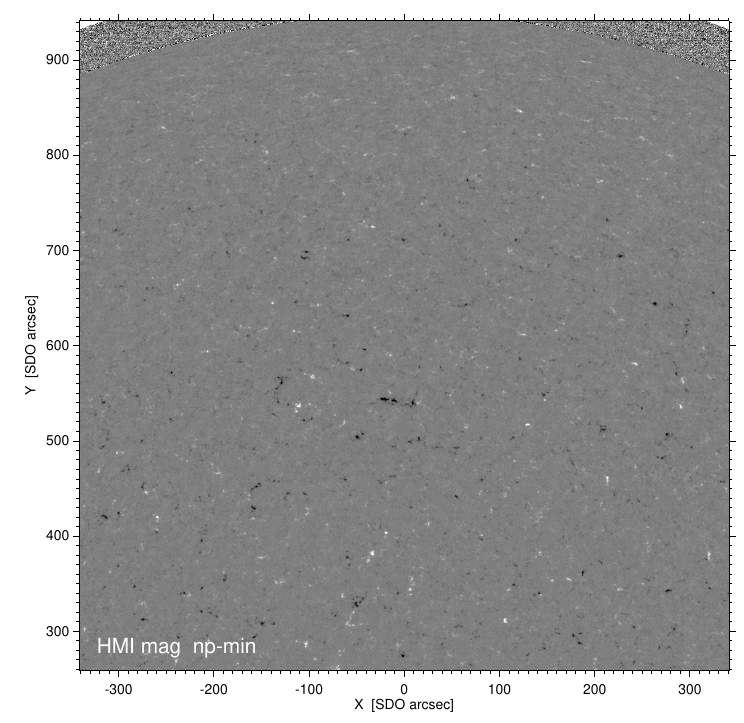}
  \caption[North-pole HMI magnetogram at minimum]
  {\label{fig:sdo-np-min-mag} %
  HMI magnetogram for the field in the two preceding figures.
  }\end{figure*}

\begin{figure*}
  \centering
  \includegraphics[width=\textwidth]{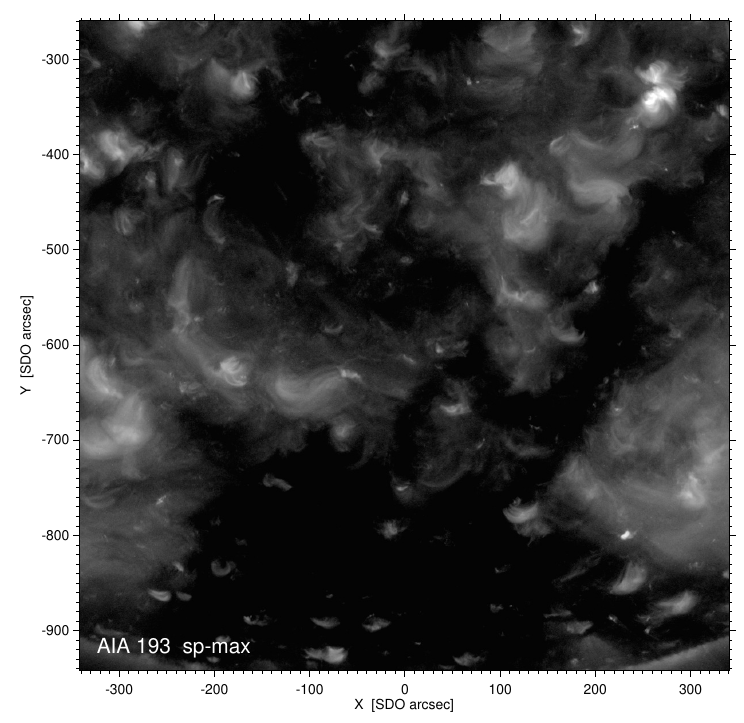}
  \caption[South-pole AIA 193\,\AA\ image at maximum]
  {\label{fig:sdo-sp-max-193} %
  AIA 193\,\AA\ cutout near the solar South pole on April 1, 2014 at
  00:00~UT.
  }\end{figure*}
\begin{figure*}
  \centering
  \includegraphics[width=\textwidth]{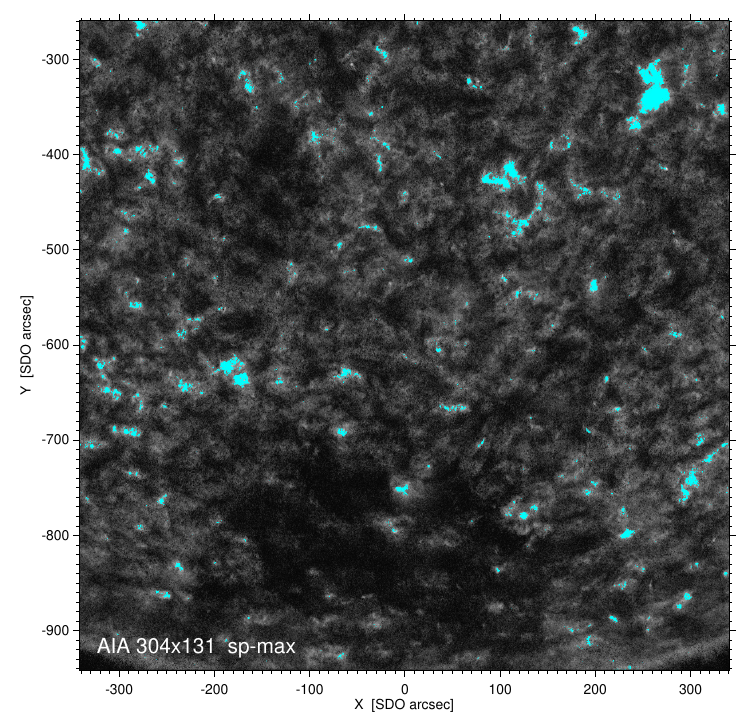}
  \caption[South-pole SDO fire detector image at maximum]
  {\label{fig:sdo-sp-max-fire} %
  SDO fire detector cutout for the same field as in the preceding
  figure. 
  High fire warning, less in the polar hole.
  }\end{figure*}
\begin{figure*}
  \centering
  \includegraphics[width=\textwidth]{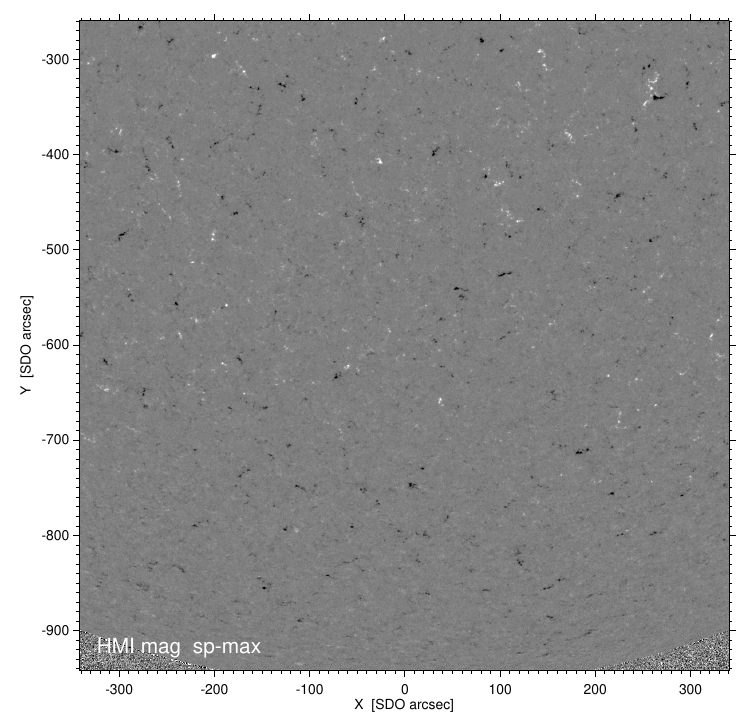}
  \caption[South-pole HMI magnetogram at maximum]
  {\label{fig:sdo-sp-max-mag} %
  HMI magnetogram for the field in the two preceding figures.
  }\end{figure*}

\begin{figure*}
  \centering
  \includegraphics[width=\textwidth]{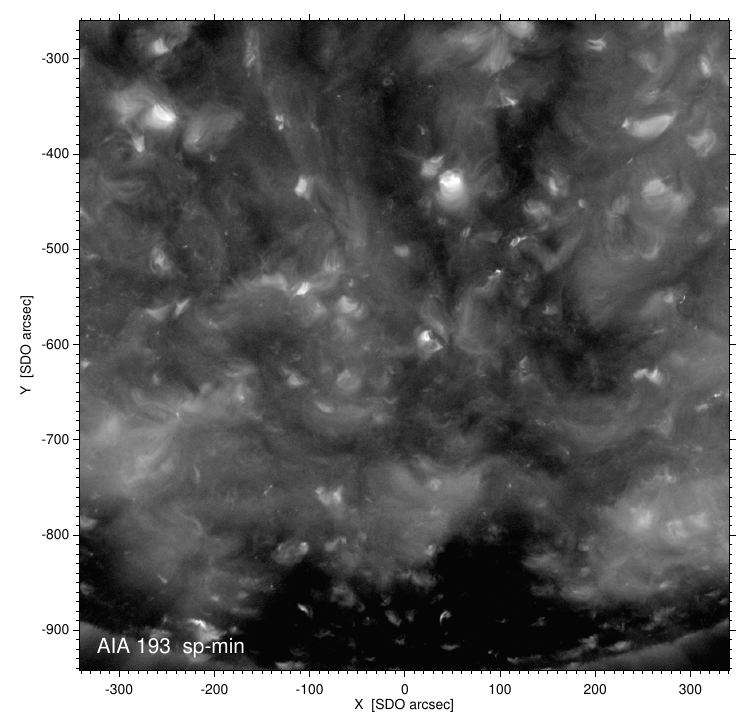}
  \caption[South-pole AIA 193\,\AA\ image at minimum]
  {\label{fig:sdo-sp-min-193} %
  AIA 193\,\AA\ cutout near the solar South pole on June 1, 2019 at
  00:00~UT.
  }\end{figure*}
\begin{figure*}
  \centering
  \includegraphics[width=\textwidth]{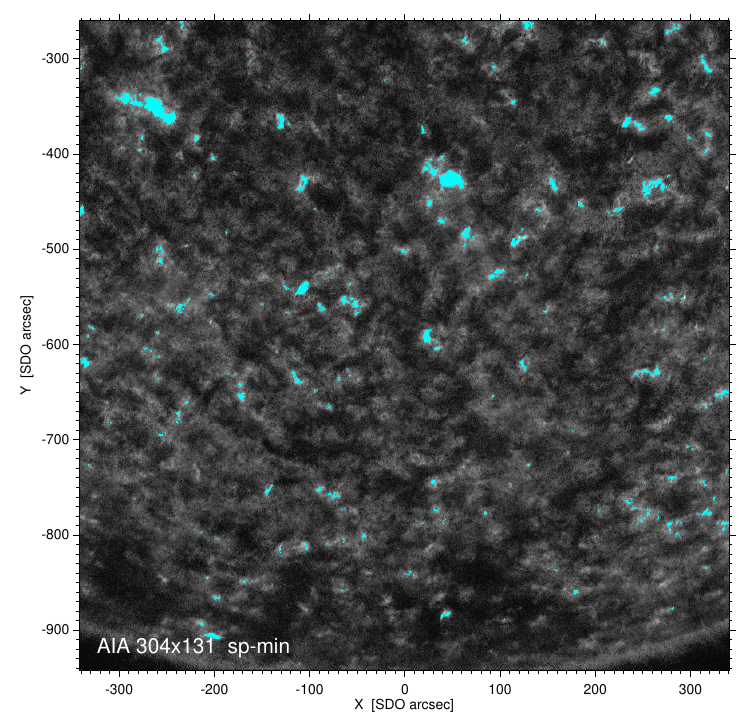}
  \caption[South-pole SDO fire detector at minimum]
  {\label{fig:sdo-sp-min-fire} %
  SDO fire detector cutout for the same field as in the preceding
  figure.
  }\end{figure*}
\begin{figure*}
  \centering
  \includegraphics[width=\textwidth]{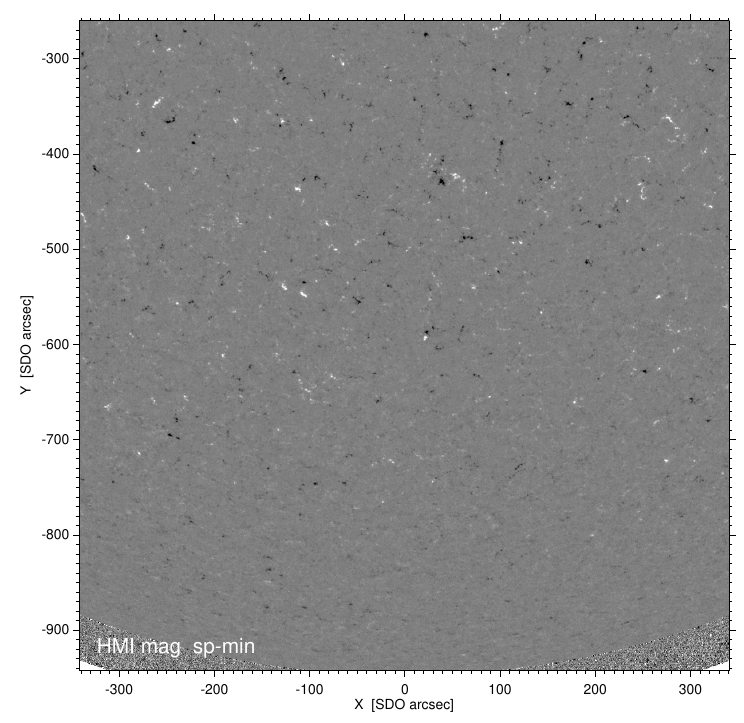}
  \caption[South-pole HMI magnetogram at minimum]
  {\label{fig:sdo-sp-min-mag} %
  HMI magnetogram for the field in the two preceding figures.
  }\end{figure*}

\clearpage 


\begin{figure*}
  \centering
  \includegraphics[width=\textwidth]{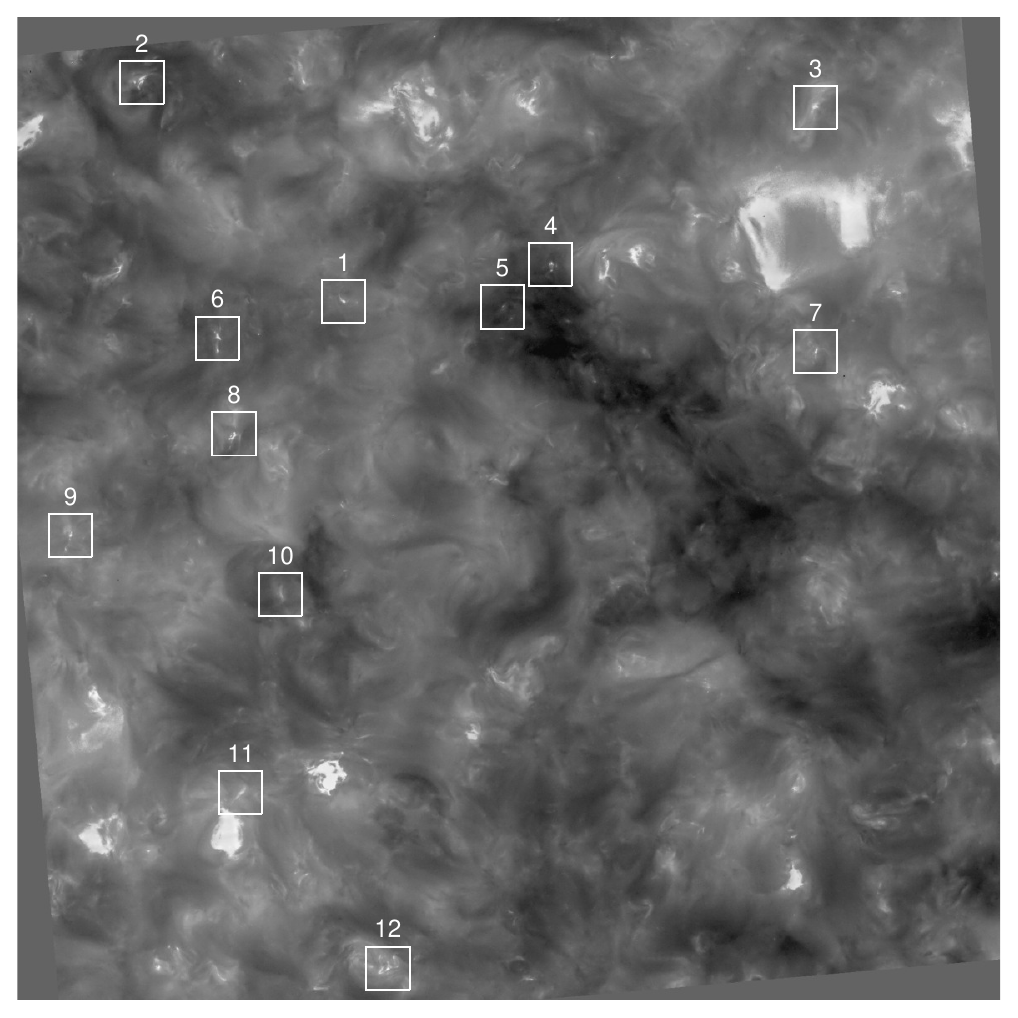}
  \caption[Campfire cutout locations in the SolO image]
  {\label{fig:cfroilocs} %
  \addtocline{SDO campfire ROI cutouts}
  The rotated SolO 174\,\AA\ image with superimposed ROI boxes for the
  campfire cutouts. 
  ROI-1 is the campfire that was marked in the annotated version in
  the press-release material.
  The others are eye-ball selected in top-to-bottom order, also
  including small bright patches with more complex than single-flame
  morphology.
  }\end{figure*}


\begin{figure*}
  \centering
  \includegraphics[width=\textwidth]{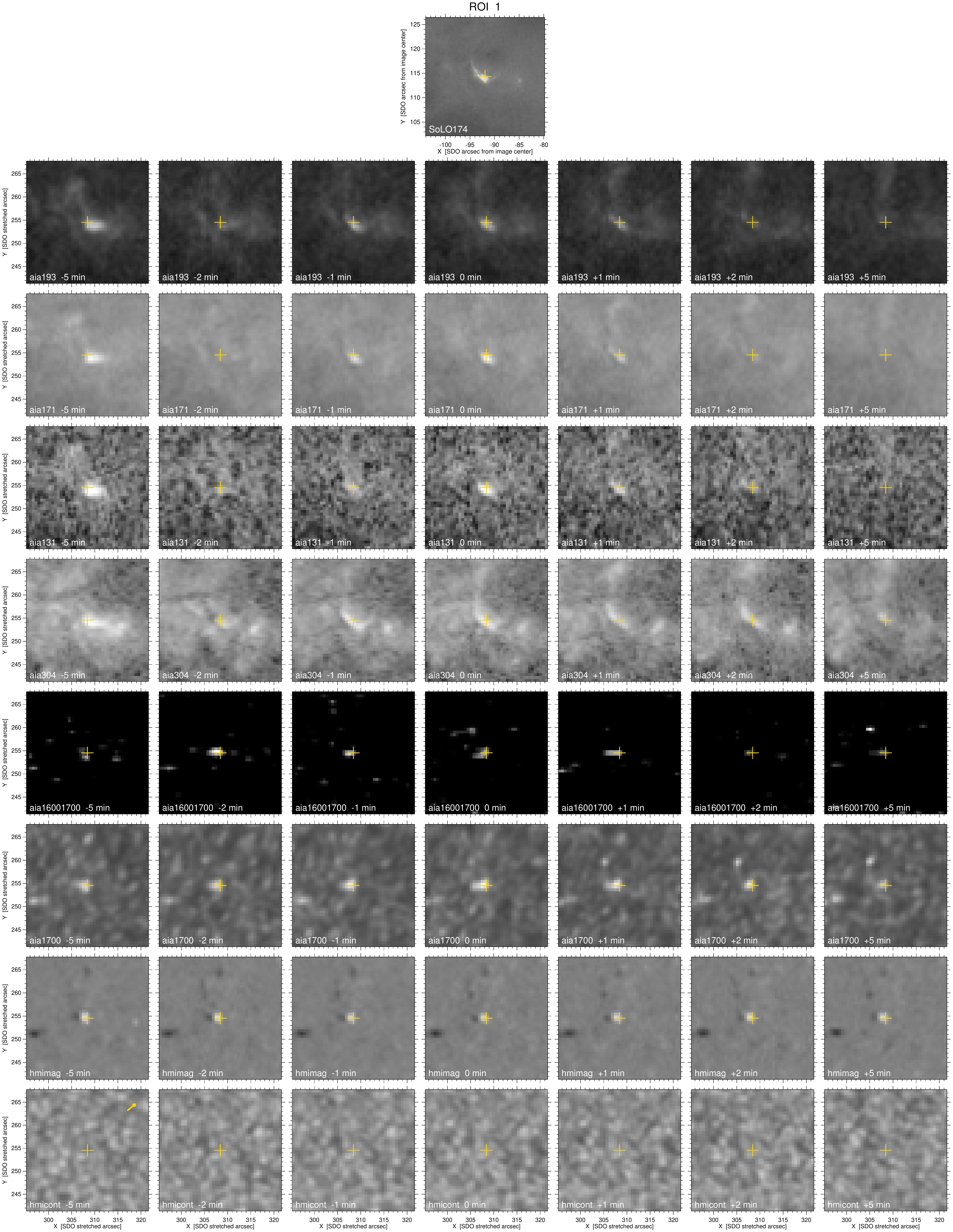}
  \caption[ROI-1 cutout assembly]{\label{fig:roi1} %
  Cutouts for ROI-1. 
  Format details are given in the main text in
  \rrref{section}{sec:campfires}.
  This SolO campfire (top) was annotated as exemplary in the press
  release.
  It was also present in the AIA EUV images, also recurrently before,
  best in AIA 304\,\AA. 
  There was a bipolar MC pair on the surface underneath, with enhanced
  brightening in the UV including excess in the 16001700 construct. 
  }\end{figure*}
\begin{figure*}
  \centering
  \includegraphics[width=0.97\textwidth]{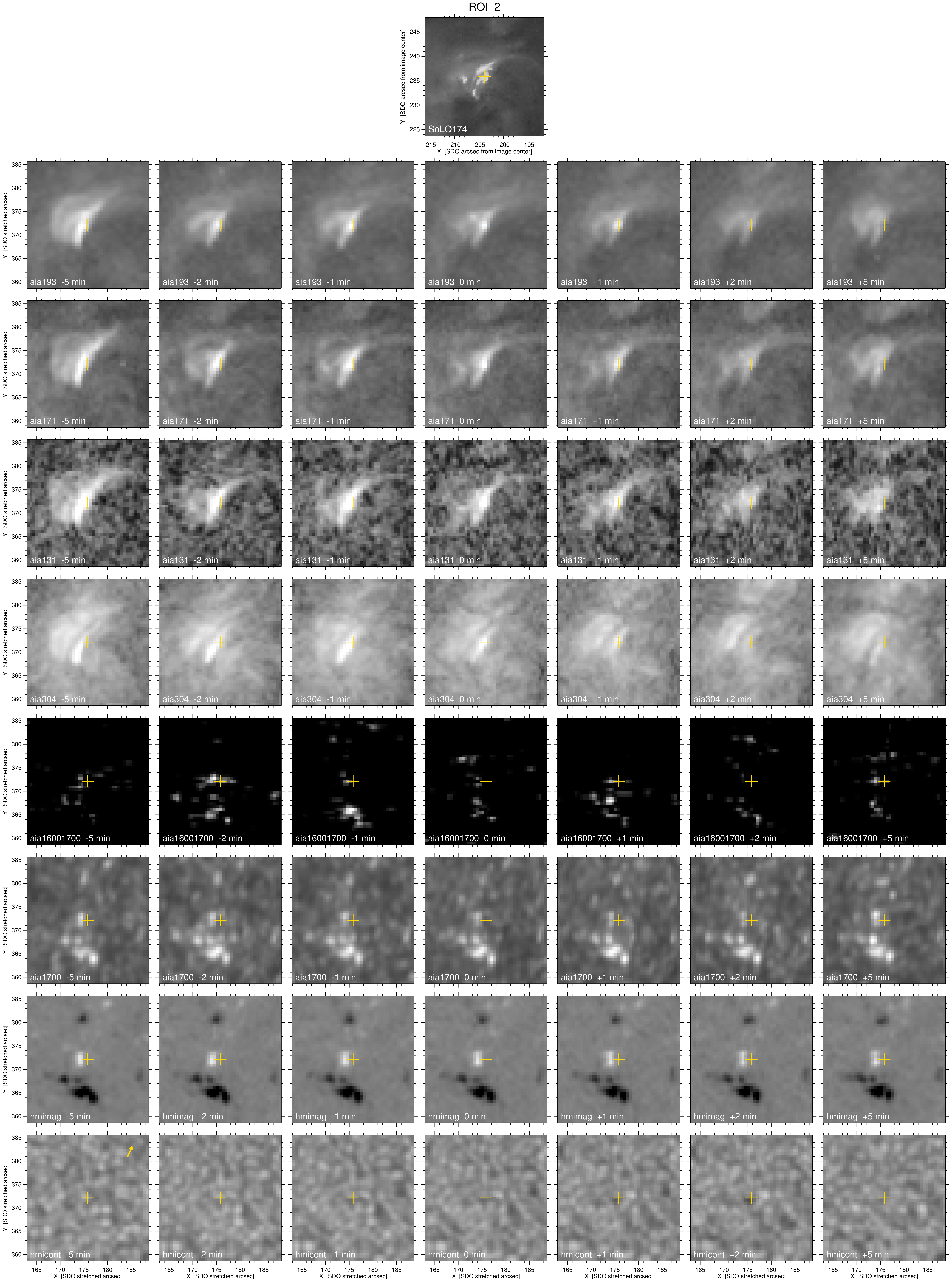} 
  \caption[ROI-2 cutout assembly]{\label{fig:roi2} %
  Cutouts for ROI-2.
  }\end{figure*}
\begin{figure*}
  \centering
 \includegraphics[width=\textwidth]{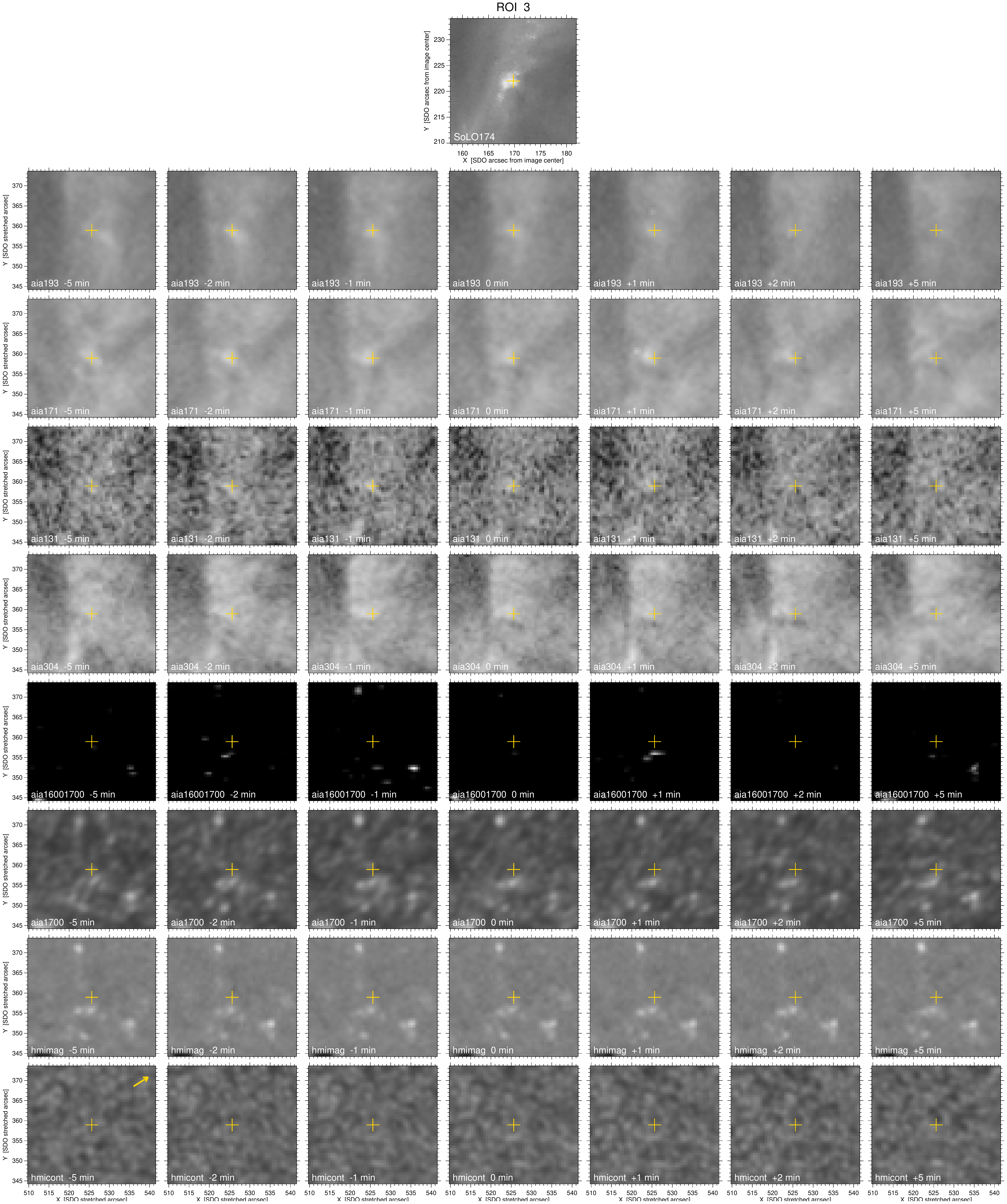}
  \caption[ROI-3 cutout assembly]{\label{fig:roi3} %
  Cutouts for ROI-3.
  }\end{figure*}
\begin{figure*}
  \centering
  \includegraphics[width=\textwidth]{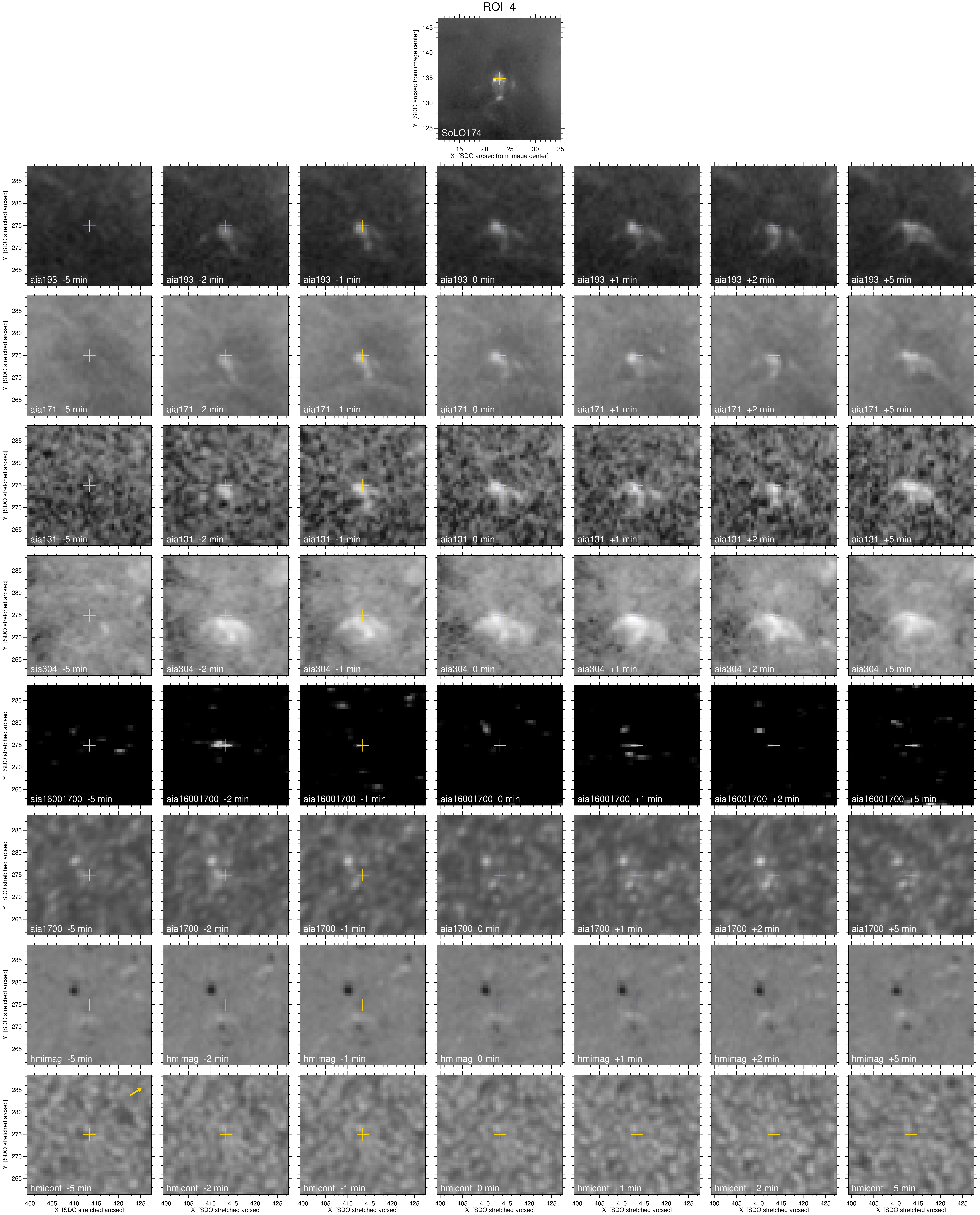}
  \caption[ROI-4 cutout assembly]{\label{fig:roi4} %
  Cutouts for ROI-4.
  }\end{figure*}
\begin{figure*}
  \centering
  \includegraphics[width=\textwidth]{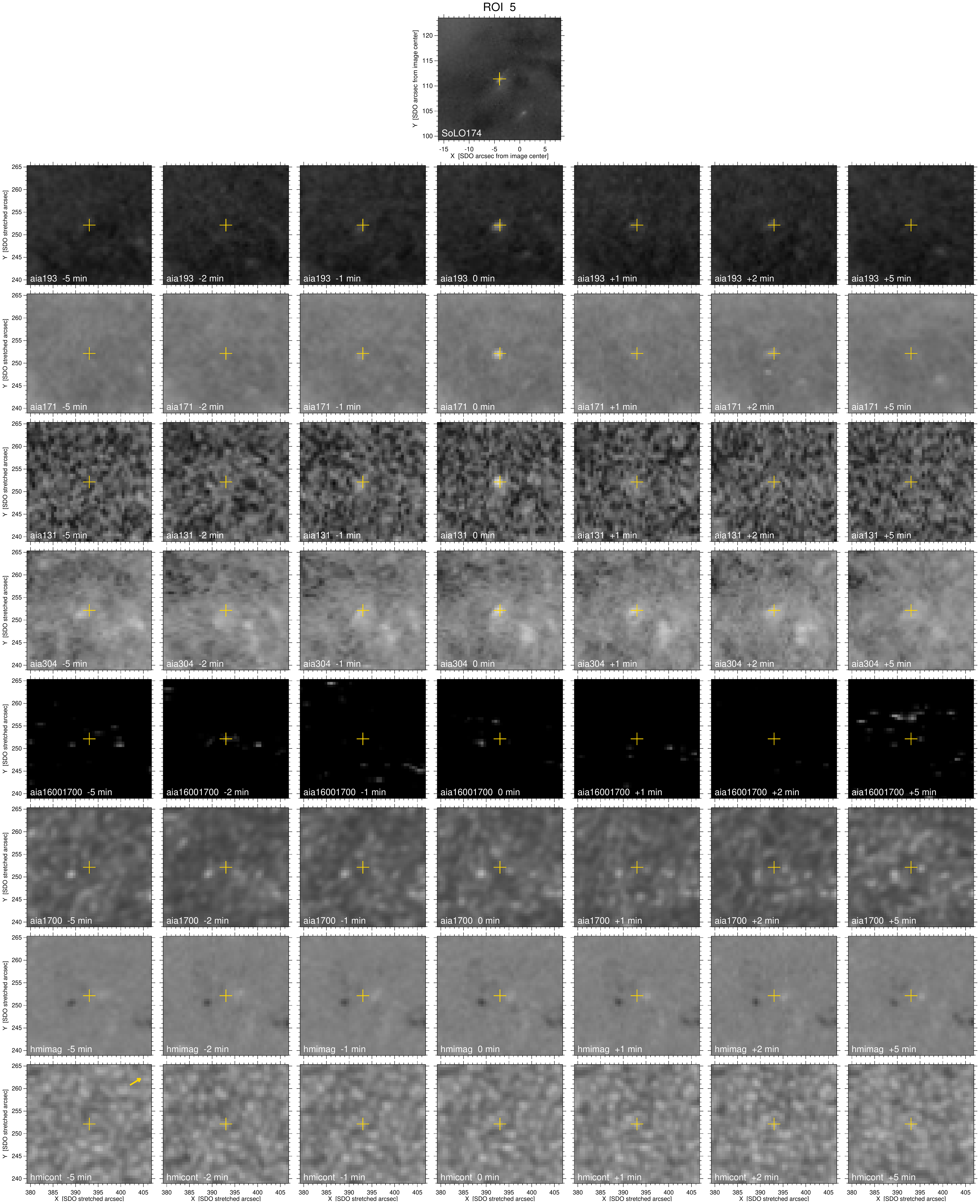}
  \caption[ROI-5 cutout assembly]{\label{fig:roi5} %
  Cutouts for ROI-5.
  }\end{figure*}
\begin{figure*}
  \centering
  \includegraphics[width=\textwidth]{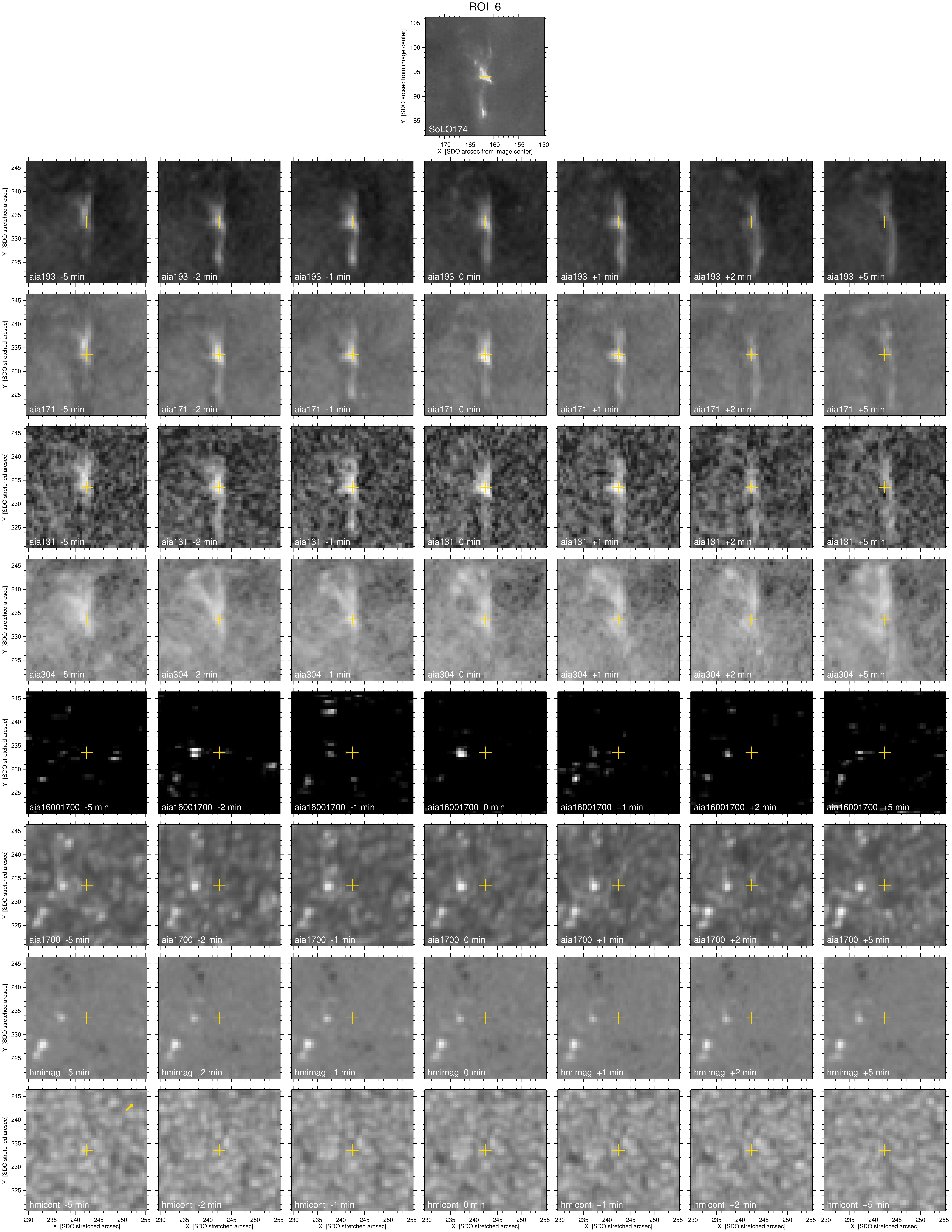}
  \caption[ROI-6 cutout assembly]{\label{fig:roi6} %
  Cutouts for ROI-6.
  }\end{figure*}
\begin{figure*}
  \centering
  \includegraphics[width=\textwidth]{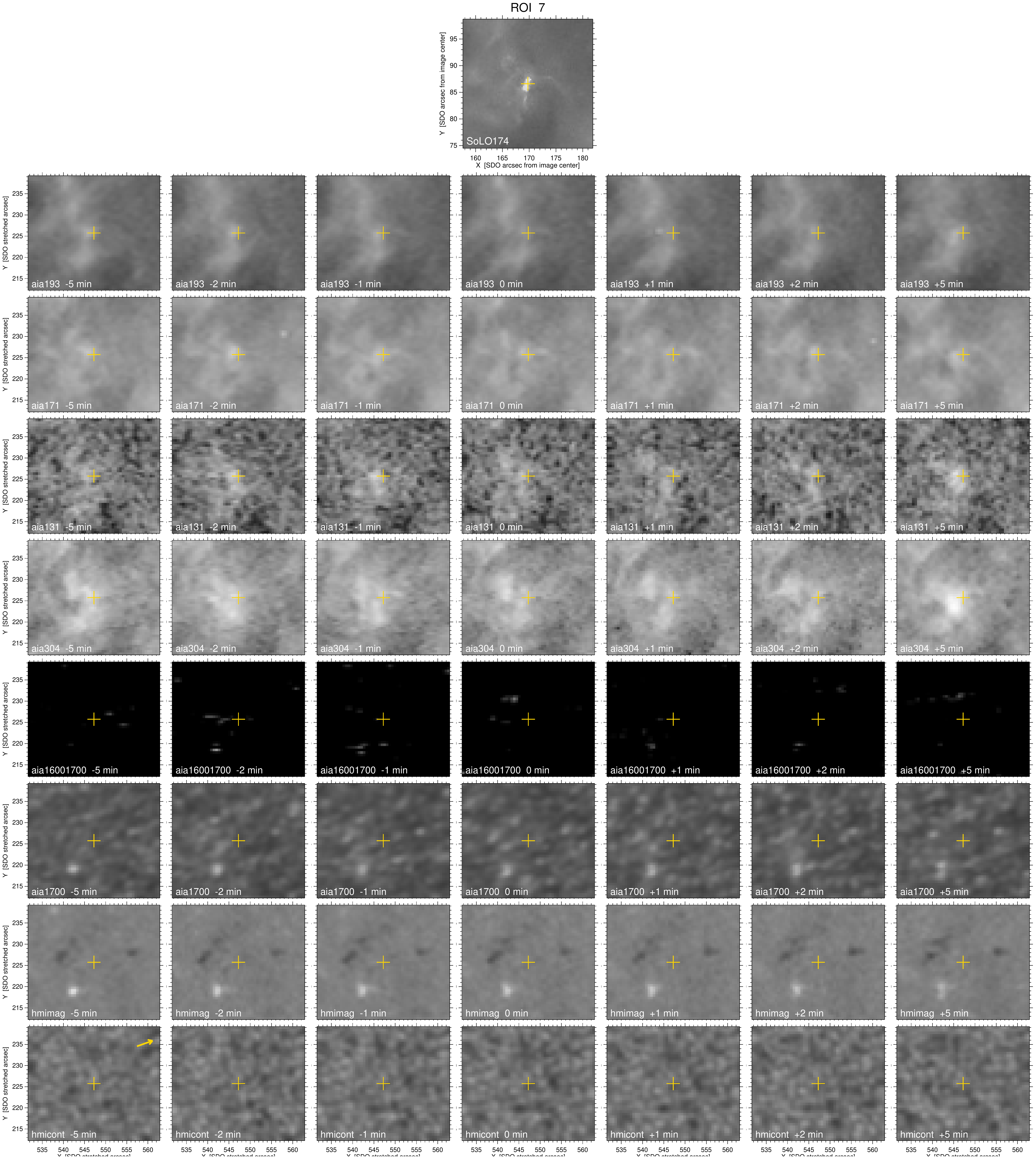}
  \caption[ROI-7 cutout assembly]{\label{fig:roi7} %
  Cutouts for ROI-7.
  }\end{figure*}
\begin{figure*}
  \centering
  \includegraphics[width=\textwidth]{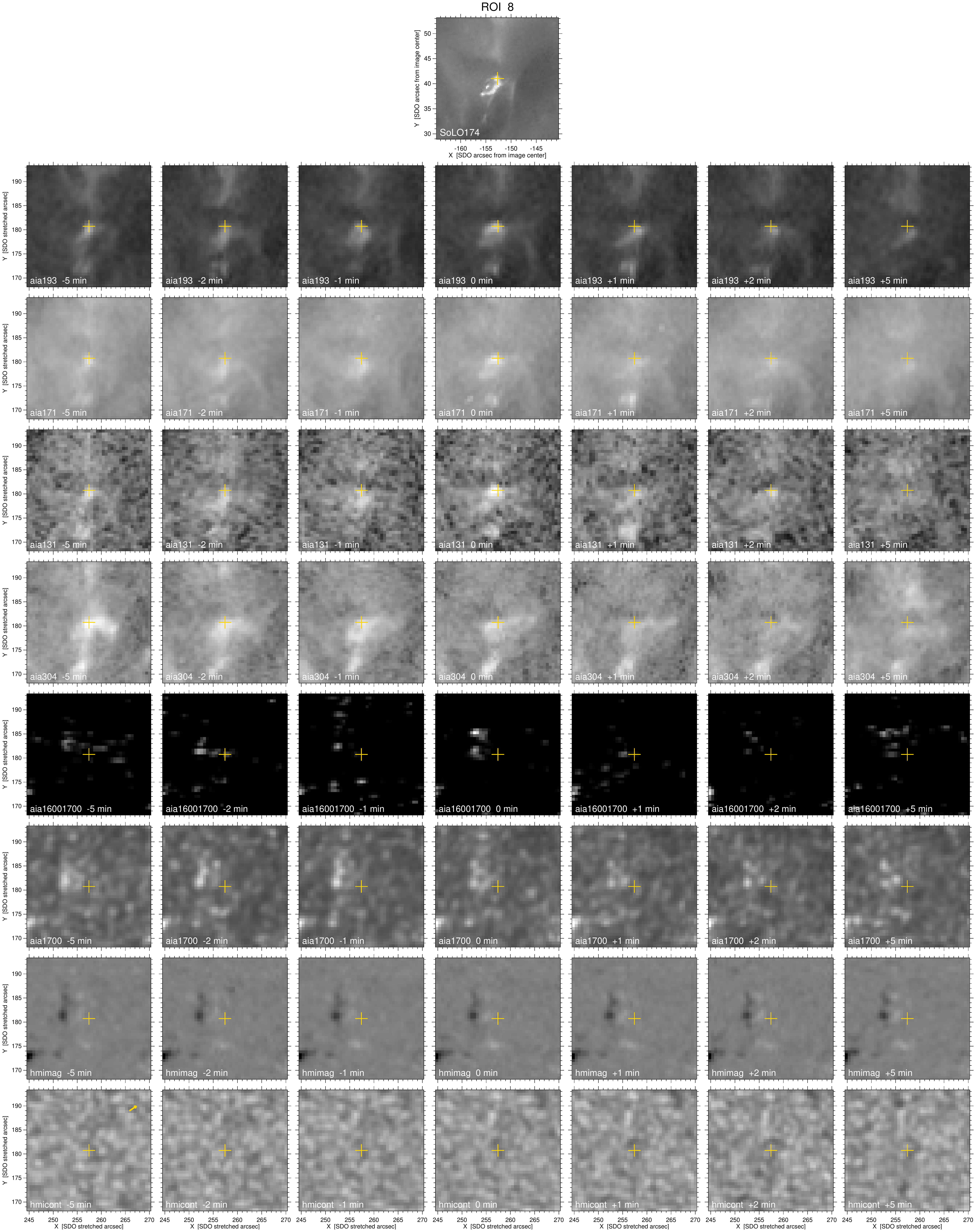}
  \caption[ROI-8 cutout assembly]{\label{fig:roi8} %
  Cutouts for ROI-8.
  }\end{figure*}
\begin{figure*}
  \centering
  \includegraphics[width=\textwidth]{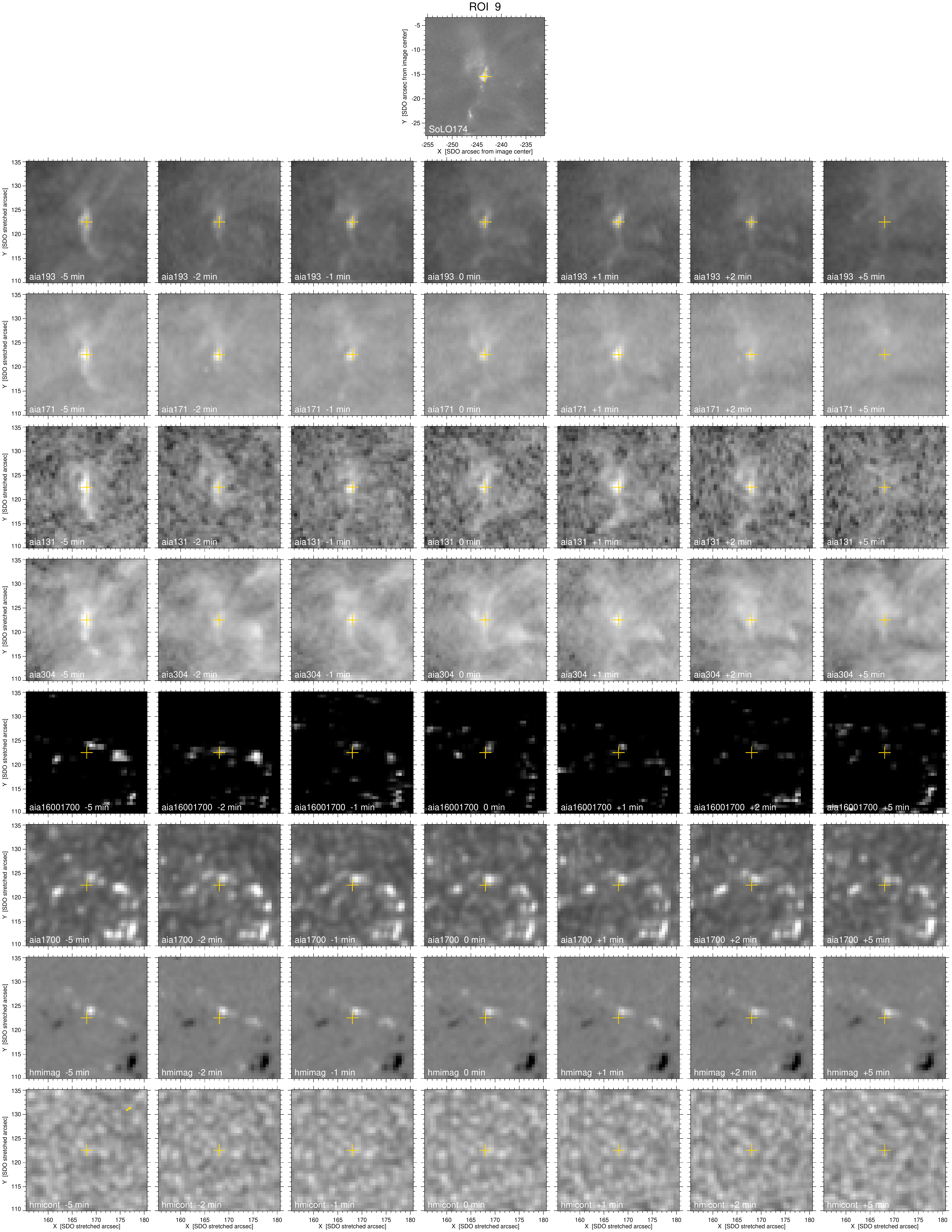}
  \caption[ROI-9 cutout assembly]{\label{fig:roi9} %
  Cutouts for ROI-9.
  }\end{figure*}
\begin{figure*}
  \centering
  \includegraphics[width=\textwidth]{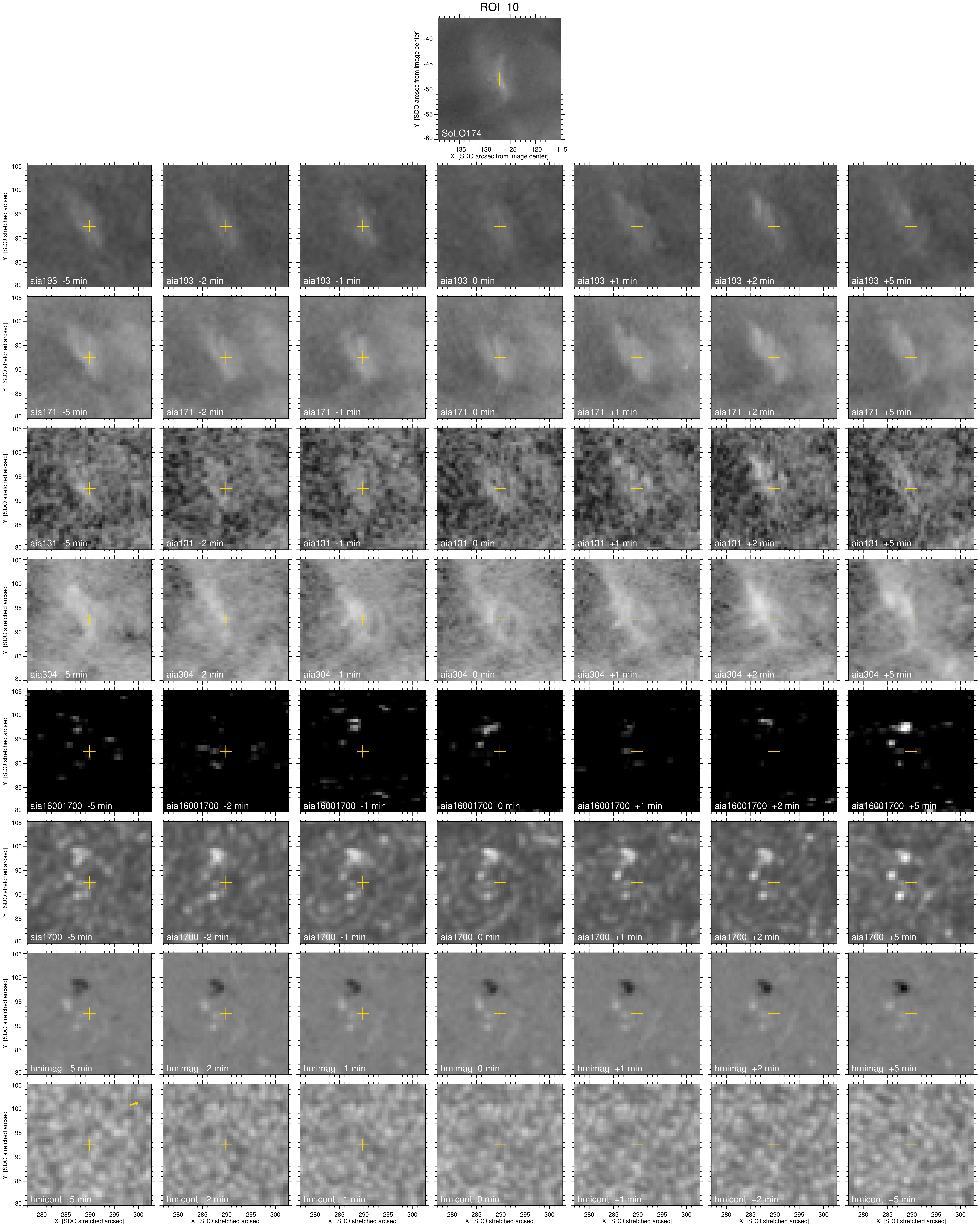}
  \caption[ROI-10 cutout assembly]{\label{fig:roi10} %
  Cutouts for ROI-10.
  }\end{figure*}
\begin{figure*}
  \centering
  \includegraphics[width=\textwidth]{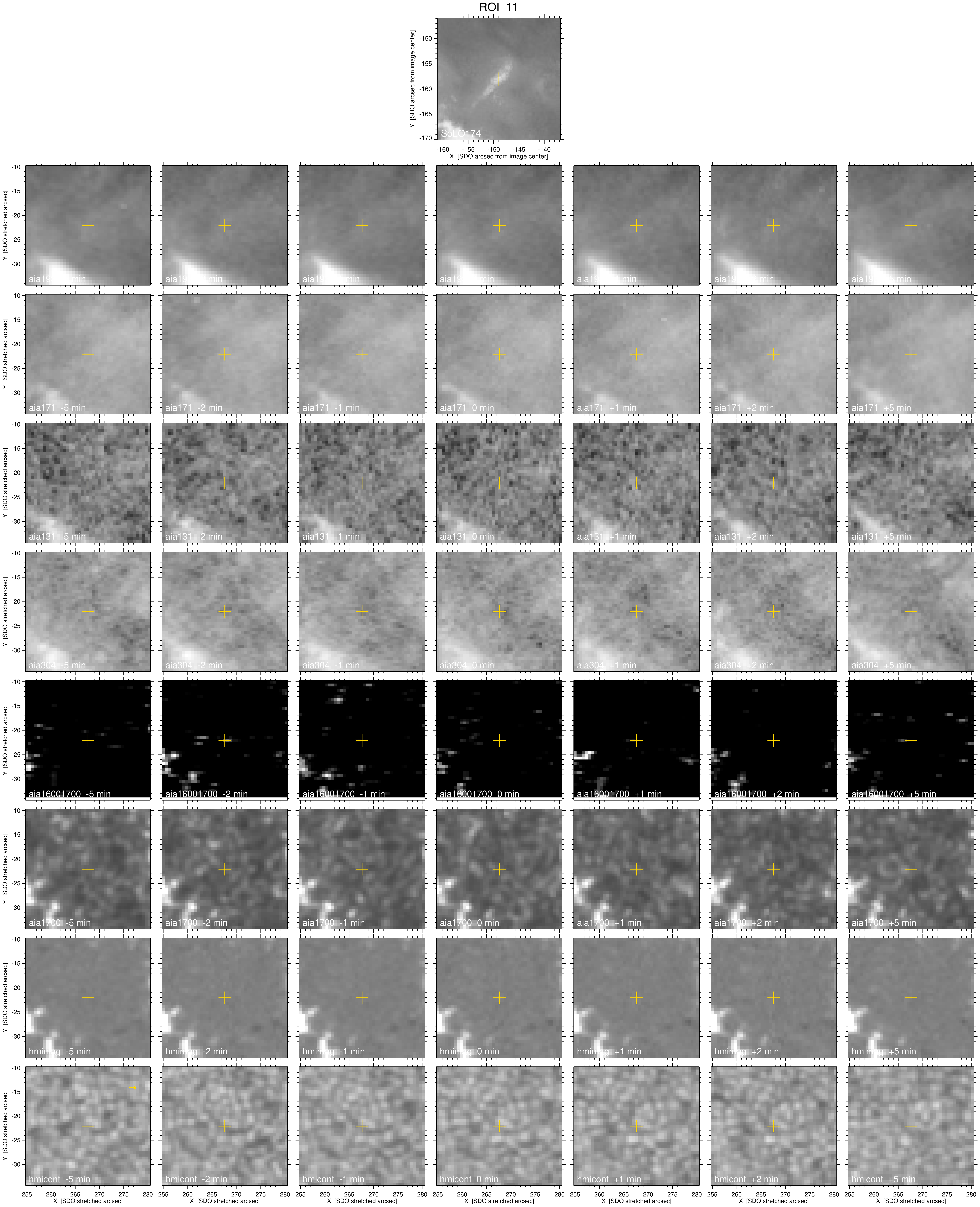}
  \caption[ROI-11 cutout assembly]{\label{fig:roi11} %
  Cutouts for ROI-11.
  }\end{figure*}
\begin{figure*}
  \centering
  \includegraphics[width=\textwidth]{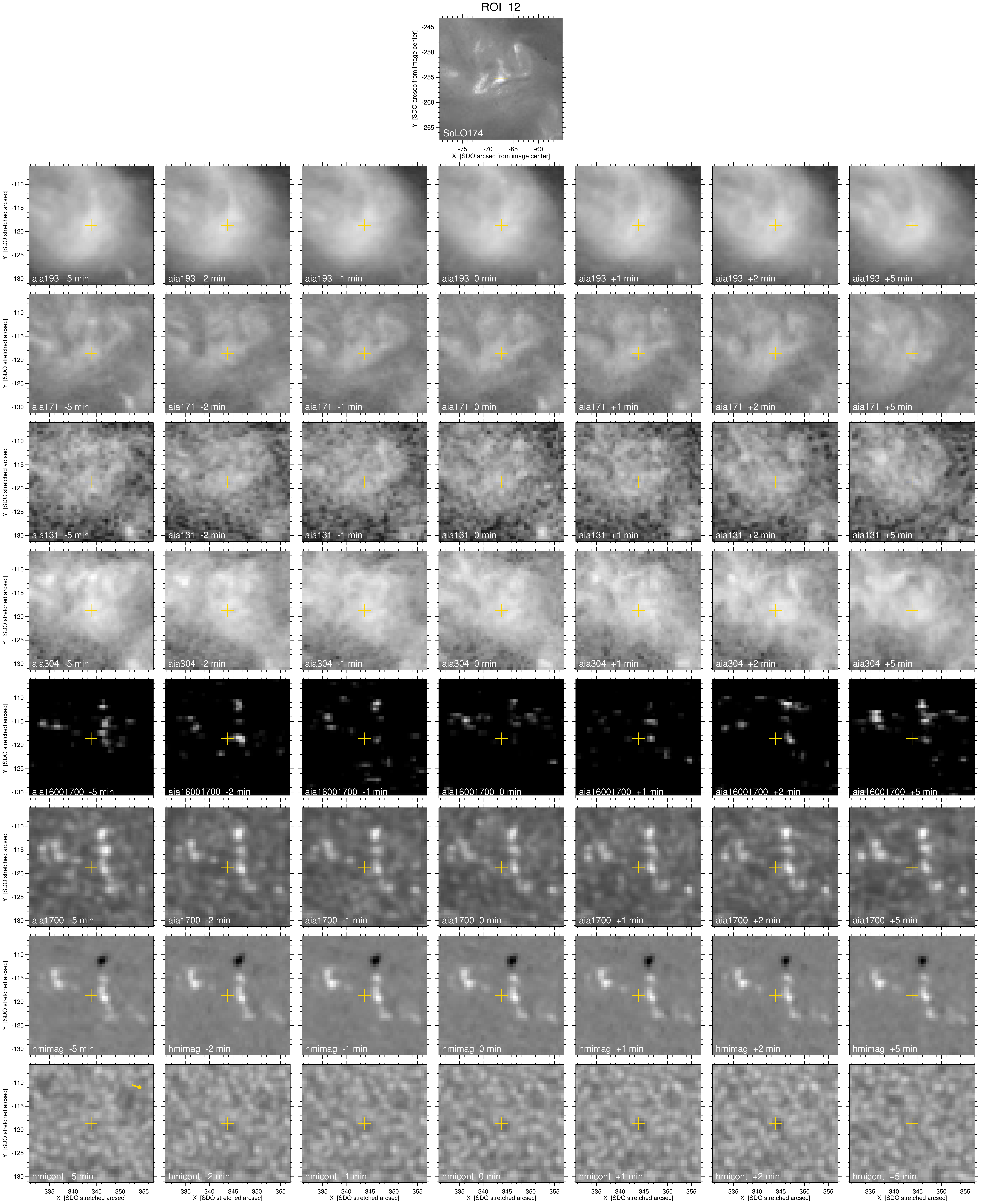}
  \caption[ROI-12 cutout assembly]{\label{fig:roi12} %
  Cutouts for ROI-12.
  }\end{figure*}

\clearpage 


\begin{figure*}
  \centering
  \includegraphics[width=\textwidth]{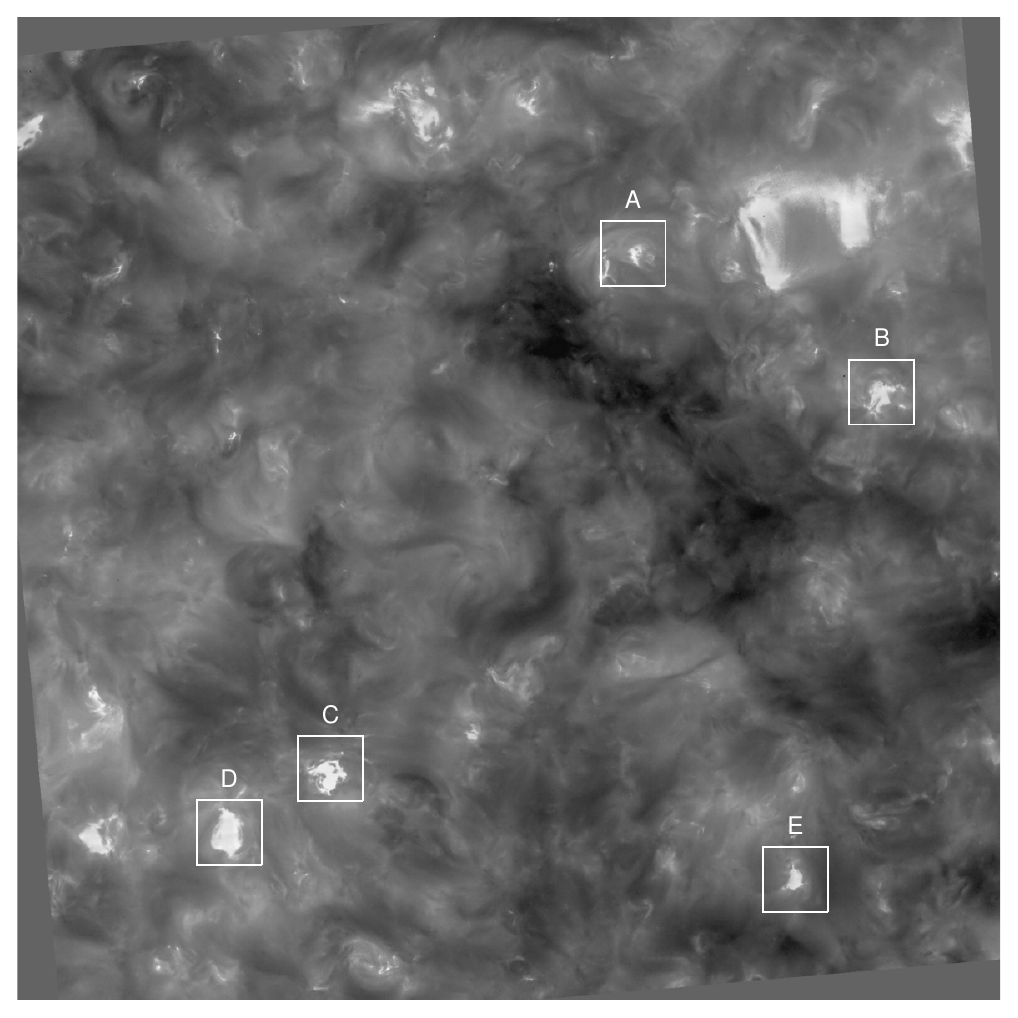}
  \caption[Brushfire cutout locations in the SolO image]
  {\label{fig:bfroilocs} %
  The rotated SolO 174\,\AA\ image with superimposed ROI boxes for the
  brushfire cutouts. 
  }\end{figure*}


\begin{figure*}
  \centering
  \includegraphics[width=0.9\textwidth]{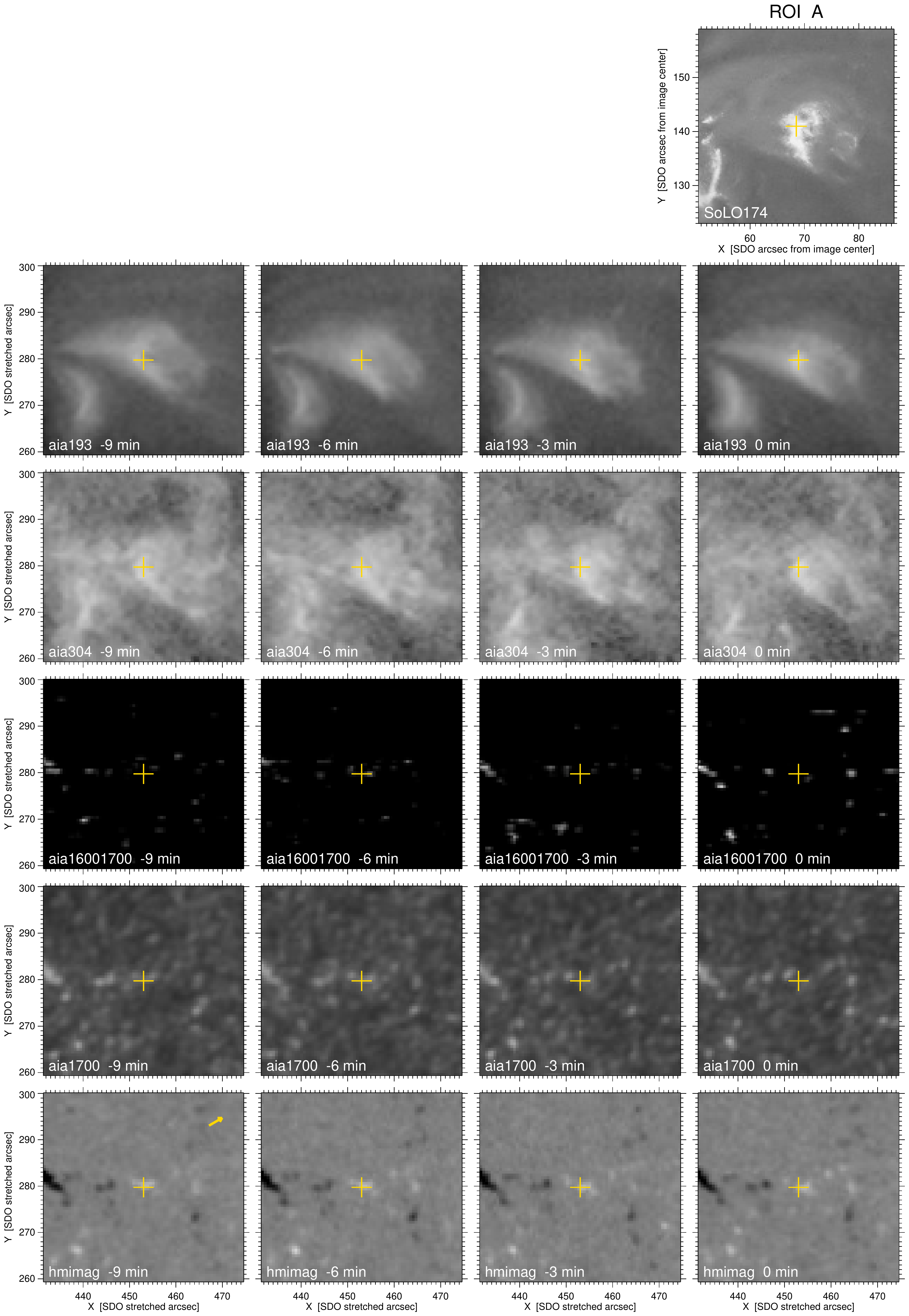}
  \caption[ROI-A cutout assembly]{\label{fig:roiA} %
  \addtocline{SDO brushfire ROI cutouts}
  Cutouts for ROI-A.  
  }\end{figure*}
\begin{figure*}
  \centering
  \includegraphics[width=0.9\textwidth]{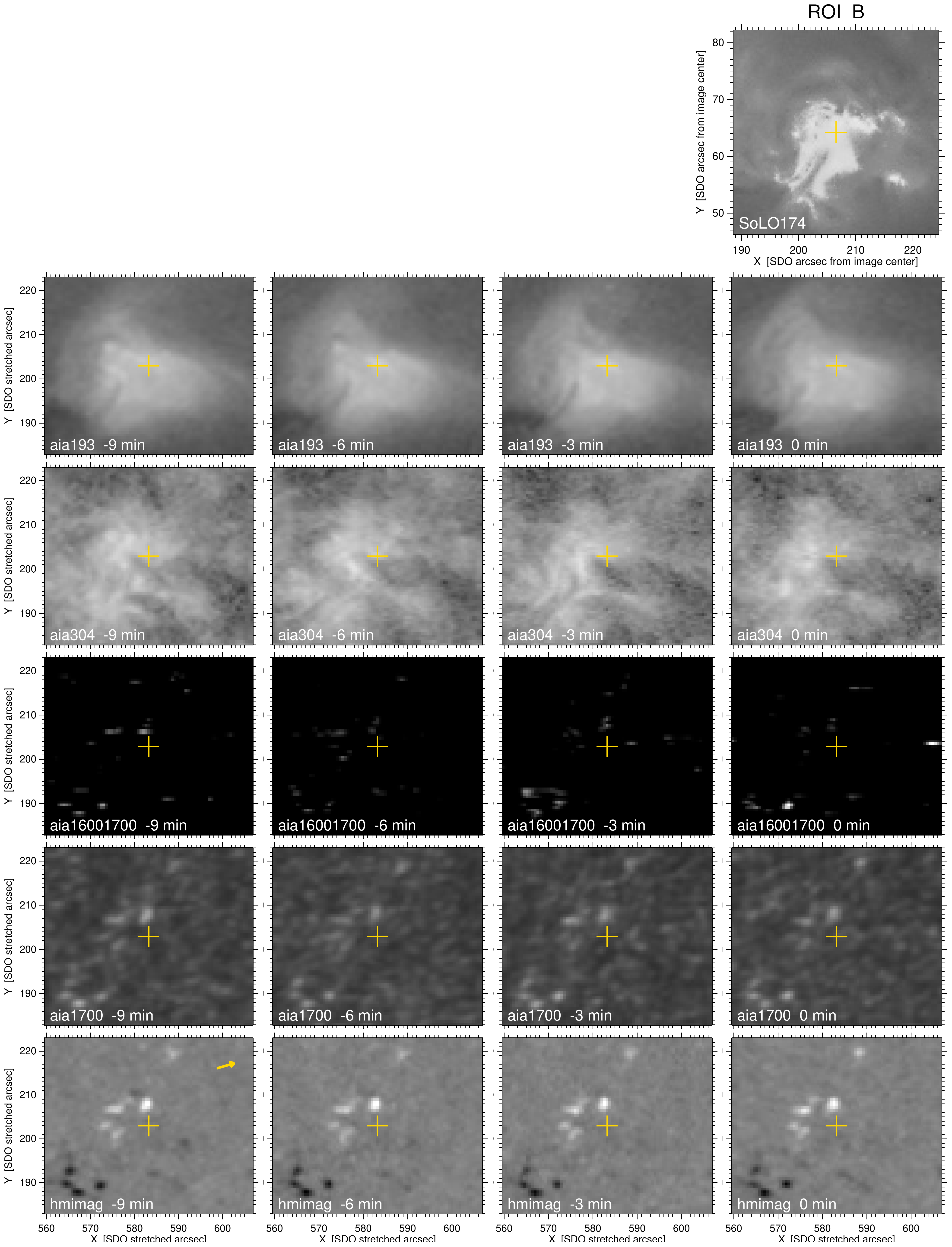}
  \caption[ROI-B cutout assembly]{\label{fig:roiB} %
  Cutouts for ROI-B.
  }\end{figure*}
\begin{figure*}
  \centering
  \includegraphics[width=0.9\textwidth]{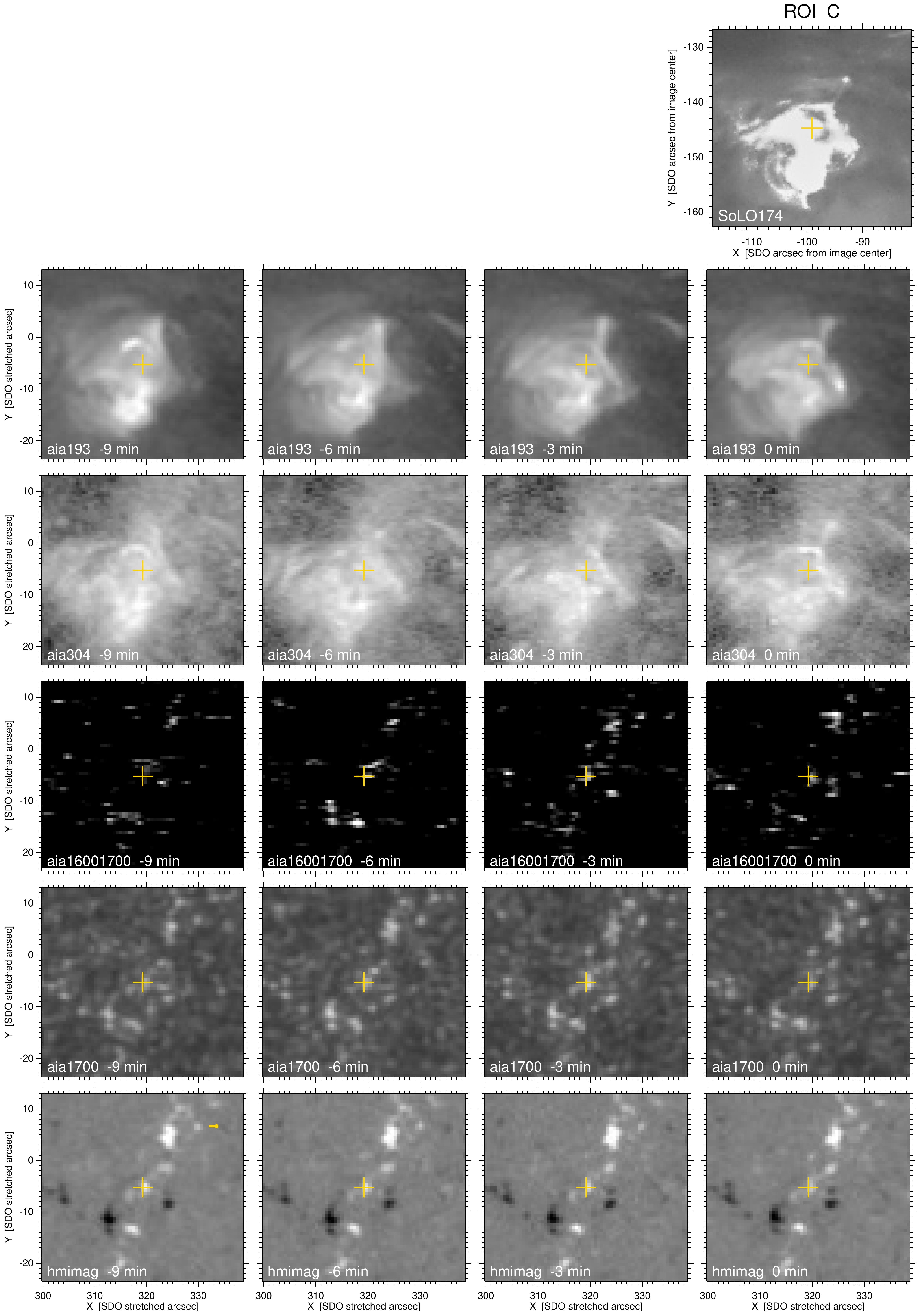}
  \caption[ROI-C cutout assembly]{\label{fig:roiC} %
  Cutouts for ROI-C.
  }\end{figure*}
\begin{figure*}
  \centering
  \includegraphics[width=0.9\textwidth]{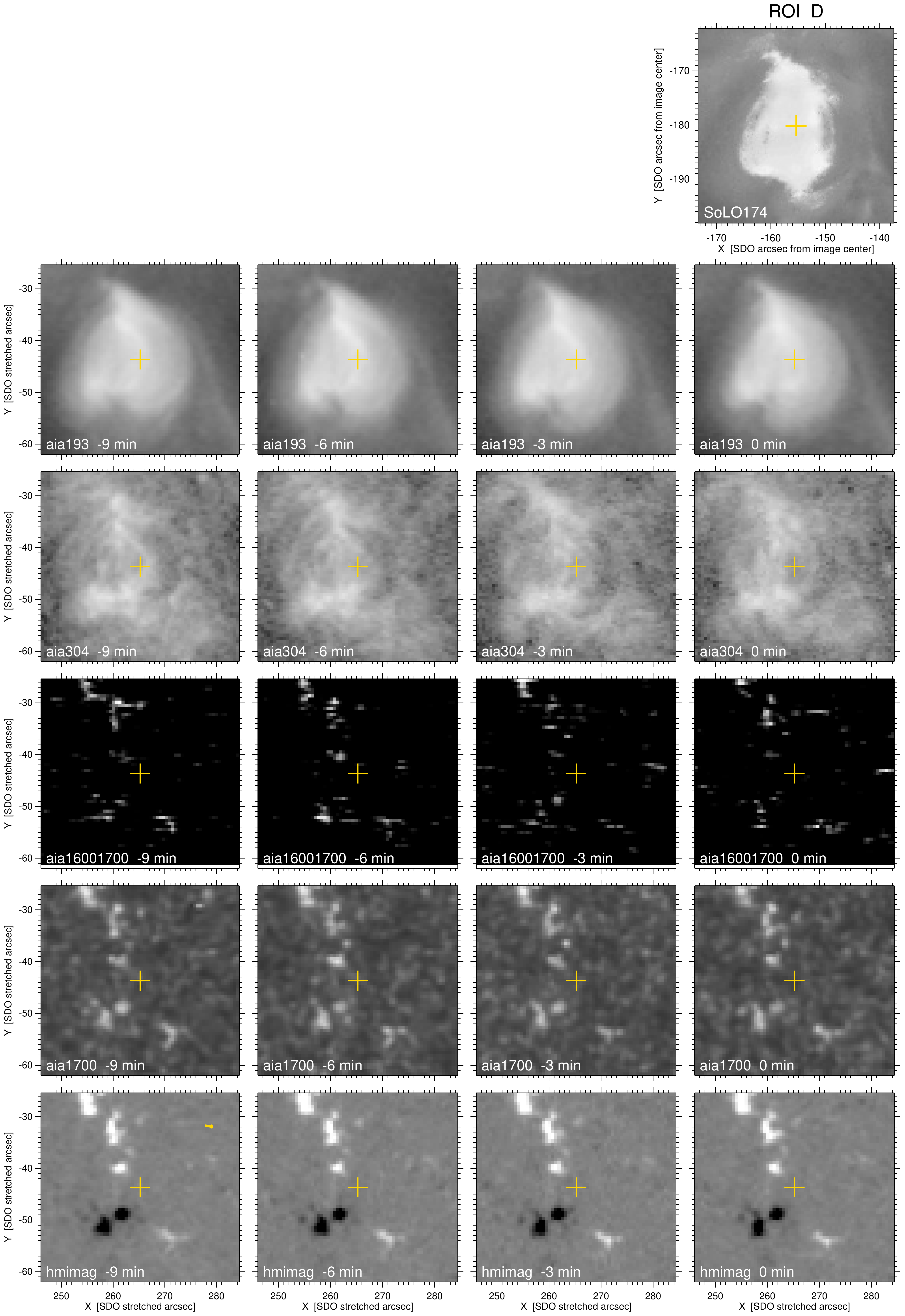}
  \caption[ROI-D cutout assembly]{\label{fig:roiD} %
  Cutouts for ROI-D.
  }\end{figure*}
\begin{figure*}
  \centering
  \includegraphics[width=0.9\textwidth]{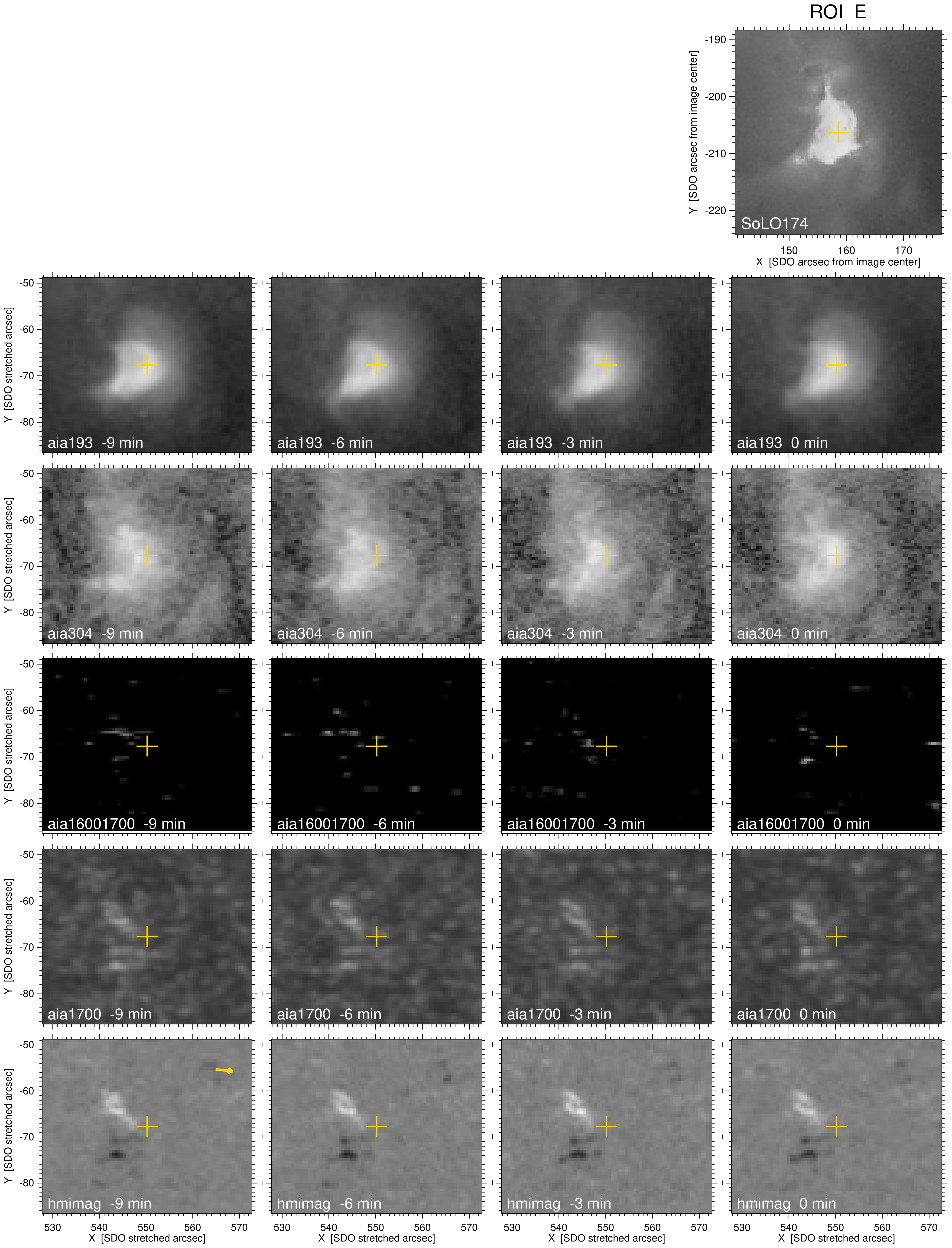}
  \caption[ROI-E cutout assembly]{\label{fig:roiE} %
  Cutouts for ROI-E.
  }\end{figure*}

\begin{figure*}
  \centering
  \includegraphics[width=\textwidth]{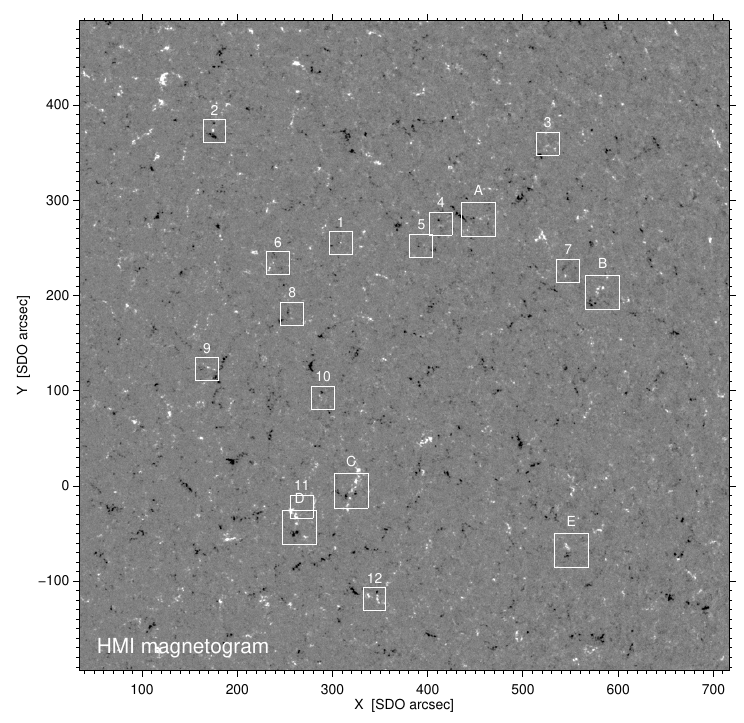}
  \caption[All cutouts in HMI magnetogram]
  {\label{fig:sdoroilocmag} %
  \addtocline{SDO ROI locations and fire detector}
  The HMI magnetogram at best-match time with superimposed ROI boxes
  for all campfire and brushfire cutouts.
  }\end{figure*}
\begin{figure*}
  \centering
  \includegraphics[width=\textwidth]{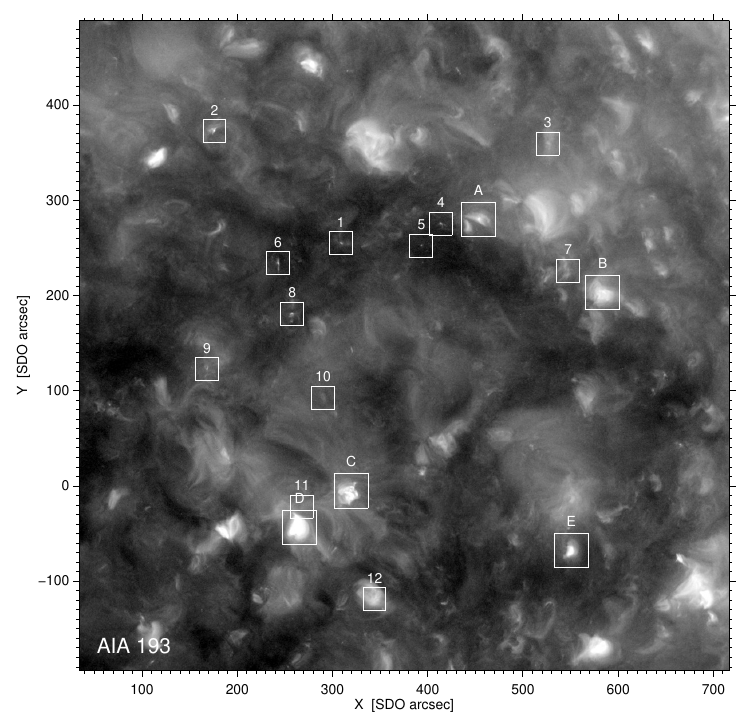}
  \caption[All cutouts in AIA 193\,\AA\ image]
  {\label{fig:sdoroiloc193} %
  The AIA 193\,\AA\ image at best-match time with superimposed ROI boxes
  for all campfire and brushfire cutouts.
  }\end{figure*}
\begin{figure*}
  \centering
  \includegraphics[width=\textwidth]{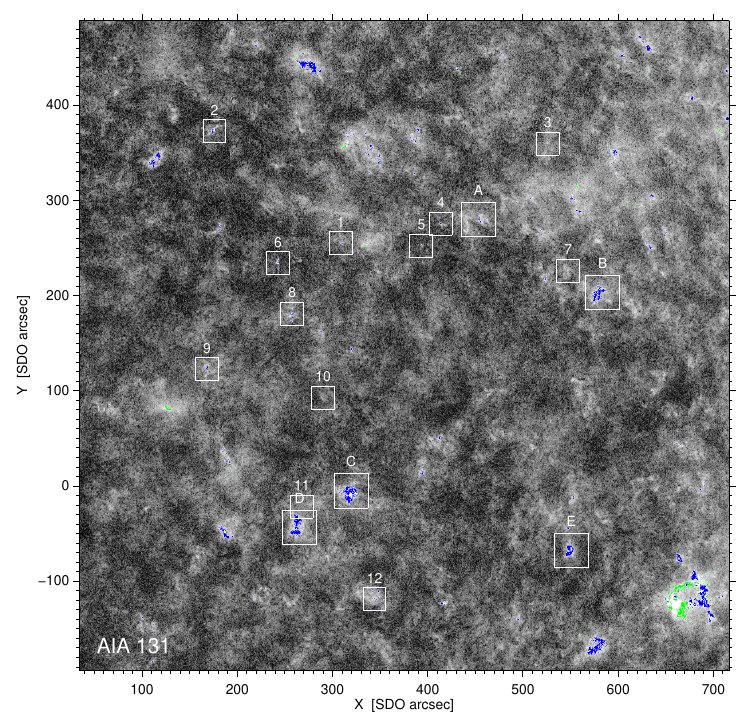}
  \caption[All cutouts in AIA 131\,\AA\ image]
  {\label{fig:sdoroiloc131} %
  The AIA 131\,\AA\ image at best-match time with superimposed ROI
  boxes for all campfire and brushfire cutouts.
  Green and blue pixels are those within the green and blue
  selection boxes in the righthand Strous diagram in
  \rrref{figure}{fig:scats}.
  Green pixels lie mostly in the bright patch in the
  lower-right corner;  blue pixels lie in small bright features.
  Blinking with the preceding AIA\,193\,\AA\ image shows that most
  are small brushfires, with bipolar MCs on the surface (blink
  one more back).  However, most campfires also contain blue pixels.
  These colored pixel selections inspired the next figure.
  }\end{figure*}
\begin{figure*}
  \centering
  \includegraphics[width=\textwidth]{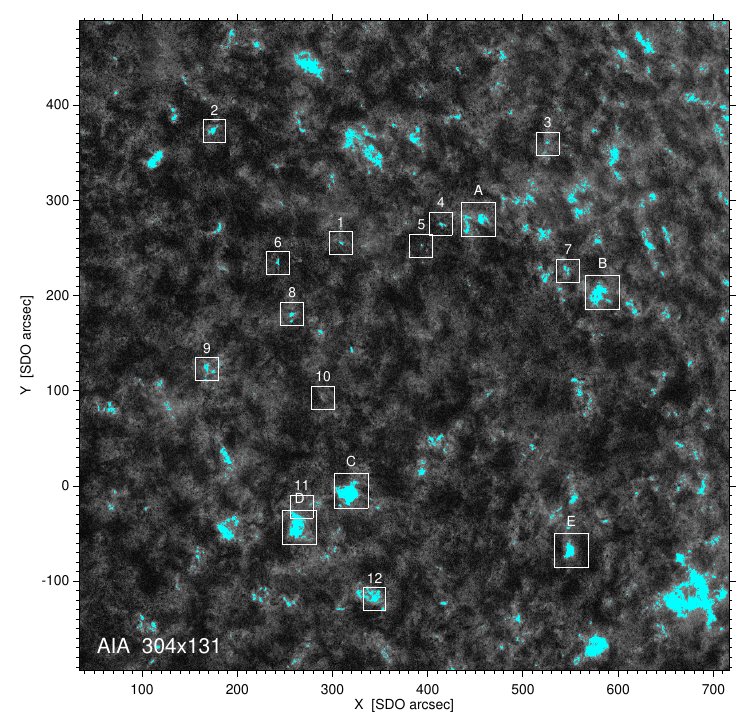}
  \caption[All cutouts in AIA 304$\times$131 construct]
  {\label{fig:sdoroiloc304x131} %
  SDO fire detector inspired by blinking AIA~304 and 131\,\AA\ in
  \rrref{figures}{fig:field304} and \ref{fig:field131}, {\tt EBFAF}
  detection (\rrref{appendix}{sec:ebfaf}) and by the blue pixels in
  the preceding figure selected with the blue box in
  \rrref{figure}{fig:scats}.
  The scene is a construct: the best-match AIA 304\,\AA\ image in
  \rrref{figure}{fig:field304} is multiplied by its 131\,\AA\
  companion in \rrref{figure}{fig:field131}.
  The check plot in \rrref{figure}{fig:drifts} for this pair shows
  that their alignment is sufficiently precise.
  The brightest product pixels are colored cyan, chosen to convey
  EUV-hot fire temperature. 
  The greyscale clip and color threshold are defined by the
  sequence-averaged quiet-network areas only, to avoid lopsiding by
  brighter active regions as present in
  \rrref{figure}{fig:sdo2018-fire}. 
  The values are determined in
  \href{https://webspace.science.uu.nl/~rutte101/rridl/sdolib/sdo_firelevel.pro}{\tt
  sdo\_firelevel.pro} by iteratively removing brightest pixels in
  average and rms computation for the full sequence.
  The values used here are 10\,$\sigma$ of left-over network above its
  average to clip active regions and
  4.5\,$\sigma$ threshold to color pixels cyan.\\
  This stringent pixel-value selection appears useful as EUV fire
  detector. 
  Spatial extent and temporal duration may then discriminate between
  small momentary campfires versus wider and more persistent
  brushfires. \\
  The grey patches everywhere else represent dynamic chromosphere
  around quiet network. 
  They are ubiquitously present with similar appearance in spreading
  around network MC concentrations, defining grey in this image
  well-suited as heated-chromosphere locator.
  It is hot since it is seen here and also as grey background in
  171\,\AA\ (\rrref{figure}{fig:field171}), AIA 221\,\AA\ and AIA
  335\,\AA, but it is also darkly visible in \Halpha\ (blink with
  \rrref{figure}{fig:fieldharev}).
  This dual visibility is discussed in \rrref{appendix}{sec:ha304}. \\
  The darkest patches between the grey ones generally match dark
  features present in most other AIA EUVs and seen sharpest in
  171\,\AA.
  However, in 193\,\AA\ (\rrref{figure}{fig:sdoroiloc193} or
  \rrref{figure}{fig:field193}) many seem covered by diffuse emission
  and others appear bright, for example the brushfire loop bundle to
  the right of ROI-A between ROI-3 and ROI-7 which only has footpoint
  pixels colored cyan here while most loops in this 193\,\AA-bright
  bundle appear dark in 131\,\AA\ (\rrref{figure}{fig:sdoroiloc131} or
  \rrref{figure}{fig:field131}) and in 171\,\AA\
  (\rrref{figure}{fig:field171}).
  Also this dual visibility is discussed in \rrref{appendix}{sec:ha304}.\\
  Ten other fire detector scenes are shown as triple member in
  \rrref{figures}{fig:sdo-minx-193}--\ref{fig:sdo-sp-min-mag}.  
  They separate well between chromospheric and coronal heating.
  }\end{figure*}


\begin{figure*}
  \centering
  \includegraphics[width=0.9\textwidth]{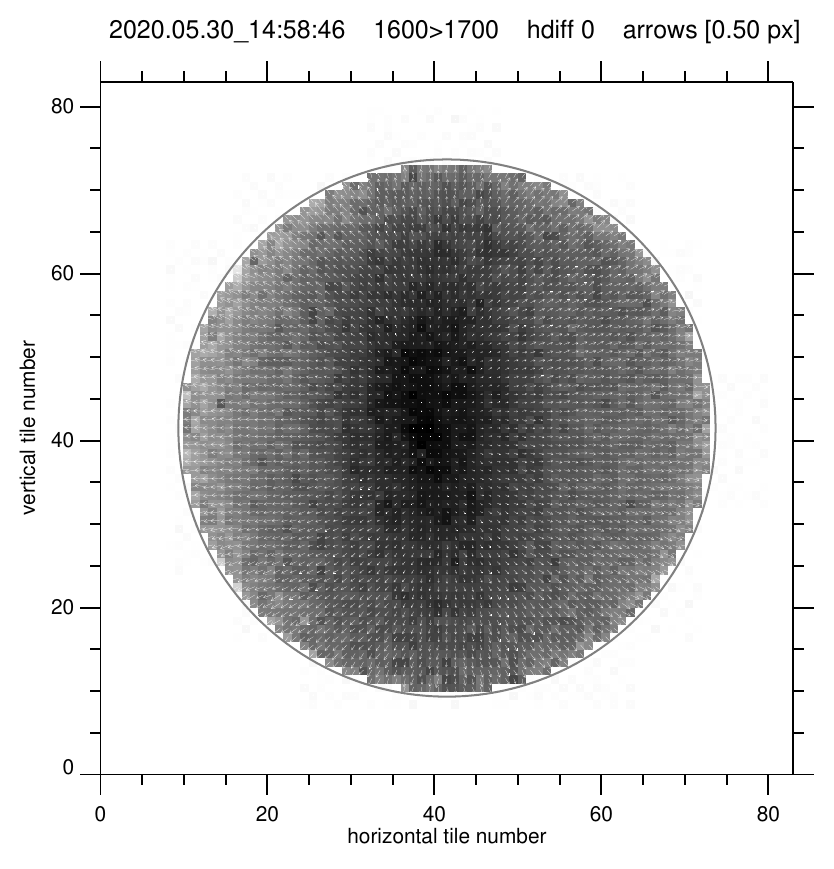}
  \caption[Tile shifts 1700--1600\,\AA] 
  {\label{fig:tileshifts-1617-0} %
  \addtocline{Appendix A}
  Vector chart of apparent shifts of features in AIA 1700\,\AA\ to
  their location in AIA 1600\,\AA, measured by cross-correlation per
  subfield tile at the best-match time. 
  The tiles are greyscaled to their vector length.  
  Similar radial-expansion charts result throughout the SDO database. 
  They represent my closest endeavor to cosmology.
  }\end{figure*}

\begin{figure*}
  \centering
  \includegraphics[width=0.6\textwidth]{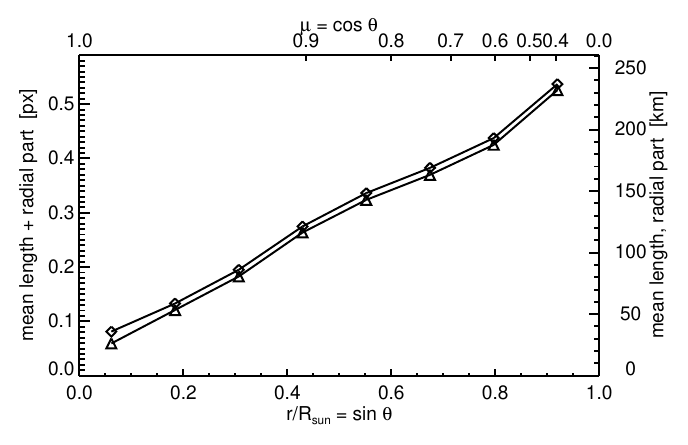}
  \caption[Zonal azimuthal tile-shift averages]
  {\label{fig:zonalshifts-1617-0} %
  Amplitudes of the shift vectors in
  \rrref{figure}{fig:tileshifts-1617-0} averaged azimuthally along
  circular zones.
  The lower curve is the zonal average for the radial components.
  I attribute the final uptilt to \CIV\ contribution in AIA 1600\,\AA\
  and chose ${\tt heightdiff} = 220$~km for
  \href{https://webspace.science.uu.nl/~rutte101/rridl/sdolib/sdo_muckimagepair.pro}{\tt
  sdo\_muckimagepair.pro} from this graph, in agreement with the
  measurement by
  \citetads{2019SoPh..294..161A}. 
  This value is not height-of-formation difference between samplings of
  vertical features as commonly thought, but a formal way of
  quantifying the apparent limbward displacements in
  \rrref{figure}{fig:tileshifts-1617-0}.
  Their actual cause is explained in \rrref{figure}{fig:uvcartoons}.
  }\end{figure*}

\begin{figure*}
  \centering
  \includegraphics[width=0.9\textwidth]{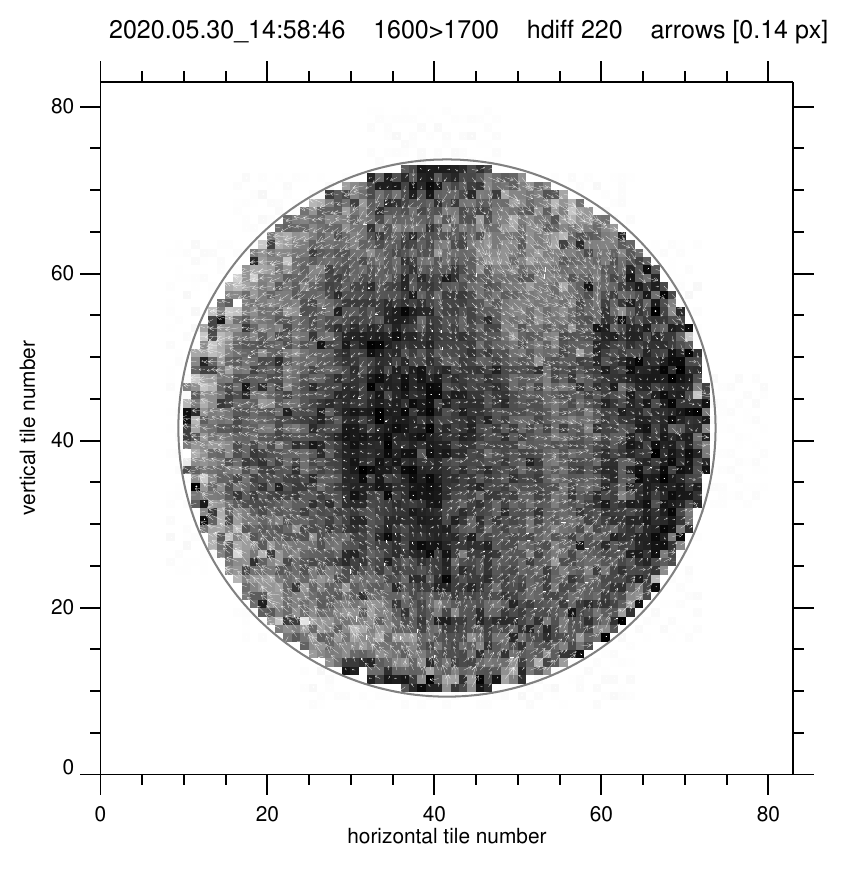}
  \caption[Tile shifts 1700--1600\,\AA\ after {\tt heightdiff}]
  {\label{fig:tileshifts-1617-220} %
  Vector chart of apparent AIA 1700 to 1600\,\AA\ shifts after
  shift-back correction per tile using {\tt heightdiff} = 220~km as
  formal limb value.
  The vector scale is expanded 3.6~times compared with
  \rrref{figure}{fig:tileshifts-1617-0}.  
  I found closely similar residue patterns throughout the SDO database
  and do not understand it. 
  }\end{figure*}

\begin{figure*}
  \centering
  \includegraphics[width=6cm]{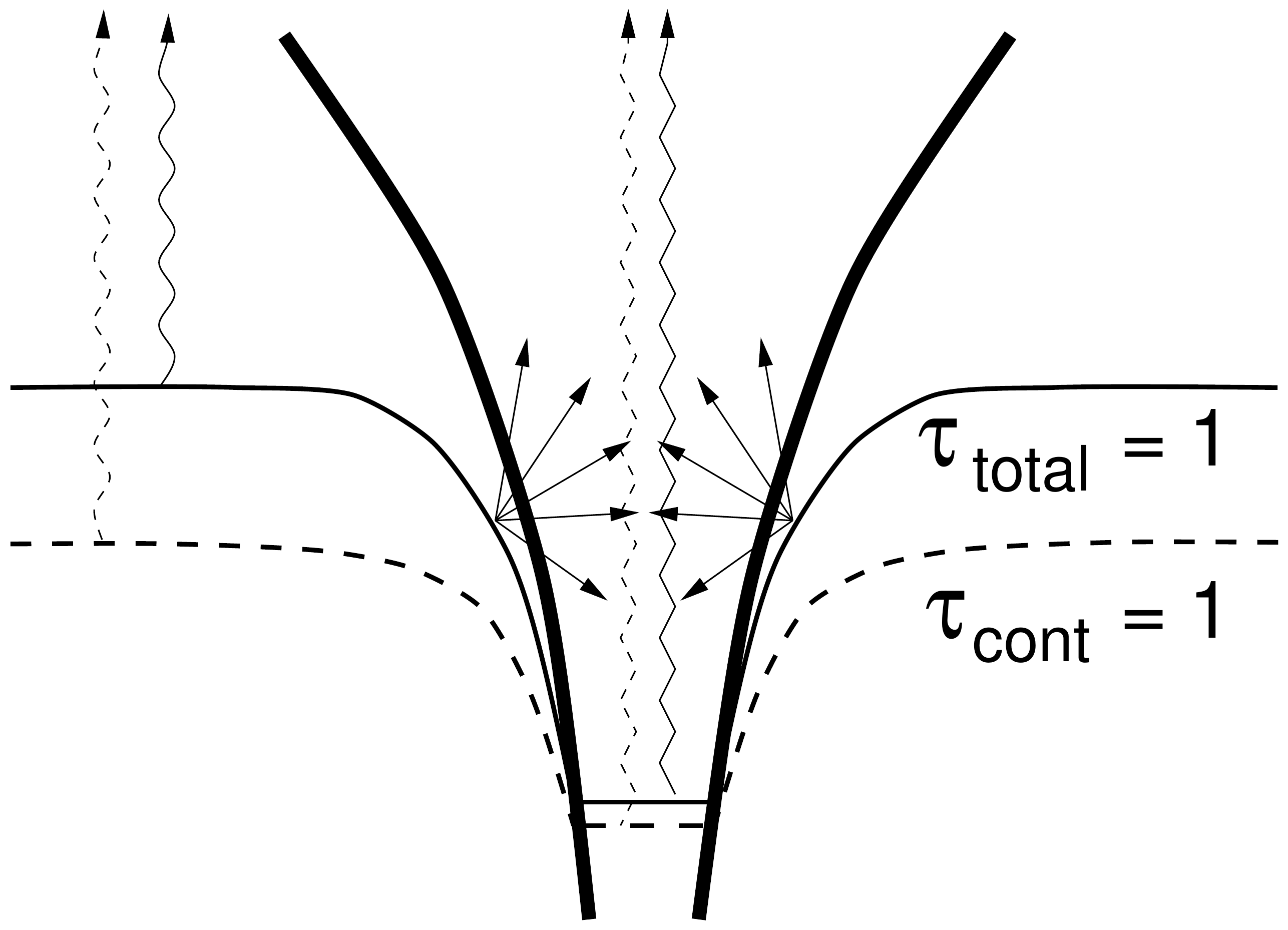} 
  \hspace{2cm}
  \includegraphics[width=6cm]{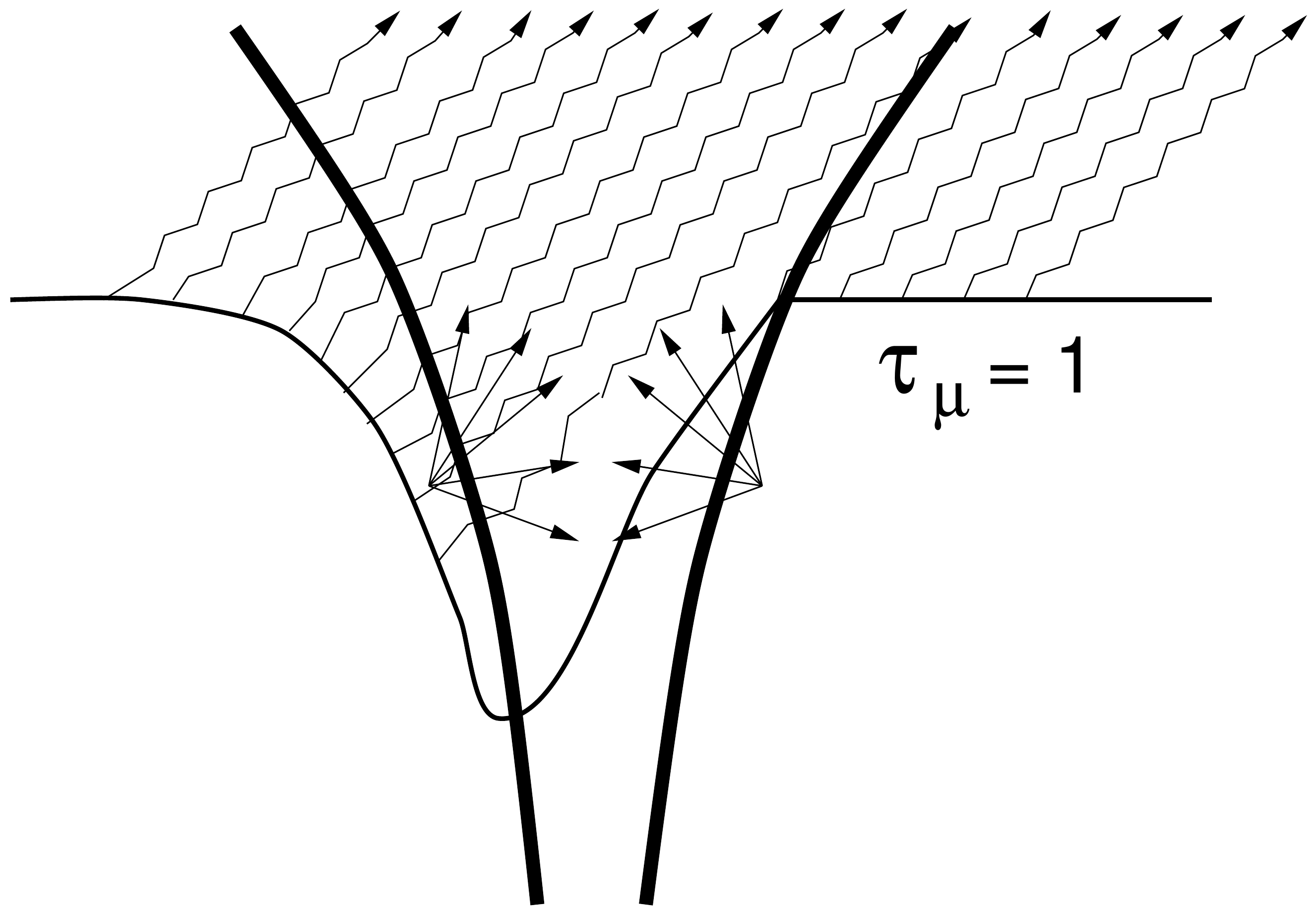}\\[4ex]
  \includegraphics[width=6cm]{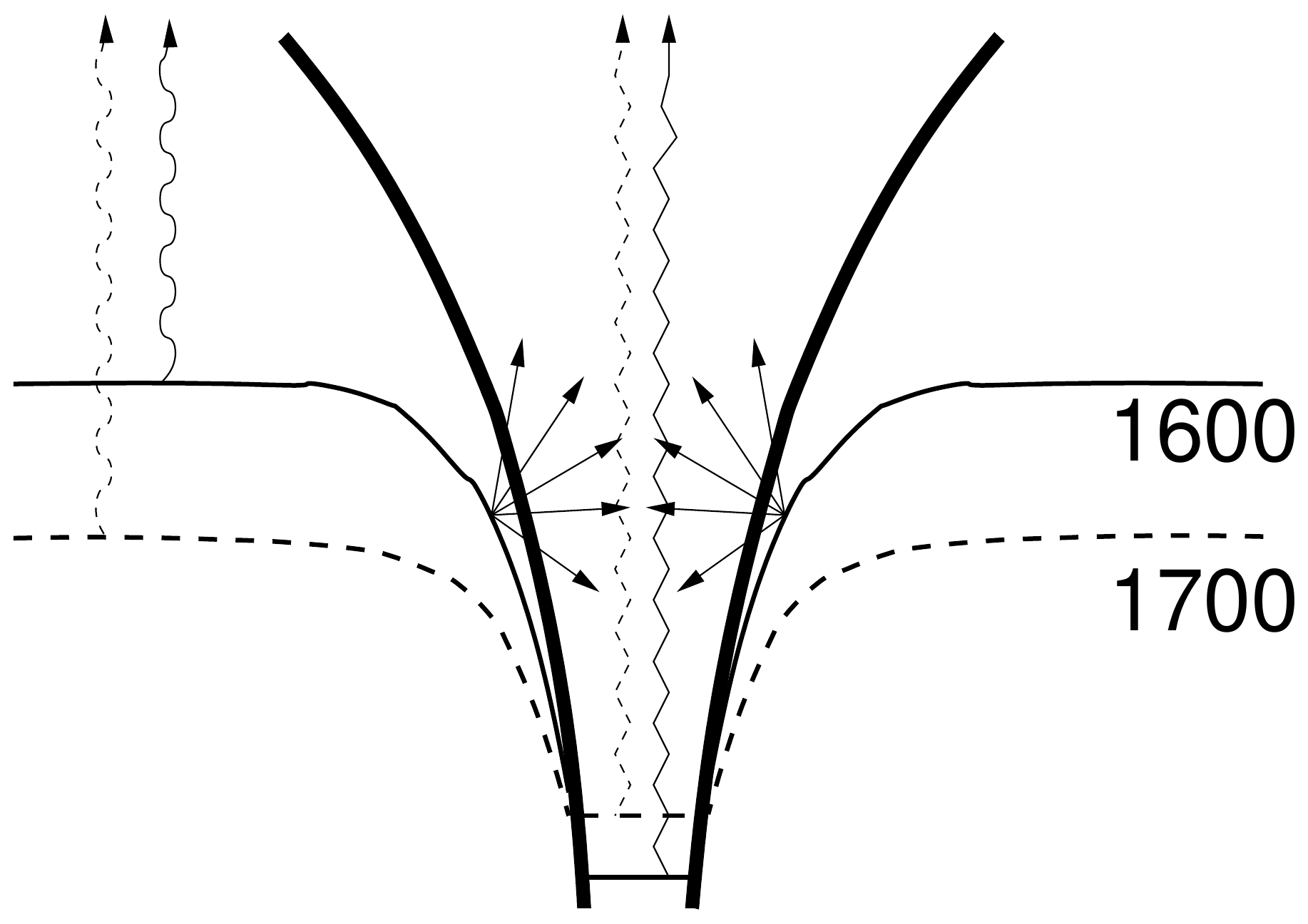}
  \hspace{2cm}
  \includegraphics[width=6cm]{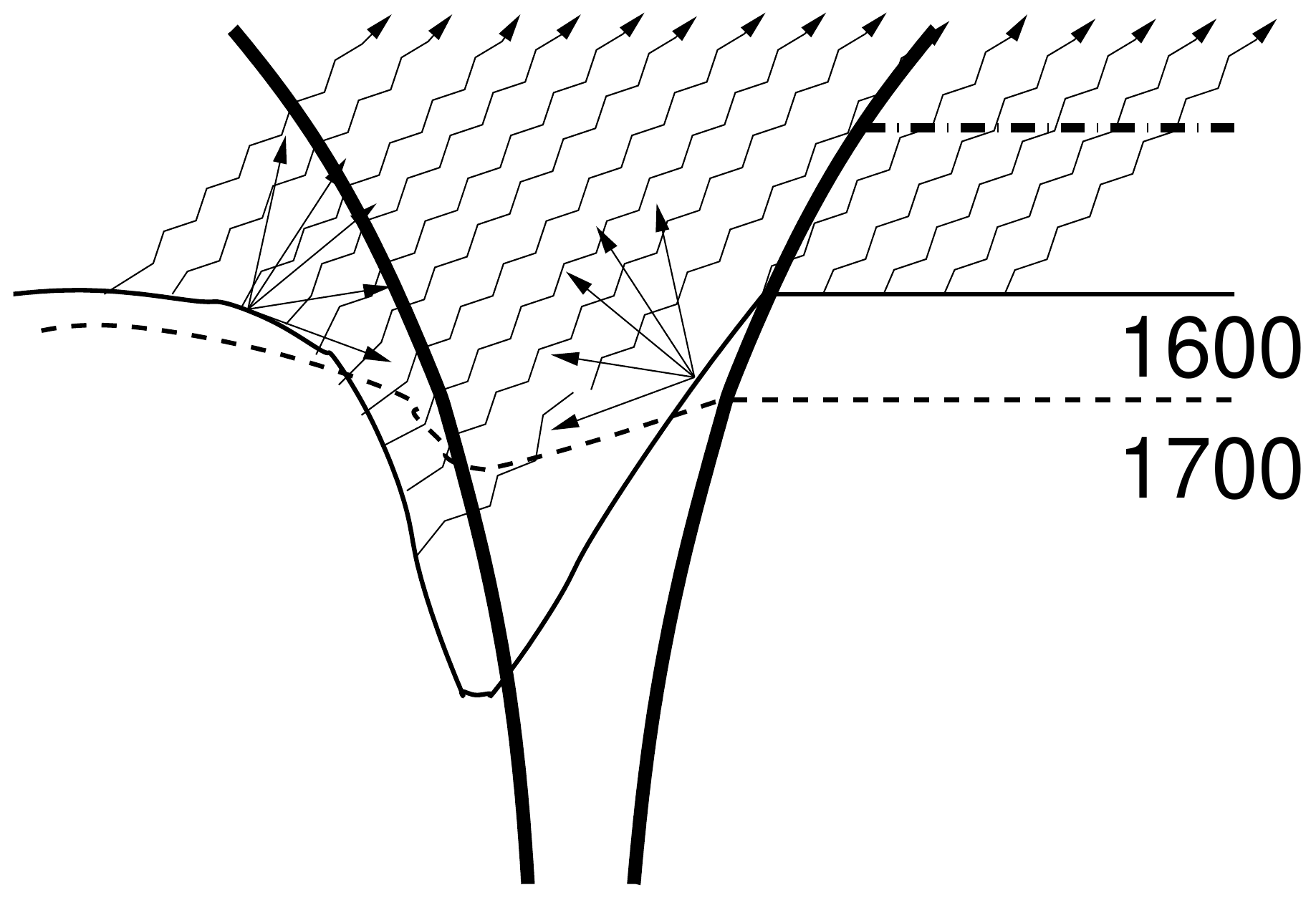}
  \caption[MC brightening at 1700 and 1600\,\AA] 
  {\label{fig:uvcartoons} %
  {\em Top row\/}: ancient sketches to explain G-band brightening of
  `''filigree grains'' (here called magnetic concentrations = MCs,
  also known as flux tubes, fluxtubes, magnetic elements, magnetic
  bright points, intergranular bright points, network points, facular
  points, faculae, plage, flowers, pseudo-EBs, line gaps,
  pseudo-moustaches) copied from
  \linkadspage{1999ASPC..184..181R}{8}{Figure 7} in
  \citetads{1999ASPC..184..181R}.\\ 
  {\em Left:\/} the radial view at disk center.  
  This cartoon was first drawn by Kees Zwaan, based on his
  thesis (\citeads{1965smss.book.....Z}) 
  and extract \citetads{1967SoPh....1..478Z}, 
  much discussed at Sterrewacht Sonnenborgh (with Hans Rosenberg on
  this
  \href{https://webspace.science.uu.nl/~rutte101/astronomershots/album1967/kees-zwaan-hans-rosenberg.jpg}{1967
  photograph}), and finally published as
  \linkadspage{1978SoPh...60..213Z}{12}{Fig.~2} of
  \citetads{1978SoPh...60..213Z}. 
  The Zwaan-style fluxtube is relatively empty because its magnetic
  pressure compensates part of the outside gas pressure in
  magnetostatic equilibrium
  (\citeads{1976SoPh...50..269S}, 
  Zwaan-inspired in contrast to this
  \linkadspage{1990IAUS..138..501R}{10}{creation diagram}). 
  Zwaan's concept was observationally quantified by Sami Solanki and
  coworkers into Z\"urich wine-glass models (\eg\
  \citeads{1993A&A...268..736B}) 
  using unresolved but sensitive multi-line spectropolarimetry
  initiated by \citetads{1984AdSpR...4h...5S} 
  (see \citeads{1984A&A...131..333S}, 
  \citeads{1993SSRv...63....1S}). 
  The concept applies to gas-dominated atmospheric stratification with
  upward fluxtube flaring set by hydrostatic density decrease; for
  non-flaring field-dominated magnetostatic coronal loops see \eg\
  \citetads{1985ApJ...293...31L}, 
  \citetads{2004A&A...417..333N}.\\ 
  {\em Visibility:\/} top-down one views deeper into the fluxtube and
  receives larger brightness from its hotter hole-in-the-surface walls
  and bottom than from the field-free gas around it. 
  The latter is sampled higher at lower temperature and also in a
  relatively dark intergranular lane (so that at
  non-subarcsec resolution these small brightenings and their
  surroundings blend into non-showing grey,
  \citeads{1996ApJ...463..797T}). 
  In the G-band the low pressure causes extra dissociation of the CH
  molecules causing this Fraunhofer-named dark spectral feature, so
  that images selecting it show enhanced brightening compared to
  continuum wavelengths.   
  Because this band is wide enough to accommodate interference-filter
  10\,\AA\ bandpass it became very popular in short-exposure
  high-resolution photosphere imaging after
  \citetads{1984apoa.conf..382M} 
  did so first.
  Thus, MC brightening comes from holes in the surface, deeper in the
  G-band, not necessarily from extra heating.
  The \CaII\ \HK\ wing study of
  \citetads{2005A&A...437.1069S} 
  found no MC heating throughout the photosphere, in conflict with the
  increasing temperature excess that is stipulated in plane-parallel plage
  models. 
  MC heating becomes visible only higher up as
  chromospheric grain brightening in the \CaII\ \HK\ cores, mostly in
  the network but occasionally above internetwork MCs as the
  ``persistent flasher'' of
  \citetads{1994ASIC..433..251B}.\\ 
  {\em Right:\/} the same sketch served to explain why MCs become
  bright stalks (``faculae'', originally in white light) towards the
  limb. 
  In slanted viewing the fluxtube foot is blocked to higher height by
  the surrounding denser gas (yet higher in the more opaque G band)
  whereas through the relatively empty fluxtube (yet emptier in G-band
  opacity) one views further into the hot = bright granule behind it
  than without a fluxtube crossing the line of sight.
  The apparent stalks represent lack of opacity along that.\\
  {\em Bottom row\/}: the same sketch but modified to illustrate
  1700--1600\,\AA\ MC brightness difference. 
  Outside the tube the 1600\,\AA\ continuum opacity is larger but
  inside it is lower because the neutral metals ionize away (they are
  already minority stage) so that only the scattering Balmer continuum
  and some Rayleigh scattering remain.\\
  {\em Left:\/} at 1600\,\AA\ the MCs are deeper holes than at
  1700\,\AA\ and relatively brighter in byte-scaled top-down images
  (the outside scene is dominated by clapotispheric shocks that
  brighten less or darken). 
  The same apparent deepening from neutral-metal ionization causes the
  classic ``line gap'' phenomenon in \FeI\ lines (\eg\
  \citeads{1991A&A...251..675S}) 
  which is seen best in \MnI\ lines
  (\citeads{1987ApJ...314..808L}) 
  from lack of microturbulent smearing in surrounding granulation
  (\citeads{2009A&A...499..301V}), 
  and also similar MC ``moustache'' brightening (not EB but pseudo-EB,
  moustaches share that confusion) of the wings of the \NaID\ and
  \MgIb\ minority-stage lines (also
  \citeads{1991A&A...251..675S}). 
  For an SST/CRISP example of the latter see the MC near the center of
  the small field in \linkadspage{2011A&A...531A..17R}{5}{Figure 4} of
  \citetads{2011A&A...531A..17R}. 
  Its brightening enhancement in the wing of the majority-stage \CaIR\
  line in that figure comes from less collisional damping at smaller
  density reducing wing opacities.
  Such wing brightening by less-damping deepening is also seen in
  \Halpha\
  (\citeads{2006A&A...449.1209L}), 
  \Hbeta\ and \CaIIK\
  (\citeads{2006A&A...452L..15L}). 
  For recent SST examples see \linkadspage{2019A&A...631L...5B}{2}{the
  first panels of Fig.~1} of
  \citetads{2019A&A...631L...5B}; 
  less-damping MC brightening is seen particularly well with the
  SST/CHROMIS wide-band filter at 3950\,\AA\ midway between \CaII\ H
  and K.\\ 
  {\em Right:\/} towards the limb the MCs have higher-up dark foot
  blocking at 1600\,\AA\ by the more opaque surroundings, but the view
  through the tube deep into the hot granule behind starts deeper down
  at 1600\,\AA\ and extends further out in reaching $\tau_\mu \tis\ 1$
  than at 1700\,\AA. 
  The 1600\,\AA\ stalk therefore differs in morphology from the
  1700\,\AA\ stalk, being brighter with apparent limbward shifts
  utterly evident in \rrref{figure}{fig:tileshifts-1617-0}.
  }\end{figure*}

 
\begin{figure*}
  \centering
  \includegraphics[width=0.33\textwidth]{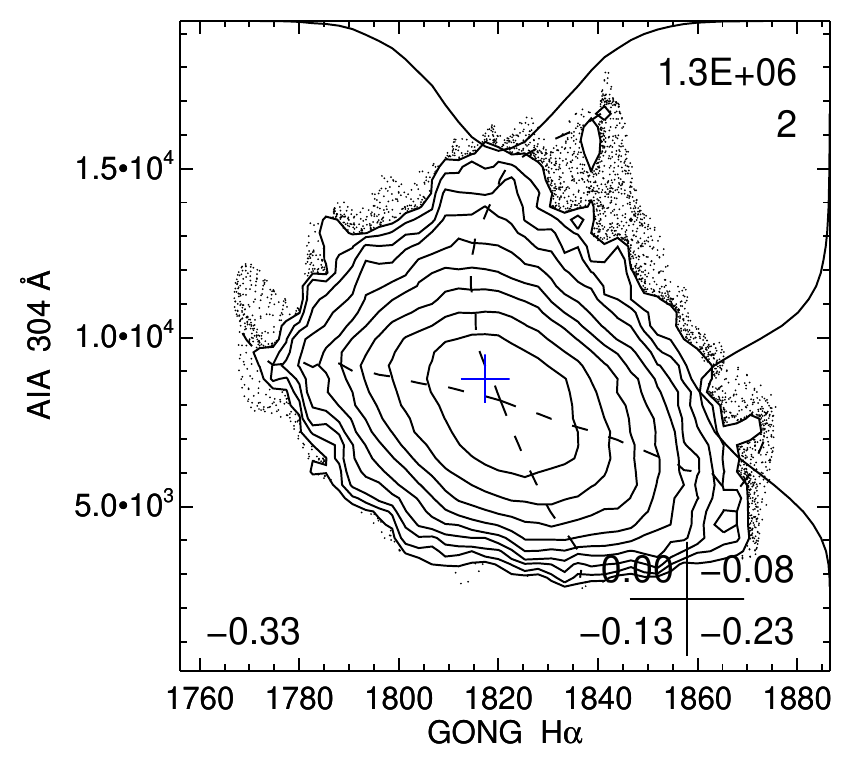}
  \hspace{-1mm}
  \includegraphics[width=0.33\textwidth]{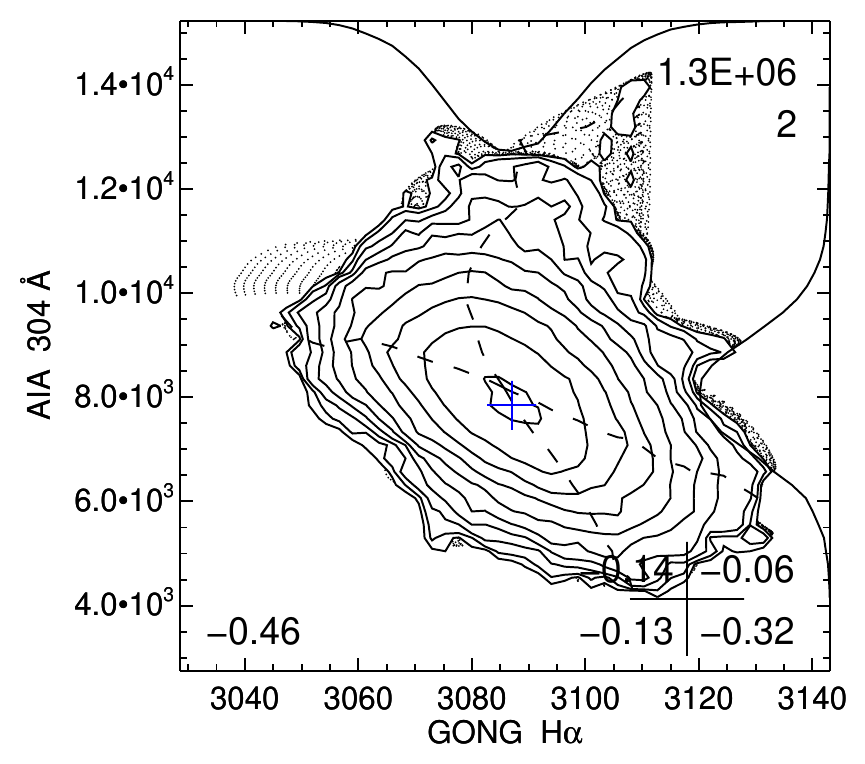}
  \hspace{-1mm}
  \includegraphics[width=0.33\textwidth]{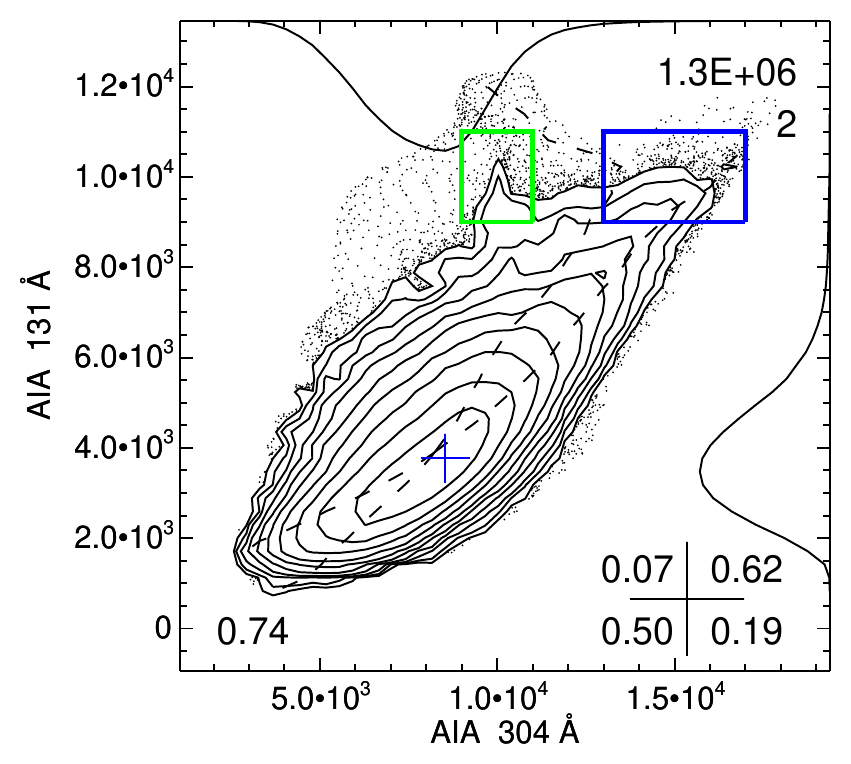}
  \caption[Strous diagrams]
  {\label{fig:scats} %
  \addtocline{Appendix B}
  Strous scatter diagrams.  
  The format and inserted numbers are explained in
  \linkadspage{2019A&A...632A..96R}{2}{Section~2} of
  \citetads{2019A&A...632A..96R} 
  with \linkadspage{2019A&A...632A..96R}{7}{Figure~5 there} a
  relatively easy to interpret example. 
  I make Strous diagrams with
  \href{https://webspace.science.uu.nl/~rutte101/rridl/cubelib/scatcont.pro}
  {\tt scatcont.pro} based on
  \href{https://webspace.science.uu.nl/~rutte101/rridl/collib/scatter_aw.pro}
  {Alfred de Wijn's version}.\\
  {\em First panel:\/}  AIA\,\,304\,\AA\ against GONG \Halpha.
  For AIA\,\,304\,\AA\ the temporal average of the
  15-min sequence is used, for \Halpha\ the non-reversed GONG image of
  \rrref{figure}{fig:fieldha} blurred over 30 pixels. 
  The apparent overall correspondence of the grey patches in
  \rrref{figures}{fig:fieldharev} and \ref{fig:field304} is quantified
  as the significant downward tilt of the contour mountain of which
  the extended summit corresponds to grey in the images. 
  Without correlation the mountain would be round with perpendicular
  first-moment curves.
  The rightward tilt of the upper end of the vertical first-moment
  curve suggests slightly brighter than normal grey for \Halpha\ at
  brushfire sites.\\
  {\em Second panel:\/} similar but now using sequence-averaged GONG
  \Halpha.  
  My success in co-aligning the single reversed \Halpha\ image of
  \rrref{figure}{fig:fieldharev} with the AIA\,304\,\AA\ image of
  \rrref{figure}{fig:field304} made me develop
  \href{https://webspace.science.uu.nl/~rutte101/rridl/gonglib/gong_sdo.pro}
  {\tt gong\_sdo.pro} to co-align each of the best in a sequence of
  GONG \Halpha\ images with simultaneous AIA\,304\,\AA\ images. 
  My motivation was that this may serve to co-align any \Halpha\
  observation with SDO by using GONG \Halpha\ as intermediary, and
  also look-alike scenes such as ALMA images. 
  Because the GONG images vary much in quality and considerably in
  scale, position and orientation automating such co-alignment is
  non-trivial, but I succeeded in obtaining a co-aligned GONG \Halpha\
  sequence for the present SDO downloads.   
  The resulting Strous diagram is similar to the one at left but shows
  tighter relationships due to the now dual 15-min temporal and 30-px
  spatial averaging.
  This improvement fits my view of the chromosphere: heating to AIA
  304\,\AA\ visibility occurs momentarily in the tips of spicules-II
  followed by darkening \Halpha-core fibrils along their tracks,
  darkest at the track onsets so that the correlation concerns fine
  structures that are not precisely synchronous nor co-spatial but
  closely adjacent in space and time.  
  Again, the tilted scatter mountain describes the grey chromospheric
  patches seen ubiquitously everywhere whereas the brightest 304\,\AA\
  pixels describe fires. 
  These appear only slightly brighter than average in \Halpha\ so that
  low-resolution \Halpha\ does not offer a viable means to locate them.\\
  {\em Third panel\/}: AIA\,131 against 304\,\AA, showing yet tighter
  spatial correlation for most-common grey pixels in
  \rrref{figures}{fig:field304} and \ref{fig:field131}. 
  For both sequences the temporal means are used.
  I was curious about the small North-ward promontory on the North
  coast of this contour island. 
  Inspection with {\tt showex}, which can plot Strous diagrams live
  while blinking and also offers box-selection with image pixel
  coloring, made me add the green and blue selection boxes. 
  Their pixels are colored correspondingly in
  \rrref{figure}{fig:sdoroiloc131}. 
  The blue ones appear to represent a viable fire locator and so
  inspired the multiplicative fire detector tested and described in
  \rrref{figure}{fig:sdoroiloc304x131}. 
  It serves as triple member in
  \rrref{figures}{fig:sdo-minx-193}--\ref{fig:sdo-sp-min-mag} to
  separate chromospheric and coronal heating.
  }\end{figure*}

\begin{figure*}
  \centering
  \includegraphics[width=0.49\textwidth]{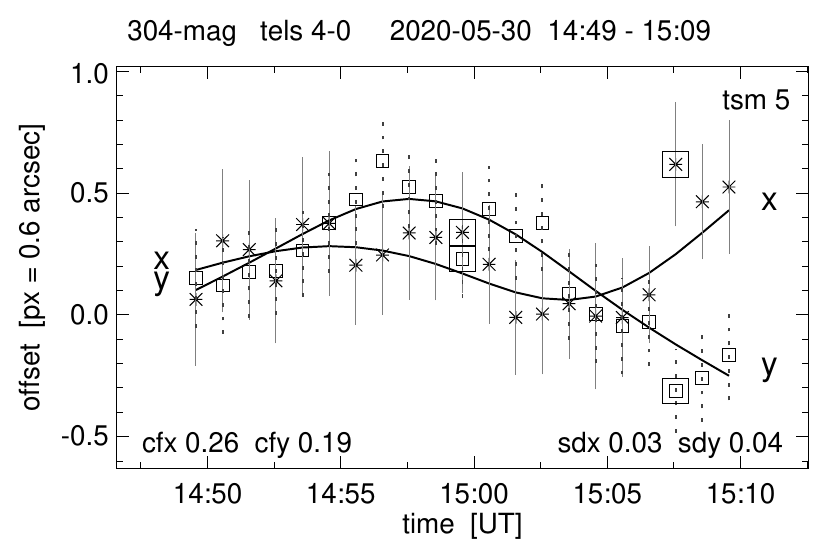}
  \includegraphics[width=0.49\textwidth]{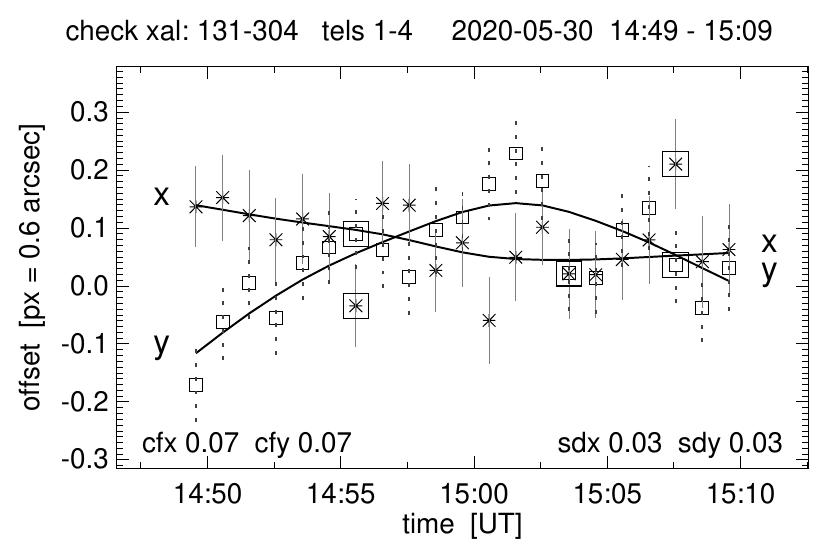}
  \caption[SDO cross-alignment results]
  {\label{fig:drifts}%
  {\em Left\/}: standard {\tt driftscenter} plot from the SDO
  cross-alignment pipeline for the present SDO download, for the AIA
  304\,\AA\ versus HMI magnetogram pair.
  These are made from low-cadence 700$\times$700 arcsec$^2$ center
  cutouts by tiling into 30$\times$30~arcsec$^2$ subfields as in
  \rrref{figures}{fig:tileshifts-1617-0} and
  \ref{fig:tileshifts-1617-220} and for every time step determining
  and averaging spatial offsets by cross-correlation per tile pair,
  after appropriate image ``mucking'' in
  \href{https://webspace.science.uu.nl/~rutte101/rridl/sdolib/sdo_muckimagepair.pro}
  {\tt sdo\_muckimagepair.pro} to make them appear more similar and
  using {\tt heightdiff} correction, here with limb value 3600~km
  determined from zonal tile-shift averaging as in
  \rrref{figure}{fig:zonalshifts-1617-0}. 
  The error bars are 96\% confidence limits for the next tile sample
  per time step.
  Their average values are specified at lower left.
  The boxed samples are outliers removed in iterative spline fitting.  
  The resulting spline curves are stored and used to cross-align the
  actual target data. 
  Their 1$\sigma$ reliability is specified at the lower right.
  The pipeline produces such graphs for all pairs it employs. 
  Generally they show time-varying drifts up to a few pixels between
  SDO diagnostics (more after eclipses or hiccups). 
  Fixing the EUVs to HMI or UV is the hardest; currently this
  304\,\AA\ -- magnetogram pair is my default anchor choice.\\
  {\em Right\/}: {\tt driftscenter} check plot after all
  cross-alignments for the AIA 131\,\AA\ versus 304\,\AA\ pair, also
  made with
  \href{https://webspace.science.uu.nl/~rutte101/rridl/sdolib/sdo_writepairspline.pro}
  {\tt sdo\_writepairspline,pro}. 
  The vertical axis range is smaller.
  Currently, the pipeline does not cross-align these directly but it
  roundabout cross-aligns 211\,\AA\ to 304\,\AA, 335\,\AA\ to
  211\,\AA, 131\,\AA\ to 335\,\AA. 
  There are more such chained cross-alignments, all anchored to the
  304\,\AA\ to HMI magnetogram alignment at left, with their orders
  defined in
  \href{https://webspace.science.uu.nl/~rutte101/rridl/sdolib/
  sdo_getsumsplineshift.pro}{\tt sdo\_getsumsplineshift.pro} from {\tt
  showex} scene comparisons to select best-matching pairs as well as
  appropriate mucking.
  Remaining errors add up statistically in such multi-step chains. 
  Here, this check from cross-aligning the pipeline disk-center
  results, not its input, shows that the final errors for this pair
  are negligible, fully so at best-match time (14:58:46~UT). 
  Before cross-alignment this {\tt aia\_prepped} input pair had
  consistent $(\Delta x,\Delta y) \approx (+0.3,-1.1)$~px offsets.
  Their reduction to 0.1~px ensures that the fire detector of
  \rrref{figure}{fig:sdoroiloc304x131} multiplies appropriate pixel
  pairs.
  }\end{figure*}

\end{document}